%% file: apssamp.tex
\documentclass[amsmath,amssymb,aps,prd,reprint,superscriptaddress,footinbib,floatfix,showkeys]{revtex4-2}

\usepackage{graphicx}
\usepackage{dcolumn}
\usepackage{bm}

\begin{document}

\preprint{APS/123-QED}

\title{\textbf{} 
Current Closure and Gamma-Ray Signatures of Spark-Gap and Current-Sheet Activities in Radiative Black Hole Magnetospheres
}%

\author{Koki Kin}
\affiliation{Academia Sinica Institute of Astronomy and Astrophysics (ASIAA), 11F of AS/NTU Astronomy-Mathematics Building, No.1, Sec. 4, Roosevelt Rd, Taipei 106319, Taiwan}
\affiliation{Astronomical Institute, Graduate School of Science, Tohoku University, Sendai 980-8578, Japan}
\email{kkin@asiaa.sinica.edu.tw}

\author{Shota Kisaka}
\affiliation{Physics Program, Graduate School of Advanced Science and Engineering, Hiroshima University, Higashi-Hiroshima, 739-8526, Japan}
\affiliation{Hiroshima Astrophysical Science Center, Hiroshima University, Higashi-Hiroshima, Hiroshima 739-8526, Japan}

\author{Kenji Toma}
\affiliation{Astronomical Institute, Graduate School of Science, Tohoku University, Sendai 980-8578, Japan}
\affiliation{Frontier Research Institute for Interdisciplinary Sciences, Tohoku University, Sendai 980-8578, Japan}

\author{John M. Mehlhaff}
\affiliation{Univ. Grenoble Alpes, CNRS, IPAG, 38000 Grenoble, France}
\affiliation{Physics Department and McDonnell Center for the Space Sciences, Washington University in St. Louis, MO, 63130, USA}

\author{Beno\^it Cerutti}
\affiliation{Univ. Grenoble Alpes, CNRS, IPAG, 38000 Grenoble, France}

\date{\today}

\begin{abstract}
We investigate the evolution of an axisymmetric black hole magnetosphere using general relativistic particle-in-cell simulations, focusing on its global current structure, variability, and radiative effects.
We include inverse-Compton (IC) scattering of ambient photons and subsequent two-photon pair production, and compare simulations with and without synchrotron-curvature (SC) radiation reaction.
Starting from an axisymmetric vertical magnetic field, the system develops a polar spark gap and a variable equatorial current sheet that together sustain the global poloidal current.
For the parameters explored in this study, we observe that the spark gap oscillates upon the onset of SC cooling, whereas the current sheet is less sensitive to radiative cooling.
Time-dependent electric fields in the return-current layer trigger magnetic-field rearrangement followed by magnetic reconnection.
Cross-correlation analysis finds no significant correlation between the two regions' variability, suggesting dynamical independence.
We further perform post-processing ray-tracing calculations to evaluate the radiative efficiencies of the IC and SC processes.
High-energy particles dominate the bolometric luminosity, which reaches up to $\sim1$\% of the Blandford-Znajek luminosity for SC emission and $\sim0.1$\% for IC emission.
The relative IC luminosity to SC is higher at lower viewing angles because particle momenta are more anisotropic near the pole than near the equator, due to efficient acceleration along the local magnetic field lines.
Together with energetic estimates, these results suggest that both regions can contribute to high-energy emission in active galactic nuclei, with spark-gap photons reaching the sub-TeV regime.
\end{abstract}


\maketitle

\section{Introduction} 
A leading theoretical framework for launching relativistic jets from accreting black holes (BHs) is the Blandford--Znajek (BZ) mechanism \citep{BZ77}, in which the rotation of the central BH is considered to be the power source of jets.
A key ingredient of the theory is the formation of a highly magnetized region, i.e., the magnetosphere, near BHs.
Electromagnetic interactions between the plasma and magnetic fields in the magnetosphere efficiently extract the BH's rotational energy as outgoing Poynting flux, which is then converted into the jet's kinetic energy.

The magnetosphere is thought to arise through the magnetic flux accumulation during the accretion of magnetized plasma onto the BH.
Dynamical properties of such a BH accretion flow with a high magnetization have been extensively investigated through general relativistic magnetohydrodynamic (GRMHD) simulations, in which a highly magnetized polar funnel is indeed formed and powerful jets are launched \citep{McKinney&Gammie04, McKinney06, Tchekho11, Nakamura18, Chael19, Porth19, Liska20, Ripperda22}.
However, the expected plasma density inside the magnetosphere can become dilute, and thus the MHD fluid approximation no longer holds.
Moreover, nonthermal plasma populations, namely those for electron-positron pairs whose energetics are essential for determining observational signatures, cannot be modeled self-consistently within the MHD framework.
A kinetic model is therefore necessary to capture magnetospheric plasma dynamics: general relativistic particle-in-cell (GRPIC) simulations have become an essential tool for investigating plasma kinetics in BH magnetospheres in recent years.
Studies have progressed from one-dimensional (1D) models focusing on local dynamics \citep{Levinson18, Chen20, Kisaka20, Kisaka22, Kin24, Kin25} to multi-D global simulations that capture the entire magnetospheric structure \citep{Parfrey19, Crin20, Crin21, Hirotani21, ElMellah23, Hirotani23, Niv23, Chen25, Vos25, Yuan25, Mehlhaff25, Meringolo25, Figueiredo25, Figueiredo26, Mehlhaff26, Meringolo26}.

One important implication from previous GRPIC simulations is the development of unscreened electric fields when the plasma amount becomes insufficient to support the current required for maintaining a force-free magnetosphere (1D: \citep[][]{Chen20, Kisaka20, Kisaka22, Kin24, Kin25}, 2D: \citep[][]{Parfrey19, Crin20, Crin21, Hirotani24, Chen25, Vos25, Yuan25, Mehlhaff26}).
Particle acceleration in these "spark gaps" triggers copious intermittent pair production, providing a potential plasma source for relativistic jets.
Furthermore, a fraction of the high-energy photons can escape and contribute to the very-high-energy (VHE) emission observed from core regions of active galactic nuclei (AGNs), such as M87 \citep{Acciari08, Albert08, Abramowski12, Algaba24, Cao24}.
Previous studies have investigated the mechanisms behind the dynamical variability of spark gaps and how their energy-dissipation efficiency depends on several parameters.

Our previous 1D GRPIC studies, which prescribed the global poloidal current \citep{Kisaka22, Kin24}, showed that variations in its amplitude can strongly affect spark-gap activity and the resulting gamma-ray emission.
This highlights the need to investigate gap dynamics alongside the global current system, including the equatorial current sheet.
Furthermore, the equatorial sheet is highly unstable and prone to magnetic reconnection \citep{Koide03, Komissarov04, Komissarov&McKinney07}.
Reconnection can dissipate a substantial fraction of the magnetic energy \citep{Ripperda20, Ripperda22, Zhou25}, produce nonthermal particles, and potentially contribute to VHE emission \citep{Hakobyan23, Hakobyan25, Solanki25}.
A detailed comparison of the energetics and variability of these two regions is essential to understand plasma dynamics in the global BH magnetosphere. 
A precise treatment of radiative cooling processes is also crucial because not only is it necessary to diagnose observables, but radiative cooling can also affect the dynamics of the accelerating regions, as suggested by local PIC simulations of radiative magnetic reconnection \citep{Werner19, Sironi20, Guo21, Schoeffler23}.
In the system considered here, inverse-Compton (IC) scattering of photons from the surrounding accretion flow and synchrotron-curvature (SC) radiation are expected to dominate the emission.
Global simulations that incorporate these processes and particle production are therefore essential for studying the magnetosphere's dynamics and radiation characteristics.

In this regard, several GRPIC studies have reproduced dynamically evolving equatorial current sheets by adopting vertical magnetic field configurations \citep{Parfrey19, Figueiredo25, Meringolo25, Meringolo26}, by incorporating the conductor disk \citep{Crin21}, or by initializing a dense magnetized plasma torus reminiscent of GRMHD simulations \cite{Vos25, Mehlhaff26}.
Some have even simulated both the development of the polar spark gap and magnetic reconnection in the current sheet by incorporating IC scattering and subsequent pair production \citep{Parfrey19, Crin21, Vos25}, with the Appendix of \citet{Vos25} reporting possible dynamical coupling between the two regions.
However, the effects of SC radiative cooling, the relative IC and SC luminosities of the gap and current-sheet regions, and the dynamical link between the two through global current closure remain unexplored in detail.

In this work, we aim to address these open questions. 
We perform 2D axisymmetric GRPIC simulations of a black hole magnetosphere, incorporating SC and IC emission processes, as well as pair production via gamma-ray absorption by the soft-photon background that seeds IC scattering.
We adopt Wald's vacuum solution as the initial magnetic field configuration to isolate the magnetosphere's intrinsic electromagnetic evolution, which complements studies that account for the dynamical influence of the accretion flow (e.g., \cite{Crin21, Vos25, Mehlhaff26}).
Section~\ref{setup} describes the numerical setup, Section~\ref{res} presents the simulation results and synthetic observables, and Section~\ref{disc} discusses implications for VHE emission from BH magnetospheres.

\section{Simulation Setup}\label{setup}
We perform 2D GRPIC simulations using $\mathit{GRZeltron}$ \citep{Parfrey19}, in which the fundamental equations are described using the 3+1 formulation \citep{Landau75, Macdonald82} in spherical Kerr--Schild (KS) coordinates.
Electromagnetic fields evolve according to the time-dependent Maxwell's equations:
\begin{equation}
\begin{aligned}\label{faraday}
\frac{1}{c}\frac{\partial\bm{B}}{\partial t}=-\nabla\times\bm{E},
\end{aligned}
\end{equation}
and
\begin{equation}\label{ampere}
\begin{aligned}
\frac{1}{c}\frac{\partial\bm{D}}{\partial t}=\nabla\times\bm{H}-\frac{4\pi}{c}\bm{J},
\end{aligned}
\end{equation}
where $\bm{J}$ is the current density.
The pair $\{\bm{E},\bm{H}\}$ and $\{\bm{D},\bm{B}\}$ can be related to each other as
\begin{equation}\label{EHDB}
\bm{E}=\alpha\bm{D}+\bm{\beta}\times\bm{B},\qquad
\bm{H}=\alpha\bm{B}-\bm{\beta}\times\bm{D}.
\end{equation}
where $\alpha$ and $\bm{\beta}$ are the lapse function and the shift vector (Appendix.~\ref{app_metric}).

The plasma consists of electrons and positrons whose motions follow Hamilton's equations \citep{Landau71, Hughes94},
\begin{equation}\label{hamilton}
\begin{aligned}
\frac{du_i}{dt}
=&-\Gamma\partial_i\alpha+u_j\partial_i\beta^j-\frac{\alpha}{2\Gamma}u_ju_k\partial_i\gamma^{jk}\\
&+\alpha\frac{q}{m}\left(D_i+\frac{\sqrt{\gamma}}{\Gamma}\epsilon_{ijk}u^jB^k\right),
\end{aligned}
\end{equation}
and
\begin{equation}
\frac{dx^i}{dt}
=\frac{\alpha}{\Gamma}\gamma^{ij}u_j-\beta^i,
\end{equation}
where Latin indices span the values $r$, $\theta$, and $\varphi$, and the Einstein summation convention is implied.
$\gamma^{ij}$ denotes the spatial 3-metric, $\sqrt{\gamma}$ the root of its determinant, $x^i$ the spatial coordinates, $u^i$ the 4-velocities, $\Gamma=\sqrt{1+\gamma^{ij}u_i u_j}$ the Lorentz factor of particles measured by the "fiducial observer" (FIDO; \citep{bardeen72, Thorne86}), $m$ and $q$ are the particle mass and the charge, respectively, and $\epsilon_{ijk}$ is the Levi-Civita symbol.

The IC and pair-production processes are treated probabilistically (cf. \cite{Levinson18, Crin20}).
Particles can upscatter photons from a homogeneous, static, isotropic, mono-energetic background radiation field with the photon number density $n_\gamma$ and the optical depth $\tau_0=n_\gamma\sigma_{\rm T}r_{\rm g}$.
Here, $\sigma_{\rm T}=8\pi e^4/3m_{\rm e}^2c^4$ is the Thomson scattering cross section, $r_g=GM/c^2$ is the gravitational radius, $G$ is the gravitational constant, $M$ is the BH mass, $m_{\rm e}$ and $e$ are the electron rest mass and charge, and $c$ is the speed of light.
If, after the IC interaction, the scattered photon lies above the pair-production threshold with the background radiation, it becomes a tracked, chargeless, massless particle with a per-timestep absorption probability set by the pair-annihilation cross section.

\begin{figure*}
\includegraphics[keepaspectratio, scale=0.7]{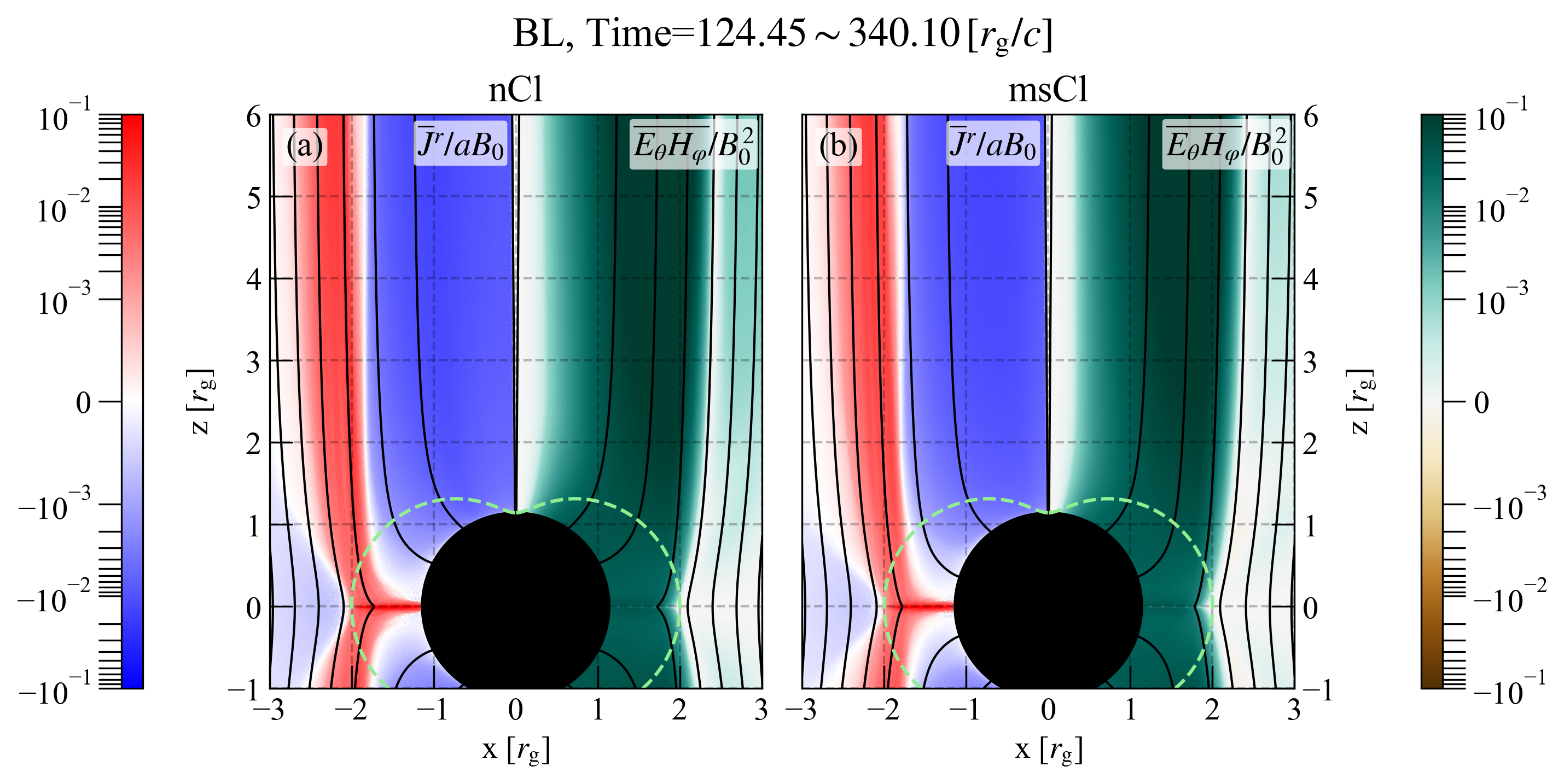}
\caption{The 2D cross-sections of $J^r/aB_0$ and $E_\theta H_\varphi/B_0^2$ averaged over $t=124.45\sim340.10\,r_{\rm g}/c$ for (a) nCl and (b) msCl models. The lime dashed line denotes the static limit.}
\label{paper_fig2}
\end{figure*}
As for the fiducial parameters, we adopt the dimensionless spin of the BH as $a=0.99$, the normalized initial magnetic field amplitude at the horizon as $\tilde B_0= eB_0r_{\rm g}/m_{\rm e}c^2=5\times10^3$, the ambient photon energy normalized by the electron rest mass as $\varepsilon_0=5\times10^{-2}$, and $\tau_0=10$, corresponding to a charge-starved regime where the macroscopic gap development is expected \citep{Crin20, Yuan25}.
(Hereafter we denote photon energies in units of the electron rest-mass energy as $\varepsilon$.)
The initial value of the cold magnetization for pair plasma is therefore $\sigma_0=B_0^2/4\pi n_0m_{\rm e}c^2\simeq1.2\times10^4$, where $n_0=n_{\rm GJ}=\rho_{\rm e, GJ}/e=\omega_{\rm h}B_0/4\pi ec=aB_0/8\pi er_{\rm h}$ denotes a typical value of the Goldreich-Julian (GJ) plasma number density (cf.~\cite{GJ69}; see Appendix.~\ref{ff}) and $\omega_{\rm h} = \omega (r_{\rm h}) = ac/2r_{\rm h}$ is the amplitude of the frame-dragging angular velocity at the horizon (see Appendix.~\ref{app_metric}).
The computational domain spans $\{r_{\rm min},\,r_{\rm max}\}=\{0.98\, r_{\rm h},\,8.0\,r_{\rm g}\}$ ($r_{\rm h}=(1+\sqrt{1-a^2})r_{\rm g}$ denotes the horizon) and $\{\theta_{\rm min},\,\theta_{\rm max}\}=\{0.001\pi,\,\pi-0.001\pi\}$, resolved by $1120\times1600$ cells uniformly spaced in $\log r$ and $\theta$.

The initial electromagnetic field is the vacuum solution of Wald \cite{Wald74}, which is fixed at the outer boundary throughout the simulation. 
To ensure a smooth transition, we employ a perfectly matched layer \cite{Berenger94} near the outer boundary, with its inner edge at $r=7.2\,r_{\rm g}$.
At the axial and inner radial boundaries, we apply reflecting and zero-gradient boundary conditions, respectively.
To initiate pair production, we uniformly initialize a cloud of photons between the horizon and $4.0\,r_{\rm g}$ at number density $n_0$, all with energy~$\varepsilon_{\rm ini}=200$.
Immediately after the simulation starts, these photons begin converting into electron-positron pairs.
This procedure results in a grid-averaged ${\sim}10$ particles per cell throughout the steady state of each run.

We include SC radiative cooling as described in \citet{Cerutti25}. 
In the ultra-relativistic ($\Gamma\gg1$) limit, the SC radiative backreaction force is (cf. \cite{Landau71})
\begin{equation}
\bm g\simeq
-\frac{2}{3}\frac{e^4}{m_{\rm e}^2c^4}
\Gamma^2\bm\beta B_{\perp,\rm eff}^2,
\end{equation}
where
\begin{equation}\label{bperp}
B_{\perp,\rm eff}=\sqrt{(\bm D+\bm\beta\times\bm B)^2
-(\bm\beta\cdot\bm B)^2}.
\end{equation}
This must be added to the RHS of Eq.~(\ref{hamilton}).
The magnitude of the reaction force $\bm{g}(\Gamma)$ can be approximated in the~$\Gamma\gg 1$ limit as $g(\Gamma)\sim-c\sigma_{\rm T}B_{\perp,\rm eff}^2\Gamma^2/6\pi$.
When further assuming that $B_{\perp,\rm eff}\sim B_0$ and considering acceleration by electric fields with amplitudes $D\sim B_{\perp,\rm eff}\sim B_0$, another energy scale, $\Gamma_{\rm SC}=\sqrt{6\pi e/B_0\sigma_{\rm T}}$, would appear.
This corresponds to a balance between the accelerating electric force and SC radiation reaction, and the SC reaction force can then be rewritten as $g(\Gamma)\sim ecB_0(\Gamma/\Gamma_{\rm SC})^2$.
Although~$\Gamma_{\rm SC}$ is fixed by~$B_0$ and physical constants, we treat it as a free parameter, allowing us to tune the relative strength of the radiative backreaction force against that produced by the already-rescaled magnetic field.
In this study, we consider two models: a no-synchrotron-cooling model (nCl; $\Gamma_{\rm SC}=\infty$) and a marginally slow-synchrotron-cooling model (msCl; $\Gamma_{\rm SC}=800$).
The latter is chosen such that SC cooling becomes efficient for only a fraction of high-energy particles: by using the same approximation as in the discussion above, we find that 
the SC cooling timescale is $t_{\rm SC}=(\Gamma m_{\rm e}c^2/g)\sim6\pi m_{\rm e}c/\sigma_{\rm T}B_0^2\Gamma=\Gamma_{\rm SC}^2r_{\rm g}/\tilde{B}_0\Gamma c$, so that $t_{\rm SC}/(r_{\rm g}/c)\leq1$ for $\Gamma\geq\Gamma_{\rm cool}\approx\Gamma_{\rm SC}^2/\tilde{B}_0\simeq128$.
The condition $\varepsilon_0^{-1}<\Gamma_{\rm cool}<\Gamma_{\rm SC}\tilde{B}_0$ ensures both the effective SC cooling for highest-energy particles and substantial pair production involving $\Gamma\sim\varepsilon_0^{-1}$ particles for the msCl model.
We note that this adopted energy hierarchy may not directly match expectations in reality: for instance, with $B_{\perp,\rm eff}\sim B_0$, $\Gamma_{\rm cool}\sim3.2$ is expected for M87 ($B_0\sim10^2\,\rm G$ \cite{Kuo14, EHT21_mag}, $M\sim6.5\times10^9M_\odot$ \cite{EHT19}), and $\sim5.2\times10^3$ for Sgr A ($B_0\sim10^2\,\rm G$ \citep{EHT_SgrA_spin}, $M\sim4.3\times10^6M_\odot$ \cite{EHT_SgrA1}).
We further discuss the caveats and limitations of the detectability assessment associated with our parameter choice in Sec.~\ref{fwhvsmnycvts}.

We perform the main time integration in KS coordinates.
In postprocessing, we transform quantities into Boyer--Lindquist (BL) coordinates when the BL metric facilitates analytic interpretation (Appendix~\ref{app_conversion}).
To reduce computational cost, the msCl run is restarted from the nCl simulation at 95000 timesteps ($t\simeq124.45\,r_{\rm g}/c$) with SC cooling enabled and evolved until $t\simeq340.10\,r_{\rm g}/c$, by which time a new quasi-steady state is established.
Plasma variables and electromagnetic fields are stored every 250 timesteps ($\sim0.33\,r_{\rm g}/c$), while synthetic observables are stored every 100 timesteps ($\sim1.3\,r_{\rm g}/c$).
 
\section{Results}\label{res}
Hereafter, figures labeled “BL” show quantities measured in and evaluated on BL constant-time slices.
\subsection{Global Structures}
\begin{figure*}
\includegraphics[keepaspectratio, scale=0.7]{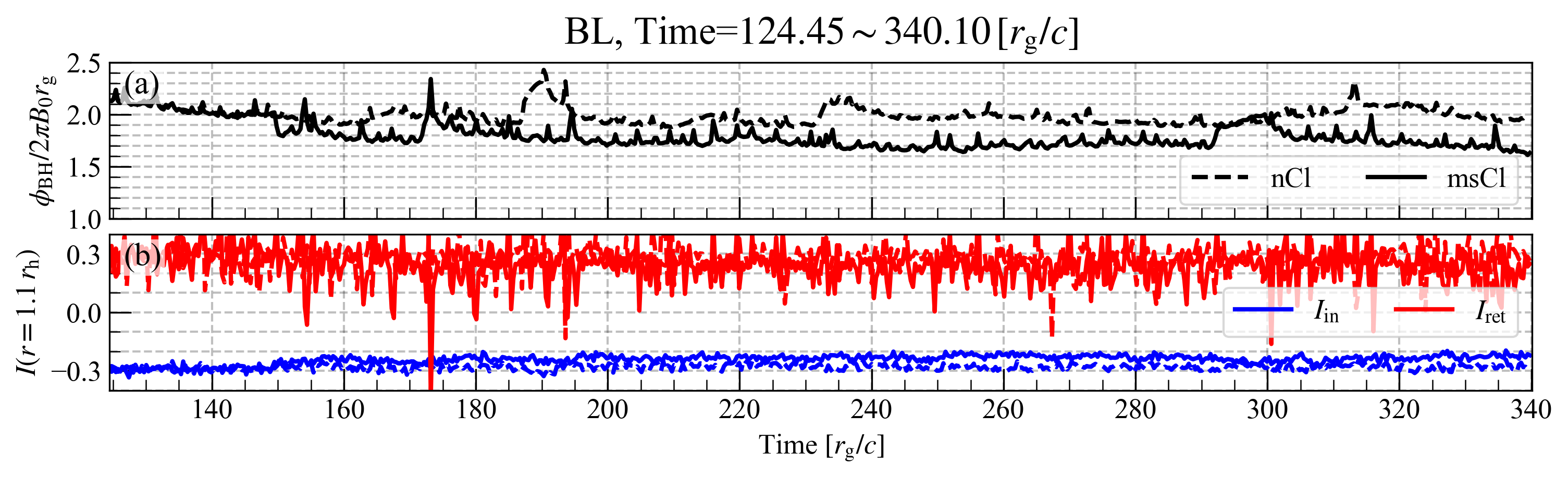}
\caption{Comparison of the temporal evolutions of (a) $\phi_{\rm BH}/2\pi r_{\rm g}B_0$ and (b) the integrated poloidal inward/return current during $t=124.45\sim340.10\,r_{\rm g}/c$. Solid and dashed lines denote the msCl and nCl models, respectively.}
\label{paper_fig1}
\end{figure*}
\begin{figure}
\hspace*{-0.5cm}
\includegraphics[keepaspectratio, scale=0.6]{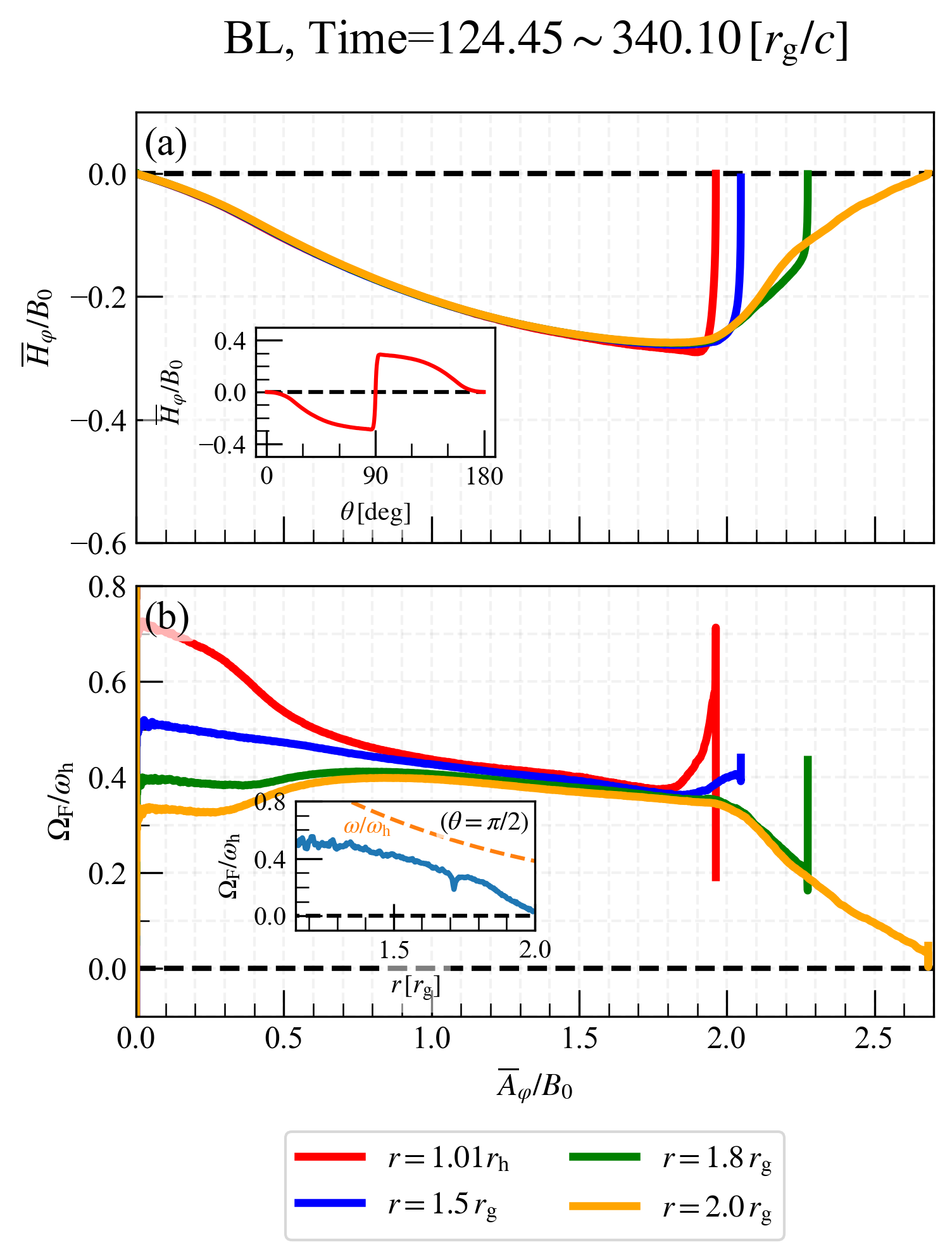}
\caption{(a) $\overline{H}_\varphi/B_0$ versus $A_\varphi$ at $r=1.01\,r_{\rm h}$, $1.5$, $1.8$, and $2.0\,r_{\rm g}$ in the northern hemisphere. The inset shows $\overline{H}_\varphi(1.01\,r_{\rm h})/B_0$ versus $\theta$, illustrating its reversal across the equator. (b) Same for $\Omega_{\rm F}/\omega_{\rm h}$. The inset shows their radial profiles along the equator, including $\omega/\omega_{\rm h}$ (orange dashed).}
\label{paper_fig3}
\end{figure}
\begin{figure*}
\includegraphics[keepaspectratio, scale=0.7]{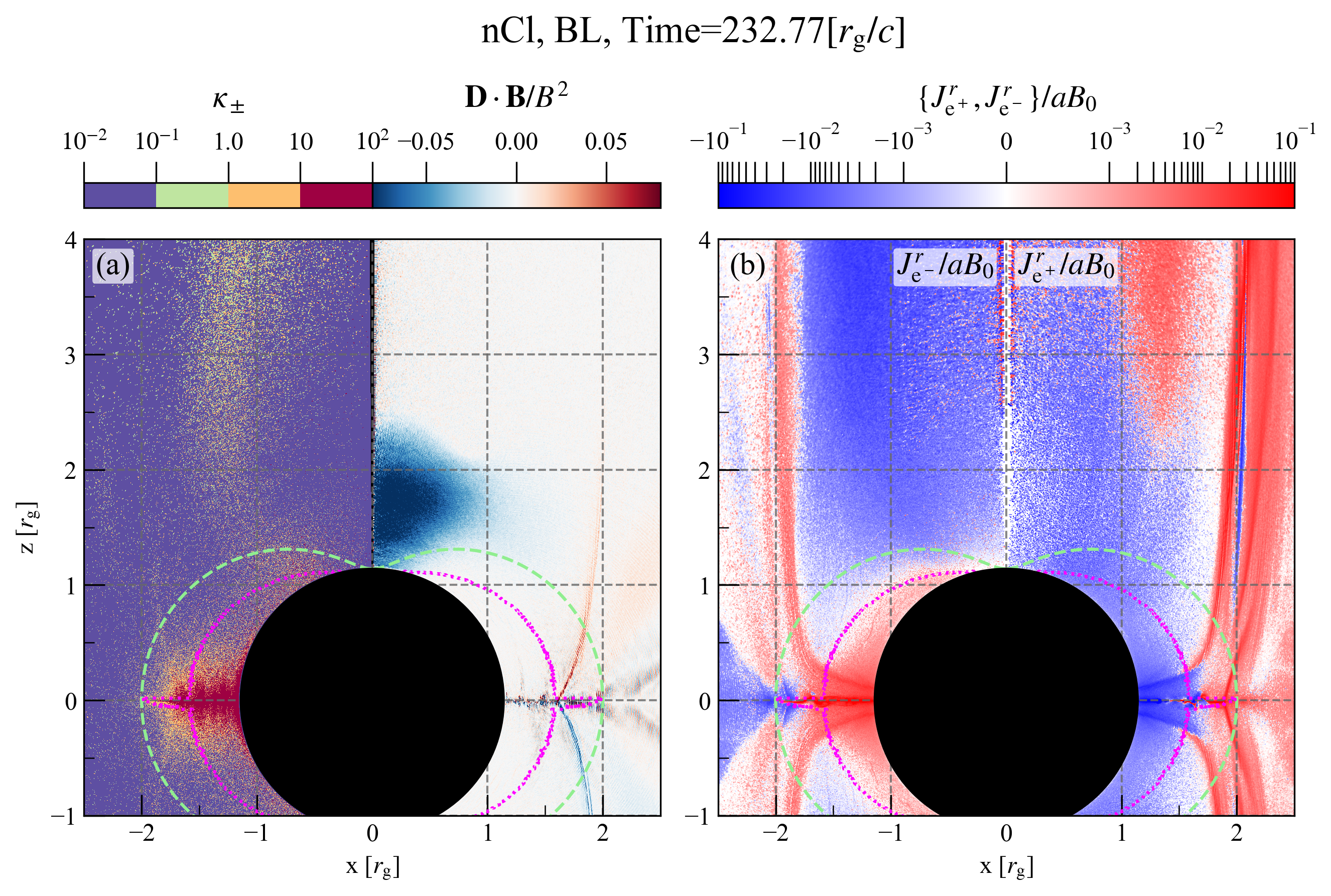}
\caption{(a) 2D snapshots of the pair-production rate, $\kappa_\pm$ (left), and the electric field parallel to the local magnetic field, $\bm{D}\cdot\bm{B}/B^2$ (right), in the northern hemisphere at $t=232.77\,r_{\rm g}/c$ for the nCl model. Lime-dashed and magenta-dotted curves denote the static limit and the inner light surface (ILS; Eq.~(\ref{lsfunc}) with $\Omega_{\rm F}=(E_\theta/\sqrt{\gamma}B^r)(c/r_{\rm g})$), respectively.
(b) 2D snapshots of the radial current densities generated by electrons, $J^r_{\rm e^-}$ (left), and that by positron populations, $J^r_{\rm e^+}$ (right).}
\label{paper_fig4}
\end{figure*}

We first describe the global properties of the magnetospheres in our simulations.
Immediately after initiation, pair production and the associated development of an equatorial current sheet reorganize the magnetic field near the horizon, transforming it from the initial configuration into a parabolic geometry.
The global electromagnetic structure of the magnetosphere is similar across the models, as seen in Fig.~\ref{paper_fig2}.
The current configuration is consistent with what is described in the steady-state force-free model for rapidly rotating BH magnetospheres \citep{Komissarov04} and with previous GRPIC simulations employing similar initial magnetic fields \citep{Parfrey19, Crin21}: poloidal currents flow inward near the pole and outward through the equatorial current sheet and the return current layer between the last magnetic field line connecting to the horizon and the outermost vertical field line threading the ergosphere, forming a closed-current system.
This ensures a hemispheric reversal of $\overline{H}_\varphi$ across the equatorial plane and sustains an outward Poynting flux.
(Hereafter, the time-averaged quantities are denoted by overlined characters.)
$\overline{H}_\varphi$ is approximately preserved along the time-averaged azimuthal vector potential, $\overline{A}_\varphi$, except near the equatorial plane (e.g., Fig.~\ref{paper_fig3}~(a)), indicating an overall satisfaction of the force-free condition for the BZ process.

Fig.~\ref {paper_fig1} shows the temporal evolution of the magnetic flux threading the horizon,
\begin{equation}
\phi_{\rm BH}=2\pi r_{\rm g}\int_0^{\pi/2}\sqrt{\gamma}B^rd\theta,
\end{equation}
and the total "inward" and "return" currents flowing through $r=1.1r_{\rm h}$, $I_{\rm in}$ and $I_{\rm ret}$.
We evaluate them by integrating over two surfaces whose boundaries are chosen as the flux surfaces of $A_\varphi(r,\theta)=A_\varphi(r_{\rm h},80^\circ)$ and $A_\varphi(r,\theta)=A_\varphi(r_{\rm h},100^\circ)$, where $A_\varphi$ denotes the $\varphi$ component of the vector potential:
\begin{equation}\label{ir}
\begin{aligned}
I_{\rm{in}}(r)&=\int_{A_\varphi(r,\theta)=A_\varphi(r_{\rm h},80^\circ)}^{A_\varphi(r,\theta)=A_\varphi(r_{\rm h},\theta_{\rm min})}\sqrt{\gamma}J^r(r) d\theta \\
&\hspace{10pt}+\int^{A_\varphi(r,\theta)=A_\varphi(r_{\rm h},100^\circ)}_{A_\varphi(r,\theta)=A_\varphi(r_{\rm h},\theta_{\rm max})}\sqrt{\gamma}J^r(r) d\theta,\\
I_{\rm{ret}}(r)&=\int_{A_\varphi(r,\theta)=A_\varphi(r_{\rm h},100^\circ)}^{A_\varphi(r,\theta)=A_\varphi(r_{\rm h},80^\circ)}\sqrt{\gamma}J^r(r)d\theta.
\end{aligned}
\end{equation}
Both the temporal evolution of the magnetic flux threading the BH and the total poloidal current saturate at broadly similar values across the models, except for an approximately 10\% decline in the saturated values.
This suggests overall consistency between the two models.

Fig.~\ref{paper_fig3} presents profiles of $\overline{H}_\varphi$ and the field-line angular velocity $\Omega_{\rm F}=-(\overline{E}_\theta/\sqrt{\gamma}\overline{B}^r)(c/r_{\rm g})$ evaluated using time-averaged fields, as functions of $\overline{A}_\varphi$.
As seen, $\Omega_{\rm F}\sim0.3\sim0.6\,\omega_{\rm h}$ remains nearly independent of radius at mid-latitudes ($1.0\lesssim \overline{A}_\varphi/B_0\lesssim1.8$ in Fig.~\ref{paper_fig3}), again consistent with the expectation from the steady-state force-free magnetosphere model.
Noticeable variations along $\overline{A}_\varphi$ appear near the pole ($\overline{A}_\varphi=0$) and the equatorial plane.
The former reflects the local breakdown of the force-free condition as the spark gap emerges.
Near the equator, $\Omega_{\rm F}$ decreases radially from $\sim0.5\omega_{\rm h}$ in the immediate vicinity of the horizon through $\sim0.3\omega_{\rm h}$ in the return current layer to almost zero around $2\,r_{\rm g}$ (see inset).
The region with $\Omega_{\rm F}\sim0$ at the equator extends along the last field line threading the ergosphere, marking the outer boundary of the return-current layer, within which the global current closes (see also \cite{Komissarov04, Toma14}).

\subsection{Local Dynamics and Current Maintenance}
Fig.~\ref{paper_fig4}~(a) shows the snapshots of the pair-production rate, $\kappa_\pm=(\dot n_\pm/n_{\rm GJ})(r_{\rm g}/c)$, and the unscreened electric field strength, $\bm D\cdot\bm B/B^2$, in the northern hemisphere at $t=232.77\,r_{\rm g}/c$, for the nCl model.
Here, $\dot{n}_\pm$ is averaged over each data dumping interval (250 timesteps~$\sim0.33\,r_{\rm g}/c$).
Particle acceleration and subsequent pair production occur around a large polar extent of the unscreened electric fields (i.e., the spark gap) and an equatorial current sheet undergoing intermittent magnetic reconnection, which maintain the global electric current structure.

\subsubsection{Polar Spark Gap}
For both models, we observe that spark gaps develop for $r\lesssim2.5\,r_{\rm g}$ and $\theta\lesssim\pi/4$, with peak amplitudes of the electric field parallel to the local magnetic field, $\bm{D}\cdot\bm{B}/B^2\sim10^{-1}$ throughout the simulation (see Fig.~\ref{paper_fig4}~(a)).

We observe self-sustained temporal evolution of the gap, similar to that reported in previous studies \cite{Levinson18, Chen20, Crin20, Kisaka20, Kisaka22, Kin24, Kin25}.
As electric fields develop parallel to the magnetic field lines due to local mismatch in $\nabla\times\bm{H}$ and the plasma current, electrons and positrons are accelerated in opposite directions to $\Gamma\gg\varepsilon_0^{-1}$, producing IC photons.
These photons, particularly those with energies near $\varepsilon_{\rm IC}\sim\varepsilon_0^{-1}$, produce secondary pairs after propagating over their mean free path, $l_{\rm mfp}(\varepsilon_0^{-1})$.
$l_{\rm mfp}(\varepsilon_0^{-1})\sim(n_\gamma\sigma_{\gamma\gamma})^{-1}\sim0.1r_{\rm g}/\tau_0\sim r_{\rm g}$, where we approximate the pair-production cross-section near the energy threshold as $\sigma_{\gamma\gamma}\sim0.1\sigma_{\rm T}$ \cite{Coppi90}, so that the extension of the electric fields beyond $\sim r_{\rm g}$ is halted.

Fig.~\ref{paper_fig4}~(b) illustrates poloidal current generation for the nCl model.
As shown, both $J^r_{\rm e^\pm}<0$ within the spark-gap extent, indicating that the bulk outgoing (ingoing) motion of electrons (positrons), produced by linear acceleration in the gap, mainly contributes to the poloidal current there.

The limited extent of the gap in the $\theta$ direction is related to the curvature of the magnetic field lines: the curvature of poloidal magnetic field lines toward the rotation axis causes IC photons produced in the polar gap to deviate from their originating field lines, leading to copious pair production at mid-latitudes.
Complete screening is therefore achieved there, while a strong electric field persists near the pole.

\subsubsection{Equatorial Current Sheet and Return Current Layer}\label{eq_sep}
\begin{figure*}
\includegraphics[keepaspectratio, scale=0.55]{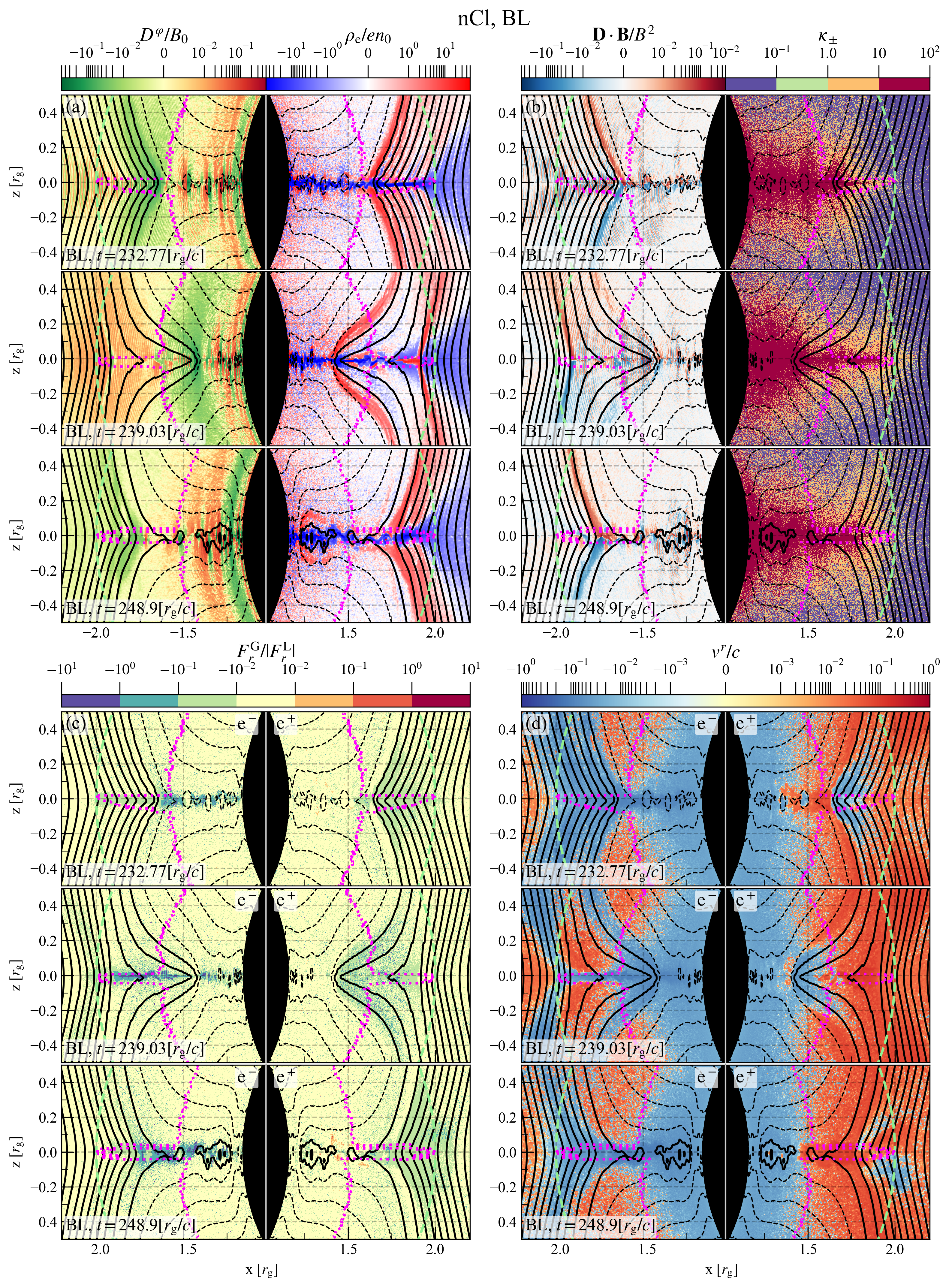}
\caption{A 2D cross-section showing (a) $D^\varphi/B_0$ and $\rho_{\rm e}/en_0$, (b) $\bm{D}\cdot\bm{B}/B^2$ and $\kappa_\pm$, (c) the ratio of the combined radial inertial force (first to third terms in Eq.~(\ref{hamilton})) to the absolute radial Lorentz force, $F^{\rm G}_r/|F^{\rm L}_r|$, and (d) radial 3-velocities, $v^r/c$, during a representative nCl reconnection event. Rows show snapshots at $t=232.77,\,239.03,\,248.9\,r_{\rm g}/c$ (top to bottom). Black solid (dashed) contours of $A_\varphi$ denote field lines disconnected from (connected to) the horizon, defined as contours not crossing (crossing) $r=1.01\,r_{\rm h}$. Panels (c) and (d) show electrons (left) and positrons (right). In each panel, Lime-dashed and magenta-dotted curves denote the static limit and the ILS, respectively.}
\label{paper_fig5}
\end{figure*}
Fig.~\ref{paper_fig4}~(b) also presents the particle populations contributing to the outward electric current flowing through the equator and the return current layer beyond.
Here, the particle population differs between the inner current sheet ($r\lesssim1.5\,r_{\rm g}$ around the equatorial plane) and the outer part ($r>1.5\,r_{\rm g}$).
Around the current sheet, $J^r_{\rm e^-}>0$ but $J^r_{\rm e^+}<0$, suggesting that both electrons and positrons move inward, but the net negative charge associated with them carries the return current.
In the return-current layer, $J^r_{\rm e^\pm}>0$ for the most part, suggesting that inward (outward) electron (positron) motion primarily carries the return current along vertical field lines.

This electric current structure is associated with the dynamical evolution of electromagnetic fields during the recurring "flux-infall" events, similar to those reported in previous studies \citep{Parfrey19, Crin21, Vos25, Mehlhaff26}.
Fig.~\ref{paper_fig5} shows the temporal evolution of one representative event, which proceeds as follows: (1) the innermost vertical field lines of the return current layer move inward with the plasma ($t=232.77\,r_{\rm g}/c$; top rows), (2) the ingoing field lines deform into a radially antiparallel configuration across the equator ($t=239.03\,r_{\rm g}/c$; middle rows), and (3) magnetic reconnection is triggered in the equatorial current sheet ($t=248.9\,r_{\rm g}/c$; bottom rows).

In the early stage of the evolution, the poloidal magnetic field evolves as an azimuthal electric field ($D^\varphi<0$) component emerges at the inner edge of the return current layer (panel~(a), left).
This time-dependent electric field initially arises from the local mismatch between the azimuthal current supported by $\bm{D}\times\bm{B}$ drift and $\left.\nabla\times\bm{H}\right|^\varphi$, and moves toward the horizon.
The resulting deformation of the magnetic field lines into a radially antiparallel configuration across the equator is related to the net negative charge along with there: the poloidal electric field associated with such an antiparallel magnetic field configuration must reverse sign across the current sheet to sustain the azimuthal drift motion, so that the net negative charge density is required to satisfy $\rho_{\rm e}=\nabla\cdot\bm{D}/4\pi<0$ in there (panel~(a), right; see also Appendix~\ref{gj}).
We also find that the radial inertial forces, $F^{\rm G}_r$ (the first-to-third RHS terms in Eq.~(\ref{hamilton})), modestly affect the inward movement of plasma compared to the Lorentz term, $F^{\rm L}_r$ (the fourth and fifth terms; see panel~(c)) in this early phase.
Instead, the drift driven by the azimuthal electric fields mainly contributes to radial motion here.

Alongside the inward transport of magnetic flux, a narrow thread of $\bm{D}\cdot\bm{B}\neq0$ develops along the innermost magnetic field line in the return-current layer (panel~(b), top left).
Its sign reversal across the equator indicates that the electric field is predominantly directed outward.
The thread emerges because the slow field-line rotation near the equator ($\Omega_{\rm F}\lesssim\omega$; see Fig.~\ref{paper_fig3}~(b)) drives an inward $\bm{D}\times\bm{B}$ drift (see Eq.~(\ref{dbdrift})), which cannot initially sustain the outward current required by the radial gradient in $H_\varphi$ across the return-current layer.
Once the thread forms, particle acceleration within it drives positrons outward and electrons inward, compensating for the current deficit and making the net charge density near the equator negative.

After the deformation, magnetic pressure pushes antiparallel poloidal magnetic fields toward the equator, driving a series of magnetic reconnection events in the sheet.
An apparent kink in the sheet indicates the onset of the drift-kink instability, as briefly mentioned in \cite{Parfrey19}.
At this stage, the inertial drag force sets in to generate the return current: the reconnection electric fields point toward the negative azimuthal direction, with amplitudes $D_{\rm acc}\sim0.1B$, accelerating positrons and decelerating electrons.
The resulting difference in angular momentum makes $F^{\rm G}_r$ significant relative to $F^{\rm L}_r$ particularly for electrons (panel~(c)), producing radial velocity shear in the sheet (panel~(d)).
This preferred electron infall maintains the radial return current.

At a certain point in the inward motion of the return-current layer, we observe that a poloidal $\bm{D}\cdot\bm{B}$ disappears, and a new thread simultaneously develops near the static limit, repeating the cycle (see Fig.~\ref{paper_fig5}~(a) left).
This cycle is associated with the growth of $H_\varphi$: as the field lines approach the horizon, frame dragging increasingly twists them, causing $H_\varphi$ at the inner edge of the return current layer to approach that of the "horizon-connected" field lines (dashed contours in Fig.~\ref{paper_fig5}).
The inward drift motion of positive charge therefore matches the sign of the current needed to support $H_\varphi$, so no nonzero $\bm{D}\cdot\bm{B}$ is needed.
Conversely, outward acceleration is necessary to generate the return current at the static limit, where the gradient of $H_\varphi$ is largest, so a new thread emerges there.
 
Pair production mainly occurs in the current sheet in the late stage (panel~(b), right), reaching $\kappa_\pm\sim10^2$, although most produced pairs fall into the horizon and contribute only to current production in the sheet.
Outside in the return current layer, electrons accelerated inward along the poloidal threads of $\bm{D}\cdot\bm{B}\neq0$ mainly trigger pair production around the equator, though the multiplicity remains lower ($\kappa_{\pm}\lesssim1$; see Fig.~\ref{paper_fig4}~(a)) than in the current sheet.
Nevertheless, the multiplicity remains lower than in the current sheet, and the cold magnetization in the return-current layer is not substantially reduced from $\sigma_0$, typically remaining at $\sigma_\pm=B^2/4\pi m_{\rm e}n_\pm c^2\gtrsim10$.
This $\sigma_\pm$ is higher than that found in previous GRPIC simulations with similar initial magnetic-field configurations but more abundant plasma supply (e.g., \cite{Parfrey19}), which may result in the ineffective inertial force in the return current layer in this study (see Sec.~\ref{rec_comp} for further discussion).

\subsection{Variability \& Particle Acceleration}
\begin{figure}
\includegraphics[keepaspectratio, scale=0.55]{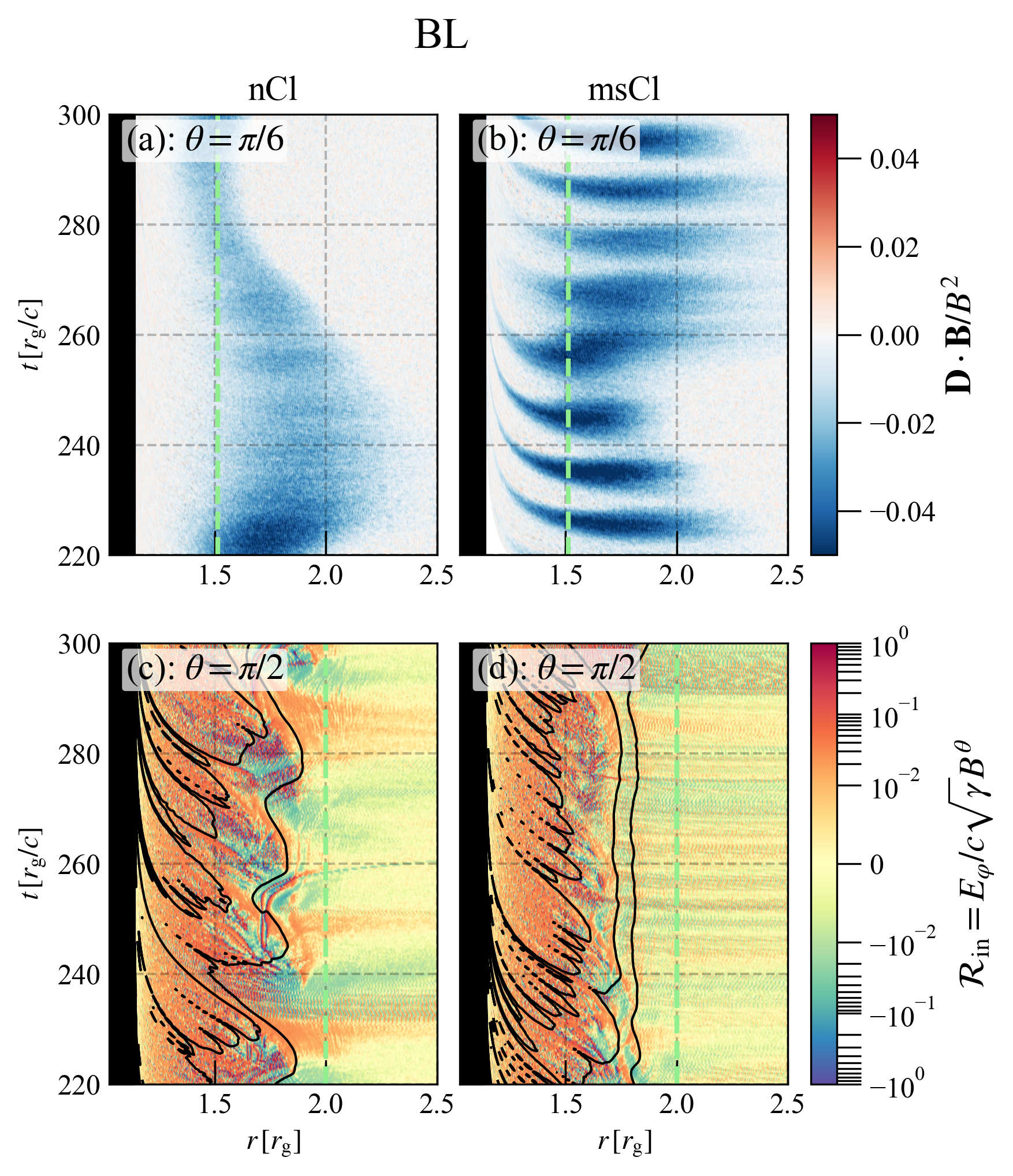}
\caption{Diagrams of the radial distribution of (a, b) $\bm{D}\cdot\bm{B}/B^2$ in $\theta=\pi/6$, and (c, d) $\mathcal{R}_{\rm in}$ in $\theta=\pi/2$, for two cooling models. Black contours for panels (c, d) denote the contour of $A_\varphi(r)/2\pi B_0r_{\rm g}$ (each line indicates 1.8, 2.0, and 2.2, respectively). Lim-dashed vertical lines indicate the static limit.}
\label{paper_fig6}
\end{figure}
\begin{figure*} 
\includegraphics[keepaspectratio, scale=0.7]{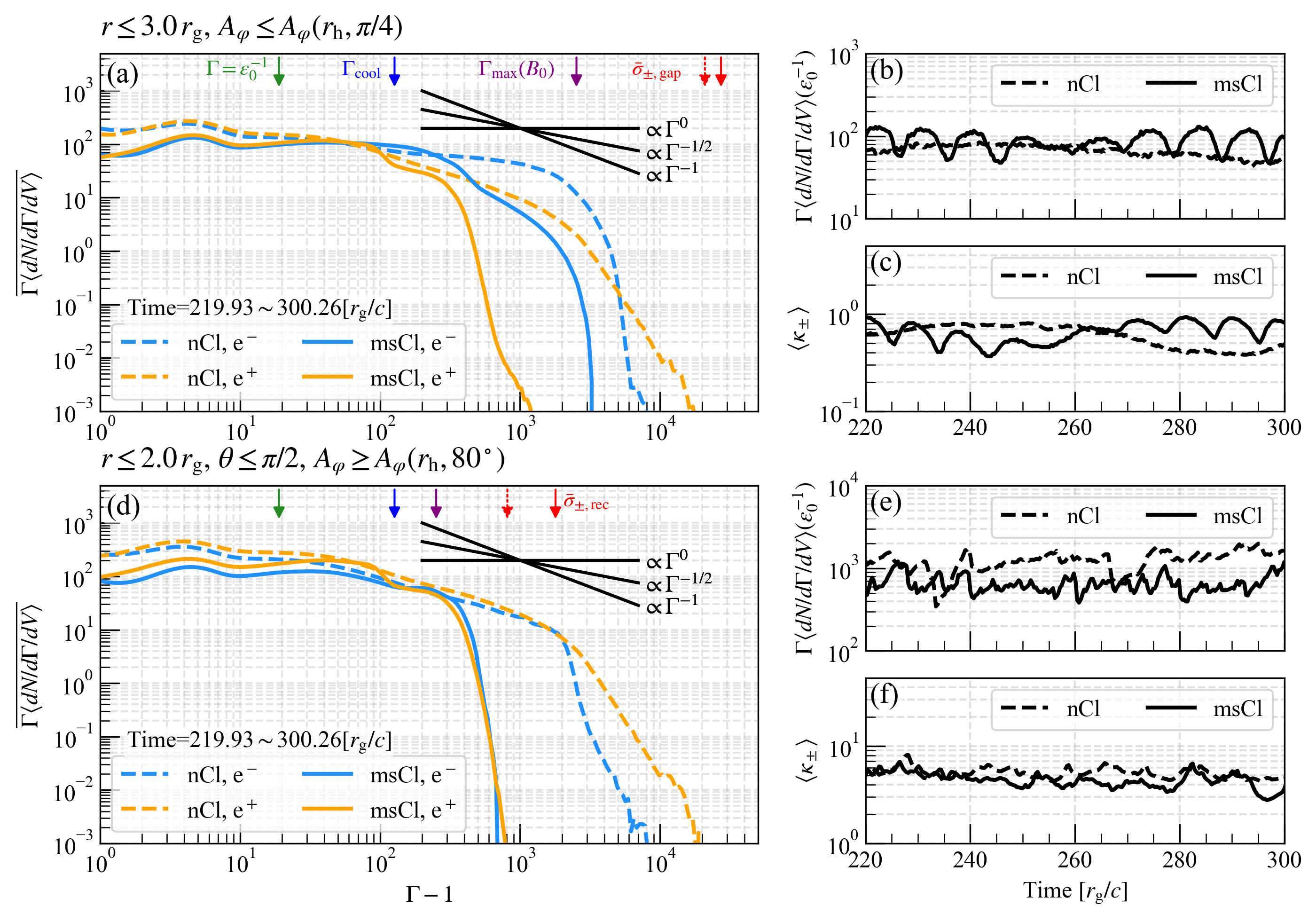}
\caption{(a) Time- and volume-averaged particle spectra, $\overline{\Gamma \langle dN/d\Gamma/dV\rangle}$, during $t=219.93\sim300.26\,r_{\rm g}/c$, for the polar-gap region ($r\le3.0\,r_{\rm g}$, $A_\varphi\le A_\varphi(r_{\rm h},\pi/4)$). Solid (dashed) curves denote the msCl (nCl) model; blue (orange) curves represent electrons (positrons). (b,c) Temporal evolution of $\Gamma \langle dN/d\Gamma/dV\rangle$ at $\Gamma=\varepsilon_0^{-1}$ and the volume-averaged pair-production rate, $\langle\kappa_\pm\rangle$. (d--f) Same as (a--c), but for the equatorial current-sheet region ($r\le2.0\,r_{\rm g}$, $\theta\le\pi/2$, $A_\varphi\ge A_\varphi(r_{\rm h},80^\circ)$). In panels (a) and (d), green, blue, purple, and red arrows indicate $\Gamma=\varepsilon_0^{-1}$, $\Gamma_{\rm cool}$, $\Gamma_{\rm max}(B_0)=\Gamma_{\rm SC}(f_D/f_{B_{\perp,\rm eff}}^2)^{1/2}$ (Eq.~(\ref{gammamax}), with $(f_D,f_{B_{\perp,\rm eff}})=(10^{-1},10^{-1})$ for (a) and $(10^{-1},1.0)$ for (d)), and the time- and volume-averaged magnetization, $\sigma_\pm=B^2/4\pi n_\pm m_{\rm e}c^2$, respectively.} 
\label{paper_fig7} 
\end{figure*}
\begin{figure}
\includegraphics[keepaspectratio, scale=0.55]{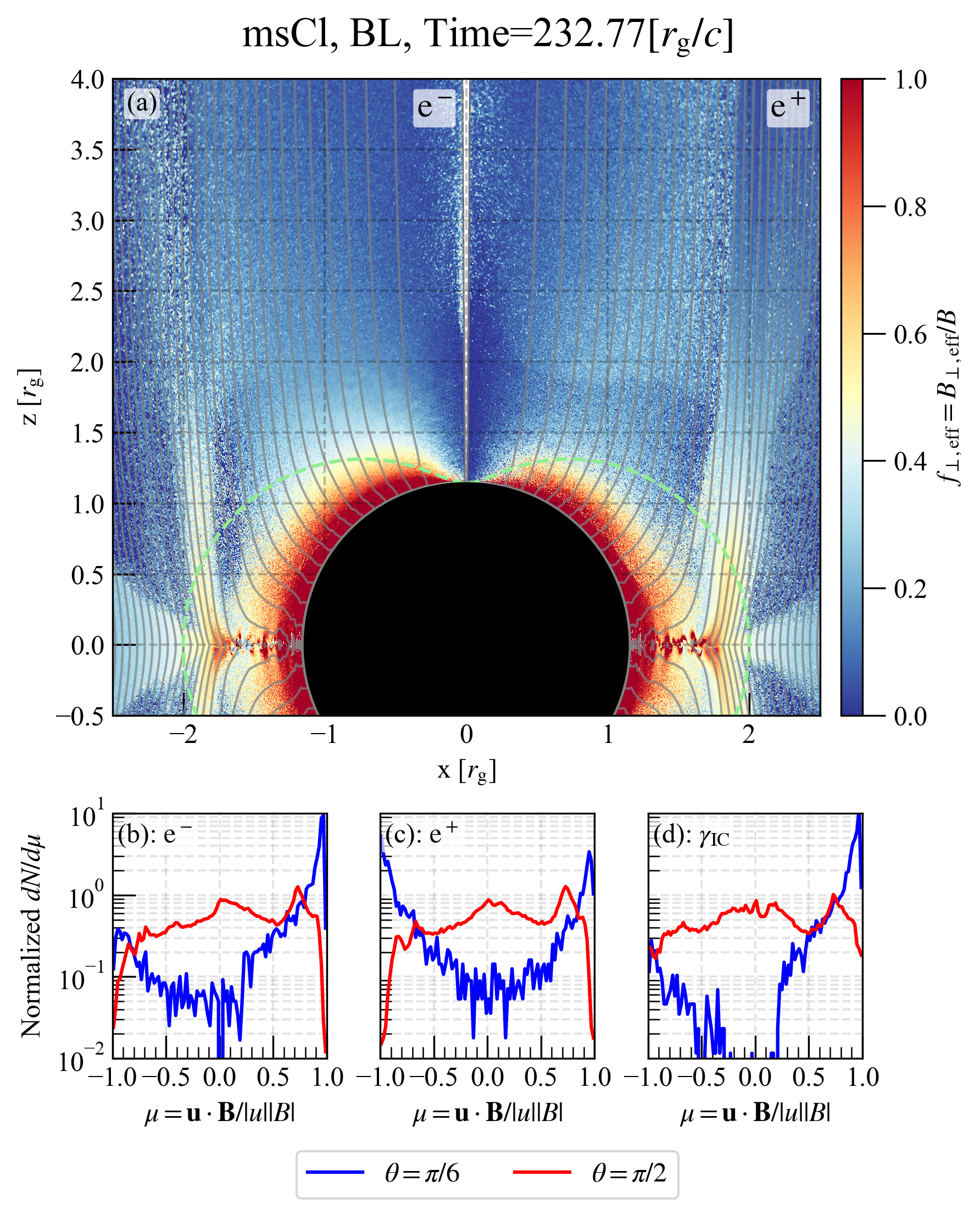}
\caption{(a) Cell-averaged effective perpendicular magnetic field, $f_{\perp,\rm eff}=B_{\perp,\rm eff}/B$, at $t=239.36\,r_{\rm g}/c$ for the msCl model (see Eq.~(\ref{bperp}) for the definition of $B_{\perp,\rm eff}$). Left and right panels show the values for electrons and positrons, respectively.
(b)--(d) Normalized distributions of the pitch-angle cosine, $\mu=\bm{u}\cdot\bm{B}/(|u||B|)$, evaluated within $1.01\,r_{\rm h}\le r<4.0\,r_{\rm g}$ along $\theta=\pi/6$ (blue) and $\pi/2$ (red). From left to right: electrons, positrons, and IC photons.}
\label{paper_fig8}
\end{figure}
Next, we examine the variability and properties of particle acceleration in the polar spark gap and equatorial current sheet, and describe their response to the onset of SC radiative cooling.

\subsubsection{Polar Spark Gap}
Fig.~\ref{paper_fig6}~(a, b) compares the radial evolution of $\bm{D}\cdot\bm{B}/B^2$ at $\theta=\pi/6$ for the nCl and msCl models.
Here, the difference in the temporal evolution of the gap electric field between the models is evident: the gap electric field remains around $r\sim1.7\,r_{\rm g}$ in the nCl model, whereas the msCl model exhibits pronounced oscillatory gap formation and screening on variability timescales of $\Delta t\lesssim10\,r_{\rm g}/c$, indicating more efficient suppression of the electric field.
The oscillating behavior with the macroscopic ($\sim r_{\rm g}$) development of unscreened electric fields in the msCl model is broadly consistent with previous 1D studies \citep{Levinson18, Chen20, Kisaka20, Kisaka22, Kin24} and with 2D simulations employing similarly low plasma injection rates or pair multiplicities \citep{Crin20, Niv23, Chen25, Yuan25}.

The difference between the two models arises from the enhanced SC radiative cooling for the highest-energy particles in the msCl model.
Fig.~\ref{paper_fig7}~(a) shows that, as the electric field develops, particles inside the gap are accelerated beyond $\Gamma_{\rm cool}$ and rapidly lose energy through SC cooling, as indicated by the spectral break above $\Gamma_{\rm cool}$ in the msCl model.
This induces a pile-up of particles below $\Gamma_{\rm cool}$.
Because $\Gamma_{\rm cool}$ is close to the pair-production threshold energy, $\varepsilon_0^{-1}$, in our setup, this also increases the number of particles around $\varepsilon\sim\varepsilon_0^{-1}$ (Fig.~\ref{paper_fig7}~(b)), thereby enhancing the production of IC photons with energies of $\varepsilon_{\rm IC}\sim\Gamma\sim\varepsilon_0^{-1}$.
Because $l_{\rm mfp}$ depends on energy, these photons travel shorter distances than higher-energy photons before annihilating.
As the typical scale of the magnetic field's curvature radius is found to be $\gtrsim r_{\rm g}$, those photons are more likely to inject pairs along the field lines where they originate, thereby screening the gap electric field and driving quasi-periodic oscillations (Fig.~\ref{paper_fig7}~(c)).

\subsubsection{Equatorial Current Sheet and Return Current Layer}
SC cooling also affects the particle spectra around the current sheet.
In the msCl model, SC cooling substantially suppresses the high-energy particle production, resulting in a clear post-break spectrum (Fig.~\ref{paper_fig7}~(d)).
A cutoff steeper than $\Gamma\,\langle dN/d\Gamma dV\rangle\propto\Gamma^{-2}$ is broadly consistent with local 2D PIC simulations of radiative magnetic reconnection \citep{Werner19, Schoeffler23}.
(Hereafter, the volume-averaged quantities are denoted by bracketed characters.)
The cooling effect appears even stronger than in the polar spark gap because the particles around the current sheet have momenta perpendicular to the local magnetic fields, as indicated by a 2D map of the relative strength of effective magnetic fields perpendicular to the particle momenta, $f_{\perp,\rm eff}=B_{\perp,\rm eff}/B$, presented in Fig.~\ref{paper_fig8}~(a). 
This reflects different preferred directions of momentum gain: the accelerating electric field, $D_{\rm acc}$, is expected to align mostly with the local magnetic field in the spark gap, whereas it is predominantly perpendicular to the magnetic field in the current sheet.
Moreover, the local magnetic field is relatively persistent in the polar region, whereas it is strongly spatially and temporally inhomogeneous in reconnection zones.
The momentum distribution of accelerated particles should therefore remain preferentially aligned with the field lines in the gap region, while becoming more isotropic in the current sheet (see Fig.~\ref{paper_fig8}~(b, c)).

Such differences in the particle spectra appear to affect the dynamics of the current sheet: the left-upward notch of the $A_\varphi$ contour in Fig.~\ref{paper_fig6}~(c, d) indicates the occurrence of major magnetic flux infall events (as in Fig.~\ref{paper_fig5}) at the equator, and the comparison of its pattern between the two models suggests their slightly higher frequency in the msCl model.
It may also be responsible for the about 10\% decline of $\phi_{\rm BH}$ for the msCl model (see Fig.~\ref{paper_fig1}~(a)), as lower plasma injection would mean the general reduction of the total azimuthal current amplitude.
The dynamical impact of SC cooling is nevertheless modest in the current comparison, possibly due to our choice of $\Gamma_{\rm SC}$ (see Sec.~\ref{fwhvsmnycvts} for further discussion).

The inflow velocity of $B^\theta$ in the return current layer can nevertheless be evaluated from 
$\theta$-component of Eq.~(\ref{faraday}), as $\mathcal{R}_{\rm in}=E_\varphi/(c\sqrt{\gamma}B^\theta)$
(notice that $E_\varphi=\alpha D_\varphi=\alpha\gamma_{\varphi\varphi}D^\varphi$ in BL coordinates).
The colors in Fig.~\ref{paper_fig6}~(c, d) show its temporal evolution at $\theta=\pi/2$.
We find $\mathcal{R}_{\rm in}\sim+0.1$ from $r\sim1.5\,r_{\rm g}$, a typical location where the current sheet intersects the return current layer, toward the horizon, in both models.
The variability timescale of the equatorial current sheet can then be estimated as
$\Delta t\sim l_{\rm rec}/(\mathcal{R}_{\rm in}c)\sim10\,r_{\rm g}/c$,
where $l_{\rm rec}\sim r_{\rm g}$ is taken as the characteristic length scale of the sheet.

\begin{figure}
 \includegraphics[keepaspectratio, scale=0.55]{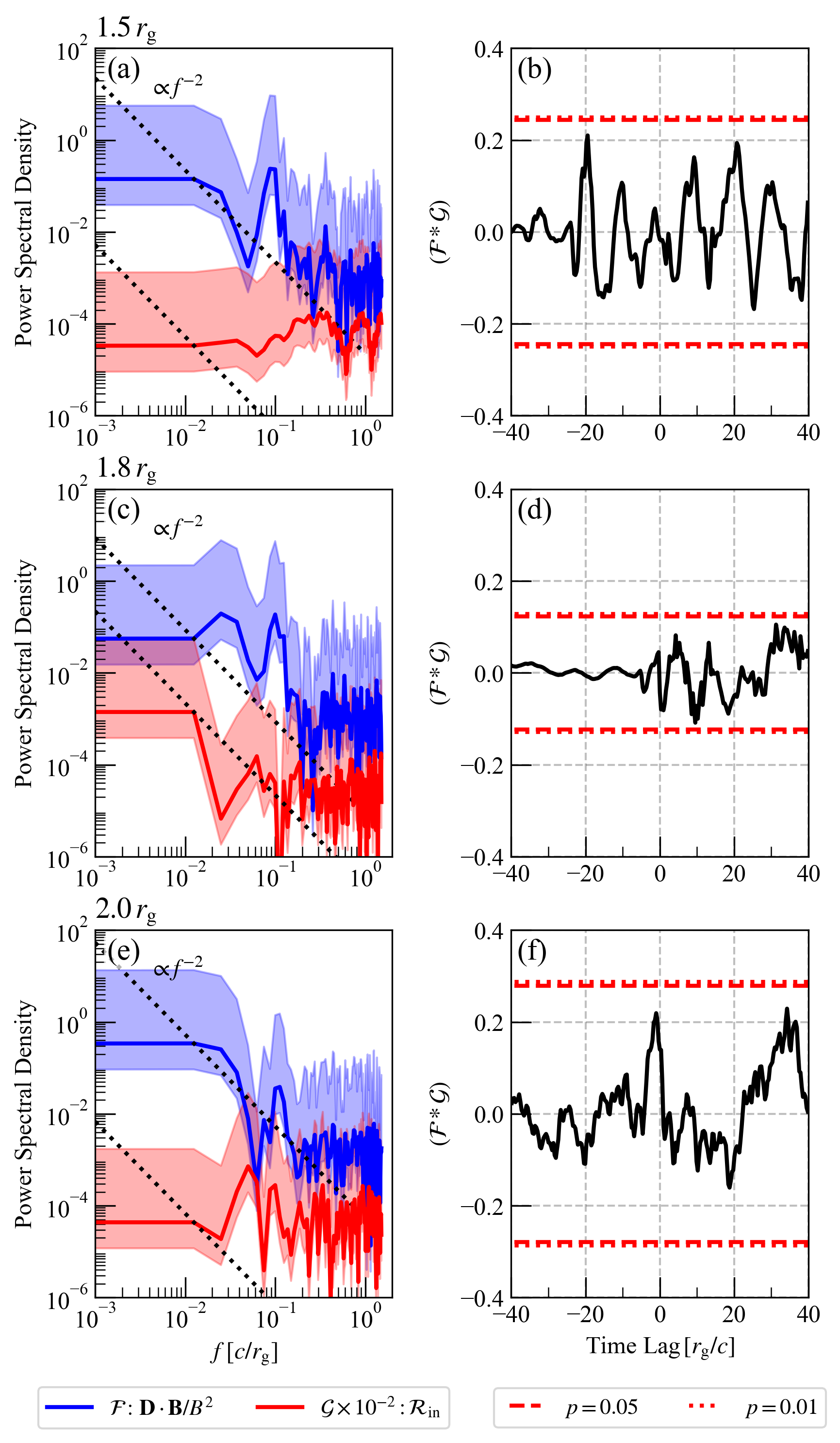}
\caption{PSDs and cross-correlation functions for $\bm{D}\cdot\bm{B}/B^2$ along $\theta=\pi/6$ and $\mathcal{R}_{\rm in}$ along $\theta=\pi/2,$ at (a, b) $r=1.5\,r_{\rm g}$, (c, d) $1.8\,r_{\rm g}$, and (e, f) $2.0\,r_{\rm g}$. Red dashed (dotted) lines on the right mark the 95 (99)\% percentiles of the circular shift test.}
\label{paper_fig9}
\end{figure}
\subsubsection{Characteristic Variability Timescales, Cross-correlation}
Fig.~\ref{paper_fig9} shows the Fourier power spectral densities (PSDs) of the same temporal sequence of quantities as in Fig.~\ref{paper_fig6} at $r=1.5,\,1.8,\,2.0\,r_{\rm g}$ for the msCl model.
As for $\bm{D}\cdot\bm{B}/B^2$, the PSDs approach a white-noise spectrum, $f^0$, at the frequncy higher than $f\sim10^{-1}\,c/r_{\rm g}$.
$\mathcal{R}_{\rm in}$ PSDs present a white-noise trend across the entire frequency range, suggesting a more stochastic nature of the current-sheet variability.
A local peak near $f\sim10^{-1}\,c/r_{\rm g}$ in $\bm{D}\cdot\bm{B}/B^2$ PSDs, most evident for $1.5\,r_{\rm g}$, would correspond to the gap oscillation timescale identified above.

To further quantify the nature of their variabilities, we compute their cross-correlation functions, defined as
\begin{equation}
(\mathcal{F}*\mathcal{G})[n]\equiv\sum^{\infty}_{m=-\infty}\mathcal{F}[m]\mathcal{G}[m+n],
\end{equation}
where $\mathcal{F}$ and $\mathcal{G}$ denote the two datasets, $m$ is the index of discrete timesteps, and $n$ is the index of a time lag of $\mathcal{G}$ relative to $\mathcal{F}$.
The right panels of Fig.~\ref{paper_fig9} show the results.
Several weak correlations appear in the sequence but never reach 95\% significance under circular time-shift testing. 

The result indicates no, or at least very weak, established cross-correlation between the equatorial and polar regions in the innermost magnetosphere, suggesting that the overall match between the ingoing and return current amplitudes (Fig.~\ref{paper_fig1}) does not arise from a direct dynamical connection near the horizon.
This appears inconsistent with discussions in earlier studies assuming that current closure occurs very close to the BH \citep[e.g.,][]{Thorne86, Toma14}, though further tests are necessary to completely rule out this possibility.

The apparent match between their current amplitudes may instead reflect the global requirement that the total ingoing and return currents balance in the time-averaged state to sustain $H_\varphi\sim0$ outside the ergosphere.
On shorter timescales, however, the currents flowing through there are sustained by different physical processes: the ingoing current is maintained primarily by charge-separated flow via particle acceleration in the polar spark gap, whereas the return current is sustained by the inward motion of negatively charged plasma in the equatorial current sheet and by nonzero $\bm{D}\cdot\bm{B}$ in the return current layer.
Their instantaneous amplitudes therefore need not coincide exactly, even though they balance on average (see also \cite{Kolos26}).

\subsection{Synthetic Observables}\label{lcs}
As shown in the previous sections, both the polar spark gap and the equatorial current sheet can produce nonthermal particles. Here, we show how their time-dependent dynamics imprint themselves on the observable radiation.
To investigate these signatures, we performed ray-tracing calculations based on the simulation outputs.

\subsubsection{Ray-tracing Procedure}
The ray-tracing procedure follows \citet{Crin21} and \citet{Mehlhaff25}.
For IC emission, we trace photons below the pair-production threshold ($\varepsilon<\varepsilon_0^{-1}$) directly to a distant observer screen at $r_{\rm obs}=10^3\,r_{\rm g}$.
For SC emission, we employ a sub-sampling procedure, in which a fraction of plasma particles is randomly selected to emit representative SC photon macroparticles with appropriately rescaled statistical weights.

Observables are evaluated in 10 viewing-angle bins spanning $0^\circ<\alpha_{\rm obs}\leq90^\circ$.
To evaluate contributions from the spark gap and the current sheet separately, we label photons originating from mid-to-high latitudes ($0<\theta\leq\pi/4$) and mid-to-low latitudes ($\pi/4<\theta\leq\pi/2$) within $r\sin\theta\leq3.0\,r_{\rm g}$ when counting.

\begin{figure}
 \includegraphics[keepaspectratio, scale=0.5]{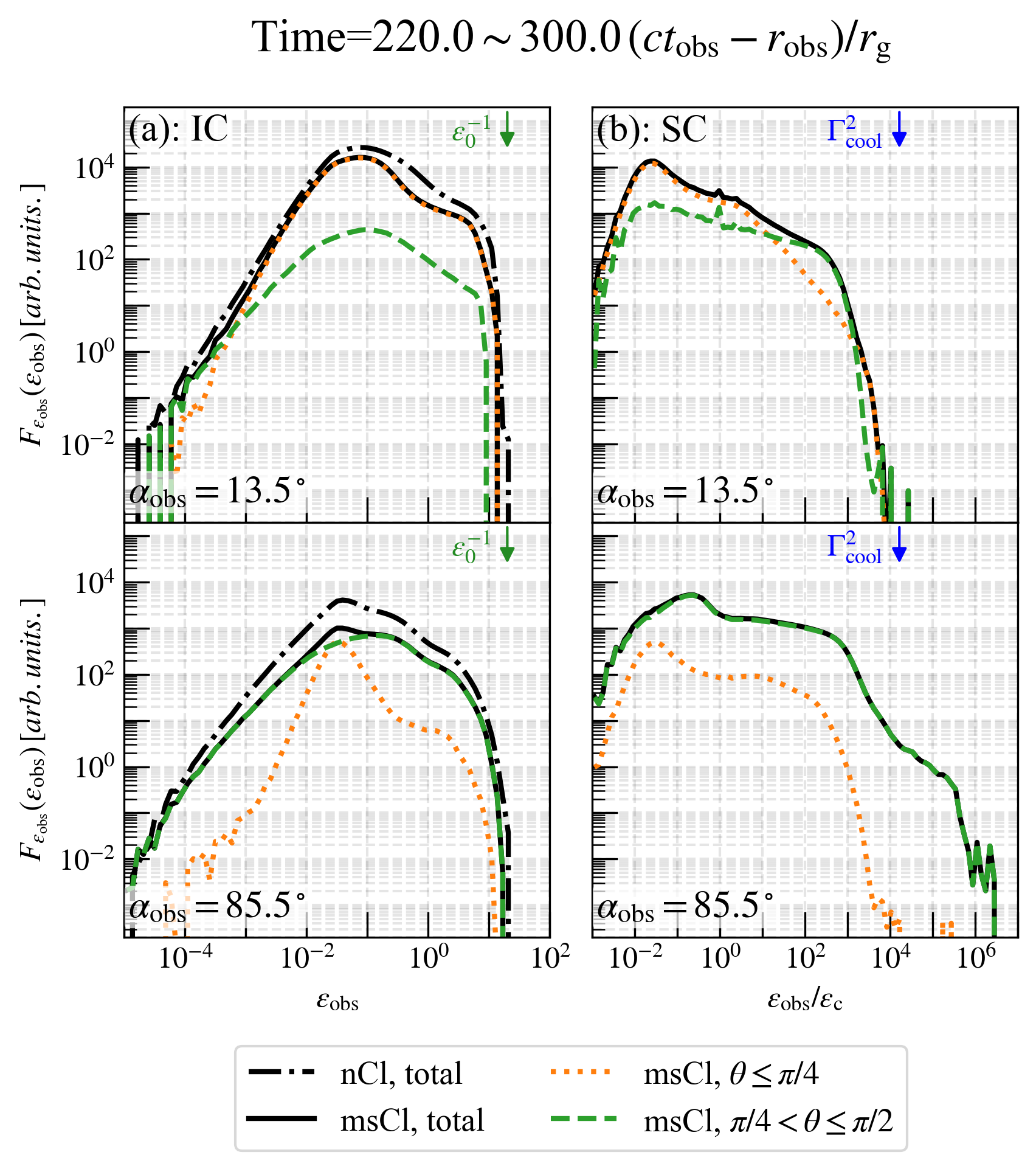}
\caption{(a) Time-averaged synthetic IC photon spectra during $t_{\rm obs}-r_{\rm obs}/c=220.0\sim300.0\,r_{\rm g}/c$. (b) Same for SC photons, shown in units of the cyclotron characteristic energy, $\varepsilon_{\rm c}=3heB_0/4\pi m_{\rm e}^2c^3$. Black solid and dash-dotted lines show the total luminosities of the msCl and nCl models. Orange dotted and lime dashed lines indicate contributions from $0<\theta\leq\pi/4$ and $\pi/4<\theta\leq\pi/2$ within $r\sin\theta\leq3.0\,r_{\rm g}$ for the msCl model.}
\label{paper_appc}
\end{figure}
\begin{figure*}
\includegraphics[keepaspectratio, scale=0.6]{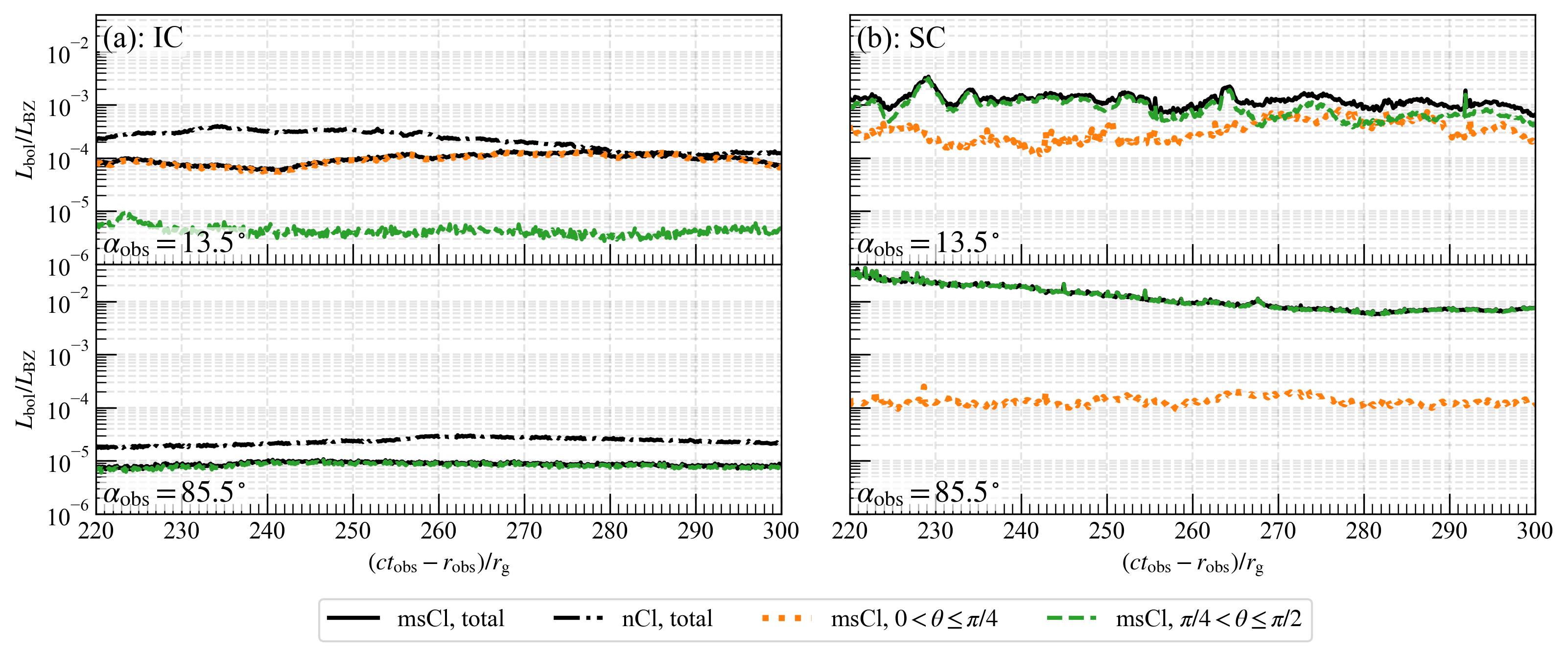}
\caption{Synthetic observed bolometric IC (left) and SC (right) light curves during $t_{\rm obs}-r_{\rm obs}/c=220\sim300\,r_{\rm g}/c$. The top and bottom rows correspond to viewing angles of $\alpha_{\rm obs}=13.5^\circ$ and $85.5^\circ$, respectively. Colors denote the same components as in Fig.~\ref{paper_appc}.}
\label{paper_fig10}
\end{figure*}
\begin{figure}
\hspace*{-0.5cm}
\includegraphics[keepaspectratio, scale=0.7]{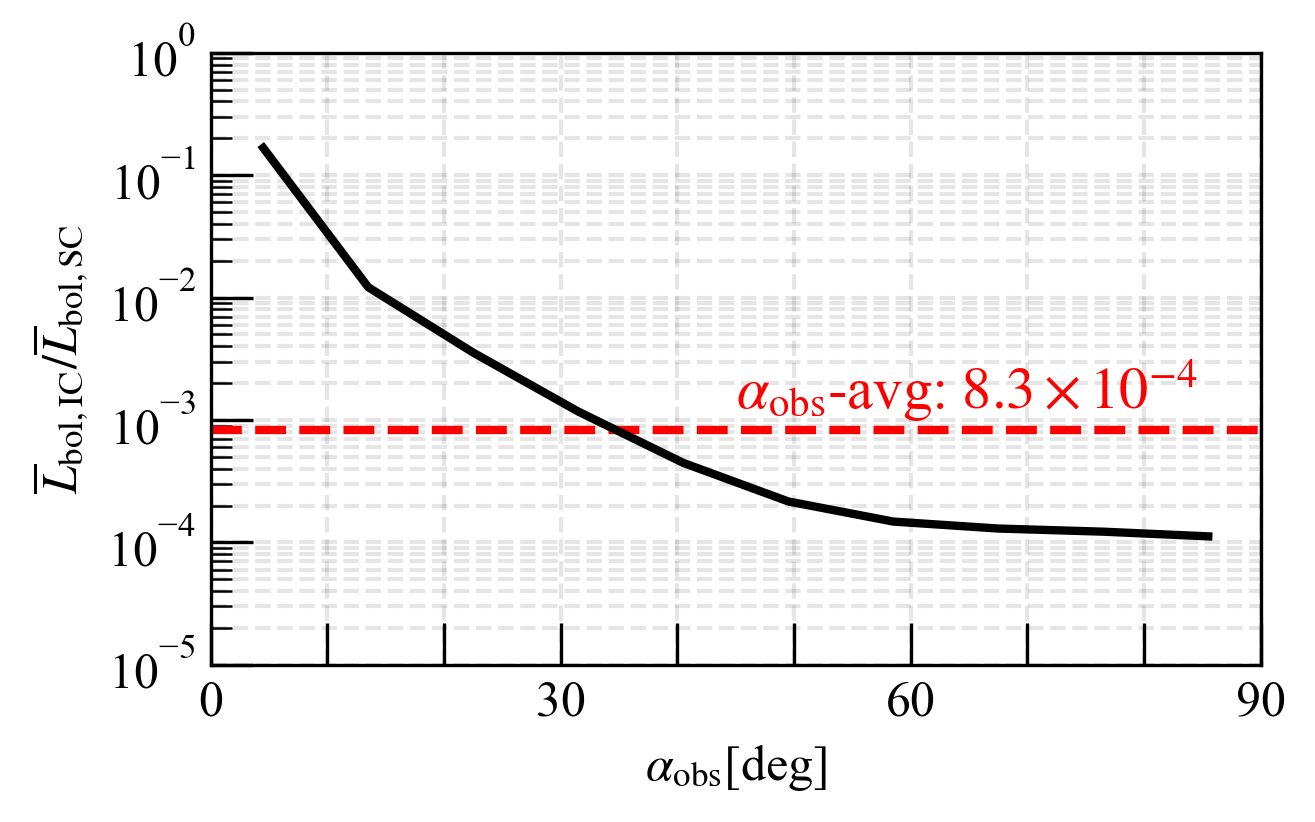}
\caption{Ratios between obtained IC and SC bolometric luminosity averaged over $t_{\rm obs}-r_{\rm obs}/c=220\sim300\,r_{\rm g}/c$, $\overline{L}_{\rm bol, IC}/\overline{L}_{\rm bol, SC}$, shown as a function of $\alpha_{\rm obs}$. The red dashed line indicates the ratio for isotropic (averaged over all $\alpha_{\rm obs}$ bins) luminosities.}
\label{paper_fig11}
\end{figure}
\subsubsection{Photon Spectra}\label{spectra}
Fig.~\ref{paper_appc} presents synthetic photon spectra temporally averaged over $t_{\rm obs}-r_{\rm obs}/c=220\sim300\,r_{\rm g}/c$ for two representative viewing angles, $\alpha_{\rm obs}=13.5^\circ$ and $85.5^\circ$.
Here, we evaluate both energy spectra using 100 bins ranging across $10^{-5}\leq\varepsilon_{\rm obs}<10^4$ for IC photons and $10^{-3}\leq\varepsilon_{\rm obs}/\varepsilon_{\rm c} < 10^7$ for SC photons, where $\varepsilon_c=3heB_0/4\pi m_{\rm e}^2c^3$ denotes the cyclotron characteristic photon energy.

The IC spectra (left) are broadly consistent between the nCl and msCl models, reflecting their similar low-energy particle populations, with pair-production cut-offs at $\varepsilon_{\rm obs}\sim\varepsilon_0^{-1}$.
In the msCl model, the high-latitude component ($\theta\leq\pi/4$) dominates for $\alpha_{\rm obs}=13.5^\circ$, whereas the low-latitude component ($\pi/4<\theta\leq\pi/2$) dominates for $\alpha_{\rm obs}=85.5^\circ$.

The SC spectra (right) show a stronger viewing-angle dependence.
A pronounced spectral break appears at $\varepsilon_{\rm obs}/\varepsilon_{\rm c}\sim10^3$ for $\alpha_{\rm obs}=13.5^\circ$, while the spectrum above the break remains harder for $\alpha_{\rm obs}=85.5^\circ$ because the low-latitude region contributes more.
This reflects the different acceleration mechanisms operating in the two emitting regions (Fig.~\ref{paper_fig8}).
The cooling break appears at about an order of magnitude lower energy than $\varepsilon/\varepsilon_{\rm c}\sim\Gamma_{\rm cool}^2$ for both spectra, likely because $B_{\perp, \rm eff}$ is lower than $B_0$.

\subsubsection{Light Curves}
We now discuss the temporal evolution of observed luminosities.
Given the broadband photon spectra, we define three energy integration ranges: \textit{low-energy} ($\varepsilon\leq10^{-2}$ for IC; $\varepsilon/\varepsilon_{\rm c}\leq10^{2}$ for SC), \textit{high-energy} ($\varepsilon\geq10^{-2}$ for IC; $\varepsilon/\varepsilon_{\rm c}\geq10^{2}$ for SC), and \textit{bolometric} (the combined energy range).
We choose the IC energy boundary near the spectral peak, whereas the SC boundary corresponds to an order of magnitude below the cooling break.
Luminosities are normalized by the analytic BZ power for a monopolar magnetic-field configuration,
\begin{equation}
L_{\rm BZ}=\frac{1}{12c}\omega_{\rm h}^2B_0^2r_{\rm g}^4.
\end{equation}

Fig.~\ref{paper_fig10} presents synthetic bolometric light curves for two viewing angles, $\alpha_{\rm obs}=13.5^\circ$ and $85.5^\circ$.
The luminosities remain relatively steady in both the nCl and msCl models, with $L_{\rm bol, IC}\sim10^{-5}$--$10^{-3}L_{\rm BZ}$ and $L_{\rm bol, SC}\sim10^{-4}$--$10^{-2}L_{\rm BZ}$, reflecting the overall stability of the system.
The values of $L_{\rm bol, IC}/L_{\rm BZ}$ in the nCl model and $L_{\rm bol, SC}/L_{\rm BZ}$ in the msCl model are consistent with those reported in previous 2D GRPIC studies including IC scattering \citep[][]{Crin20, Kisaka20, Crin21, Yuan25} and 1D studies using simplified analytic prescriptions for curvature losses \citep{Chen20, Kisaka20, Kisaka22, Kin24}.

The angle-averaged SC luminosity exceeds the IC luminosity, with $\overline{L}_{\rm bol, IC}/\overline{L}_{\rm bol, SC}\sim8.3\times10^{-4}$, consistent with the system's magnetically dominated nature.
However, the two components exhibit distinct angular distributions: IC emission is more prominent toward the rotational axis, whereas SC emission dominates for equatorial observers (Fig.~\ref{paper_fig11}). Consequently, $\overline{L}_{\rm bol, IC}/\overline{L}_{\rm bol, SC}$ decreases with increasing $\alpha_{\rm obs}$.

This anisotropy reflects the different acceleration mechanisms in the two emission regions.
As shown by the colored curves in Figs.~\ref{paper_appc} and \ref{paper_fig10}, the polar spark gap preferentially contributes to low-$\alpha_{\rm obs}$ emission, whereas the equatorial current sheet dominates at high $\alpha_{\rm obs}$.
Particles are accelerated mainly along magnetic field lines in the spark gap but across them in the reconnection layer, yielding $f_{\perp,\rm eff}=B_{\perp,\rm eff}/B\ll1$ at mid-to-high latitudes, but $f_{\perp,\rm eff}\sim1$ at low latitudes (Fig.~\ref{paper_fig8}~(a)), and thus less efficient SC emission.
IC emission is insensitive to field orientation, while its photon momentum distribution closely follows that of the scattering particles. Their momenta are beamed along poloidal field lines toward polar observers (Fig.~\ref{paper_fig8}~(b--d)), explaining the enhanced IC luminosity at low $\alpha_{\rm obs}$.

\begin{figure}
\hspace*{-0.5cm}
\includegraphics[keepaspectratio, scale=0.55]{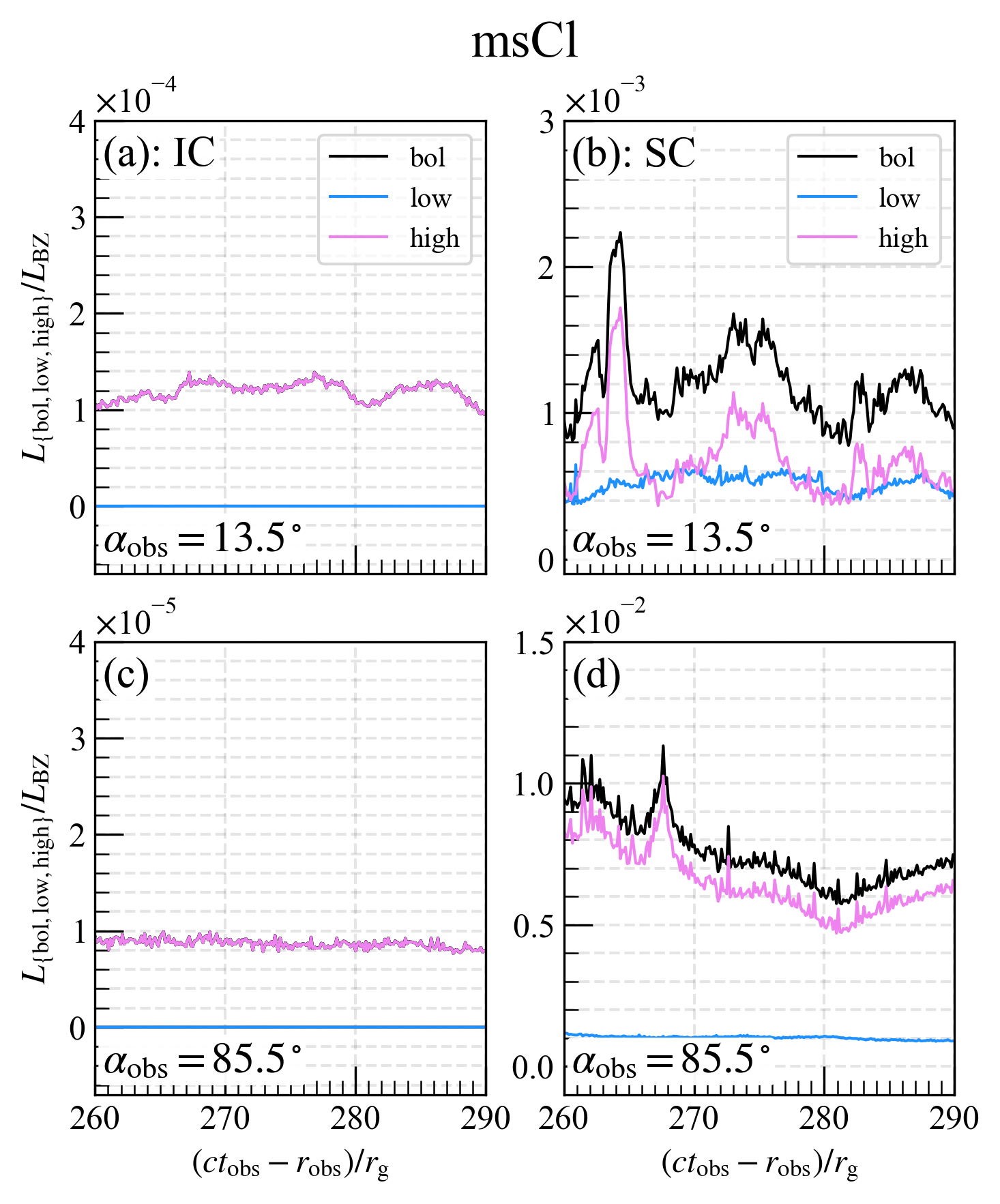}
\caption{The msCl light curves for $\alpha_{\rm obs}=13.5^\circ,\,85.5^\circ$ in $t_{\rm obs}-r_{\rm obs}/c=260\sim290\,r_{\rm g}/c$, evaluated for three different energy ranges. "bol" denotes the bolometric light curves, while "low" ("high") curves are the light curves for the \textit{low (high)-energy} ranges, respectively (see the text for the energy-range definition).}
\label{paper_lc_blh}
\end{figure}
Fig.~\ref{paper_lc_blh} shows the light curves for low-energy (sky-blue), high-energy (violet), and bolometric (black) emission, over $t_{\rm obs}-r_{\rm obs}/c=260.0\sim290.0\,r_{\rm g}/c$ on a linear scale, for the msCl model.
As shown, total luminosities are dominated by high-energy photons for IC emission and by SC emission at lower viewing angles, while in the SC $\alpha_{\rm obs}=13.5^\circ$ light curve both low- and high-energy ranges contribute almost equally.

For luminosity variability, the SC $\alpha_{\rm obs}=13.5^\circ$ curve in Fig.~\ref{paper_fig10} shows the strongest oscillations with a typical timescale of $\Delta t\sim10\,r_{\rm g}/c$.
This reflects the dynamical fluctuations of the spark gap: $\Gamma>\Gamma_{\rm cool}$ particle populations are intermittently produced during the peak of gap electric-field development, which predominantly contributes to SC emission.
Escaping IC emission, by contrast, is primarily produced by more persistent $\Gamma\lesssim\varepsilon_0^{-1}$ populations.
Moreover, IC emission can also arise from particles outside the primary acceleration regions, potentially smoothing temporal variations (see also \cite{Kin24}).
As a result, IC luminosity variations are generally modest and slower than those of SC emission.

\section{Discussions}\label{disc}
\subsection{Energetics}\label{applic}
The regional differences in $f_{\perp,\rm eff}$ presented in Fig.~\ref{paper_fig10} suggest that particles' maximum energies would differ between the polar and equatorial regions.
Equating the SC radiative cooling timescale $t_{\rm SC}=6\pi m_{\rm e}c/(\sigma_{\rm T}f_{\perp,\rm eff}^2B_0^2\Gamma)$ and the particle acceleration timescale via the electric field of the amplitude $D_{\rm acc}=f_{D}B_0$ is $t_{\rm acc}\approx\Gamma m_{\rm e}c/ef_{D}B_0$ yields the maximum Lorentz factor, 
\begin{equation}\label{gammamax}
\Gamma_{\rm max}=\sqrt{\frac{6\pi e f_D}{\sigma_{\rm T}f_{\perp,\rm eff}^2B_0}}=\sqrt{\frac{f_{D}}{f_{\perp,\rm eff}^2}} \Gamma_{\rm SC}.
\end{equation}
Estimating $B_0\sim10^2\rm G$, we obtain $\Gamma_{\rm max}\simeq3.7\times10^{6}f_{\perp,\rm eff,0}^{-1}\,f_{D,-1}^{1/2}B_{0,2}^{-1/2}$, where we have plugged in~$(f_{D},\,f_{\perp,\rm eff})=(10^{-1},\,1.0)$, representative of the equatorial current sheet of our simulations. In the polar region, though, where $(f_{D},\,f_{\perp,\rm eff})=(10^{-1},\,10^{-1})$ is more typical,~$\Gamma_{\rm max}$ increases by an order of magnitude.
As above, and as throughout this section, we adopt the convention that $Q_X$ denotes $Q=10^X$ in cgs units, except for BH masses, for which $M_X$ denotes $M=10^XM_\odot$, with $ M_\odot$ the Solar mass.
Following this argument, the corresponding characteristic photon energies for the polar spark gap and the current sheet are (cf. \cite{Cerutti25})
\begin{equation}
\begin{aligned}
E_{\gamma,\rm SC}&\approx\frac{3heB_{\perp,\rm eff}}{4\pi m_{\rm e}c}\Gamma_{\rm max}^2\\
&\simeq1.5\times\left\{\begin{aligned}
&10^1\,f_{\perp,\rm eff,0}^{-1}\,f_{D,-1}\,\mathrm{MeV}\\
&10^2\,f_{\perp,\rm eff,-1}^{-1}\,f_{D,-1}\,\mathrm{MeV}
\end{aligned}\right.,
\end{aligned}
\end{equation}
where $h$ denotes the Planck constant.
The IC scattering is in the Thomson regime for $\Gamma_{\rm max}\varepsilon_0\sim\{10^{-3}, 10^{-2}\}$, assuming the ambient photon energy to be in the meV range ($\varepsilon_{\rm 0}\sim10^{-9}$; e.g., \cite{Prieto16}), so that
\begin{equation}
\begin{aligned}
E_{\gamma,\rm IC}&\sim\Gamma_{\rm max}^2\varepsilon_0m_{\rm e}c^2\\
&\simeq7.0\times\left\{\begin{aligned}
&1\,\,f_{\perp,\rm eff,0}^{-2}f_{\rm D,-1}B_{0,2}^{-1}\varepsilon_{0,-9}\,\rm GeV\\
&10^2\,f_{\perp,\rm eff,-1}^{-2}f_{\rm D,-1}B_{0,2}^{-1}\varepsilon_{0,-9}\,\rm GeV
\end{aligned}\right.,
\end{aligned}
\end{equation}
for the polar gap and for the current sheet, respectively.
Overall, energy dissipation inside the magnetosphere could produce photons from tens of MeV to sub-TeV, since the polar region with small $f_{\perp,\rm eff}$ yields higher photon energies.
The angular dependence of the IC/SC luminosity ratio also favors a polar origin for the higher-energy gamma-ray emission at viewing angles of a few 10s of degrees \citep{Biretta99, Perlman11, Hada13}.

The rapid ($\Delta t\lesssim 10\,r_{\rm g}/c$) luminosity variations could arise from dynamical fluctuations in the spark gap.
However, the luminosity-variation amplitudes of the synthetic light curves are smaller than those observed in some of the most prominent VHE flares, which show order-of-magnitude enhancements.
Reproducing such large-amplitude flares likely requires stronger external perturbations, such as variations in the accretion rate that modify the ambient photon field \citep[e.g.][]{Jia23, Zhou25}.

Our estimate of the energetics is broadly consistent with several previous studies of spark-gap radiation efficiency \citep[e.g.][]{Levinson11, Hirotani16, Katsoulakos18, Yuan25}, while equatorial magnetic reconnection has recently been proposed as a possible origin of the VHE emission from the M87 core region \citep{Hakobyan23, Chernoglazov23, Solanki25}.
These studies rely on local PIC simulations of relativistic, high-$\sigma$ radiative magnetic reconnection and suggest that a fraction of electrons and positrons can be accelerated to $\Gamma>\sigma_\pm$, exceeding the SC cooling limit ($\Gamma_{\rm max}$) used in their analyses.
However, several dynamical feedbacks expected in our situation of interest, including plasma shear flow in both in-plane and out-of-plane directions (see Fig.~\ref{paper_fig5}), may disturb the sheet and suppress reconnection (e.g., \cite{French23, Figueiredo24, Peery24}).
Furthermore, recent studies have shown that pair injection into the reconnection region can substantially reduce the upstream magnetization \citep{Chen23, Kuze25, Oikawa26}, potentially yielding a saturated value well below $\sigma_\pm\sim10^7$ adopted in \citep{Hakobyan23, Chernoglazov23, Solanki25}.
The attainable particle energies may therefore be comparable to or lower than those assumed here, even if selective acceleration up to $\Gamma\sim\sigma_\pm$ were to operate.
Indeed, the low-latitude particle spectrum in the msCl model (Fig.~\ref{paper_fig7}~(d)) exhibits a steep cutoff near the SC burn-off limit, $\Gamma_{\rm max}(B_0)$, suggesting that highly efficient reconnection-driven acceleration beyond~$\Gamma_{\rm max}$ does not operate in our simulations.
Our axisymmetric 2D setup would, however, underestimate acceleration efficiency, since 3D meandering motions around the reconnection layer can enhance particle acceleration \citep{Guo21, Chernoglazov23}, which are absent in axisymmetry.

\subsection{Comparison of Equatorial Dynamics to Previous Studies}\label{rec_comp}
We observe several qualitative differences in the current-sheet dynamics compared to previous GRPIC studies with similar initial magnetic-field configurations \citep{Parfrey19, Hirotani23, Vos25, Mehlhaff26}: the intervals between successive flux-infall events in our simulations are typically shorter ($\lesssim10\,r_{\rm g}/c$), while the radial extent of the current sheet is much smaller, remaining within the ergosphere.
Furthermore, the contribution of inertial drag, often attributed to triggering flux infall, is minor, whereas acceleration by the time-dependent azimuthal electric field appears more effective.
These differences would stem from a limited plasma supply in our setup: the low plasma supply leaves the charge density in the return-current layer prone to deviations from the GJ value, inducing substantial electric fields.
Conversely, a large plasma reservoir there would reduce the magnetization to order unity, much lower than in our studies, which should enhance the inertial effect.

The fixed outer boundary condition for the magnetic field may also help maintain its nearly vertical configuration within the ergosphere even after plasma is replenished, potentially determining the location of the outer boundary of the return current layer where the largest current amplitude is required.
\citet{Hirotani23} performed a series of 2D GRPIC simulations using the Wald solution as the initial electromagnetic-field configuration, similarly to our study, and found that the nearly vertical shape of the poloidal magnetic field is maintained and that recurring magnetic reconnection occurs only within the ergosphere.
They also reported that local charge excesses along the poloidal magnetic field lines just above and below the reconnection point lead to the development of $\bm{D}\cdot\bm{B}$ during the reconnection event (see Figs.~2 and 3 of their paper), which appears to have certain similarity to the dynamical behavior observed in our simulations.

Given these possible effects of our numerical setup on the results of this study, the external accretion flow would significantly impact the magnetospheric dynamics.
For instance, if the accretion flow is steadily sustained, its abundant plasma supply may efficiently screen the electric fields in the return-current layer.
The circulating electric current in the accretion flow could also reorganize the magnetic-field configuration, so the vertical configuration of the magnetic field lines within the ergosphere, as in the present study, would not be maintained.
If the separatrix point is pushed back to the outer magnetosphere, $\Omega_{\rm F}$ could be faster than $\omega$, and then the $\bm{D}\times\bm{B}$ drift motion could then go outward (see Eq.~(\ref{dbdrift})) to sustain the return current even without developing displacement current.
A charge-starved state may still be achieved if the accretion flow is time-variable: when accretion is temporarily halted, the main plasma injection channel would be two-photon pair production, whose injection rate could be as low as $n_\pm\lesssim n_{\rm GJ}$ for several low-luminosity AGNs with highly sub-Eddington mass accretion rates (\cite{Levinson11, Hirotani16}, but see also \cite{Wong21}).
Further studies systematically varying pair multiplicity and/or the supply rate at the simulation boundary are needed to clarify their effects on the dynamics of the return-current layer.

\subsection{Caveats and Limitations}\label{fwhvsmnycvts}
In this part, we mention caveats and limitations of our study.
\subsubsection{Energy Rescaling}
The adopted rescaling of the magnetic-field strength may affect the acceleration process by artificially enlarging the Larmor radius and shortening the particle escape timescale.
In particular, longer-duration entrainment of particles in the acceleration region may further emphasize contrasts in $f_{\perp,\rm eff}$ between the gap and the current sheet.
If present, the hierarchy of observed IC and SC photon energies from the two respective regions could also be enhanced.
Future diagnostics of magnetic reconnection dynamics in BH magnetospheres using full 3D GRPIC simulations, together with more extensive parameter surveys, will therefore be needed to determine the maximum attainable particle energies more precisely.

\subsubsection{Radiation Microphysics Parameters}
One major caveat in the direct comparison discussed in Sec.~\ref{applic} is a possible mismatch between our adopted optical depth, $\tau_0=10$, and the value expected for low-luminosity AGNs.
For instance, the typical energy flux density at 1.3 mm in the M87 core is $F_{\nu,\rm obs}\simeq0.64\,\rm Jy$ \cite{EHT19}, and thus we can estimate $\tau_0$ (see also \cite{Kin24, Kin25, Yuan25}) as
$\tau_0\sim[4F_{\nu,\rm obs}d_{\rm L}^2/(1+z)R^2]\sigma_{\rm T}r_{\rm g}\simeq2.4\times10^3$, where we use $1+z\simeq1.00428$ and $d_{\rm L}\simeq16.8\,\rm Mpc$ for the redshift and the luminosity distance of M87, and $R=10\,r_{\rm g}$ is the size of the photon-emitting region (i.e., the accretion flow).

Although a larger $\tau_0$ (i.e., higher multiplicity) would favor shorter gap variability timescales comparable to those of VHE flares, it would also reduce the electric field strength and suppress the overall gamma-ray luminosity as well as the observed photon energies (\cite{Crin20, Kisaka20, Kisaka22, Kin24}, but see also \cite{Yuan25}).
The highest-energy IC emission may be less sensitive to $\tau_0$, since steady magnetospheric operation requires a persistent plasma density of order $n_{\rm GJ}$, whose contribution would sustain the IC luminosity (see also \cite{Kin24}).
Conversely, temporal reductions in the ambient photon field due to accretion-flow variability \citep{Jia23, Zhou25} could lower $\tau_0$ and trigger intermittent gap formation, potentially leading to large-amplitude flares \citep{Kisaka22}. 

Another caveat is the discrepancy between the characteristic energy of the SC process and the pair-production energy threshold, $\Gamma\sim\varepsilon_0^{-1}$.
We choose the SC rescaling energy parameter for the msCl model as $\Gamma_{\rm SC}=800$, so that $\varepsilon_0^{-1}<\Gamma_{\rm cool}<\Gamma_{\rm max}(B_0)<\Gamma_{\rm SC}$ throughout the course of our runs, but much stronger SC backreaction may persist in AGN magnetospheres: in the case of M87, we obtain $\Gamma_{\rm cool}\sim2.3\, f_{\perp,\rm eff,0}^{-2}B_{0,2}^{-2}M_{10}^{-1}$, and thus $\varepsilon_0^{-1}<\Gamma_{\rm cool}<\Gamma_{\rm max}$ can only be achieved for the extremely beamed particle populations with $f_{\perp,\rm eff}\gtrsim5\times10^4$.
The reversed energy hierarchy would affect the dynamics, namely the spark gap oscillations observed in the present study, as it is closely related to the SC cooling efficiency.
It is not entirely clear whether the rapid oscillations would be a persistent characteristic of the spark gap regardless of SC radiation efficiency, although there are several previous examples that apply only to IC emission \cite{Chen20, Kisaka20, Kisaka22, Kin24, Yuan25}.
The current sheet dynamics may also be affected.
Although the impact on the overall morphology and variability around the equator is minor in the present comparison between the nCl and msCl models, further suppression of pair production is expected for the $\Gamma_{\rm cool}\ll\epsilon_0^{-1}$ case, which may result in a larger interval between each flux-infall event, or their complete termination (see also \cite{Mehlhaff26}).
A future broader parameter survey, particularly of $\tau_0$, the SC cooling efficiency $\Gamma_{\rm SC}$, and the combination $\varepsilon_0 B_0$, will be necessary for more precise constraints.

\section{Conclusions}
In this work, we performed 2D GRPIC simulations of the BH magnetosphere incorporating fundamental interactions associated with the production of high-energy photons and plasma particles, initialized with Wald’s vacuum solution.

Main results are summarized as:
\begin{itemize}
\item
Once plasma is supplied, the electromagnetic field relaxes into a state qualitatively consistent with the expectations from a force-free magnetosphere model and previous GRMHD/GRPIC studies, launching a steady Poynting-dominated polar outflow via the BZ mechanism.

\item
The polar spark gap and a time-dependent equatorial current sheet develop simultaneously, with particle acceleration and pair production sustaining the global current structure, as is consistent with previous studies of similar numerical setups \cite{Crin21, Vos25}.

\item
With our low-pair-multiplicity parameter choice, SC cooling leaves the overall structure largely unchanged but induces significant spark-gap electric-field oscillations through variations in pair-production efficiency.
\item
Both regions exhibit rapid ($\lesssim10\,r_{\rm g}/c$) variability for both models.
Fourier PSDs show a distinct gap-oscillation peak in the polar region but no clear peak in the current sheet, reflecting its stochastic variability.

\item
The two regions show no significant cross-correlation, indicating that the variabilities at the innermost magnetosphere in this setup are mutually independent.

\item
Post-processed ray tracing yields bolometric luminosities of $\sim10^{-4}$ and $\sim10^{-2}$ of the BZ power for the IC and SC components, respectively.
Because particle pitch-angle distributions vary regionally, the IC-to-SC luminosity ratio is higher at lower viewing angles, and vice versa.

\item
Particles' momentum distributions are highly beamed along the local magnetic field lines in the polar region because of linear acceleration in the gap, while those around the equatorial plane are more isotropized.
We discuss the magnetosphere's radiation efficiency and find that the polar spark gap can produce gamma rays up to the sub-TeV regime, providing a viable VHE emission scenario.
\end{itemize}

Improved sensitivity and time resolution in the TeV band from the Cherenkov Telescope Array Observatory's large-sized telescopes will provide key diagnostics for testing the magnetospheric VHE flare scenario.
Efficient gap acceleration ($\Gamma dN/d\Gamma\propto\Gamma^0$ in the slow-cooling regime and $\propto\Gamma^{-1}$ in the fast-cooling regime; see Fig.~\ref{paper_fig6}) predicts a harder TeV spectrum than alternative scenarios assuming the conventional Fermi-type acceleration \citep{Spada01, Baring17, Bottcher19}.
In addition, reduced IC efficiency in the Klein-Nishina regime implies a sharp cutoff at the highest energies.
Careful diagnostics of the photon spectral index in future VHE observations of AGNs will therefore be crucial.
A larger VHE flare sample associated with nearby radio galaxies may enable a population study of the correlation between jet inclination and TeV luminosity, further testing the viewing-angle relation of $L_{\rm IC}$ obtained from our simulations.
Joint multiwavelength analyses can provide additional insights.
The relative contributions across the GeV--TeV bands constrain magnetospheric properties; the synchrotron-to-IC luminosity ratio probes the ratio of magnetic to photon energy densities, while the synchrotron-to-curvature ratio informs the magnetic field geometry, with curvature-dominated emission indicating a large-scale ordered field.

The possible influence of plasma injection channels not addressed in our simulation is worth consideration.
One possibility is two-photon pair production from GeV--X-ray photon interactions: proton acceleration in the current sheet may produce GeV synchrotron photons \citep{Kimura20, Kuze24, Hakobyan25}, though it remains unclear whether sufficient protons are present in the accelerating region \cite{Chow26}.
Alternatively, enhanced MeV photon emission from nonthermal electrons during magnetic reconnection \citep{Kimura22, Kuze24} could increase the pair multiplicity by several orders of magnitude, even near the pole \citep{Oikawa26}.
Spark gaps may still re-form after plasma evacuation, given the long ($\Delta t_{\rm dur}\sim10^2$--$10^3 r_{\rm g}/c$) intervals between large-scale reconnection events.
In such cases, a time lag from the plasma evacuation timescale may persist between the VHE signals and radio-X counterparts. 
Nevertheless, the magnetosphere's dynamical response to episodic plasma injection remains an important open problem: rapid magnetic flux decay and transient starvation may induce large-amplitude variability (see also \cite{Nalewajko26}), potentially affecting the polar electromagnetic structure.

\begin{acknowledgements}
K. K. thanks Enzo Figueiredo for support in numerical computations for this study.
K. K. also thanks Morifumi Mizuno and Rin Oikawa for valuable comments on interpreting the dynamics in the simulation results.
This work was in part supported by the Graduate Program
on Physics for the Universe (GP-PU) (K. K.), JSPS KAKENHI Grant Number JP26K00696 (K. T., S. K.), and JP26K07066 (S. K.).
This project has received funding from the European Research Council (ERC) under the European Union’s Horizon 2020 research and innovation program (Grant Agreement No. 863412).
\end{acknowledgements}
\appendix
\section{Kerr Spacetime}\label{app_metric}
The Kerr metric describes a time-independent, axisymmetric gravitational field outside the BH with finite angular momentum.
In the 3+1 formalism \cite{Landau75, Thorne82, Komissarov04} in BL coordinates $(t,r,\theta,\varphi)$, it is written as
\begin{equation}\label{3p1ds}
ds^2=\alpha^2dt^2+\gamma_{ij}(\beta^idt+dx^i)(\beta^jdt+dx^j),
\end{equation}
where $\gamma_{ij}\,(i,j=r,\theta,\varphi)$ is the spatial metric, $\alpha$ and $\beta^i$ are the lapse function and the shift vector, and
\begin{equation}
\begin{aligned}
&\rho^2=r^2+a^2\cos^2\theta,\;\Delta=r^2+a^2-2r,\\
&\hspace{20pt}\Sigma=(r^2+a^2)^2-a^2\Delta\sin^2\theta.
\end{aligned}
\end{equation}
The determinant of the spatial metric is $\sqrt{\gamma}=\rho^2\sin\theta/\alpha$.
We may also define the frame-dragging angular velocity as $\omega=2ar/\Sigma$.

We may define the local observer normal to the constant-time hypersurface for each coordinate, often denoted as the "fiducial observer" (FIDO; \citep{bardeen72, Thorne86}).
Its four-vector, $n^\mu$, is defined as 
\begin{equation}
n^\mu=\left(\frac{1}{\alpha},-\frac{\beta^i}{\alpha} \right),\,n_\mu=(-\alpha,0,0,0).
\end{equation}
In the BL coordinates, FIDO's local orthonormal basis reads
\begin{equation}
\begin{aligned}
&\omega^{(t)}=\alpha dt,
&\omega^{(r)}&=\sqrt{\gamma_{rr}}dr,\\
&\omega^{(\theta)}=\sqrt{\gamma_{\theta\theta}}d\theta,
&\omega^{(\varphi)}&=\sqrt{\gamma_{\varphi\varphi}}(\beta^\varphi dt+d\varphi),
\end{aligned}
\end{equation}
and the corresponding basis vectors are
\begin{equation}
\begin{aligned}
&e_{(t)}=\dfrac{1}{\alpha}(\partial_t-\beta^\varphi\partial_\varphi),
&e_{(r)}=\dfrac{1}{\sqrt{\gamma_{rr}}}\partial_r,\\
&e_{(\theta)}=\dfrac{1}{\sqrt{\gamma_{\theta\theta}}}\partial_\theta,
&e_{(\varphi)}=\dfrac{1}{\sqrt{\gamma_{\varphi\varphi}}}\partial_\varphi.
\end{aligned}
\end{equation}
One could then evaluate the vector with respect to the FIDO orthonormal basis, $C^{\hat{\mu}}=g_{lm}C^le^m_{(\mu)}$, 
as the "FIDO-measured" vector.
Note that the FIDO in Kerr spacetime is often called the zero angular momentum observer (ZAMO; \citep[][]{Thorne86}), as its angular momentum is $n_\varphi=0$.

\begin{widetext}
\section{Conversion between Coordinates}\label{app_conversion}
Time/spatial displacements in KS ($\tilde{t},\tilde{r},\tilde{\theta}, \tilde{\varphi}$) and BL ($t,r,\theta,\varphi$) coordinates are related with each other as
\begin{equation}\label{kbint}
\left\{\begin{aligned}
&d\tilde{t}=dt+\dfrac{2r}{\Delta}dr;\\
&d\tilde{r}=dr;\\
&d\tilde{\theta}=d\theta;\\
&d\tilde{\varphi}=d\varphi+\dfrac{a}{\Delta}dr,
\end{aligned}\right.
\end{equation}
and thus, by integrating this, we obtain the coordinate conversion equations as (cf., e.g., \cite{Garofalo10, Chen25})
\begin{equation}\label{ktob_eqs}
\left\{\begin{aligned}
&\tilde{t} = t
+ \frac{1}{\sqrt{1-a^{2}}}
    \left[
        (1+\sqrt{1-a^{2}})\,\ln\!\left|\frac{r}{1+\sqrt{1-a^{2}}}-1\right|\right.
        \left.- (1-\sqrt{1-a^{2}})\,\ln\!\left|\frac{r}{1-\sqrt{1-a^{2}}}-1\right|
    \right]; \\
&\tilde{r} = r; \\
&\tilde{\theta} = \theta; \\
&\tilde{\varphi}= \varphi
+ \frac{a}{2\sqrt{1-a^{2}}}
    \ln\!\left(\frac{r-(1+\sqrt{1-a^{2}})}{r-(1-\sqrt{1-a^{2}})}\right).
\end{aligned}\right.
\end{equation}
The Jacobian matrix $J^{\tilde{\mu}}_{\;\nu}=(\partial \tilde{x}^{\tilde{\mu}}/\partial x^\nu)$ and its inverse are
\begin{equation}\label{convdet}
J^{\tilde{\mu}}_{\;\nu}=\left(\begin{array}{lccr}
1 \ -2r/\Delta \ \:0 \ \;0 \\
0 \ \;\;\;\;\;\;\,\,1 \ \;\;\;\:\,0 \ \;0 \\
0 \ \;\;\;\;\;\;\,\,0 \ \;\;\;\:\,1 \ \;0 \\
0 \ \;-a/\Delta \ \:\,0 \ \;1
\end{array}\right);\;
(J^{-1})^{\mu}_{\;\tilde{\nu}}=\left(\begin{array}{lccr}
1 \ \;2r/\Delta \ \;0 \ \;0 \\
0 \ \;\;\;\;\,1 \ \;\;\;\:0 \ \;0 \\
0 \ \;\;\;\;\,0 \ \;\;\;\:1 \ \;0 \\
0 \ \;\;a/\Delta \ \:\,0 \ \;1
\end{array}\right).
\end{equation}
Any 4-vector can be translated as $C^{\tilde{\mu}}=J^{\tilde{\mu}}_{\;\nu}C^\nu$, or $C_{\tilde{\mu}}=(J^{-1})^{\mu}_{\;\tilde{\nu}}C_\nu$.

All time integrations in our simulations are performed in KS coordinates, so the datasets represent physical quantities defined on KS constant-time hypersurfaces.
To construct snapshots on constant-time hypersurfaces in BL coordinates, as shown in the main text, we combine and rearrange time-sequenced datasets using the coordinate transformation in Eq.~(\ref{ktob_eqs}), then transform four-vectors using the Jacobian in Eq.~(\ref{convdet}).
In the main text, figures labeled “BL” show quantities evaluated on BL constant-time slices.
\end{widetext}

\begin{figure}
\includegraphics[keepaspectratio, scale=0.7]{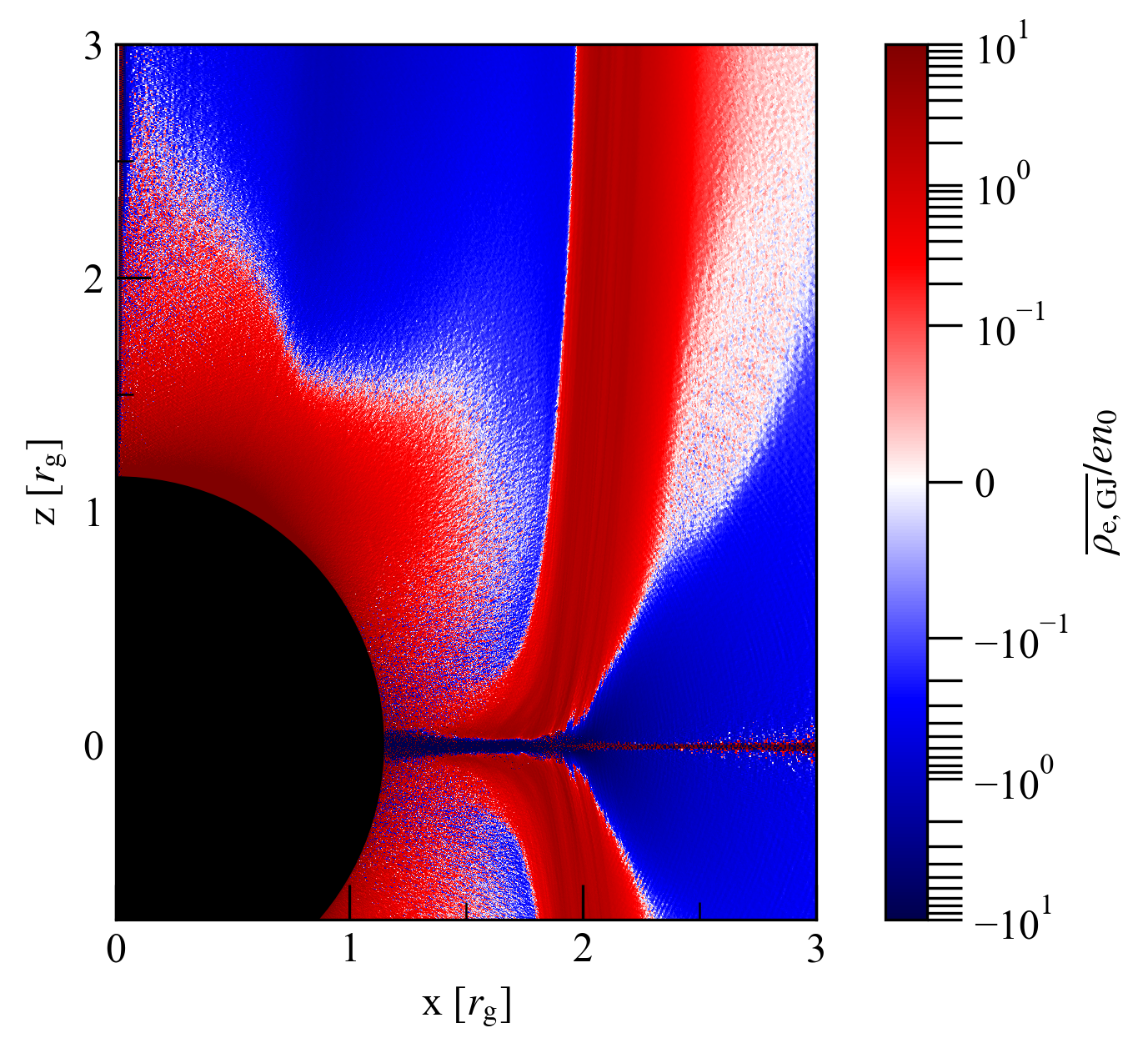}
\caption{The time-averaged GJ density, $\overline{\rho_{\rm GJ}}$, calculated for the nCl model.}
\label{paper_appd}
\end{figure}
\section{A Brief Review of the Axisymmetric Force-Free Kerr BH Magnetosphere}\label{ff}
We hereby briefly review the theoretical model of the axisymmetric force-free Kerr BH magnetosphere in the 3+1 formulation of electrodynamics to derive several equations used in the main text (for more details, see, e.g., \cite{Komissarov04}).
Note that we use the expressions in BL coordinates in this section.

The magnetic field can be written in terms of the vector potential $\bm{A}$ as $\bm{B}=\nabla\times\bm{A}$, so that
\begin{equation}\label{b_as_a}
B^r=\dfrac{1}{\sqrt{\gamma}}\partial_\theta A_\varphi,\quad B^\theta=-\dfrac{1}{\sqrt{\gamma}}\partial_r A_\varphi.
\end{equation}
Thus, $\bm{B}\cdot\nabla A_\varphi=0$, so $A_\varphi$ is constant along magnetic field lines and defines surfaces of constant magnetic flux.

From the force-free condition, $\bm{D}\cdot\bm{B}=\bm{E}\cdot\bm{B}=0$, combined with the axisymmetry, we can write
\begin{equation}\label{omegaf}
\bm{E}=-\Omega_{\rm F}\bm{m}\times\bm{B},
\end{equation}
and
\begin{equation}\label{dff}
\bm{D}=-\frac{1}{\alpha}(\Omega_{\rm F}-\omega)\bm{m}\times\bm{B},
\end{equation}
where $\bm{m}=\partial_\varphi$.
One can see from Eq.~(\ref{faraday}) and $\nabla\cdot \bm{B}=0$ with $\partial_t=0$ that $\Omega_{\rm F}$ satisfies the condition $\bm{B}\cdot\nabla\Omega_{\rm F}=0$,
i.e., $\Omega_{\rm F}$ is constant along the field lines, often referred to as the "angular velocity of the magnetic field line".

The charge density that satisfies the force-free condition is obtained by substituting Eq.~(\ref{dff}) into Gauss's equation, $\nabla\cdot\bm{D}=4\pi\rho$, which is often referred to as the Goldreich-Julian charge density, $\rho_{\rm GJ}$, as
\begin{equation}\label{gjff}
\rho_{\rm GJ}=\frac{1}{4\pi}\nabla\cdot\left[-\frac{1}{\alpha}(\Omega_{\rm F}-\omega)\bm{m}\times\bm{B}\right].
\end{equation}

The plasma motion in the force-free state is dominated by the $\bm{D}\times \bm{B}$ drift, whose 3-velocities, $\bm{v_{\rm d}} = \alpha(\bm{D} \times \bm{B})/B^2-\bm{\beta}$, are given by
\begin{equation}
\begin{aligned}
\label{dbdrift}
v_{\rm d}^{r} &= \dfrac{D_{\theta} B_{\varphi}}{\sqrt{\gamma}B^2}
= -(\Omega_{\rm F} - \omega)\dfrac{B^r B_\varphi}{B^2}, \\
v_{\rm d}^{\theta} &= -\dfrac{D_{r} B_{\varphi}}{\sqrt{\gamma}B^2}
= -(\Omega_{\rm F} - \omega)\dfrac{B^\theta B_\varphi}{B^2}, \\
v_{\rm d}^{\varphi} &= \dfrac{D_{r} B_{\theta} - D_{\theta} B_{r}}{\sqrt{\gamma}B^2}-\beta^\varphi
= (\Omega_{\rm F} - \omega)\dfrac{B_p^2}{B^2}-\beta^\varphi,
\end{aligned}
\end{equation}
where $B_p^2=B^rB_r+B^\theta B_\theta$.
One can see that the direction of the poloidal drift reverses at $\Omega_{\rm F}=\omega$: in the northern hemisphere (where $B^\varphi<0$), $v_{\rm d}^{r} \lessgtr 0$, $v_{\rm d}^{\theta} \gtrless 0$ for $\Omega_{\rm F} \lessgtr \omega $, respectively.

\begin{figure} 
\includegraphics[keepaspectratio, scale=0.6]{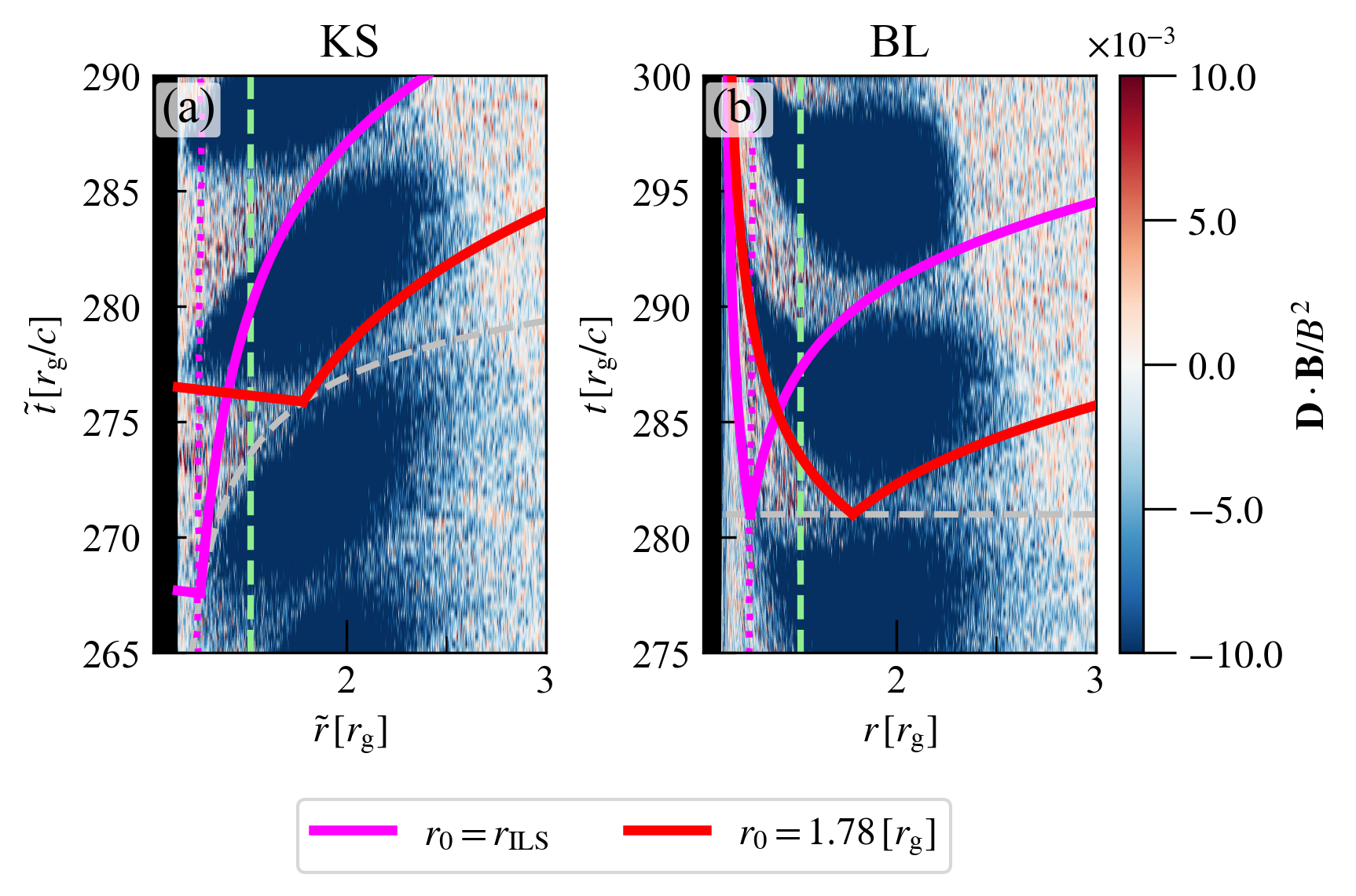} 
\caption{Comparisons of saturated $\bm{D}\cdot\bm{B}/B^2$ diagrams at $\theta=\pi/6$ between (a) KS and (b) BL coordinates. Magenta and lime green vertical lines mark the ILS and the static limit, respectively. The magenta and red V-shaped curves show the null light cones for radial propagation originating from $t_0=281.50\,r_{\rm g}/c$ and $r_0=\{r_{\rm ILS},1.78r_{\rm g}\}$, respectively.}
\label{paper_fig12}
\end{figure}
\section{Light Surfaces}\label{LSs}
Inner/outer light surfaces are defined as the surface for the observer rotating with constant angular velocity $\Omega$ at constant $(r,\theta)$ to be null, $g_{tt}dt^2+2g_{t\varphi}dtd\varphi+g_{\varphi\varphi}d\varphi^2\leq0$, which can be rewritten as 
\begin{equation}\label{lsfunc}
\begin{aligned}
f(\Omega,r,\theta,a)&=g_{tt}+2\Omega g_{t\varphi}+\Omega^2g_{\varphi\varphi}\\
&=-\alpha^2+\gamma_{\varphi\varphi}(\Omega-\omega)^2\leq0,
\end{aligned}
\end{equation}
where we define $d\varphi/dt=\Omega$.
In the main text, we substitute $\Omega_{\rm F}$ defined by Eq.~(\ref{omegaf}) into the equation $f(\Omega,r,\theta, a)=0$ to find the inner light surface.

\section{The Approximate GJ Density Profile in Simulations}\label{gj}
The overall force-free nature of our simulation results motivates us to evaluate $\rho_{\rm GJ}$ from the obtained electromagnetic fields using Eq.~(\ref{gjff}).
Fig.~\ref{paper_appd} presents the 2D map of $\overline{\rho_{\rm GJ}}$ for the nCl model (the msCl model is overall similar).
At high latitudes, $\rho_{\rm GJ}$ resembles that of monopole/parabolic magnetic-field configurations \citep{Hirotani16, Chen20}, decreasing from positive values near the horizon to negative values beyond $r\sim2\,r_{\rm g}$.
At lower latitudes, the negative-charge region pushes inward.
In the return-current layer, however, positive charge dominates again.

The narrow corridor of negative charge along the equator results from the sign reversal of $D^\theta$ across the equator, which is caused by the antiparallel configuration of the poloidal magnetic field lines.
This connection may indicate that forming an antiparallel field configuration across the equator is an inevitable consequence of return-current maintenance, rather than an inertial effect, as has often been discussed (cf., e.g., \cite{Komissarov05}).
Testing the robustness of this interpretation with GRPIC simulations using different setups remains for future work.

\section{The comparion of the $\bm{D}\cdot\bm{B}/B^2$ diagram morphologies in KS and BL coordinates}
In Fig.~\ref{paper_fig12}, we compare $\bm{D}\cdot\bm{B}/B^2$ along $\theta=\pi/6$ in KS and BL coordinates.
Although the apparent morphology differs, with outward propagation in KS and both inward and outward propagation in BL coordinates, the radial light cones (V-shaped curves) indicate that both edges are more closely associated with inward and outward propagation from $r_0\sim1.78\,r_{\rm g}$ than with the ILS (magenta), regardless of coordinates (see also \cite{Chen25, Yuan25}).

Gap expansion from $r\sim1.7\,r_{\rm g}$ is commonly observed in 1D GRPIC simulations including curvature-radiation losses \citep{Kisaka20, Kisaka22, Kin24}, where plasma evacuation near the null-charge surface ($\rho_{\rm GJ}=0$) is considered to initiate the expansion, resembling outer-gap models of pulsar emission \citep{Cheng86, Romani96}.
In contrast, previous 2D studies report apparent outward propagation from around the ILS \citep{Parfrey19, Crin20}. Our msCl results more closely resemble the 1D models, with $r_0$ near $\rho_{\rm GJ}=0$ (Fig.~\ref{paper_appd}).

This discrepancy may reflect differences in numerical setups.
\citet{Parfrey19} injected plasma immediately when $|\bm{D}\cdot\bm{B}|/B^2$ exceeded a threshold ($\lesssim10^{-2}$), neglecting photon mean free paths, which would have suppressed gap expansion beyond the grid scale.
\citet{Crin20} showed results only for the high-multiplicity case ($\tau_0=30$), where the gap dynamics may differ from the lower-multiplicity regime explored here: higher multiplicity would generally bring the system closer to complete screening of $\bm{D}\cdot\bm{B}$, and local deficiency may occur only at the nearest horizon, where inertial drag could force flow separation.
Curvature in the polar magnetic field lines may still create a mismatch between the required poloidal current amplitude and the local plasma supply for the outer region, leading to outer-gap opening, as is observed in the equatorial return-current layer of the present study (see Sec.~\ref{eq_sep}).
Verifying such speculations about gap dynamics in multi-D models would be important in the future.

\input{apssamp.bbl}

\end{document}

%% file: apssamp.bbl
%

%% file: apssamp.bbl
\begin{thebibliography}{96}%
\makeatletter
\providecommand \@ifxundefined [1]{%
 \@ifx{#1\undefined}
}%
\providecommand \@ifnum [1]{%
 \ifnum #1\expandafter \@firstoftwo
 \else \expandafter \@secondoftwo
 \fi
}%
\providecommand \@ifx [1]{%
 \ifx #1\expandafter \@firstoftwo
 \else \expandafter \@secondoftwo
 \fi
}%
\providecommand \natexlab [1]{#1}%
\providecommand \enquote  [1]{``#1''}%
\providecommand \bibnamefont  [1]{#1}%
\providecommand \bibfnamefont [1]{#1}%
\providecommand \citenamefont [1]{#1}%
\providecommand \href@noop [0]{\@secondoftwo}%
\providecommand \href [0]{\begingroup \@sanitize@url \@href}%
\providecommand \@href[1]{\@@startlink{#1}\@@href}%
\providecommand \@@href[1]{\endgroup#1\@@endlink}%
\providecommand \@sanitize@url [0]{\catcode `\\12\catcode `\$12\catcode `\&12\catcode `\#12\catcode `\^12\catcode `\_12\catcode `\%12\relax}%
\providecommand \@@startlink[1]{}%
\providecommand \@@endlink[0]{}%
\providecommand \url  [0]{\begingroup\@sanitize@url \@url }%
\providecommand \@url [1]{\endgroup\@href {#1}{\urlprefix }}%
\providecommand \urlprefix  [0]{URL }%
\providecommand \Eprint [0]{\href }%
\providecommand \doibase [0]{https://doi.org/}%
\providecommand \selectlanguage [0]{\@gobble}%
\providecommand \bibinfo  [0]{\@secondoftwo}%
\providecommand \bibfield  [0]{\@secondoftwo}%
\providecommand \translation [1]{[#1]}%
\providecommand \BibitemOpen [0]{}%
\providecommand \bibitemStop [0]{}%
\providecommand \bibitemNoStop [0]{.\EOS\space}%
\providecommand \EOS [0]{\spacefactor3000\relax}%
\providecommand \BibitemShut  [1]{\csname bibitem#1\endcsname}%
\let\auto@bib@innerbib\@empty
\bibitem [{\citenamefont {{Blandford}}\ and\ \citenamefont {{Znajek}}(1977)}]{BZ77}%
  \BibitemOpen
  \bibfield  {author} {\bibinfo {author} {\bibfnamefont {R.~D.}\ \bibnamefont {{Blandford}}}\ and\ \bibinfo {author} {\bibfnamefont {R.~L.}\ \bibnamefont {{Znajek}}},\ }\href {https://doi.org/10.1093/mnras/179.3.433} {\bibfield  {journal} {\bibinfo  {journal} {MNRAS}\ }\textbf {\bibinfo {volume} {179}},\ \bibinfo {pages} {433} (\bibinfo {year} {1977})}\BibitemShut {NoStop}%
\bibitem [{\citenamefont {{McKinney}}\ and\ \citenamefont {{Gammie}}(2004)}]{McKinney&Gammie04}%
  \BibitemOpen
  \bibfield  {author} {\bibinfo {author} {\bibfnamefont {J.~C.}\ \bibnamefont {{McKinney}}}\ and\ \bibinfo {author} {\bibfnamefont {C.~F.}\ \bibnamefont {{Gammie}}},\ }\href {https://doi.org/10.1086/422244} {\bibfield  {journal} {\bibinfo  {journal} {ApJ}\ }\textbf {\bibinfo {volume} {611}},\ \bibinfo {pages} {977} (\bibinfo {year} {2004})},\ \Eprint {https://arxiv.org/abs/astro-ph/0404512} {arXiv:astro-ph/0404512 [astro-ph]} \BibitemShut {NoStop}%
\bibitem [{\citenamefont {{McKinney}}(2006)}]{McKinney06}%
  \BibitemOpen
  \bibfield  {author} {\bibinfo {author} {\bibfnamefont {J.~C.}\ \bibnamefont {{McKinney}}},\ }\href {https://doi.org/10.1111/j.1365-2966.2006.10256.x} {\bibfield  {journal} {\bibinfo  {journal} {MNRAS}\ }\textbf {\bibinfo {volume} {368}},\ \bibinfo {pages} {1561} (\bibinfo {year} {2006})},\ \Eprint {https://arxiv.org/abs/astro-ph/0603045} {arXiv:astro-ph/0603045 [astro-ph]} \BibitemShut {NoStop}%
\bibitem [{\citenamefont {{Tchekhovskoy}}\ \emph {et~al.}(2011)\citenamefont {{Tchekhovskoy}}, \citenamefont {{Narayan}},\ and\ \citenamefont {{McKinney}}}]{Tchekho11}%
  \BibitemOpen
  \bibfield  {author} {\bibinfo {author} {\bibfnamefont {A.}~\bibnamefont {{Tchekhovskoy}}}, \bibinfo {author} {\bibfnamefont {R.}~\bibnamefont {{Narayan}}},\ and\ \bibinfo {author} {\bibfnamefont {J.~C.}\ \bibnamefont {{McKinney}}},\ }\href {https://doi.org/10.1111/j.1745-3933.2011.01147.x} {\bibfield  {journal} {\bibinfo  {journal} {MNRAS}\ }\textbf {\bibinfo {volume} {418}},\ \bibinfo {pages} {L79} (\bibinfo {year} {2011})},\ \Eprint {https://arxiv.org/abs/1108.0412} {arXiv:1108.0412 [astro-ph.HE]} \BibitemShut {NoStop}%
\bibitem [{\citenamefont {{Nakamura}}\ \emph {et~al.}(2018)\citenamefont {{Nakamura}}, \citenamefont {{Asada}}, \citenamefont {{Hada}}, \citenamefont {{Pu}}, \citenamefont {{Noble}}, \citenamefont {{Tseng}}, \citenamefont {{Toma}}, \citenamefont {{Kino}}, \citenamefont {{Nagai}}, \citenamefont {{Takahashi}}, \citenamefont {{Algaba}}, \citenamefont {{Orienti}}, \citenamefont {{Akiyama}}, \citenamefont {{Doi}}, \citenamefont {{Giovannini}}, \citenamefont {{Giroletti}}, \citenamefont {{Honma}}, \citenamefont {{Koyama}}, \citenamefont {{Lico}}, \citenamefont {{Niinuma}},\ and\ \citenamefont {{Tazaki}}}]{Nakamura18}%
  \BibitemOpen
  \bibfield  {author} {\bibinfo {author} {\bibfnamefont {M.}~\bibnamefont {{Nakamura}}}, \bibinfo {author} {\bibfnamefont {K.}~\bibnamefont {{Asada}}}, \bibinfo {author} {\bibfnamefont {K.}~\bibnamefont {{Hada}}}, \bibinfo {author} {\bibfnamefont {H.-Y.}\ \bibnamefont {{Pu}}}, \bibinfo {author} {\bibfnamefont {S.}~\bibnamefont {{Noble}}}, \bibinfo {author} {\bibfnamefont {C.}~\bibnamefont {{Tseng}}}, \bibinfo {author} {\bibfnamefont {K.}~\bibnamefont {{Toma}}}, \bibinfo {author} {\bibfnamefont {M.}~\bibnamefont {{Kino}}}, \bibinfo {author} {\bibfnamefont {H.}~\bibnamefont {{Nagai}}}, \bibinfo {author} {\bibfnamefont {K.}~\bibnamefont {{Takahashi}}}, \bibinfo {author} {\bibfnamefont {J.-C.}\ \bibnamefont {{Algaba}}}, \bibinfo {author} {\bibfnamefont {M.}~\bibnamefont {{Orienti}}}, \bibinfo {author} {\bibfnamefont {K.}~\bibnamefont {{Akiyama}}}, \bibinfo {author} {\bibfnamefont {A.}~\bibnamefont {{Doi}}}, \bibinfo {author} {\bibfnamefont {G.}~\bibnamefont {{Giovannini}}}, \bibinfo {author} {\bibfnamefont
  {M.}~\bibnamefont {{Giroletti}}}, \bibinfo {author} {\bibfnamefont {M.}~\bibnamefont {{Honma}}}, \bibinfo {author} {\bibfnamefont {S.}~\bibnamefont {{Koyama}}}, \bibinfo {author} {\bibfnamefont {R.}~\bibnamefont {{Lico}}}, \bibinfo {author} {\bibfnamefont {K.}~\bibnamefont {{Niinuma}}},\ and\ \bibinfo {author} {\bibfnamefont {F.}~\bibnamefont {{Tazaki}}},\ }\href {https://doi.org/10.3847/1538-4357/aaeb2d} {\bibfield  {journal} {\bibinfo  {journal} {ApJ}\ }\textbf {\bibinfo {volume} {868}},\ \bibinfo {eid} {146} (\bibinfo {year} {2018})},\ \Eprint {https://arxiv.org/abs/1810.09963} {arXiv:1810.09963 [astro-ph.HE]} \BibitemShut {NoStop}%
\bibitem [{\citenamefont {{Chael}}\ \emph {et~al.}(2019)\citenamefont {{Chael}}, \citenamefont {{Narayan}},\ and\ \citenamefont {{Johnson}}}]{Chael19}%
  \BibitemOpen
  \bibfield  {author} {\bibinfo {author} {\bibfnamefont {A.}~\bibnamefont {{Chael}}}, \bibinfo {author} {\bibfnamefont {R.}~\bibnamefont {{Narayan}}},\ and\ \bibinfo {author} {\bibfnamefont {M.~D.}\ \bibnamefont {{Johnson}}},\ }\href {https://doi.org/10.1093/mnras/stz988} {\bibfield  {journal} {\bibinfo  {journal} {MNRAS}\ }\textbf {\bibinfo {volume} {486}},\ \bibinfo {pages} {2873} (\bibinfo {year} {2019})},\ \Eprint {https://arxiv.org/abs/1810.01983} {arXiv:1810.01983 [astro-ph.HE]} \BibitemShut {NoStop}%
\bibitem [{\citenamefont {{Porth}}\ \emph {et~al.}(2019)\citenamefont {{Porth}}, \citenamefont {{Chatterjee}}, \citenamefont {{Narayan}}, \citenamefont {{Gammie}}, \citenamefont {{Mizuno}}, \citenamefont {{Anninos}}, \citenamefont {{Baker}}, \citenamefont {{Bugli}}, \citenamefont {{Chan}}, \citenamefont {{Davelaar}}, \citenamefont {{Del Zanna}}, \citenamefont {{Etienne}}, \citenamefont {{Fragile}}, \citenamefont {{Kelly}}, \citenamefont {{Liska}}, \citenamefont {{Markoff}}, \citenamefont {{McKinney}}, \citenamefont {{Mishra}}, \citenamefont {{Noble}}, \citenamefont {{Olivares}}, \citenamefont {{Prather}}, \citenamefont {{Rezzolla}}, \citenamefont {{Ryan}}, \citenamefont {{Stone}}, \citenamefont {{Tomei}}, \citenamefont {{White}}, \citenamefont {{Younsi}}, \citenamefont {{Akiyama}}, \citenamefont {{Alberdi}}, \citenamefont {{Alef}}, \citenamefont {{Asada}}, \citenamefont {{Azulay}}, \citenamefont {{Baczko}}, \citenamefont {{Ball}}, \citenamefont {{Balokovi{\'c}}}, \citenamefont {{Barrett}}, \citenamefont
  {{Bintley}}, \citenamefont {{Blackburn}}, \citenamefont {{Boland}}, \citenamefont {{Bouman}}, \citenamefont {{Bower}}, \citenamefont {{Bremer}}, \citenamefont {{Brinkerink}}, \citenamefont {{Brissenden}}, \citenamefont {{Britzen}}, \citenamefont {{Broderick}}, \citenamefont {{Broguiere}}, \citenamefont {{Bronzwaer}}, \citenamefont {{Byun}}, \citenamefont {{Carlstrom}}, \citenamefont {{Chael}}, \citenamefont {{Chatterjee}}, \citenamefont {{Chen}}, \citenamefont {{Chen}}, \citenamefont {{Cho}}, \citenamefont {{Christian}}, \citenamefont {{Conway}}, \citenamefont {{Cordes}}, \citenamefont {{Geoffrey}}, \citenamefont {{Crew}}, \citenamefont {{Cui}}, \citenamefont {{De Laurentis}}, \citenamefont {{Deane}}, \citenamefont {{Dempsey}}, \citenamefont {{Desvignes}}, \citenamefont {{Doeleman}}, \citenamefont {{Eatough}}, \citenamefont {{Falcke}}, \citenamefont {{Fish}}, \citenamefont {{Fomalont}}, \citenamefont {{Fraga-Encinas}}, \citenamefont {{Freeman}}, \citenamefont {{Friberg}}, \citenamefont {{Fromm}},
  \citenamefont {{G{\'o}mez}}, \citenamefont {{Galison}}, \citenamefont {{Garc{\'\i}a}}, \citenamefont {{Gentaz}}, \citenamefont {{Georgiev}}, \citenamefont {{Goddi}}, \citenamefont {{Gold}}, \citenamefont {{Gu}}, \citenamefont {{Gurwell}}, \citenamefont {{Hada}}, \citenamefont {{Hecht}}, \citenamefont {{Hesper}}, \citenamefont {{Ho}}, \citenamefont {{Ho}}, \citenamefont {{Honma}}, \citenamefont {{Huang}}, \citenamefont {{Huang}}, \citenamefont {{Hughes}}, \citenamefont {{Ikeda}}, \citenamefont {{Inoue}}, \citenamefont {{Issaoun}}, \citenamefont {{James}}, \citenamefont {{Jannuzi}}, \citenamefont {{Janssen}}, \citenamefont {{Jeter}}, \citenamefont {{Jiang}}, \citenamefont {{Johnson}}, \citenamefont {{Jorstad}}, \citenamefont {{Jung}}, \citenamefont {{Karami}}, \citenamefont {{Karuppusamy}}, \citenamefont {{Kawashima}}, \citenamefont {{Keating}}, \citenamefont {{Kettenis}}, \citenamefont {{Kim}}, \citenamefont {{Kim}}, \citenamefont {{Kim}}, \citenamefont {{Kino}}, \citenamefont {{Koay}}, \citenamefont
  {{Patrick}}, \citenamefont {{Koch}}, \citenamefont {{Koyama}}, \citenamefont {{Kramer}}, \citenamefont {{Kramer}}, \citenamefont {{Krichbaum}}, \citenamefont {{Kuo}}, \citenamefont {{Lauer}}, \citenamefont {{Lee}}, \citenamefont {{Li}}, \citenamefont {{Li}}, \citenamefont {{Lindqvist}}, \citenamefont {{Liu}}, \citenamefont {{Liuzzo}}, \citenamefont {{Lo}}, \citenamefont {{Lobanov}}, \citenamefont {{Loinard}}, \citenamefont {{Lonsdale}}, \citenamefont {{Lu}}, \citenamefont {{MacDonald}}, \citenamefont {{Mao}}, \citenamefont {{Marrone}}, \citenamefont {{Marscher}}, \citenamefont {{Mart{\'\i}-Vidal}}, \citenamefont {{Matsushita}}, \citenamefont {{Matthews}}, \citenamefont {{Medeiros}}, \citenamefont {{Menten}}, \citenamefont {{Mizuno}}, \citenamefont {{Moran}}, \citenamefont {{Moriyama}}, \citenamefont {{Moscibrodzka}}, \citenamefont {{M{\"u}ller}}, \citenamefont {{Nagai}}, \citenamefont {{Nagar}}, \citenamefont {{Nakamura}}, \citenamefont {{Narayanan}}, \citenamefont {{Natarajan}}, \citenamefont {{Neri}},
  \citenamefont {{Ni}}, \citenamefont {{Noutsos}}, \citenamefont {{Okino}}, \citenamefont {{Oyama}}, \citenamefont {{{\"O}zel}}, \citenamefont {{Palumbo}}, \citenamefont {{Patel}}, \citenamefont {{Pen}}, \citenamefont {{Pesce}}, \citenamefont {{Pi{\'e}tu}}, \citenamefont {{Plambeck}}, \citenamefont {{PopStefanija}}, \citenamefont {{Preciado-L{\'o}pez}}, \citenamefont {{Psaltis}}, \citenamefont {{Pu}}, \citenamefont {{Ramakrishnan}}, \citenamefont {{Rao}}, \citenamefont {{Rawlings}}, \citenamefont {{Raymond}}, \citenamefont {{Ripperda}}, \citenamefont {{Roelofs}}, \citenamefont {{Rogers}}, \citenamefont {{Ros}}, \citenamefont {{Rose}}, \citenamefont {{Roshanineshat}}, \citenamefont {{Rottmann}}, \citenamefont {{Roy}}, \citenamefont {{Ruszczyk}}, \citenamefont {{Rygl}}, \citenamefont {{S{\'a}nchez}}, \citenamefont {{S{\'a}nchez-Arguelles}}, \citenamefont {{Sasada}}, \citenamefont {{Savolainen}}, \citenamefont {{Schloerb}}, \citenamefont {{Schuster}}, \citenamefont {{Shao}}, \citenamefont {{Shen}}, \citenamefont
  {{Small}}, \citenamefont {{Sohn}}, \citenamefont {{SooHoo}}, \citenamefont {{Tazaki}}, \citenamefont {{Tiede}}, \citenamefont {{Tilanus}}, \citenamefont {{Titus}}, \citenamefont {{Toma}}, \citenamefont {{Torne}}, \citenamefont {{Trent}}, \citenamefont {{Trippe}}, \citenamefont {{Tsuda}}, \citenamefont {{van Bemmel}}, \citenamefont {{van Langevelde}}, \citenamefont {{van Rossum}}, \citenamefont {{Wagner}}, \citenamefont {{Wardle}}, \citenamefont {{Weintroub}}, \citenamefont {{Wex}}, \citenamefont {{Wharton}}, \citenamefont {{Wielgus}}, \citenamefont {{Wong}}, \citenamefont {{Wu}}, \citenamefont {{Young}}, \citenamefont {{Young}}, \citenamefont {{Yuan}}, \citenamefont {{Yuan}}, \citenamefont {{Zensus}}, \citenamefont {{Zhao}}, \citenamefont {{Zhao}}, \citenamefont {{Zhu}},\ and\ \citenamefont {{Event Horizon Telescope Collaboration}}}]{Porth19}%
  \BibitemOpen
  \bibfield  {author} {\bibinfo {author} {\bibfnamefont {O.}~\bibnamefont {{Porth}}}, \bibinfo {author} {\bibfnamefont {K.}~\bibnamefont {{Chatterjee}}}, \bibinfo {author} {\bibfnamefont {R.}~\bibnamefont {{Narayan}}}, \bibinfo {author} {\bibfnamefont {C.~F.}\ \bibnamefont {{Gammie}}}, \bibinfo {author} {\bibfnamefont {Y.}~\bibnamefont {{Mizuno}}}, \bibinfo {author} {\bibfnamefont {P.}~\bibnamefont {{Anninos}}}, \bibinfo {author} {\bibfnamefont {J.~G.}\ \bibnamefont {{Baker}}}, \bibinfo {author} {\bibfnamefont {M.}~\bibnamefont {{Bugli}}}, \bibinfo {author} {\bibfnamefont {C.-k.}\ \bibnamefont {{Chan}}}, \bibinfo {author} {\bibfnamefont {J.}~\bibnamefont {{Davelaar}}}, \bibinfo {author} {\bibfnamefont {L.}~\bibnamefont {{Del Zanna}}}, \bibinfo {author} {\bibfnamefont {Z.~B.}\ \bibnamefont {{Etienne}}}, \bibinfo {author} {\bibfnamefont {P.~C.}\ \bibnamefont {{Fragile}}}, \bibinfo {author} {\bibfnamefont {B.~J.}\ \bibnamefont {{Kelly}}}, \bibinfo {author} {\bibfnamefont {M.}~\bibnamefont {{Liska}}}, \bibinfo
  {author} {\bibfnamefont {S.}~\bibnamefont {{Markoff}}}, \bibinfo {author} {\bibfnamefont {J.~C.}\ \bibnamefont {{McKinney}}}, \bibinfo {author} {\bibfnamefont {B.}~\bibnamefont {{Mishra}}}, \bibinfo {author} {\bibfnamefont {S.~C.}\ \bibnamefont {{Noble}}}, \bibinfo {author} {\bibfnamefont {H.}~\bibnamefont {{Olivares}}}, \bibinfo {author} {\bibfnamefont {B.}~\bibnamefont {{Prather}}}, \bibinfo {author} {\bibfnamefont {L.}~\bibnamefont {{Rezzolla}}}, \bibinfo {author} {\bibfnamefont {B.~R.}\ \bibnamefont {{Ryan}}}, \bibinfo {author} {\bibfnamefont {J.~M.}\ \bibnamefont {{Stone}}}, \bibinfo {author} {\bibfnamefont {N.}~\bibnamefont {{Tomei}}}, \bibinfo {author} {\bibfnamefont {C.~J.}\ \bibnamefont {{White}}}, \bibinfo {author} {\bibfnamefont {Z.}~\bibnamefont {{Younsi}}}, \bibinfo {author} {\bibfnamefont {K.}~\bibnamefont {{Akiyama}}}, \bibinfo {author} {\bibfnamefont {A.}~\bibnamefont {{Alberdi}}}, \bibinfo {author} {\bibfnamefont {W.}~\bibnamefont {{Alef}}}, \bibinfo {author} {\bibfnamefont
  {K.}~\bibnamefont {{Asada}}}, \bibinfo {author} {\bibfnamefont {R.}~\bibnamefont {{Azulay}}}, \bibinfo {author} {\bibfnamefont {A.-K.}\ \bibnamefont {{Baczko}}}, \bibinfo {author} {\bibfnamefont {D.}~\bibnamefont {{Ball}}}, \bibinfo {author} {\bibfnamefont {M.}~\bibnamefont {{Balokovi{\'c}}}}, \bibinfo {author} {\bibfnamefont {J.}~\bibnamefont {{Barrett}}}, \bibinfo {author} {\bibfnamefont {D.}~\bibnamefont {{Bintley}}}, \bibinfo {author} {\bibfnamefont {L.}~\bibnamefont {{Blackburn}}}, \bibinfo {author} {\bibfnamefont {W.}~\bibnamefont {{Boland}}}, \bibinfo {author} {\bibfnamefont {K.~L.}\ \bibnamefont {{Bouman}}}, \bibinfo {author} {\bibfnamefont {G.~C.}\ \bibnamefont {{Bower}}}, \bibinfo {author} {\bibfnamefont {M.}~\bibnamefont {{Bremer}}}, \bibinfo {author} {\bibfnamefont {C.~D.}\ \bibnamefont {{Brinkerink}}}, \bibinfo {author} {\bibfnamefont {R.}~\bibnamefont {{Brissenden}}}, \bibinfo {author} {\bibfnamefont {S.}~\bibnamefont {{Britzen}}}, \bibinfo {author} {\bibfnamefont {A.~E.}\ \bibnamefont
  {{Broderick}}}, \bibinfo {author} {\bibfnamefont {D.}~\bibnamefont {{Broguiere}}}, \bibinfo {author} {\bibfnamefont {T.}~\bibnamefont {{Bronzwaer}}}, \bibinfo {author} {\bibfnamefont {D.-Y.}\ \bibnamefont {{Byun}}}, \bibinfo {author} {\bibfnamefont {J.~E.}\ \bibnamefont {{Carlstrom}}}, \bibinfo {author} {\bibfnamefont {A.}~\bibnamefont {{Chael}}}, \bibinfo {author} {\bibfnamefont {S.}~\bibnamefont {{Chatterjee}}}, \bibinfo {author} {\bibfnamefont {M.-T.}\ \bibnamefont {{Chen}}}, \bibinfo {author} {\bibfnamefont {Y.}~\bibnamefont {{Chen}}}, \bibinfo {author} {\bibfnamefont {I.}~\bibnamefont {{Cho}}}, \bibinfo {author} {\bibfnamefont {P.}~\bibnamefont {{Christian}}}, \bibinfo {author} {\bibfnamefont {J.~E.}\ \bibnamefont {{Conway}}}, \bibinfo {author} {\bibfnamefont {J.~M.}\ \bibnamefont {{Cordes}}}, \bibinfo {author} {\bibnamefont {{Geoffrey}}}, \bibinfo {author} {\bibfnamefont {B.}~\bibnamefont {{Crew}}}, \bibinfo {author} {\bibfnamefont {Y.}~\bibnamefont {{Cui}}}, \bibinfo {author} {\bibfnamefont
  {M.}~\bibnamefont {{De Laurentis}}}, \bibinfo {author} {\bibfnamefont {R.}~\bibnamefont {{Deane}}}, \bibinfo {author} {\bibfnamefont {J.}~\bibnamefont {{Dempsey}}}, \bibinfo {author} {\bibfnamefont {G.}~\bibnamefont {{Desvignes}}}, \bibinfo {author} {\bibfnamefont {S.~S.}\ \bibnamefont {{Doeleman}}}, \bibinfo {author} {\bibfnamefont {R.~P.}\ \bibnamefont {{Eatough}}}, \bibinfo {author} {\bibfnamefont {H.}~\bibnamefont {{Falcke}}}, \bibinfo {author} {\bibfnamefont {V.~L.}\ \bibnamefont {{Fish}}}, \bibinfo {author} {\bibfnamefont {E.}~\bibnamefont {{Fomalont}}}, \bibinfo {author} {\bibfnamefont {R.}~\bibnamefont {{Fraga-Encinas}}}, \bibinfo {author} {\bibfnamefont {B.}~\bibnamefont {{Freeman}}}, \bibinfo {author} {\bibfnamefont {P.}~\bibnamefont {{Friberg}}}, \bibinfo {author} {\bibfnamefont {C.~M.}\ \bibnamefont {{Fromm}}}, \bibinfo {author} {\bibfnamefont {J.~L.}\ \bibnamefont {{G{\'o}mez}}}, \bibinfo {author} {\bibfnamefont {P.}~\bibnamefont {{Galison}}}, \bibinfo {author} {\bibfnamefont {R.}~\bibnamefont
  {{Garc{\'\i}a}}}, \bibinfo {author} {\bibfnamefont {O.}~\bibnamefont {{Gentaz}}}, \bibinfo {author} {\bibfnamefont {B.}~\bibnamefont {{Georgiev}}}, \bibinfo {author} {\bibfnamefont {C.}~\bibnamefont {{Goddi}}}, \bibinfo {author} {\bibfnamefont {R.}~\bibnamefont {{Gold}}}, \bibinfo {author} {\bibfnamefont {M.}~\bibnamefont {{Gu}}}, \bibinfo {author} {\bibfnamefont {M.}~\bibnamefont {{Gurwell}}}, \bibinfo {author} {\bibfnamefont {K.}~\bibnamefont {{Hada}}}, \bibinfo {author} {\bibfnamefont {M.~H.}\ \bibnamefont {{Hecht}}}, \bibinfo {author} {\bibfnamefont {R.}~\bibnamefont {{Hesper}}}, \bibinfo {author} {\bibfnamefont {L.~C.}\ \bibnamefont {{Ho}}}, \bibinfo {author} {\bibfnamefont {P.}~\bibnamefont {{Ho}}}, \bibinfo {author} {\bibfnamefont {M.}~\bibnamefont {{Honma}}}, \bibinfo {author} {\bibfnamefont {C.-W.~L.}\ \bibnamefont {{Huang}}}, \bibinfo {author} {\bibfnamefont {L.}~\bibnamefont {{Huang}}}, \bibinfo {author} {\bibfnamefont {D.~H.}\ \bibnamefont {{Hughes}}}, \bibinfo {author} {\bibfnamefont
  {S.}~\bibnamefont {{Ikeda}}}, \bibinfo {author} {\bibfnamefont {M.}~\bibnamefont {{Inoue}}}, \bibinfo {author} {\bibfnamefont {S.}~\bibnamefont {{Issaoun}}}, \bibinfo {author} {\bibfnamefont {D.~J.}\ \bibnamefont {{James}}}, \bibinfo {author} {\bibfnamefont {B.~T.}\ \bibnamefont {{Jannuzi}}}, \bibinfo {author} {\bibfnamefont {M.}~\bibnamefont {{Janssen}}}, \bibinfo {author} {\bibfnamefont {B.}~\bibnamefont {{Jeter}}}, \bibinfo {author} {\bibfnamefont {W.}~\bibnamefont {{Jiang}}}, \bibinfo {author} {\bibfnamefont {M.~D.}\ \bibnamefont {{Johnson}}}, \bibinfo {author} {\bibfnamefont {S.}~\bibnamefont {{Jorstad}}}, \bibinfo {author} {\bibfnamefont {T.}~\bibnamefont {{Jung}}}, \bibinfo {author} {\bibfnamefont {M.}~\bibnamefont {{Karami}}}, \bibinfo {author} {\bibfnamefont {R.}~\bibnamefont {{Karuppusamy}}}, \bibinfo {author} {\bibfnamefont {T.}~\bibnamefont {{Kawashima}}}, \bibinfo {author} {\bibfnamefont {G.~K.}\ \bibnamefont {{Keating}}}, \bibinfo {author} {\bibfnamefont {M.}~\bibnamefont {{Kettenis}}},
  \bibinfo {author} {\bibfnamefont {J.-Y.}\ \bibnamefont {{Kim}}}, \bibinfo {author} {\bibfnamefont {J.}~\bibnamefont {{Kim}}}, \bibinfo {author} {\bibfnamefont {J.}~\bibnamefont {{Kim}}}, \bibinfo {author} {\bibfnamefont {M.}~\bibnamefont {{Kino}}}, \bibinfo {author} {\bibfnamefont {J.~Y.}\ \bibnamefont {{Koay}}}, \bibinfo {author} {\bibnamefont {{Patrick}}}, \bibinfo {author} {\bibfnamefont {M.}~\bibnamefont {{Koch}}}, \bibinfo {author} {\bibfnamefont {S.}~\bibnamefont {{Koyama}}}, \bibinfo {author} {\bibfnamefont {M.}~\bibnamefont {{Kramer}}}, \bibinfo {author} {\bibfnamefont {C.}~\bibnamefont {{Kramer}}}, \bibinfo {author} {\bibfnamefont {T.~P.}\ \bibnamefont {{Krichbaum}}}, \bibinfo {author} {\bibfnamefont {C.-Y.}\ \bibnamefont {{Kuo}}}, \bibinfo {author} {\bibfnamefont {T.~R.}\ \bibnamefont {{Lauer}}}, \bibinfo {author} {\bibfnamefont {S.-S.}\ \bibnamefont {{Lee}}}, \bibinfo {author} {\bibfnamefont {Y.-R.}\ \bibnamefont {{Li}}}, \bibinfo {author} {\bibfnamefont {Z.}~\bibnamefont {{Li}}}, \bibinfo
  {author} {\bibfnamefont {M.}~\bibnamefont {{Lindqvist}}}, \bibinfo {author} {\bibfnamefont {K.}~\bibnamefont {{Liu}}}, \bibinfo {author} {\bibfnamefont {E.}~\bibnamefont {{Liuzzo}}}, \bibinfo {author} {\bibfnamefont {W.-P.}\ \bibnamefont {{Lo}}}, \bibinfo {author} {\bibfnamefont {A.~P.}\ \bibnamefont {{Lobanov}}}, \bibinfo {author} {\bibfnamefont {L.}~\bibnamefont {{Loinard}}}, \bibinfo {author} {\bibfnamefont {C.}~\bibnamefont {{Lonsdale}}}, \bibinfo {author} {\bibfnamefont {R.-S.}\ \bibnamefont {{Lu}}}, \bibinfo {author} {\bibfnamefont {N.~R.}\ \bibnamefont {{MacDonald}}}, \bibinfo {author} {\bibfnamefont {J.}~\bibnamefont {{Mao}}}, \bibinfo {author} {\bibfnamefont {D.~P.}\ \bibnamefont {{Marrone}}}, \bibinfo {author} {\bibfnamefont {A.~P.}\ \bibnamefont {{Marscher}}}, \bibinfo {author} {\bibfnamefont {I.}~\bibnamefont {{Mart{\'\i}-Vidal}}}, \bibinfo {author} {\bibfnamefont {S.}~\bibnamefont {{Matsushita}}}, \bibinfo {author} {\bibfnamefont {L.~D.}\ \bibnamefont {{Matthews}}}, \bibinfo {author}
  {\bibfnamefont {L.}~\bibnamefont {{Medeiros}}}, \bibinfo {author} {\bibfnamefont {K.~M.}\ \bibnamefont {{Menten}}}, \bibinfo {author} {\bibfnamefont {I.}~\bibnamefont {{Mizuno}}}, \bibinfo {author} {\bibfnamefont {J.~M.}\ \bibnamefont {{Moran}}}, \bibinfo {author} {\bibfnamefont {K.}~\bibnamefont {{Moriyama}}}, \bibinfo {author} {\bibfnamefont {M.}~\bibnamefont {{Moscibrodzka}}}, \bibinfo {author} {\bibfnamefont {C.}~\bibnamefont {{M{\"u}ller}}}, \bibinfo {author} {\bibfnamefont {H.}~\bibnamefont {{Nagai}}}, \bibinfo {author} {\bibfnamefont {N.~M.}\ \bibnamefont {{Nagar}}}, \bibinfo {author} {\bibfnamefont {M.}~\bibnamefont {{Nakamura}}}, \bibinfo {author} {\bibfnamefont {G.}~\bibnamefont {{Narayanan}}}, \bibinfo {author} {\bibfnamefont {I.}~\bibnamefont {{Natarajan}}}, \bibinfo {author} {\bibfnamefont {R.}~\bibnamefont {{Neri}}}, \bibinfo {author} {\bibfnamefont {C.}~\bibnamefont {{Ni}}}, \bibinfo {author} {\bibfnamefont {A.}~\bibnamefont {{Noutsos}}}, \bibinfo {author} {\bibfnamefont {H.}~\bibnamefont
  {{Okino}}}, \bibinfo {author} {\bibfnamefont {T.}~\bibnamefont {{Oyama}}}, \bibinfo {author} {\bibfnamefont {F.}~\bibnamefont {{{\"O}zel}}}, \bibinfo {author} {\bibfnamefont {D.~C.~M.}\ \bibnamefont {{Palumbo}}}, \bibinfo {author} {\bibfnamefont {N.}~\bibnamefont {{Patel}}}, \bibinfo {author} {\bibfnamefont {U.-L.}\ \bibnamefont {{Pen}}}, \bibinfo {author} {\bibfnamefont {D.~W.}\ \bibnamefont {{Pesce}}}, \bibinfo {author} {\bibfnamefont {V.}~\bibnamefont {{Pi{\'e}tu}}}, \bibinfo {author} {\bibfnamefont {R.}~\bibnamefont {{Plambeck}}}, \bibinfo {author} {\bibfnamefont {A.}~\bibnamefont {{PopStefanija}}}, \bibinfo {author} {\bibfnamefont {J.~A.}\ \bibnamefont {{Preciado-L{\'o}pez}}}, \bibinfo {author} {\bibfnamefont {D.}~\bibnamefont {{Psaltis}}}, \bibinfo {author} {\bibfnamefont {H.-Y.}\ \bibnamefont {{Pu}}}, \bibinfo {author} {\bibfnamefont {V.}~\bibnamefont {{Ramakrishnan}}}, \bibinfo {author} {\bibfnamefont {R.}~\bibnamefont {{Rao}}}, \bibinfo {author} {\bibfnamefont {M.~G.}\ \bibnamefont {{Rawlings}}},
  \bibinfo {author} {\bibfnamefont {A.~W.}\ \bibnamefont {{Raymond}}}, \bibinfo {author} {\bibfnamefont {B.}~\bibnamefont {{Ripperda}}}, \bibinfo {author} {\bibfnamefont {F.}~\bibnamefont {{Roelofs}}}, \bibinfo {author} {\bibfnamefont {A.}~\bibnamefont {{Rogers}}}, \bibinfo {author} {\bibfnamefont {E.}~\bibnamefont {{Ros}}}, \bibinfo {author} {\bibfnamefont {M.}~\bibnamefont {{Rose}}}, \bibinfo {author} {\bibfnamefont {A.}~\bibnamefont {{Roshanineshat}}}, \bibinfo {author} {\bibfnamefont {H.}~\bibnamefont {{Rottmann}}}, \bibinfo {author} {\bibfnamefont {A.~L.}\ \bibnamefont {{Roy}}}, \bibinfo {author} {\bibfnamefont {C.}~\bibnamefont {{Ruszczyk}}}, \bibinfo {author} {\bibfnamefont {K.~L.~J.}\ \bibnamefont {{Rygl}}}, \bibinfo {author} {\bibfnamefont {S.}~\bibnamefont {{S{\'a}nchez}}}, \bibinfo {author} {\bibfnamefont {D.}~\bibnamefont {{S{\'a}nchez-Arguelles}}}, \bibinfo {author} {\bibfnamefont {M.}~\bibnamefont {{Sasada}}}, \bibinfo {author} {\bibfnamefont {T.}~\bibnamefont {{Savolainen}}}, \bibinfo {author}
  {\bibfnamefont {F.~P.}\ \bibnamefont {{Schloerb}}}, \bibinfo {author} {\bibfnamefont {K.-F.}\ \bibnamefont {{Schuster}}}, \bibinfo {author} {\bibfnamefont {L.}~\bibnamefont {{Shao}}}, \bibinfo {author} {\bibfnamefont {Z.}~\bibnamefont {{Shen}}}, \bibinfo {author} {\bibfnamefont {D.}~\bibnamefont {{Small}}}, \bibinfo {author} {\bibfnamefont {B.~W.}\ \bibnamefont {{Sohn}}}, \bibinfo {author} {\bibfnamefont {J.}~\bibnamefont {{SooHoo}}}, \bibinfo {author} {\bibfnamefont {F.}~\bibnamefont {{Tazaki}}}, \bibinfo {author} {\bibfnamefont {P.}~\bibnamefont {{Tiede}}}, \bibinfo {author} {\bibfnamefont {R.~P.~J.}\ \bibnamefont {{Tilanus}}}, \bibinfo {author} {\bibfnamefont {M.}~\bibnamefont {{Titus}}}, \bibinfo {author} {\bibfnamefont {K.}~\bibnamefont {{Toma}}}, \bibinfo {author} {\bibfnamefont {P.}~\bibnamefont {{Torne}}}, \bibinfo {author} {\bibfnamefont {T.}~\bibnamefont {{Trent}}}, \bibinfo {author} {\bibfnamefont {S.}~\bibnamefont {{Trippe}}}, \bibinfo {author} {\bibfnamefont {S.}~\bibnamefont {{Tsuda}}},
  \bibinfo {author} {\bibfnamefont {I.}~\bibnamefont {{van Bemmel}}}, \bibinfo {author} {\bibfnamefont {H.~J.}\ \bibnamefont {{van Langevelde}}}, \bibinfo {author} {\bibfnamefont {D.~R.}\ \bibnamefont {{van Rossum}}}, \bibinfo {author} {\bibfnamefont {J.}~\bibnamefont {{Wagner}}}, \bibinfo {author} {\bibfnamefont {J.}~\bibnamefont {{Wardle}}}, \bibinfo {author} {\bibfnamefont {J.}~\bibnamefont {{Weintroub}}}, \bibinfo {author} {\bibfnamefont {N.}~\bibnamefont {{Wex}}}, \bibinfo {author} {\bibfnamefont {R.}~\bibnamefont {{Wharton}}}, \bibinfo {author} {\bibfnamefont {M.}~\bibnamefont {{Wielgus}}}, \bibinfo {author} {\bibfnamefont {G.~N.}\ \bibnamefont {{Wong}}}, \bibinfo {author} {\bibfnamefont {Q.}~\bibnamefont {{Wu}}}, \bibinfo {author} {\bibfnamefont {K.}~\bibnamefont {{Young}}}, \bibinfo {author} {\bibfnamefont {A.}~\bibnamefont {{Young}}}, \bibinfo {author} {\bibfnamefont {F.}~\bibnamefont {{Yuan}}}, \bibinfo {author} {\bibfnamefont {Y.-F.}\ \bibnamefont {{Yuan}}}, \bibinfo {author} {\bibfnamefont
  {J.~A.}\ \bibnamefont {{Zensus}}}, \bibinfo {author} {\bibfnamefont {G.}~\bibnamefont {{Zhao}}}, \bibinfo {author} {\bibfnamefont {S.-S.}\ \bibnamefont {{Zhao}}}, \bibinfo {author} {\bibfnamefont {Z.}~\bibnamefont {{Zhu}}},\ and\ \bibinfo {author} {\bibnamefont {{Event Horizon Telescope Collaboration}}},\ }\href {https://doi.org/10.3847/1538-4365/ab29fd} {\bibfield  {journal} {\bibinfo  {journal} {ApJS}\ }\textbf {\bibinfo {volume} {243}},\ \bibinfo {eid} {26} (\bibinfo {year} {2019})},\ \Eprint {https://arxiv.org/abs/1904.04923} {arXiv:1904.04923 [astro-ph.HE]} \BibitemShut {NoStop}%
\bibitem [{\citenamefont {{Liska}}\ \emph {et~al.}(2020)\citenamefont {{Liska}}, \citenamefont {{Tchekhovskoy}},\ and\ \citenamefont {{Quataert}}}]{Liska20}%
  \BibitemOpen
  \bibfield  {author} {\bibinfo {author} {\bibfnamefont {M.}~\bibnamefont {{Liska}}}, \bibinfo {author} {\bibfnamefont {A.}~\bibnamefont {{Tchekhovskoy}}},\ and\ \bibinfo {author} {\bibfnamefont {E.}~\bibnamefont {{Quataert}}},\ }\href {https://doi.org/10.1093/mnras/staa955} {\bibfield  {journal} {\bibinfo  {journal} {MNRAS}\ }\textbf {\bibinfo {volume} {494}},\ \bibinfo {pages} {3656} (\bibinfo {year} {2020})},\ \Eprint {https://arxiv.org/abs/1809.04608} {arXiv:1809.04608 [astro-ph.HE]} \BibitemShut {NoStop}%
\bibitem [{\citenamefont {{Ripperda}}\ \emph {et~al.}(2022)\citenamefont {{Ripperda}}, \citenamefont {{Liska}}, \citenamefont {{Chatterjee}}, \citenamefont {{Musoke}}, \citenamefont {{Philippov}}, \citenamefont {{Markoff}}, \citenamefont {{Tchekhovskoy}},\ and\ \citenamefont {{Younsi}}}]{Ripperda22}%
  \BibitemOpen
  \bibfield  {author} {\bibinfo {author} {\bibfnamefont {B.}~\bibnamefont {{Ripperda}}}, \bibinfo {author} {\bibfnamefont {M.}~\bibnamefont {{Liska}}}, \bibinfo {author} {\bibfnamefont {K.}~\bibnamefont {{Chatterjee}}}, \bibinfo {author} {\bibfnamefont {G.}~\bibnamefont {{Musoke}}}, \bibinfo {author} {\bibfnamefont {A.~A.}\ \bibnamefont {{Philippov}}}, \bibinfo {author} {\bibfnamefont {S.~B.}\ \bibnamefont {{Markoff}}}, \bibinfo {author} {\bibfnamefont {A.}~\bibnamefont {{Tchekhovskoy}}},\ and\ \bibinfo {author} {\bibfnamefont {Z.}~\bibnamefont {{Younsi}}},\ }\href {https://doi.org/10.3847/2041-8213/ac46a1} {\bibfield  {journal} {\bibinfo  {journal} {ApJL}\ }\textbf {\bibinfo {volume} {924}},\ \bibinfo {eid} {L32} (\bibinfo {year} {2022})},\ \Eprint {https://arxiv.org/abs/2109.15115} {arXiv:2109.15115 [astro-ph.HE]} \BibitemShut {NoStop}%
\bibitem [{\citenamefont {{Levinson}}\ and\ \citenamefont {{Cerutti}}(2018)}]{Levinson18}%
  \BibitemOpen
  \bibfield  {author} {\bibinfo {author} {\bibfnamefont {A.}~\bibnamefont {{Levinson}}}\ and\ \bibinfo {author} {\bibfnamefont {B.}~\bibnamefont {{Cerutti}}},\ }\href {https://doi.org/10.1051/0004-6361/201832915} {\bibfield  {journal} {\bibinfo  {journal} {A\&A}\ }\textbf {\bibinfo {volume} {616}},\ \bibinfo {eid} {A184} (\bibinfo {year} {2018})},\ \Eprint {https://arxiv.org/abs/1803.04427} {arXiv:1803.04427 [astro-ph.HE]} \BibitemShut {NoStop}%
\bibitem [{\citenamefont {{Chen}}\ and\ \citenamefont {{Yuan}}(2020)}]{Chen20}%
  \BibitemOpen
  \bibfield  {author} {\bibinfo {author} {\bibfnamefont {A.~Y.}\ \bibnamefont {{Chen}}}\ and\ \bibinfo {author} {\bibfnamefont {Y.}~\bibnamefont {{Yuan}}},\ }\href {https://doi.org/10.3847/1538-4357/ab8c46} {\bibfield  {journal} {\bibinfo  {journal} {ApJ}\ }\textbf {\bibinfo {volume} {895}},\ \bibinfo {eid} {121} (\bibinfo {year} {2020})},\ \Eprint {https://arxiv.org/abs/1908.06919} {arXiv:1908.06919 [astro-ph.HE]} \BibitemShut {NoStop}%
\bibitem [{\citenamefont {{Kisaka}}\ \emph {et~al.}(2020)\citenamefont {{Kisaka}}, \citenamefont {{Levinson}},\ and\ \citenamefont {{Toma}}}]{Kisaka20}%
  \BibitemOpen
  \bibfield  {author} {\bibinfo {author} {\bibfnamefont {S.}~\bibnamefont {{Kisaka}}}, \bibinfo {author} {\bibfnamefont {A.}~\bibnamefont {{Levinson}}},\ and\ \bibinfo {author} {\bibfnamefont {K.}~\bibnamefont {{Toma}}},\ }\href {https://doi.org/10.3847/1538-4357/abb46c} {\bibfield  {journal} {\bibinfo  {journal} {ApJ}\ }\textbf {\bibinfo {volume} {902}},\ \bibinfo {eid} {80} (\bibinfo {year} {2020})},\ \Eprint {https://arxiv.org/abs/2007.02838} {arXiv:2007.02838 [astro-ph.HE]} \BibitemShut {NoStop}%
\bibitem [{\citenamefont {{Kisaka}}\ \emph {et~al.}(2022)\citenamefont {{Kisaka}}, \citenamefont {{Levinson}}, \citenamefont {{Toma}},\ and\ \citenamefont {{Niv}}}]{Kisaka22}%
  \BibitemOpen
  \bibfield  {author} {\bibinfo {author} {\bibfnamefont {S.}~\bibnamefont {{Kisaka}}}, \bibinfo {author} {\bibfnamefont {A.}~\bibnamefont {{Levinson}}}, \bibinfo {author} {\bibfnamefont {K.}~\bibnamefont {{Toma}}},\ and\ \bibinfo {author} {\bibfnamefont {I.}~\bibnamefont {{Niv}}},\ }\href {https://doi.org/10.3847/1538-4357/ac35da} {\bibfield  {journal} {\bibinfo  {journal} {ApJ}\ }\textbf {\bibinfo {volume} {924}},\ \bibinfo {eid} {28} (\bibinfo {year} {2022})},\ \Eprint {https://arxiv.org/abs/2108.02971} {arXiv:2108.02971 [astro-ph.HE]} \BibitemShut {NoStop}%
\bibitem [{\citenamefont {{Kin}}\ \emph {et~al.}(2024)\citenamefont {{Kin}}, \citenamefont {{Kisaka}}, \citenamefont {{Toma}}, \citenamefont {{Kimura}},\ and\ \citenamefont {{Levinson}}}]{Kin24}%
  \BibitemOpen
  \bibfield  {author} {\bibinfo {author} {\bibfnamefont {K.}~\bibnamefont {{Kin}}}, \bibinfo {author} {\bibfnamefont {S.}~\bibnamefont {{Kisaka}}}, \bibinfo {author} {\bibfnamefont {K.}~\bibnamefont {{Toma}}}, \bibinfo {author} {\bibfnamefont {S.~S.}\ \bibnamefont {{Kimura}}},\ and\ \bibinfo {author} {\bibfnamefont {A.}~\bibnamefont {{Levinson}}},\ }\href {https://doi.org/10.3847/1538-4357/ad20cd} {\bibfield  {journal} {\bibinfo  {journal} {ApJ}\ }\textbf {\bibinfo {volume} {964}},\ \bibinfo {eid} {78} (\bibinfo {year} {2024})},\ \Eprint {https://arxiv.org/abs/2310.12532} {arXiv:2310.12532 [astro-ph.HE]} \BibitemShut {NoStop}%
\bibitem [{\citenamefont {{Kin}}\ \emph {et~al.}(2025)\citenamefont {{Kin}}, \citenamefont {{Kuze}},\ and\ \citenamefont {{Kimura}}}]{Kin25}%
  \BibitemOpen
  \bibfield  {author} {\bibinfo {author} {\bibfnamefont {K.}~\bibnamefont {{Kin}}}, \bibinfo {author} {\bibfnamefont {R.}~\bibnamefont {{Kuze}}},\ and\ \bibinfo {author} {\bibfnamefont {S.~S.}\ \bibnamefont {{Kimura}}},\ }\href {https://doi.org/10.3847/1538-4357/adcb3d} {\bibfield  {journal} {\bibinfo  {journal} {ApJ}\ }\textbf {\bibinfo {volume} {985}},\ \bibinfo {eid} {251} (\bibinfo {year} {2025})},\ \Eprint {https://arxiv.org/abs/2502.09181} {arXiv:2502.09181 [astro-ph.HE]} \BibitemShut {NoStop}%
\bibitem [{\citenamefont {{Parfrey}}\ \emph {et~al.}(2019)\citenamefont {{Parfrey}}, \citenamefont {{Philippov}},\ and\ \citenamefont {{Cerutti}}}]{Parfrey19}%
  \BibitemOpen
  \bibfield  {author} {\bibinfo {author} {\bibfnamefont {K.}~\bibnamefont {{Parfrey}}}, \bibinfo {author} {\bibfnamefont {A.}~\bibnamefont {{Philippov}}},\ and\ \bibinfo {author} {\bibfnamefont {B.}~\bibnamefont {{Cerutti}}},\ }\href {https://doi.org/10.1103/PhysRevLett.122.035101} {\bibfield  {journal} {\bibinfo  {journal} {\prl}\ }\textbf {\bibinfo {volume} {122}},\ \bibinfo {eid} {035101} (\bibinfo {year} {2019})},\ \Eprint {https://arxiv.org/abs/1810.03613} {arXiv:1810.03613 [astro-ph.HE]} \BibitemShut {NoStop}%
\bibitem [{\citenamefont {{Crinquand}}\ \emph {et~al.}(2020)\citenamefont {{Crinquand}}, \citenamefont {{Cerutti}}, \citenamefont {{Philippov}}, \citenamefont {{Parfrey}},\ and\ \citenamefont {{Dubus}}}]{Crin20}%
  \BibitemOpen
  \bibfield  {author} {\bibinfo {author} {\bibfnamefont {B.}~\bibnamefont {{Crinquand}}}, \bibinfo {author} {\bibfnamefont {B.}~\bibnamefont {{Cerutti}}}, \bibinfo {author} {\bibfnamefont {A.}~\bibnamefont {{Philippov}}}, \bibinfo {author} {\bibfnamefont {K.}~\bibnamefont {{Parfrey}}},\ and\ \bibinfo {author} {\bibfnamefont {G.}~\bibnamefont {{Dubus}}},\ }\href {https://doi.org/10.1103/PhysRevLett.124.145101} {\bibfield  {journal} {\bibinfo  {journal} {\prl}\ }\textbf {\bibinfo {volume} {124}},\ \bibinfo {eid} {145101} (\bibinfo {year} {2020})},\ \Eprint {https://arxiv.org/abs/2003.03548} {arXiv:2003.03548 [astro-ph.HE]} \BibitemShut {NoStop}%
\bibitem [{\citenamefont {{Crinquand}}\ \emph {et~al.}(2021)\citenamefont {{Crinquand}}, \citenamefont {{Cerutti}}, \citenamefont {{Dubus}}, \citenamefont {{Parfrey}},\ and\ \citenamefont {{Philippov}}}]{Crin21}%
  \BibitemOpen
  \bibfield  {author} {\bibinfo {author} {\bibfnamefont {B.}~\bibnamefont {{Crinquand}}}, \bibinfo {author} {\bibfnamefont {B.}~\bibnamefont {{Cerutti}}}, \bibinfo {author} {\bibfnamefont {G.}~\bibnamefont {{Dubus}}}, \bibinfo {author} {\bibfnamefont {K.}~\bibnamefont {{Parfrey}}},\ and\ \bibinfo {author} {\bibfnamefont {A.}~\bibnamefont {{Philippov}}},\ }\href {https://doi.org/10.1051/0004-6361/202040158} {\bibfield  {journal} {\bibinfo  {journal} {A\&A}\ }\textbf {\bibinfo {volume} {650}},\ \bibinfo {eid} {A163} (\bibinfo {year} {2021})},\ \Eprint {https://arxiv.org/abs/2012.09733} {arXiv:2012.09733 [astro-ph.HE]} \BibitemShut {NoStop}%
\bibitem [{\citenamefont {{Hirotani}}\ \emph {et~al.}(2021)\citenamefont {{Hirotani}}, \citenamefont {{Krasnopolsky}}, \citenamefont {{Shang}}, \citenamefont {{Nishikawa}},\ and\ \citenamefont {{Watson}}}]{Hirotani21}%
  \BibitemOpen
  \bibfield  {author} {\bibinfo {author} {\bibfnamefont {K.}~\bibnamefont {{Hirotani}}}, \bibinfo {author} {\bibfnamefont {R.}~\bibnamefont {{Krasnopolsky}}}, \bibinfo {author} {\bibfnamefont {H.}~\bibnamefont {{Shang}}}, \bibinfo {author} {\bibfnamefont {K.-i.}\ \bibnamefont {{Nishikawa}}},\ and\ \bibinfo {author} {\bibfnamefont {M.}~\bibnamefont {{Watson}}},\ }\href {https://doi.org/10.3847/1538-4357/abd3a6} {\bibfield  {journal} {\bibinfo  {journal} {ApJ}\ }\textbf {\bibinfo {volume} {908}},\ \bibinfo {eid} {88} (\bibinfo {year} {2021})},\ \Eprint {https://arxiv.org/abs/2012.07229} {arXiv:2012.07229 [astro-ph.HE]} \BibitemShut {NoStop}%
\bibitem [{\citenamefont {{El Mellah}}\ \emph {et~al.}(2023)\citenamefont {{El Mellah}}, \citenamefont {{Cerutti}},\ and\ \citenamefont {{Crinquand}}}]{ElMellah23}%
  \BibitemOpen
  \bibfield  {author} {\bibinfo {author} {\bibfnamefont {I.}~\bibnamefont {{El Mellah}}}, \bibinfo {author} {\bibfnamefont {B.}~\bibnamefont {{Cerutti}}},\ and\ \bibinfo {author} {\bibfnamefont {B.}~\bibnamefont {{Crinquand}}},\ }\href {https://doi.org/10.1051/0004-6361/202346781} {\bibfield  {journal} {\bibinfo  {journal} {A\&A}\ }\textbf {\bibinfo {volume} {677}},\ \bibinfo {eid} {A67} (\bibinfo {year} {2023})},\ \Eprint {https://arxiv.org/abs/2305.01689} {arXiv:2305.01689 [astro-ph.HE]} \BibitemShut {NoStop}%
\bibitem [{\citenamefont {{Hirotani}}\ \emph {et~al.}(2023)\citenamefont {{Hirotani}}, \citenamefont {{Shang}}, \citenamefont {{Krasnopolsky}},\ and\ \citenamefont {{Nishikawa}}}]{Hirotani23}%
  \BibitemOpen
  \bibfield  {author} {\bibinfo {author} {\bibfnamefont {K.}~\bibnamefont {{Hirotani}}}, \bibinfo {author} {\bibfnamefont {H.}~\bibnamefont {{Shang}}}, \bibinfo {author} {\bibfnamefont {R.}~\bibnamefont {{Krasnopolsky}}},\ and\ \bibinfo {author} {\bibfnamefont {K.}~\bibnamefont {{Nishikawa}}},\ }\href {https://doi.org/10.3847/1538-4357/aca8b0} {\bibfield  {journal} {\bibinfo  {journal} {ApJ}\ }\textbf {\bibinfo {volume} {943}},\ \bibinfo {eid} {164} (\bibinfo {year} {2023})},\ \Eprint {https://arxiv.org/abs/2209.13140} {arXiv:2209.13140 [astro-ph.HE]} \BibitemShut {NoStop}%
\bibitem [{\citenamefont {{Niv}}\ \emph {et~al.}(2023)\citenamefont {{Niv}}, \citenamefont {{Bromberg}}, \citenamefont {{Levinson}}, \citenamefont {{Cerutti}},\ and\ \citenamefont {{Crinquand}}}]{Niv23}%
  \BibitemOpen
  \bibfield  {author} {\bibinfo {author} {\bibfnamefont {I.}~\bibnamefont {{Niv}}}, \bibinfo {author} {\bibfnamefont {O.}~\bibnamefont {{Bromberg}}}, \bibinfo {author} {\bibfnamefont {A.}~\bibnamefont {{Levinson}}}, \bibinfo {author} {\bibfnamefont {B.}~\bibnamefont {{Cerutti}}},\ and\ \bibinfo {author} {\bibfnamefont {B.}~\bibnamefont {{Crinquand}}},\ }\href {https://doi.org/10.1093/mnras/stad2904} {\bibfield  {journal} {\bibinfo  {journal} {MNRAS}\ }\textbf {\bibinfo {volume} {526}},\ \bibinfo {pages} {2709} (\bibinfo {year} {2023})},\ \Eprint {https://arxiv.org/abs/2306.09161} {arXiv:2306.09161 [astro-ph.HE]} \BibitemShut {NoStop}%
\bibitem [{\citenamefont {{Chen}}\ \emph {et~al.}(2025)\citenamefont {{Chen}}, \citenamefont {{Luepker}},\ and\ \citenamefont {{Yuan}}}]{Chen25}%
  \BibitemOpen
  \bibfield  {author} {\bibinfo {author} {\bibfnamefont {A.~Y.}\ \bibnamefont {{Chen}}}, \bibinfo {author} {\bibfnamefont {M.}~\bibnamefont {{Luepker}}},\ and\ \bibinfo {author} {\bibfnamefont {Y.}~\bibnamefont {{Yuan}}},\ }\href {https://doi.org/10.48550/arXiv.2503.04558} {\bibfield  {journal} {\bibinfo  {journal} {arXiv e-prints}\ ,\ \bibinfo {eid} {arXiv:2503.04558}} (\bibinfo {year} {2025})},\ \Eprint {https://arxiv.org/abs/2503.04558} {arXiv:2503.04558 [astro-ph.HE]} \BibitemShut {NoStop}%
\bibitem [{\citenamefont {{Vos}}\ \emph {et~al.}(2025)\citenamefont {{Vos}}, \citenamefont {{Cerutti}}, \citenamefont {{Mo{\'s}cibrodzka}},\ and\ \citenamefont {{Parfrey}}}]{Vos25}%
  \BibitemOpen
  \bibfield  {author} {\bibinfo {author} {\bibfnamefont {J.}~\bibnamefont {{Vos}}}, \bibinfo {author} {\bibfnamefont {B.}~\bibnamefont {{Cerutti}}}, \bibinfo {author} {\bibfnamefont {M.}~\bibnamefont {{Mo{\'s}cibrodzka}}},\ and\ \bibinfo {author} {\bibfnamefont {K.}~\bibnamefont {{Parfrey}}},\ }\href {https://doi.org/10.1103/5wwb-7t2n} {\bibfield  {journal} {\bibinfo  {journal} {\prl}\ }\textbf {\bibinfo {volume} {135}},\ \bibinfo {eid} {015201} (\bibinfo {year} {2025})},\ \Eprint {https://arxiv.org/abs/2410.19061} {arXiv:2410.19061 [astro-ph.HE]} \BibitemShut {NoStop}%
\bibitem [{\citenamefont {{Yuan}}\ \emph {et~al.}(2025)\citenamefont {{Yuan}}, \citenamefont {{Chen}},\ and\ \citenamefont {{Luepker}}}]{Yuan25}%
  \BibitemOpen
  \bibfield  {author} {\bibinfo {author} {\bibfnamefont {Y.}~\bibnamefont {{Yuan}}}, \bibinfo {author} {\bibfnamefont {A.~Y.}\ \bibnamefont {{Chen}}},\ and\ \bibinfo {author} {\bibfnamefont {M.}~\bibnamefont {{Luepker}}},\ }\href {https://doi.org/10.3847/1538-4357/adce79} {\bibfield  {journal} {\bibinfo  {journal} {ApJ}\ }\textbf {\bibinfo {volume} {985}},\ \bibinfo {eid} {159} (\bibinfo {year} {2025})},\ \Eprint {https://arxiv.org/abs/2503.08487} {arXiv:2503.08487 [astro-ph.HE]} \BibitemShut {NoStop}%
\bibitem [{\citenamefont {{Mehlhaff}}\ \emph {et~al.}(2025)\citenamefont {{Mehlhaff}}, \citenamefont {{Cerutti}},\ and\ \citenamefont {{Crinquand}}}]{Mehlhaff25}%
  \BibitemOpen
  \bibfield  {author} {\bibinfo {author} {\bibfnamefont {J.}~\bibnamefont {{Mehlhaff}}}, \bibinfo {author} {\bibfnamefont {B.}~\bibnamefont {{Cerutti}}},\ and\ \bibinfo {author} {\bibfnamefont {B.}~\bibnamefont {{Crinquand}}},\ }\href {https://doi.org/10.1051/0004-6361/202453561} {\bibfield  {journal} {\bibinfo  {journal} {A\&A}\ }\textbf {\bibinfo {volume} {701}},\ \bibinfo {eid} {A62} (\bibinfo {year} {2025})},\ \Eprint {https://arxiv.org/abs/2504.01062} {arXiv:2504.01062 [astro-ph.HE]} \BibitemShut {NoStop}%
\bibitem [{\citenamefont {{Meringolo}}\ \emph {et~al.}(2025)\citenamefont {{Meringolo}}, \citenamefont {{Camilloni}},\ and\ \citenamefont {{Rezzolla}}}]{Meringolo25}%
  \BibitemOpen
  \bibfield  {author} {\bibinfo {author} {\bibfnamefont {C.}~\bibnamefont {{Meringolo}}}, \bibinfo {author} {\bibfnamefont {F.}~\bibnamefont {{Camilloni}}},\ and\ \bibinfo {author} {\bibfnamefont {L.}~\bibnamefont {{Rezzolla}}},\ }\href {https://doi.org/10.3847/2041-8213/ae06a6} {\bibfield  {journal} {\bibinfo  {journal} {ApJL}\ }\textbf {\bibinfo {volume} {992}},\ \bibinfo {eid} {L8} (\bibinfo {year} {2025})},\ \Eprint {https://arxiv.org/abs/2507.08942} {arXiv:2507.08942 [gr-qc]} \BibitemShut {NoStop}%
\bibitem [{\citenamefont {{Figueiredo}}\ \emph {et~al.}(2025)\citenamefont {{Figueiredo}}, \citenamefont {{Cerutti}},\ and\ \citenamefont {{Parfrey}}}]{Figueiredo25}%
  \BibitemOpen
  \bibfield  {author} {\bibinfo {author} {\bibfnamefont {E.}~\bibnamefont {{Figueiredo}}}, \bibinfo {author} {\bibfnamefont {B.}~\bibnamefont {{Cerutti}}},\ and\ \bibinfo {author} {\bibfnamefont {K.}~\bibnamefont {{Parfrey}}},\ }\href {https://doi.org/10.1051/0004-6361/202555826} {\bibfield  {journal} {\bibinfo  {journal} {A\&A}\ }\textbf {\bibinfo {volume} {700}},\ \bibinfo {eid} {L19} (\bibinfo {year} {2025})},\ \Eprint {https://arxiv.org/abs/2508.02211} {arXiv:2508.02211 [astro-ph.HE]} \BibitemShut {NoStop}%
\bibitem [{\citenamefont {{Figueiredo}}\ \emph {et~al.}(2026)\citenamefont {{Figueiredo}}, \citenamefont {{Mehlhaff}}, \citenamefont {{Soudais}},\ and\ \citenamefont {{Cerutti}}}]{Figueiredo26}%
  \BibitemOpen
  \bibfield  {author} {\bibinfo {author} {\bibfnamefont {E.}~\bibnamefont {{Figueiredo}}}, \bibinfo {author} {\bibfnamefont {J.}~\bibnamefont {{Mehlhaff}}}, \bibinfo {author} {\bibfnamefont {A.}~\bibnamefont {{Soudais}}},\ and\ \bibinfo {author} {\bibfnamefont {B.}~\bibnamefont {{Cerutti}}},\ }\href {https://doi.org/10.1051/0004-6361/202659313} {\bibfield  {journal} {\bibinfo  {journal} {A\&A}\ }\textbf {\bibinfo {volume} {711}},\ \bibinfo {eid} {A230} (\bibinfo {year} {2026})},\ \Eprint {https://arxiv.org/abs/2602.04519} {arXiv:2602.04519 [astro-ph.HE]} \BibitemShut {NoStop}%
\bibitem [{\citenamefont {{Mehlhaff}}\ \emph {et~al.}(2026)\citenamefont {{Mehlhaff}}, \citenamefont {{Chen}}, \citenamefont {{Luepker}},\ and\ \citenamefont {{Yuan}}}]{Mehlhaff26}%
  \BibitemOpen
  \bibfield  {author} {\bibinfo {author} {\bibfnamefont {J.~M.}\ \bibnamefont {{Mehlhaff}}}, \bibinfo {author} {\bibfnamefont {A.~Y.}\ \bibnamefont {{Chen}}}, \bibinfo {author} {\bibfnamefont {M.}~\bibnamefont {{Luepker}}},\ and\ \bibinfo {author} {\bibfnamefont {Y.}~\bibnamefont {{Yuan}}},\ }\href {https://doi.org/10.1103/n92m-jvpj} {\bibfield  {journal} {\bibinfo  {journal} {\prl}\ }\textbf {\bibinfo {volume} {137}},\ \bibinfo {eid} {155201} (\bibinfo {year} {2026})},\ \Eprint {https://arxiv.org/abs/2602.22168} {arXiv:2602.22168 [astro-ph.HE]} \BibitemShut {NoStop}%
\bibitem [{\citenamefont {{Meringolo}}\ and\ \citenamefont {{Rezzolla}}(2026)}]{Meringolo26}%
  \BibitemOpen
  \bibfield  {author} {\bibinfo {author} {\bibfnamefont {C.}~\bibnamefont {{Meringolo}}}\ and\ \bibinfo {author} {\bibfnamefont {L.}~\bibnamefont {{Rezzolla}}},\ }\href {https://doi.org/10.1016/j.ascom.2026.101154} {\bibfield  {journal} {\bibinfo  {journal} {Astronomy and Computing}\ }\textbf {\bibinfo {volume} {57}},\ \bibinfo {eid} {101154} (\bibinfo {year} {2026})}\BibitemShut {NoStop}%
\bibitem [{\citenamefont {{Hirotani}}\ \emph {et~al.}(2024)\citenamefont {{Hirotani}}, \citenamefont {{Shang}}, \citenamefont {{Krasnopolsky}},\ and\ \citenamefont {{Nishikawa}}}]{Hirotani24}%
  \BibitemOpen
  \bibfield  {author} {\bibinfo {author} {\bibfnamefont {K.}~\bibnamefont {{Hirotani}}}, \bibinfo {author} {\bibfnamefont {H.}~\bibnamefont {{Shang}}}, \bibinfo {author} {\bibfnamefont {R.}~\bibnamefont {{Krasnopolsky}}},\ and\ \bibinfo {author} {\bibfnamefont {K.}~\bibnamefont {{Nishikawa}}},\ }\href {https://doi.org/10.3847/1538-4357/ad28c7} {\bibfield  {journal} {\bibinfo  {journal} {ApJ}\ }\textbf {\bibinfo {volume} {965}},\ \bibinfo {eid} {50} (\bibinfo {year} {2024})},\ \Eprint {https://arxiv.org/abs/2403.03574} {arXiv:2403.03574 [astro-ph.HE]} \BibitemShut {NoStop}%
\bibitem [{\citenamefont {{Acciari}}\ \emph {et~al.}(2008)\citenamefont {{Acciari}}, \citenamefont {{Beilicke}}, \citenamefont {{Blaylock}}, \citenamefont {{Bradbury}}, \citenamefont {{Buckley}}, \citenamefont {{Bugaev}}, \citenamefont {{Butt}}, \citenamefont {{Celik}}, \citenamefont {{Cesarini}}, \citenamefont {{Ciupik}}, \citenamefont {{Cogan}}, \citenamefont {{Colin}}, \citenamefont {{Cui}}, \citenamefont {{Daniel}}, \citenamefont {{Duke}}, \citenamefont {{Ergin}}, \citenamefont {{Falcone}}, \citenamefont {{Fegan}}, \citenamefont {{Finley}}, \citenamefont {{Finnegan}}, \citenamefont {{Fortin}}, \citenamefont {{Fortson}}, \citenamefont {{Gibbs}}, \citenamefont {{Gillanders}}, \citenamefont {{Grube}}, \citenamefont {{Guenette}}, \citenamefont {{Gyuk}}, \citenamefont {{Hanna}}, \citenamefont {{Hays}}, \citenamefont {{Holder}}, \citenamefont {{Horan}}, \citenamefont {{Hughes}}, \citenamefont {{Hui}}, \citenamefont {{Humensky}}, \citenamefont {{Imran}}, \citenamefont {{Kaaret}}, \citenamefont {{Kertzman}},
  \citenamefont {{Kieda}}, \citenamefont {{Kildea}}, \citenamefont {{Konopelko}}, \citenamefont {{Krawczynski}}, \citenamefont {{Krennrich}}, \citenamefont {{Lang}}, \citenamefont {{LeBohec}}, \citenamefont {{Lee}}, \citenamefont {{Maier}}, \citenamefont {{McCann}}, \citenamefont {{McCutcheon}}, \citenamefont {{Millis}}, \citenamefont {{Moriarty}}, \citenamefont {{Mukherjee}}, \citenamefont {{Nagai}}, \citenamefont {{Ong}}, \citenamefont {{Pandel}}, \citenamefont {{Perkins}}, \citenamefont {{Pohl}}, \citenamefont {{Quinn}}, \citenamefont {{Ragan}}, \citenamefont {{Reynolds}}, \citenamefont {{Rose}}, \citenamefont {{Schroedter}}, \citenamefont {{Sembroski}}, \citenamefont {{Smith}}, \citenamefont {{Steele}}, \citenamefont {{Swordy}}, \citenamefont {{Syson}}, \citenamefont {{Toner}}, \citenamefont {{Valcarcel}}, \citenamefont {{Vassiliev}}, \citenamefont {{Wakely}}, \citenamefont {{Ward}}, \citenamefont {{Weekes}}, \citenamefont {{Weinstein}}, \citenamefont {{White}}, \citenamefont {{Williams}}, \citenamefont
  {{Wissel}}, \citenamefont {{Wood}},\ and\ \citenamefont {{Zitzer}}}]{Acciari08}%
  \BibitemOpen
  \bibfield  {author} {\bibinfo {author} {\bibfnamefont {V.~A.}\ \bibnamefont {{Acciari}}}, \bibinfo {author} {\bibfnamefont {M.}~\bibnamefont {{Beilicke}}}, \bibinfo {author} {\bibfnamefont {G.}~\bibnamefont {{Blaylock}}}, \bibinfo {author} {\bibfnamefont {S.~M.}\ \bibnamefont {{Bradbury}}}, \bibinfo {author} {\bibfnamefont {J.~H.}\ \bibnamefont {{Buckley}}}, \bibinfo {author} {\bibfnamefont {V.}~\bibnamefont {{Bugaev}}}, \bibinfo {author} {\bibfnamefont {Y.}~\bibnamefont {{Butt}}}, \bibinfo {author} {\bibfnamefont {O.}~\bibnamefont {{Celik}}}, \bibinfo {author} {\bibfnamefont {A.}~\bibnamefont {{Cesarini}}}, \bibinfo {author} {\bibfnamefont {L.}~\bibnamefont {{Ciupik}}}, \bibinfo {author} {\bibfnamefont {P.}~\bibnamefont {{Cogan}}}, \bibinfo {author} {\bibfnamefont {P.}~\bibnamefont {{Colin}}}, \bibinfo {author} {\bibfnamefont {W.}~\bibnamefont {{Cui}}}, \bibinfo {author} {\bibfnamefont {M.~K.}\ \bibnamefont {{Daniel}}}, \bibinfo {author} {\bibfnamefont {C.}~\bibnamefont {{Duke}}}, \bibinfo {author}
  {\bibfnamefont {T.}~\bibnamefont {{Ergin}}}, \bibinfo {author} {\bibfnamefont {A.~D.}\ \bibnamefont {{Falcone}}}, \bibinfo {author} {\bibfnamefont {S.~J.}\ \bibnamefont {{Fegan}}}, \bibinfo {author} {\bibfnamefont {J.~P.}\ \bibnamefont {{Finley}}}, \bibinfo {author} {\bibfnamefont {G.}~\bibnamefont {{Finnegan}}}, \bibinfo {author} {\bibfnamefont {P.}~\bibnamefont {{Fortin}}}, \bibinfo {author} {\bibfnamefont {L.~F.}\ \bibnamefont {{Fortson}}}, \bibinfo {author} {\bibfnamefont {K.}~\bibnamefont {{Gibbs}}}, \bibinfo {author} {\bibfnamefont {G.~H.}\ \bibnamefont {{Gillanders}}}, \bibinfo {author} {\bibfnamefont {J.}~\bibnamefont {{Grube}}}, \bibinfo {author} {\bibfnamefont {R.}~\bibnamefont {{Guenette}}}, \bibinfo {author} {\bibfnamefont {G.}~\bibnamefont {{Gyuk}}}, \bibinfo {author} {\bibfnamefont {D.}~\bibnamefont {{Hanna}}}, \bibinfo {author} {\bibfnamefont {E.}~\bibnamefont {{Hays}}}, \bibinfo {author} {\bibfnamefont {J.}~\bibnamefont {{Holder}}}, \bibinfo {author} {\bibfnamefont {D.}~\bibnamefont
  {{Horan}}}, \bibinfo {author} {\bibfnamefont {S.~B.}\ \bibnamefont {{Hughes}}}, \bibinfo {author} {\bibfnamefont {M.~C.}\ \bibnamefont {{Hui}}}, \bibinfo {author} {\bibfnamefont {T.~B.}\ \bibnamefont {{Humensky}}}, \bibinfo {author} {\bibfnamefont {A.}~\bibnamefont {{Imran}}}, \bibinfo {author} {\bibfnamefont {P.}~\bibnamefont {{Kaaret}}}, \bibinfo {author} {\bibfnamefont {M.}~\bibnamefont {{Kertzman}}}, \bibinfo {author} {\bibfnamefont {D.~B.}\ \bibnamefont {{Kieda}}}, \bibinfo {author} {\bibfnamefont {J.}~\bibnamefont {{Kildea}}}, \bibinfo {author} {\bibfnamefont {A.}~\bibnamefont {{Konopelko}}}, \bibinfo {author} {\bibfnamefont {H.}~\bibnamefont {{Krawczynski}}}, \bibinfo {author} {\bibfnamefont {F.}~\bibnamefont {{Krennrich}}}, \bibinfo {author} {\bibfnamefont {M.~J.}\ \bibnamefont {{Lang}}}, \bibinfo {author} {\bibfnamefont {S.}~\bibnamefont {{LeBohec}}}, \bibinfo {author} {\bibfnamefont {K.}~\bibnamefont {{Lee}}}, \bibinfo {author} {\bibfnamefont {G.}~\bibnamefont {{Maier}}}, \bibinfo {author}
  {\bibfnamefont {A.}~\bibnamefont {{McCann}}}, \bibinfo {author} {\bibfnamefont {M.}~\bibnamefont {{McCutcheon}}}, \bibinfo {author} {\bibfnamefont {J.}~\bibnamefont {{Millis}}}, \bibinfo {author} {\bibfnamefont {P.}~\bibnamefont {{Moriarty}}}, \bibinfo {author} {\bibfnamefont {R.}~\bibnamefont {{Mukherjee}}}, \bibinfo {author} {\bibfnamefont {T.}~\bibnamefont {{Nagai}}}, \bibinfo {author} {\bibfnamefont {R.~A.}\ \bibnamefont {{Ong}}}, \bibinfo {author} {\bibfnamefont {D.}~\bibnamefont {{Pandel}}}, \bibinfo {author} {\bibfnamefont {J.~S.}\ \bibnamefont {{Perkins}}}, \bibinfo {author} {\bibfnamefont {M.}~\bibnamefont {{Pohl}}}, \bibinfo {author} {\bibfnamefont {J.}~\bibnamefont {{Quinn}}}, \bibinfo {author} {\bibfnamefont {K.}~\bibnamefont {{Ragan}}}, \bibinfo {author} {\bibfnamefont {P.~T.}\ \bibnamefont {{Reynolds}}}, \bibinfo {author} {\bibfnamefont {H.~J.}\ \bibnamefont {{Rose}}}, \bibinfo {author} {\bibfnamefont {M.}~\bibnamefont {{Schroedter}}}, \bibinfo {author} {\bibfnamefont {G.~H.}\ \bibnamefont
  {{Sembroski}}}, \bibinfo {author} {\bibfnamefont {A.~W.}\ \bibnamefont {{Smith}}}, \bibinfo {author} {\bibfnamefont {D.}~\bibnamefont {{Steele}}}, \bibinfo {author} {\bibfnamefont {S.~P.}\ \bibnamefont {{Swordy}}}, \bibinfo {author} {\bibfnamefont {A.}~\bibnamefont {{Syson}}}, \bibinfo {author} {\bibfnamefont {J.~A.}\ \bibnamefont {{Toner}}}, \bibinfo {author} {\bibfnamefont {L.}~\bibnamefont {{Valcarcel}}}, \bibinfo {author} {\bibfnamefont {V.~V.}\ \bibnamefont {{Vassiliev}}}, \bibinfo {author} {\bibfnamefont {S.~P.}\ \bibnamefont {{Wakely}}}, \bibinfo {author} {\bibfnamefont {J.~E.}\ \bibnamefont {{Ward}}}, \bibinfo {author} {\bibfnamefont {T.~C.}\ \bibnamefont {{Weekes}}}, \bibinfo {author} {\bibfnamefont {A.}~\bibnamefont {{Weinstein}}}, \bibinfo {author} {\bibfnamefont {R.~J.}\ \bibnamefont {{White}}}, \bibinfo {author} {\bibfnamefont {D.~A.}\ \bibnamefont {{Williams}}}, \bibinfo {author} {\bibfnamefont {S.~A.}\ \bibnamefont {{Wissel}}}, \bibinfo {author} {\bibfnamefont {M.~D.}\ \bibnamefont
  {{Wood}}},\ and\ \bibinfo {author} {\bibfnamefont {B.}~\bibnamefont {{Zitzer}}},\ }\href {https://doi.org/10.1086/587458} {\bibfield  {journal} {\bibinfo  {journal} {ApJ}\ }\textbf {\bibinfo {volume} {679}},\ \bibinfo {pages} {397} (\bibinfo {year} {2008})},\ \Eprint {https://arxiv.org/abs/0802.1951} {arXiv:0802.1951 [astro-ph]} \BibitemShut {NoStop}%
\bibitem [{\citenamefont {{Albert}}\ \emph {et~al.}(2008)\citenamefont {{Albert}}, \citenamefont {{Aliu}}, \citenamefont {{Anderhub}}, \citenamefont {{Antonelli}}, \citenamefont {{Antoranz}}, \citenamefont {{Backes}}, \citenamefont {{Baixeras}}, \citenamefont {{Barrio}}, \citenamefont {{Bartko}}, \citenamefont {{Bastieri}}, \citenamefont {{Becker}}, \citenamefont {{Bednarek}}, \citenamefont {{Berger}}, \citenamefont {{Bernardini}}, \citenamefont {{Bigongiari}}, \citenamefont {{Biland}}, \citenamefont {{Bock}}, \citenamefont {{Bonnoli}}, \citenamefont {{Bordas}}, \citenamefont {{Bosch-Ramon}}, \citenamefont {{Bretz}}, \citenamefont {{Britvitch}}, \citenamefont {{Camara}}, \citenamefont {{Carmona}}, \citenamefont {{Chilingarian}}, \citenamefont {{Commichau}}, \citenamefont {{Contreras}}, \citenamefont {{Cortina}}, \citenamefont {{Costado}}, \citenamefont {{Covino}}, \citenamefont {{Curtef}}, \citenamefont {{Dazzi}}, \citenamefont {{De Angelis}}, \citenamefont {{De Cea del Pozo}}, \citenamefont {{de los Reyes}},
  \citenamefont {{De Lotto}}, \citenamefont {{De Maria}}, \citenamefont {{De Sabata}}, \citenamefont {{Delgado Mendez}}, \citenamefont {{Dominguez}}, \citenamefont {{Dorner}}, \citenamefont {{Doro}}, \citenamefont {{Errando}}, \citenamefont {{Fagiolini}}, \citenamefont {{Ferenc}}, \citenamefont {{Fern{\'a}ndez}}, \citenamefont {{Firpo}}, \citenamefont {{Fonseca}}, \citenamefont {{Font}}, \citenamefont {{Galante}}, \citenamefont {{Garc{\'\i}a L{\'o}pez}}, \citenamefont {{Garczarczyk}}, \citenamefont {{Gaug}}, \citenamefont {{Goebel}}, \citenamefont {{Hayashida}}, \citenamefont {{Herrero}}, \citenamefont {{H{\"o}hne}}, \citenamefont {{Hose}}, \citenamefont {{Hsu}}, \citenamefont {{Huber}}, \citenamefont {{Jogler}}, \citenamefont {{Kranich}}, \citenamefont {{La Barbera}}, \citenamefont {{Laille}}, \citenamefont {{Leonardo}}, \citenamefont {{Lindfors}}, \citenamefont {{Lombardi}}, \citenamefont {{Longo}}, \citenamefont {{L{\'o}pez}}, \citenamefont {{Lorenz}}, \citenamefont {{Majumdar}}, \citenamefont {{Maneva}},
  \citenamefont {{Mankuzhiyil}}, \citenamefont {{Mannheim}}, \citenamefont {{Maraschi}}, \citenamefont {{Mariotti}}, \citenamefont {{Mart{\'\i}nez}}, \citenamefont {{Mazin}}, \citenamefont {{Meucci}}, \citenamefont {{Meyer}}, \citenamefont {{Miranda}}, \citenamefont {{Mirzoyan}}, \citenamefont {{Mizobuchi}}, \citenamefont {{Moles}}, \citenamefont {{Moralejo}}, \citenamefont {{Nieto}}, \citenamefont {{Nilsson}}, \citenamefont {{Ninkovic}}, \citenamefont {{Otte}}, \citenamefont {{Oya}}, \citenamefont {{Panniello}}, \citenamefont {{Paoletti}}, \citenamefont {{Paredes}}, \citenamefont {{Pasanen}}, \citenamefont {{Pascoli}}, \citenamefont {{Pauss}}, \citenamefont {{Pegna}}, \citenamefont {{Perez-Torres}}, \citenamefont {{Persic}}, \citenamefont {{Peruzzo}}, \citenamefont {{Piccioli}}, \citenamefont {{Prada}}, \citenamefont {{Prandini}}, \citenamefont {{Puchades}}, \citenamefont {{Raymers}}, \citenamefont {{Rhode}}, \citenamefont {{Rib{\'o}}}, \citenamefont {{Rico}}, \citenamefont {{Rissi}}, \citenamefont
  {{Robert}}, \citenamefont {{R{\"u}gamer}}, \citenamefont {{Saggion}}, \citenamefont {{Saito}}, \citenamefont {{Salvati}}, \citenamefont {{Sanchez-Conde}}, \citenamefont {{Sartori}}, \citenamefont {{Satalecka}}, \citenamefont {{Scalzotto}}, \citenamefont {{Scapin}}, \citenamefont {{Schweizer}}, \citenamefont {{Shayduk}}, \citenamefont {{Shinozaki}}, \citenamefont {{Shore}}, \citenamefont {{Sidro}}, \citenamefont {{Sierpowska-Bartosik}}, \citenamefont {{Sillanp{\"a}{\"a}}}, \citenamefont {{Sobczynska}}, \citenamefont {{Spanier}}, \citenamefont {{Stamerra}}, \citenamefont {{Stark}}, \citenamefont {{Takalo}}, \citenamefont {{Tavecchio}}, \citenamefont {{Temnikov}}, \citenamefont {{Tescaro}}, \citenamefont {{Teshima}}, \citenamefont {{Tluczykont}}, \citenamefont {{Torres}}, \citenamefont {{Turini}}, \citenamefont {{Vankov}}, \citenamefont {{Venturini}}, \citenamefont {{Vitale}}, \citenamefont {{Wagner}}, \citenamefont {{Wittek}}, \citenamefont {{Zabalza}}, \citenamefont {{Zandanel}}, \citenamefont {{Zanin}},\
  and\ \citenamefont {{Zapatero}}}]{Albert08}%
  \BibitemOpen
  \bibfield  {author} {\bibinfo {author} {\bibfnamefont {J.}~\bibnamefont {{Albert}}}, \bibinfo {author} {\bibfnamefont {E.}~\bibnamefont {{Aliu}}}, \bibinfo {author} {\bibfnamefont {H.}~\bibnamefont {{Anderhub}}}, \bibinfo {author} {\bibfnamefont {L.~A.}\ \bibnamefont {{Antonelli}}}, \bibinfo {author} {\bibfnamefont {P.}~\bibnamefont {{Antoranz}}}, \bibinfo {author} {\bibfnamefont {M.}~\bibnamefont {{Backes}}}, \bibinfo {author} {\bibfnamefont {C.}~\bibnamefont {{Baixeras}}}, \bibinfo {author} {\bibfnamefont {J.~A.}\ \bibnamefont {{Barrio}}}, \bibinfo {author} {\bibfnamefont {H.}~\bibnamefont {{Bartko}}}, \bibinfo {author} {\bibfnamefont {D.}~\bibnamefont {{Bastieri}}}, \bibinfo {author} {\bibfnamefont {J.~K.}\ \bibnamefont {{Becker}}}, \bibinfo {author} {\bibfnamefont {W.}~\bibnamefont {{Bednarek}}}, \bibinfo {author} {\bibfnamefont {K.}~\bibnamefont {{Berger}}}, \bibinfo {author} {\bibfnamefont {E.}~\bibnamefont {{Bernardini}}}, \bibinfo {author} {\bibfnamefont {C.}~\bibnamefont {{Bigongiari}}}, \bibinfo
  {author} {\bibfnamefont {A.}~\bibnamefont {{Biland}}}, \bibinfo {author} {\bibfnamefont {R.~K.}\ \bibnamefont {{Bock}}}, \bibinfo {author} {\bibfnamefont {G.}~\bibnamefont {{Bonnoli}}}, \bibinfo {author} {\bibfnamefont {P.}~\bibnamefont {{Bordas}}}, \bibinfo {author} {\bibfnamefont {V.}~\bibnamefont {{Bosch-Ramon}}}, \bibinfo {author} {\bibfnamefont {T.}~\bibnamefont {{Bretz}}}, \bibinfo {author} {\bibfnamefont {I.}~\bibnamefont {{Britvitch}}}, \bibinfo {author} {\bibfnamefont {M.}~\bibnamefont {{Camara}}}, \bibinfo {author} {\bibfnamefont {E.}~\bibnamefont {{Carmona}}}, \bibinfo {author} {\bibfnamefont {A.}~\bibnamefont {{Chilingarian}}}, \bibinfo {author} {\bibfnamefont {S.}~\bibnamefont {{Commichau}}}, \bibinfo {author} {\bibfnamefont {J.~L.}\ \bibnamefont {{Contreras}}}, \bibinfo {author} {\bibfnamefont {J.}~\bibnamefont {{Cortina}}}, \bibinfo {author} {\bibfnamefont {M.~T.}\ \bibnamefont {{Costado}}}, \bibinfo {author} {\bibfnamefont {S.}~\bibnamefont {{Covino}}}, \bibinfo {author} {\bibfnamefont
  {V.}~\bibnamefont {{Curtef}}}, \bibinfo {author} {\bibfnamefont {F.}~\bibnamefont {{Dazzi}}}, \bibinfo {author} {\bibfnamefont {A.}~\bibnamefont {{De Angelis}}}, \bibinfo {author} {\bibfnamefont {E.}~\bibnamefont {{De Cea del Pozo}}}, \bibinfo {author} {\bibfnamefont {R.}~\bibnamefont {{de los Reyes}}}, \bibinfo {author} {\bibfnamefont {B.}~\bibnamefont {{De Lotto}}}, \bibinfo {author} {\bibfnamefont {M.}~\bibnamefont {{De Maria}}}, \bibinfo {author} {\bibfnamefont {F.}~\bibnamefont {{De Sabata}}}, \bibinfo {author} {\bibfnamefont {C.}~\bibnamefont {{Delgado Mendez}}}, \bibinfo {author} {\bibfnamefont {A.}~\bibnamefont {{Dominguez}}}, \bibinfo {author} {\bibfnamefont {D.}~\bibnamefont {{Dorner}}}, \bibinfo {author} {\bibfnamefont {M.}~\bibnamefont {{Doro}}}, \bibinfo {author} {\bibfnamefont {M.}~\bibnamefont {{Errando}}}, \bibinfo {author} {\bibfnamefont {M.}~\bibnamefont {{Fagiolini}}}, \bibinfo {author} {\bibfnamefont {D.}~\bibnamefont {{Ferenc}}}, \bibinfo {author} {\bibfnamefont {E.}~\bibnamefont
  {{Fern{\'a}ndez}}}, \bibinfo {author} {\bibfnamefont {R.}~\bibnamefont {{Firpo}}}, \bibinfo {author} {\bibfnamefont {M.~V.}\ \bibnamefont {{Fonseca}}}, \bibinfo {author} {\bibfnamefont {L.}~\bibnamefont {{Font}}}, \bibinfo {author} {\bibfnamefont {N.}~\bibnamefont {{Galante}}}, \bibinfo {author} {\bibfnamefont {R.~J.}\ \bibnamefont {{Garc{\'\i}a L{\'o}pez}}}, \bibinfo {author} {\bibfnamefont {M.}~\bibnamefont {{Garczarczyk}}}, \bibinfo {author} {\bibfnamefont {M.}~\bibnamefont {{Gaug}}}, \bibinfo {author} {\bibfnamefont {F.}~\bibnamefont {{Goebel}}}, \bibinfo {author} {\bibfnamefont {M.}~\bibnamefont {{Hayashida}}}, \bibinfo {author} {\bibfnamefont {A.}~\bibnamefont {{Herrero}}}, \bibinfo {author} {\bibfnamefont {D.}~\bibnamefont {{H{\"o}hne}}}, \bibinfo {author} {\bibfnamefont {J.}~\bibnamefont {{Hose}}}, \bibinfo {author} {\bibfnamefont {C.~C.}\ \bibnamefont {{Hsu}}}, \bibinfo {author} {\bibfnamefont {S.}~\bibnamefont {{Huber}}}, \bibinfo {author} {\bibfnamefont {T.}~\bibnamefont {{Jogler}}}, \bibinfo
  {author} {\bibfnamefont {D.}~\bibnamefont {{Kranich}}}, \bibinfo {author} {\bibfnamefont {A.}~\bibnamefont {{La Barbera}}}, \bibinfo {author} {\bibfnamefont {A.}~\bibnamefont {{Laille}}}, \bibinfo {author} {\bibfnamefont {E.}~\bibnamefont {{Leonardo}}}, \bibinfo {author} {\bibfnamefont {E.}~\bibnamefont {{Lindfors}}}, \bibinfo {author} {\bibfnamefont {S.}~\bibnamefont {{Lombardi}}}, \bibinfo {author} {\bibfnamefont {F.}~\bibnamefont {{Longo}}}, \bibinfo {author} {\bibfnamefont {M.}~\bibnamefont {{L{\'o}pez}}}, \bibinfo {author} {\bibfnamefont {E.}~\bibnamefont {{Lorenz}}}, \bibinfo {author} {\bibfnamefont {P.}~\bibnamefont {{Majumdar}}}, \bibinfo {author} {\bibfnamefont {G.}~\bibnamefont {{Maneva}}}, \bibinfo {author} {\bibfnamefont {N.}~\bibnamefont {{Mankuzhiyil}}}, \bibinfo {author} {\bibfnamefont {K.}~\bibnamefont {{Mannheim}}}, \bibinfo {author} {\bibfnamefont {L.}~\bibnamefont {{Maraschi}}}, \bibinfo {author} {\bibfnamefont {M.}~\bibnamefont {{Mariotti}}}, \bibinfo {author} {\bibfnamefont
  {M.}~\bibnamefont {{Mart{\'\i}nez}}}, \bibinfo {author} {\bibfnamefont {D.}~\bibnamefont {{Mazin}}}, \bibinfo {author} {\bibfnamefont {M.}~\bibnamefont {{Meucci}}}, \bibinfo {author} {\bibfnamefont {M.}~\bibnamefont {{Meyer}}}, \bibinfo {author} {\bibfnamefont {J.~M.}\ \bibnamefont {{Miranda}}}, \bibinfo {author} {\bibfnamefont {R.}~\bibnamefont {{Mirzoyan}}}, \bibinfo {author} {\bibfnamefont {S.}~\bibnamefont {{Mizobuchi}}}, \bibinfo {author} {\bibfnamefont {M.}~\bibnamefont {{Moles}}}, \bibinfo {author} {\bibfnamefont {A.}~\bibnamefont {{Moralejo}}}, \bibinfo {author} {\bibfnamefont {D.}~\bibnamefont {{Nieto}}}, \bibinfo {author} {\bibfnamefont {K.}~\bibnamefont {{Nilsson}}}, \bibinfo {author} {\bibfnamefont {J.}~\bibnamefont {{Ninkovic}}}, \bibinfo {author} {\bibfnamefont {N.}~\bibnamefont {{Otte}}}, \bibinfo {author} {\bibfnamefont {I.}~\bibnamefont {{Oya}}}, \bibinfo {author} {\bibfnamefont {M.}~\bibnamefont {{Panniello}}}, \bibinfo {author} {\bibfnamefont {R.}~\bibnamefont {{Paoletti}}}, \bibinfo
  {author} {\bibfnamefont {J.~M.}\ \bibnamefont {{Paredes}}}, \bibinfo {author} {\bibfnamefont {M.}~\bibnamefont {{Pasanen}}}, \bibinfo {author} {\bibfnamefont {D.}~\bibnamefont {{Pascoli}}}, \bibinfo {author} {\bibfnamefont {F.}~\bibnamefont {{Pauss}}}, \bibinfo {author} {\bibfnamefont {R.~G.}\ \bibnamefont {{Pegna}}}, \bibinfo {author} {\bibfnamefont {M.~A.}\ \bibnamefont {{Perez-Torres}}}, \bibinfo {author} {\bibfnamefont {M.}~\bibnamefont {{Persic}}}, \bibinfo {author} {\bibfnamefont {L.}~\bibnamefont {{Peruzzo}}}, \bibinfo {author} {\bibfnamefont {A.}~\bibnamefont {{Piccioli}}}, \bibinfo {author} {\bibfnamefont {F.}~\bibnamefont {{Prada}}}, \bibinfo {author} {\bibfnamefont {E.}~\bibnamefont {{Prandini}}}, \bibinfo {author} {\bibfnamefont {N.}~\bibnamefont {{Puchades}}}, \bibinfo {author} {\bibfnamefont {A.}~\bibnamefont {{Raymers}}}, \bibinfo {author} {\bibfnamefont {W.}~\bibnamefont {{Rhode}}}, \bibinfo {author} {\bibfnamefont {M.}~\bibnamefont {{Rib{\'o}}}}, \bibinfo {author} {\bibfnamefont
  {J.}~\bibnamefont {{Rico}}}, \bibinfo {author} {\bibfnamefont {M.}~\bibnamefont {{Rissi}}}, \bibinfo {author} {\bibfnamefont {A.}~\bibnamefont {{Robert}}}, \bibinfo {author} {\bibfnamefont {S.}~\bibnamefont {{R{\"u}gamer}}}, \bibinfo {author} {\bibfnamefont {A.}~\bibnamefont {{Saggion}}}, \bibinfo {author} {\bibfnamefont {T.~Y.}\ \bibnamefont {{Saito}}}, \bibinfo {author} {\bibfnamefont {M.}~\bibnamefont {{Salvati}}}, \bibinfo {author} {\bibfnamefont {M.}~\bibnamefont {{Sanchez-Conde}}}, \bibinfo {author} {\bibfnamefont {P.}~\bibnamefont {{Sartori}}}, \bibinfo {author} {\bibfnamefont {K.}~\bibnamefont {{Satalecka}}}, \bibinfo {author} {\bibfnamefont {V.}~\bibnamefont {{Scalzotto}}}, \bibinfo {author} {\bibfnamefont {V.}~\bibnamefont {{Scapin}}}, \bibinfo {author} {\bibfnamefont {T.}~\bibnamefont {{Schweizer}}}, \bibinfo {author} {\bibfnamefont {M.}~\bibnamefont {{Shayduk}}}, \bibinfo {author} {\bibfnamefont {K.}~\bibnamefont {{Shinozaki}}}, \bibinfo {author} {\bibfnamefont {S.~N.}\ \bibnamefont {{Shore}}},
  \bibinfo {author} {\bibfnamefont {N.}~\bibnamefont {{Sidro}}}, \bibinfo {author} {\bibfnamefont {A.}~\bibnamefont {{Sierpowska-Bartosik}}}, \bibinfo {author} {\bibfnamefont {A.}~\bibnamefont {{Sillanp{\"a}{\"a}}}}, \bibinfo {author} {\bibfnamefont {D.}~\bibnamefont {{Sobczynska}}}, \bibinfo {author} {\bibfnamefont {F.}~\bibnamefont {{Spanier}}}, \bibinfo {author} {\bibfnamefont {A.}~\bibnamefont {{Stamerra}}}, \bibinfo {author} {\bibfnamefont {L.~S.}\ \bibnamefont {{Stark}}}, \bibinfo {author} {\bibfnamefont {L.}~\bibnamefont {{Takalo}}}, \bibinfo {author} {\bibfnamefont {F.}~\bibnamefont {{Tavecchio}}}, \bibinfo {author} {\bibfnamefont {P.}~\bibnamefont {{Temnikov}}}, \bibinfo {author} {\bibfnamefont {D.}~\bibnamefont {{Tescaro}}}, \bibinfo {author} {\bibfnamefont {M.}~\bibnamefont {{Teshima}}}, \bibinfo {author} {\bibfnamefont {M.}~\bibnamefont {{Tluczykont}}}, \bibinfo {author} {\bibfnamefont {D.~F.}\ \bibnamefont {{Torres}}}, \bibinfo {author} {\bibfnamefont {N.}~\bibnamefont {{Turini}}}, \bibinfo
  {author} {\bibfnamefont {H.}~\bibnamefont {{Vankov}}}, \bibinfo {author} {\bibfnamefont {A.}~\bibnamefont {{Venturini}}}, \bibinfo {author} {\bibfnamefont {V.}~\bibnamefont {{Vitale}}}, \bibinfo {author} {\bibfnamefont {R.~M.}\ \bibnamefont {{Wagner}}}, \bibinfo {author} {\bibfnamefont {W.}~\bibnamefont {{Wittek}}}, \bibinfo {author} {\bibfnamefont {V.}~\bibnamefont {{Zabalza}}}, \bibinfo {author} {\bibfnamefont {F.}~\bibnamefont {{Zandanel}}}, \bibinfo {author} {\bibfnamefont {R.}~\bibnamefont {{Zanin}}},\ and\ \bibinfo {author} {\bibfnamefont {J.}~\bibnamefont {{Zapatero}}},\ }\href {https://doi.org/10.1086/592348} {\bibfield  {journal} {\bibinfo  {journal} {ApJL}\ }\textbf {\bibinfo {volume} {685}},\ \bibinfo {pages} {L23} (\bibinfo {year} {2008})},\ \Eprint {https://arxiv.org/abs/0806.0988} {arXiv:0806.0988 [astro-ph]} \BibitemShut {NoStop}%
\bibitem [{\citenamefont {{Abramowski}}\ \emph {et~al.}(2012)\citenamefont {{Abramowski}}, \citenamefont {{Acero}}, \citenamefont {{Aharonian}}, \citenamefont {{Akhperjanian}}, \citenamefont {{Anton}}, \citenamefont {{Balzer}}, \citenamefont {{Barnacka}}, \citenamefont {{Barres de Almeida}}, \citenamefont {{Becherini}}, \citenamefont {{Becker}}, \citenamefont {{Behera}}, \citenamefont {{Bernl{\"o}hr}}, \citenamefont {{Birsin}}, \citenamefont {{Biteau}}, \citenamefont {{Bochow}}, \citenamefont {{Boisson}}, \citenamefont {{Bolmont}}, \citenamefont {{Bordas}}, \citenamefont {{Brucker}}, \citenamefont {{Brun}}, \citenamefont {{Brun}}, \citenamefont {{Bulik}}, \citenamefont {{B{\"u}sching}}, \citenamefont {{Carrigan}}, \citenamefont {{Casanova}}, \citenamefont {{Cerruti}}, \citenamefont {{Chadwick}}, \citenamefont {{Charbonnier}}, \citenamefont {{Chaves}}, \citenamefont {{Cheesebrough}}, \citenamefont {{Clapson}}, \citenamefont {{Coignet}}, \citenamefont {{Cologna}}, \citenamefont {{Conrad}}, \citenamefont
  {{Dalton}}, \citenamefont {{Daniel}}, \citenamefont {{Davids}}, \citenamefont {{Degrange}}, \citenamefont {{Deil}}, \citenamefont {{Dickinson}}, \citenamefont {{Djannati-Ata{\"\i}}}, \citenamefont {{Domainko}}, \citenamefont {{Drury}}, \citenamefont {{Dubus}}, \citenamefont {{Dutson}}, \citenamefont {{Dyks}}, \citenamefont {{Dyrda}}, \citenamefont {{Egberts}}, \citenamefont {{Eger}}, \citenamefont {{Espigat}}, \citenamefont {{Fallon}}, \citenamefont {{Farnier}}, \citenamefont {{Fegan}}, \citenamefont {{Feinstein}}, \citenamefont {{Fernandes}}, \citenamefont {{Fiasson}}, \citenamefont {{Fontaine}}, \citenamefont {{F{\"o}rster}}, \citenamefont {{F{\"u}{\ss}ling}}, \citenamefont {{Gallant}}, \citenamefont {{Gast}}, \citenamefont {{G{\'e}rard}}, \citenamefont {{Gerbig}}, \citenamefont {{Giebels}}, \citenamefont {{Glicenstein}}, \citenamefont {{Gl{\"u}ck}}, \citenamefont {{Goret}}, \citenamefont {{G{\"o}ring}}, \citenamefont {{H{\"a}ffner}}, \citenamefont {{Hague}}, \citenamefont {{Hampf}}, \citenamefont
  {{Hauser}}, \citenamefont {{Heinz}}, \citenamefont {{Heinzelmann}}, \citenamefont {{Henri}}, \citenamefont {{Hermann}}, \citenamefont {{Hinton}}, \citenamefont {{Hoffmann}}, \citenamefont {{Hofmann}}, \citenamefont {{Hofverberg}}, \citenamefont {{Holler}}, \citenamefont {{Horns}}, \citenamefont {{Jacholkowska}}, \citenamefont {{de Jager}}, \citenamefont {{Jahn}}, \citenamefont {{Jamrozy}}, \citenamefont {{Jung}}, \citenamefont {{Kastendieck}}, \citenamefont {{Katarzy{\'n}ski}}, \citenamefont {{Katz}}, \citenamefont {{Kaufmann}}, \citenamefont {{Keogh}}, \citenamefont {{Khangulyan}}, \citenamefont {{Kh{\'e}lifi}}, \citenamefont {{Klochkov}}, \citenamefont {{Klu{\'z}niak}}, \citenamefont {{Kneiske}}, \citenamefont {{Komin}}, \citenamefont {{Kosack}}, \citenamefont {{Kossakowski}}, \citenamefont {{Laffon}}, \citenamefont {{Lamanna}}, \citenamefont {{Lennarz}}, \citenamefont {{Lohse}}, \citenamefont {{Lopatin}}, \citenamefont {{Lu}}, \citenamefont {{Marandon}}, \citenamefont {{Marcowith}}, \citenamefont
  {{Masbou}}, \citenamefont {{Maurin}}, \citenamefont {{Maxted}}, \citenamefont {{Mayer}}, \citenamefont {{McComb}}, \citenamefont {{Medina}}, \citenamefont {{M{\'e}hault}}, \citenamefont {{Moderski}}, \citenamefont {{Moulin}}, \citenamefont {{Naumann}}, \citenamefont {{Naumann-Godo}}, \citenamefont {{de Naurois}}, \citenamefont {{Nedbal}}, \citenamefont {{Nekrassov}}, \citenamefont {{Nguyen}}, \citenamefont {{Nicholas}}, \citenamefont {{Niemiec}}, \citenamefont {{Nolan}}, \citenamefont {{Ohm}}, \citenamefont {{de O{\~n}a Wilhelmi}}, \citenamefont {{Opitz}}, \citenamefont {{Ostrowski}}, \citenamefont {{Oya}}, \citenamefont {{Panter}}, \citenamefont {{Paz Arribas}}, \citenamefont {{Pedaletti}}, \citenamefont {{Pelletier}}, \citenamefont {{Petrucci}}, \citenamefont {{Pita}}, \citenamefont {{P{\"u}hlhofer}}, \citenamefont {{Punch}}, \citenamefont {{Quirrenbach}}, \citenamefont {{Raue}}, \citenamefont {{Rayner}}, \citenamefont {{Reimer}}, \citenamefont {{Reimer}}, \citenamefont {{Renaud}}, \citenamefont {{de los
  Reyes}}, \citenamefont {{Rieger}}, \citenamefont {{Ripken}}, \citenamefont {{Rob}}, \citenamefont {{Rosier-Lees}}, \citenamefont {{Rowell}}, \citenamefont {{Rudak}}, \citenamefont {{Rulten}}, \citenamefont {{Ruppel}}, \citenamefont {{Sahakian}}, \citenamefont {{Sanchez}}, \citenamefont {{Santangelo}}, \citenamefont {{Schlickeiser}}, \citenamefont {{Sch{\"o}ck}}, \citenamefont {{Schulz}}, \citenamefont {{Schwanke}}, \citenamefont {{Schwarzburg}}, \citenamefont {{Schwemmer}}, \citenamefont {{Sheidaei}}, \citenamefont {{Skilton}}, \citenamefont {{Sol}}, \citenamefont {{Spengler}}, \citenamefont {{Stawarz}}, \citenamefont {{Steenkamp}}, \citenamefont {{Stegmann}}, \citenamefont {{Stinzing}}, \citenamefont {{Stycz}}, \citenamefont {{Sushch}}, \citenamefont {{Szostek}}, \citenamefont {{Tavernet}}, \citenamefont {{Terrier}}, \citenamefont {{Tluczykont}}, \citenamefont {{Valerius}}, \citenamefont {{van Eldik}}, \citenamefont {{Vasileiadis}}, \citenamefont {{Venter}}, \citenamefont {{Vialle}}, \citenamefont
  {{Viana}}, \citenamefont {{Vincent}}, \citenamefont {{V{\"o}lk}}, \citenamefont {{Volpe}}, \citenamefont {{Vorobiov}}, \citenamefont {{Vorster}}, \citenamefont {{Wagner}}, \citenamefont {{Ward}}, \citenamefont {{White}}, \citenamefont {{Wierzcholska}}, \citenamefont {{Zacharias}}, \citenamefont {{Zajczyk}}, \citenamefont {{Zdziarski}}, \citenamefont {{Zech}}, \citenamefont {{Zechlin}}, \citenamefont {{H.~E.~S.~S. Collaboration}}, \citenamefont {{Aleksi{\'c}}}, \citenamefont {{Antonelli}}, \citenamefont {{Antoranz}}, \citenamefont {{Backes}}, \citenamefont {{Barrio}}, \citenamefont {{Bastieri}}, \citenamefont {{Becerra Gonz{\'a}lez}}, \citenamefont {{Bednarek}}, \citenamefont {{Berdyugin}}, \citenamefont {{Berger}}, \citenamefont {{Bernardini}}, \citenamefont {{Biland}}, \citenamefont {{Blanch}}, \citenamefont {{Bock}}, \citenamefont {{Boller}}, \citenamefont {{Bonnoli}}, \citenamefont {{Borla Tridon}}, \citenamefont {{Braun}}, \citenamefont {{Bretz}}, \citenamefont {{Ca{\~n}ellas}}, \citenamefont
  {{Carmona}}, \citenamefont {{Carosi}}, \citenamefont {{Colin}}, \citenamefont {{Colombo}}, \citenamefont {{Contreras}}, \citenamefont {{Cortina}}, \citenamefont {{Cossio}}, \citenamefont {{Covino}}, \citenamefont {{Dazzi}}, \citenamefont {{De Angelis}}, \citenamefont {{De Cea del Pozo}}, \citenamefont {{De Lotto}}, \citenamefont {{Delgado Mendez}}, \citenamefont {{Diago Ortega}}, \citenamefont {{Doert}}, \citenamefont {{Dom{\'\i}nguez}}, \citenamefont {{Dominis Prester}}, \citenamefont {{Dorner}}, \citenamefont {{Doro}}, \citenamefont {{Elsaesser}}, \citenamefont {{Ferenc}}, \citenamefont {{Fonseca}}, \citenamefont {{Font}}, \citenamefont {{Fruck}}, \citenamefont {{Garc{\'\i}a L{\'o}pez}}, \citenamefont {{Garczarczyk}}, \citenamefont {{Garrido}}, \citenamefont {{Giavitto}}, \citenamefont {{Godinovi{\'c}}}, \citenamefont {{Hadasch}}, \citenamefont {{H{\"a}fner}}, \citenamefont {{Herrero}}, \citenamefont {{Hildebrand}}, \citenamefont {{H{\"o}hne-M{\"o}nch}}, \citenamefont {{Hose}}, \citenamefont {{Hrupec}},
  \citenamefont {{Huber}}, \citenamefont {{Jogler}}, \citenamefont {{Klepser}}, \citenamefont {{Kr{\"a}henb{\"u}hl}}, \citenamefont {{Krause}}, \citenamefont {{La Barbera}}, \citenamefont {{Lelas}}, \citenamefont {{Leonardo}}, \citenamefont {{Lindfors}}, \citenamefont {{Lombardi}}, \citenamefont {{L{\'o}pez}}, \citenamefont {{Lorenz}}, \citenamefont {{Makariev}}, \citenamefont {{Maneva}}, \citenamefont {{Mankuzhiyil}}, \citenamefont {{Mannheim}}, \citenamefont {{Maraschi}}, \citenamefont {{Mariotti}}, \citenamefont {{Mart{\'\i}nez}}, \citenamefont {{Mazin}}, \citenamefont {{Meucci}}, \citenamefont {{Miranda}}, \citenamefont {{Mirzoyan}}, \citenamefont {{Miyamoto}}, \citenamefont {{Mold{\'o}n}}, \citenamefont {{Moralejo}}, \citenamefont {{Munar}}, \citenamefont {{Nieto}}, \citenamefont {{Nilsson}}, \citenamefont {{Orito}}, \citenamefont {{Oya}}, \citenamefont {{Paneque}}, \citenamefont {{Paoletti}}, \citenamefont {{Pardo}}, \citenamefont {{Paredes}}, \citenamefont {{Partini}}, \citenamefont {{Pasanen}},
  \citenamefont {{Pauss}}, \citenamefont {{Perez-Torres}}, \citenamefont {{Persic}}, \citenamefont {{Peruzzo}}, \citenamefont {{Pilia}}, \citenamefont {{Pochon}}, \citenamefont {{Prada}}, \citenamefont {{Prada Moroni}}, \citenamefont {{Prandini}}, \citenamefont {{Puljak}}, \citenamefont {{Reichardt}}, \citenamefont {{Reinthal}}, \citenamefont {{Rhode}}, \citenamefont {{Rib{\'o}}}, \citenamefont {{Rico}}, \citenamefont {{R{\"u}gamer}}, \citenamefont {{Saggion}}, \citenamefont {{Saito}}, \citenamefont {{Saito}}, \citenamefont {{Salvati}}, \citenamefont {{Satalecka}}, \citenamefont {{Scalzotto}}, \citenamefont {{Scapin}}, \citenamefont {{Schultz}}, \citenamefont {{Schweizer}}, \citenamefont {{Shayduk}}, \citenamefont {{Shore}}, \citenamefont {{Sillanp{\"a}{\"a}}}, \citenamefont {{Sitarek}}, \citenamefont {{Sobczynska}}, \citenamefont {{Spanier}}, \citenamefont {{Spiro}}, \citenamefont {{Stamerra}}, \citenamefont {{Steinke}}, \citenamefont {{Storz}}, \citenamefont {{Strah}}, \citenamefont {{Suri{\'c}}},
  \citenamefont {{Takalo}}, \citenamefont {{Takami}}, \citenamefont {{Tavecchio}}, \citenamefont {{Temnikov}}, \citenamefont {{Terzi{\'c}}}, \citenamefont {{Tescaro}}, \citenamefont {{Teshima}}, \citenamefont {{Thom}}, \citenamefont {{Tibolla}}, \citenamefont {{Torres}}, \citenamefont {{Treves}}, \citenamefont {{Vankov}}, \citenamefont {{Vogler}}, \citenamefont {{Wagner}}, \citenamefont {{Weitzel}}, \citenamefont {{Zabalza}}, \citenamefont {{Zandanel}}, \citenamefont {{Zanin}}, \citenamefont {{MAGIC Collaboration}}, \citenamefont {{Arlen}}, \citenamefont {{Aune}}, \citenamefont {{Beilicke}}, \citenamefont {{Benbow}}, \citenamefont {{Bouvier}}, \citenamefont {{Bradbury}}, \citenamefont {{Buckley}}, \citenamefont {{Bugaev}}, \citenamefont {{Byrum}}, \citenamefont {{Cannon}}, \citenamefont {{Cesarini}}, \citenamefont {{Ciupik}}, \citenamefont {{Connolly}}, \citenamefont {{Cui}}, \citenamefont {{Dickherber}}, \citenamefont {{Duke}}, \citenamefont {{Errando}}, \citenamefont {{Falcone}}, \citenamefont {{Finley}},
  \citenamefont {{Finnegan}}, \citenamefont {{Fortson}}, \citenamefont {{Furniss}}, \citenamefont {{Galante}}, \citenamefont {{Gall}}, \citenamefont {{Godambe}}, \citenamefont {{Griffin}}, \citenamefont {{Grube}}, \citenamefont {{Gyuk}}, \citenamefont {{Hanna}}, \citenamefont {{Holder}}, \citenamefont {{Huan}}, \citenamefont {{Hui}}, \citenamefont {{Kaaret}}, \citenamefont {{Karlsson}}, \citenamefont {{Kertzman}}, \citenamefont {{Khassen}}, \citenamefont {{Kieda}}, \citenamefont {{Krawczynski}}, \citenamefont {{Krennrich}}, \citenamefont {{Lang}}, \citenamefont {{LeBohec}}, \citenamefont {{Maier}}, \citenamefont {{McArthur}}, \citenamefont {{McCann}}, \citenamefont {{Moriarty}}, \citenamefont {{Mukherjee}}, \citenamefont {{Nu{\~n}ez}}, \citenamefont {{Ong}}, \citenamefont {{Orr}}, \citenamefont {{Otte}}, \citenamefont {{Park}}, \citenamefont {{Perkins}}, \citenamefont {{Pichel}}, \citenamefont {{Pohl}}, \citenamefont {{Prokoph}}, \citenamefont {{Ragan}}, \citenamefont {{Reyes}}, \citenamefont {{Reynolds}},
  \citenamefont {{Roache}}, \citenamefont {{Rose}}, \citenamefont {{Ruppel}}, \citenamefont {{Schroedter}}, \citenamefont {{Sembroski}}, \citenamefont {{{\c{S}}ent{\"u}rk}}, \citenamefont {{Telezhinsky}}, \citenamefont {{Te{\v{s}}i{\'c}}}, \citenamefont {{Theiling}}, \citenamefont {{Thibadeau}}, \citenamefont {{Varlotta}}, \citenamefont {{Vassiliev}}, \citenamefont {{Vivier}}, \citenamefont {{Wakely}}, \citenamefont {{Weekes}}, \citenamefont {{Williams}}, \citenamefont {{Zitzer}}, \citenamefont {{VERITAS Collaboration}}, \citenamefont {{Barres de Almeida}}, \citenamefont {{Cara}}, \citenamefont {{Casadio}}, \citenamefont {{Cheung}}, \citenamefont {{McConville}}, \citenamefont {{Davies}}, \citenamefont {{Doi}}, \citenamefont {{Giovannini}}, \citenamefont {{Giroletti}}, \citenamefont {{Hada}}, \citenamefont {{Hardee}}, \citenamefont {{Harris}}, \citenamefont {{Junor}}, \citenamefont {{Kino}}, \citenamefont {{Lee}}, \citenamefont {{Ly}}, \citenamefont {{Madrid}}, \citenamefont {{Massaro}}, \citenamefont
  {{Mundell}}, \citenamefont {{Nagai}}, \citenamefont {{Perlman}}, \citenamefont {{Steele}}, \citenamefont {{Walker}},\ and\ \citenamefont {{Wood}}}]{Abramowski12}%
  \BibitemOpen
  \bibfield  {author} {\bibinfo {author} {\bibfnamefont {A.}~\bibnamefont {{Abramowski}}}, \bibinfo {author} {\bibfnamefont {F.}~\bibnamefont {{Acero}}}, \bibinfo {author} {\bibfnamefont {F.}~\bibnamefont {{Aharonian}}}, \bibinfo {author} {\bibfnamefont {A.~G.}\ \bibnamefont {{Akhperjanian}}}, \bibinfo {author} {\bibfnamefont {G.}~\bibnamefont {{Anton}}}, \bibinfo {author} {\bibfnamefont {A.}~\bibnamefont {{Balzer}}}, \bibinfo {author} {\bibfnamefont {A.}~\bibnamefont {{Barnacka}}}, \bibinfo {author} {\bibfnamefont {U.}~\bibnamefont {{Barres de Almeida}}}, \bibinfo {author} {\bibfnamefont {Y.}~\bibnamefont {{Becherini}}}, \bibinfo {author} {\bibfnamefont {J.}~\bibnamefont {{Becker}}}, \bibinfo {author} {\bibfnamefont {B.}~\bibnamefont {{Behera}}}, \bibinfo {author} {\bibfnamefont {K.}~\bibnamefont {{Bernl{\"o}hr}}}, \bibinfo {author} {\bibfnamefont {E.}~\bibnamefont {{Birsin}}}, \bibinfo {author} {\bibfnamefont {J.}~\bibnamefont {{Biteau}}}, \bibinfo {author} {\bibfnamefont {A.}~\bibnamefont {{Bochow}}},
  \bibinfo {author} {\bibfnamefont {C.}~\bibnamefont {{Boisson}}}, \bibinfo {author} {\bibfnamefont {J.}~\bibnamefont {{Bolmont}}}, \bibinfo {author} {\bibfnamefont {P.}~\bibnamefont {{Bordas}}}, \bibinfo {author} {\bibfnamefont {J.}~\bibnamefont {{Brucker}}}, \bibinfo {author} {\bibfnamefont {F.}~\bibnamefont {{Brun}}}, \bibinfo {author} {\bibfnamefont {P.}~\bibnamefont {{Brun}}}, \bibinfo {author} {\bibfnamefont {T.}~\bibnamefont {{Bulik}}}, \bibinfo {author} {\bibfnamefont {I.}~\bibnamefont {{B{\"u}sching}}}, \bibinfo {author} {\bibfnamefont {S.}~\bibnamefont {{Carrigan}}}, \bibinfo {author} {\bibfnamefont {S.}~\bibnamefont {{Casanova}}}, \bibinfo {author} {\bibfnamefont {M.}~\bibnamefont {{Cerruti}}}, \bibinfo {author} {\bibfnamefont {P.~M.}\ \bibnamefont {{Chadwick}}}, \bibinfo {author} {\bibfnamefont {A.}~\bibnamefont {{Charbonnier}}}, \bibinfo {author} {\bibfnamefont {R.~C.~G.}\ \bibnamefont {{Chaves}}}, \bibinfo {author} {\bibfnamefont {A.}~\bibnamefont {{Cheesebrough}}}, \bibinfo {author}
  {\bibfnamefont {A.~C.}\ \bibnamefont {{Clapson}}}, \bibinfo {author} {\bibfnamefont {G.}~\bibnamefont {{Coignet}}}, \bibinfo {author} {\bibfnamefont {G.}~\bibnamefont {{Cologna}}}, \bibinfo {author} {\bibfnamefont {J.}~\bibnamefont {{Conrad}}}, \bibinfo {author} {\bibfnamefont {M.}~\bibnamefont {{Dalton}}}, \bibinfo {author} {\bibfnamefont {M.~K.}\ \bibnamefont {{Daniel}}}, \bibinfo {author} {\bibfnamefont {I.~D.}\ \bibnamefont {{Davids}}}, \bibinfo {author} {\bibfnamefont {B.}~\bibnamefont {{Degrange}}}, \bibinfo {author} {\bibfnamefont {C.}~\bibnamefont {{Deil}}}, \bibinfo {author} {\bibfnamefont {H.~J.}\ \bibnamefont {{Dickinson}}}, \bibinfo {author} {\bibfnamefont {A.}~\bibnamefont {{Djannati-Ata{\"\i}}}}, \bibinfo {author} {\bibfnamefont {W.}~\bibnamefont {{Domainko}}}, \bibinfo {author} {\bibfnamefont {L.~O.}\ \bibnamefont {{Drury}}}, \bibinfo {author} {\bibfnamefont {G.}~\bibnamefont {{Dubus}}}, \bibinfo {author} {\bibfnamefont {K.}~\bibnamefont {{Dutson}}}, \bibinfo {author} {\bibfnamefont
  {J.}~\bibnamefont {{Dyks}}}, \bibinfo {author} {\bibfnamefont {M.}~\bibnamefont {{Dyrda}}}, \bibinfo {author} {\bibfnamefont {K.}~\bibnamefont {{Egberts}}}, \bibinfo {author} {\bibfnamefont {P.}~\bibnamefont {{Eger}}}, \bibinfo {author} {\bibfnamefont {P.}~\bibnamefont {{Espigat}}}, \bibinfo {author} {\bibfnamefont {L.}~\bibnamefont {{Fallon}}}, \bibinfo {author} {\bibfnamefont {C.}~\bibnamefont {{Farnier}}}, \bibinfo {author} {\bibfnamefont {S.}~\bibnamefont {{Fegan}}}, \bibinfo {author} {\bibfnamefont {F.}~\bibnamefont {{Feinstein}}}, \bibinfo {author} {\bibfnamefont {M.~V.}\ \bibnamefont {{Fernandes}}}, \bibinfo {author} {\bibfnamefont {A.}~\bibnamefont {{Fiasson}}}, \bibinfo {author} {\bibfnamefont {G.}~\bibnamefont {{Fontaine}}}, \bibinfo {author} {\bibfnamefont {A.}~\bibnamefont {{F{\"o}rster}}}, \bibinfo {author} {\bibfnamefont {M.}~\bibnamefont {{F{\"u}{\ss}ling}}}, \bibinfo {author} {\bibfnamefont {Y.~A.}\ \bibnamefont {{Gallant}}}, \bibinfo {author} {\bibfnamefont {H.}~\bibnamefont {{Gast}}},
  \bibinfo {author} {\bibfnamefont {L.}~\bibnamefont {{G{\'e}rard}}}, \bibinfo {author} {\bibfnamefont {D.}~\bibnamefont {{Gerbig}}}, \bibinfo {author} {\bibfnamefont {B.}~\bibnamefont {{Giebels}}}, \bibinfo {author} {\bibfnamefont {J.~F.}\ \bibnamefont {{Glicenstein}}}, \bibinfo {author} {\bibfnamefont {B.}~\bibnamefont {{Gl{\"u}ck}}}, \bibinfo {author} {\bibfnamefont {P.}~\bibnamefont {{Goret}}}, \bibinfo {author} {\bibfnamefont {D.}~\bibnamefont {{G{\"o}ring}}}, \bibinfo {author} {\bibfnamefont {S.}~\bibnamefont {{H{\"a}ffner}}}, \bibinfo {author} {\bibfnamefont {J.~D.}\ \bibnamefont {{Hague}}}, \bibinfo {author} {\bibfnamefont {D.}~\bibnamefont {{Hampf}}}, \bibinfo {author} {\bibfnamefont {M.}~\bibnamefont {{Hauser}}}, \bibinfo {author} {\bibfnamefont {S.}~\bibnamefont {{Heinz}}}, \bibinfo {author} {\bibfnamefont {G.}~\bibnamefont {{Heinzelmann}}}, \bibinfo {author} {\bibfnamefont {G.}~\bibnamefont {{Henri}}}, \bibinfo {author} {\bibfnamefont {G.}~\bibnamefont {{Hermann}}}, \bibinfo {author}
  {\bibfnamefont {J.~A.}\ \bibnamefont {{Hinton}}}, \bibinfo {author} {\bibfnamefont {A.}~\bibnamefont {{Hoffmann}}}, \bibinfo {author} {\bibfnamefont {W.}~\bibnamefont {{Hofmann}}}, \bibinfo {author} {\bibfnamefont {P.}~\bibnamefont {{Hofverberg}}}, \bibinfo {author} {\bibfnamefont {M.}~\bibnamefont {{Holler}}}, \bibinfo {author} {\bibfnamefont {D.}~\bibnamefont {{Horns}}}, \bibinfo {author} {\bibfnamefont {A.}~\bibnamefont {{Jacholkowska}}}, \bibinfo {author} {\bibfnamefont {O.~C.}\ \bibnamefont {{de Jager}}}, \bibinfo {author} {\bibfnamefont {C.}~\bibnamefont {{Jahn}}}, \bibinfo {author} {\bibfnamefont {M.}~\bibnamefont {{Jamrozy}}}, \bibinfo {author} {\bibfnamefont {I.}~\bibnamefont {{Jung}}}, \bibinfo {author} {\bibfnamefont {M.~A.}\ \bibnamefont {{Kastendieck}}}, \bibinfo {author} {\bibfnamefont {K.}~\bibnamefont {{Katarzy{\'n}ski}}}, \bibinfo {author} {\bibfnamefont {U.}~\bibnamefont {{Katz}}}, \bibinfo {author} {\bibfnamefont {S.}~\bibnamefont {{Kaufmann}}}, \bibinfo {author} {\bibfnamefont
  {D.}~\bibnamefont {{Keogh}}}, \bibinfo {author} {\bibfnamefont {D.}~\bibnamefont {{Khangulyan}}}, \bibinfo {author} {\bibfnamefont {B.}~\bibnamefont {{Kh{\'e}lifi}}}, \bibinfo {author} {\bibfnamefont {D.}~\bibnamefont {{Klochkov}}}, \bibinfo {author} {\bibfnamefont {W.}~\bibnamefont {{Klu{\'z}niak}}}, \bibinfo {author} {\bibfnamefont {T.}~\bibnamefont {{Kneiske}}}, \bibinfo {author} {\bibfnamefont {N.}~\bibnamefont {{Komin}}}, \bibinfo {author} {\bibfnamefont {K.}~\bibnamefont {{Kosack}}}, \bibinfo {author} {\bibfnamefont {R.}~\bibnamefont {{Kossakowski}}}, \bibinfo {author} {\bibfnamefont {H.}~\bibnamefont {{Laffon}}}, \bibinfo {author} {\bibfnamefont {G.}~\bibnamefont {{Lamanna}}}, \bibinfo {author} {\bibfnamefont {D.}~\bibnamefont {{Lennarz}}}, \bibinfo {author} {\bibfnamefont {T.}~\bibnamefont {{Lohse}}}, \bibinfo {author} {\bibfnamefont {A.}~\bibnamefont {{Lopatin}}}, \bibinfo {author} {\bibfnamefont {C.~C.}\ \bibnamefont {{Lu}}}, \bibinfo {author} {\bibfnamefont {V.}~\bibnamefont {{Marandon}}},
  \bibinfo {author} {\bibfnamefont {A.}~\bibnamefont {{Marcowith}}}, \bibinfo {author} {\bibfnamefont {J.}~\bibnamefont {{Masbou}}}, \bibinfo {author} {\bibfnamefont {D.}~\bibnamefont {{Maurin}}}, \bibinfo {author} {\bibfnamefont {N.}~\bibnamefont {{Maxted}}}, \bibinfo {author} {\bibfnamefont {M.}~\bibnamefont {{Mayer}}}, \bibinfo {author} {\bibfnamefont {T.~J.~L.}\ \bibnamefont {{McComb}}}, \bibinfo {author} {\bibfnamefont {M.~C.}\ \bibnamefont {{Medina}}}, \bibinfo {author} {\bibfnamefont {J.}~\bibnamefont {{M{\'e}hault}}}, \bibinfo {author} {\bibfnamefont {R.}~\bibnamefont {{Moderski}}}, \bibinfo {author} {\bibfnamefont {E.}~\bibnamefont {{Moulin}}}, \bibinfo {author} {\bibfnamefont {C.~L.}\ \bibnamefont {{Naumann}}}, \bibinfo {author} {\bibfnamefont {M.}~\bibnamefont {{Naumann-Godo}}}, \bibinfo {author} {\bibfnamefont {M.}~\bibnamefont {{de Naurois}}}, \bibinfo {author} {\bibfnamefont {D.}~\bibnamefont {{Nedbal}}}, \bibinfo {author} {\bibfnamefont {D.}~\bibnamefont {{Nekrassov}}}, \bibinfo {author}
  {\bibfnamefont {N.}~\bibnamefont {{Nguyen}}}, \bibinfo {author} {\bibfnamefont {B.}~\bibnamefont {{Nicholas}}}, \bibinfo {author} {\bibfnamefont {J.}~\bibnamefont {{Niemiec}}}, \bibinfo {author} {\bibfnamefont {S.~J.}\ \bibnamefont {{Nolan}}}, \bibinfo {author} {\bibfnamefont {S.}~\bibnamefont {{Ohm}}}, \bibinfo {author} {\bibfnamefont {E.}~\bibnamefont {{de O{\~n}a Wilhelmi}}}, \bibinfo {author} {\bibfnamefont {B.}~\bibnamefont {{Opitz}}}, \bibinfo {author} {\bibfnamefont {M.}~\bibnamefont {{Ostrowski}}}, \bibinfo {author} {\bibfnamefont {I.}~\bibnamefont {{Oya}}}, \bibinfo {author} {\bibfnamefont {M.}~\bibnamefont {{Panter}}}, \bibinfo {author} {\bibfnamefont {M.}~\bibnamefont {{Paz Arribas}}}, \bibinfo {author} {\bibfnamefont {G.}~\bibnamefont {{Pedaletti}}}, \bibinfo {author} {\bibfnamefont {G.}~\bibnamefont {{Pelletier}}}, \bibinfo {author} {\bibfnamefont {P.~O.}\ \bibnamefont {{Petrucci}}}, \bibinfo {author} {\bibfnamefont {S.}~\bibnamefont {{Pita}}}, \bibinfo {author} {\bibfnamefont {G.}~\bibnamefont
  {{P{\"u}hlhofer}}}, \bibinfo {author} {\bibfnamefont {M.}~\bibnamefont {{Punch}}}, \bibinfo {author} {\bibfnamefont {A.}~\bibnamefont {{Quirrenbach}}}, \bibinfo {author} {\bibfnamefont {M.}~\bibnamefont {{Raue}}}, \bibinfo {author} {\bibfnamefont {S.~M.}\ \bibnamefont {{Rayner}}}, \bibinfo {author} {\bibfnamefont {A.}~\bibnamefont {{Reimer}}}, \bibinfo {author} {\bibfnamefont {O.}~\bibnamefont {{Reimer}}}, \bibinfo {author} {\bibfnamefont {M.}~\bibnamefont {{Renaud}}}, \bibinfo {author} {\bibfnamefont {R.}~\bibnamefont {{de los Reyes}}}, \bibinfo {author} {\bibfnamefont {F.}~\bibnamefont {{Rieger}}}, \bibinfo {author} {\bibfnamefont {J.}~\bibnamefont {{Ripken}}}, \bibinfo {author} {\bibfnamefont {L.}~\bibnamefont {{Rob}}}, \bibinfo {author} {\bibfnamefont {S.}~\bibnamefont {{Rosier-Lees}}}, \bibinfo {author} {\bibfnamefont {G.}~\bibnamefont {{Rowell}}}, \bibinfo {author} {\bibfnamefont {B.}~\bibnamefont {{Rudak}}}, \bibinfo {author} {\bibfnamefont {C.~B.}\ \bibnamefont {{Rulten}}}, \bibinfo {author}
  {\bibfnamefont {J.}~\bibnamefont {{Ruppel}}}, \bibinfo {author} {\bibfnamefont {V.}~\bibnamefont {{Sahakian}}}, \bibinfo {author} {\bibfnamefont {D.~A.}\ \bibnamefont {{Sanchez}}}, \bibinfo {author} {\bibfnamefont {A.}~\bibnamefont {{Santangelo}}}, \bibinfo {author} {\bibfnamefont {R.}~\bibnamefont {{Schlickeiser}}}, \bibinfo {author} {\bibfnamefont {F.~M.}\ \bibnamefont {{Sch{\"o}ck}}}, \bibinfo {author} {\bibfnamefont {A.}~\bibnamefont {{Schulz}}}, \bibinfo {author} {\bibfnamefont {U.}~\bibnamefont {{Schwanke}}}, \bibinfo {author} {\bibfnamefont {S.}~\bibnamefont {{Schwarzburg}}}, \bibinfo {author} {\bibfnamefont {S.}~\bibnamefont {{Schwemmer}}}, \bibinfo {author} {\bibfnamefont {F.}~\bibnamefont {{Sheidaei}}}, \bibinfo {author} {\bibfnamefont {J.~L.}\ \bibnamefont {{Skilton}}}, \bibinfo {author} {\bibfnamefont {H.}~\bibnamefont {{Sol}}}, \bibinfo {author} {\bibfnamefont {G.}~\bibnamefont {{Spengler}}}, \bibinfo {author} {\bibfnamefont {{\L}.}~\bibnamefont {{Stawarz}}}, \bibinfo {author} {\bibfnamefont
  {R.}~\bibnamefont {{Steenkamp}}}, \bibinfo {author} {\bibfnamefont {C.}~\bibnamefont {{Stegmann}}}, \bibinfo {author} {\bibfnamefont {F.}~\bibnamefont {{Stinzing}}}, \bibinfo {author} {\bibfnamefont {K.}~\bibnamefont {{Stycz}}}, \bibinfo {author} {\bibfnamefont {I.}~\bibnamefont {{Sushch}}}, \bibinfo {author} {\bibfnamefont {A.}~\bibnamefont {{Szostek}}}, \bibinfo {author} {\bibfnamefont {J.~P.}\ \bibnamefont {{Tavernet}}}, \bibinfo {author} {\bibfnamefont {R.}~\bibnamefont {{Terrier}}}, \bibinfo {author} {\bibfnamefont {M.}~\bibnamefont {{Tluczykont}}}, \bibinfo {author} {\bibfnamefont {K.}~\bibnamefont {{Valerius}}}, \bibinfo {author} {\bibfnamefont {C.}~\bibnamefont {{van Eldik}}}, \bibinfo {author} {\bibfnamefont {G.}~\bibnamefont {{Vasileiadis}}}, \bibinfo {author} {\bibfnamefont {C.}~\bibnamefont {{Venter}}}, \bibinfo {author} {\bibfnamefont {J.~P.}\ \bibnamefont {{Vialle}}}, \bibinfo {author} {\bibfnamefont {A.}~\bibnamefont {{Viana}}}, \bibinfo {author} {\bibfnamefont {P.}~\bibnamefont {{Vincent}}},
  \bibinfo {author} {\bibfnamefont {H.~J.}\ \bibnamefont {{V{\"o}lk}}}, \bibinfo {author} {\bibfnamefont {F.}~\bibnamefont {{Volpe}}}, \bibinfo {author} {\bibfnamefont {S.}~\bibnamefont {{Vorobiov}}}, \bibinfo {author} {\bibfnamefont {M.}~\bibnamefont {{Vorster}}}, \bibinfo {author} {\bibfnamefont {S.~J.}\ \bibnamefont {{Wagner}}}, \bibinfo {author} {\bibfnamefont {M.}~\bibnamefont {{Ward}}}, \bibinfo {author} {\bibfnamefont {R.}~\bibnamefont {{White}}}, \bibinfo {author} {\bibfnamefont {A.}~\bibnamefont {{Wierzcholska}}}, \bibinfo {author} {\bibfnamefont {M.}~\bibnamefont {{Zacharias}}}, \bibinfo {author} {\bibfnamefont {A.}~\bibnamefont {{Zajczyk}}}, \bibinfo {author} {\bibfnamefont {A.~A.}\ \bibnamefont {{Zdziarski}}}, \bibinfo {author} {\bibfnamefont {A.}~\bibnamefont {{Zech}}}, \bibinfo {author} {\bibfnamefont {H.~S.}\ \bibnamefont {{Zechlin}}}, \bibinfo {author} {\bibnamefont {{H.~E.~S.~S. Collaboration}}}, \bibinfo {author} {\bibfnamefont {J.}~\bibnamefont {{Aleksi{\'c}}}}, \bibinfo {author}
  {\bibfnamefont {L.~A.}\ \bibnamefont {{Antonelli}}}, \bibinfo {author} {\bibfnamefont {P.}~\bibnamefont {{Antoranz}}}, \bibinfo {author} {\bibfnamefont {M.}~\bibnamefont {{Backes}}}, \bibinfo {author} {\bibfnamefont {J.~A.}\ \bibnamefont {{Barrio}}}, \bibinfo {author} {\bibfnamefont {D.}~\bibnamefont {{Bastieri}}}, \bibinfo {author} {\bibfnamefont {J.}~\bibnamefont {{Becerra Gonz{\'a}lez}}}, \bibinfo {author} {\bibfnamefont {W.}~\bibnamefont {{Bednarek}}}, \bibinfo {author} {\bibfnamefont {A.}~\bibnamefont {{Berdyugin}}}, \bibinfo {author} {\bibfnamefont {K.}~\bibnamefont {{Berger}}}, \bibinfo {author} {\bibfnamefont {E.}~\bibnamefont {{Bernardini}}}, \bibinfo {author} {\bibfnamefont {A.}~\bibnamefont {{Biland}}}, \bibinfo {author} {\bibfnamefont {O.}~\bibnamefont {{Blanch}}}, \bibinfo {author} {\bibfnamefont {R.~K.}\ \bibnamefont {{Bock}}}, \bibinfo {author} {\bibfnamefont {A.}~\bibnamefont {{Boller}}}, \bibinfo {author} {\bibfnamefont {G.}~\bibnamefont {{Bonnoli}}}, \bibinfo {author} {\bibfnamefont
  {D.}~\bibnamefont {{Borla Tridon}}}, \bibinfo {author} {\bibfnamefont {I.}~\bibnamefont {{Braun}}}, \bibinfo {author} {\bibfnamefont {T.}~\bibnamefont {{Bretz}}}, \bibinfo {author} {\bibfnamefont {A.}~\bibnamefont {{Ca{\~n}ellas}}}, \bibinfo {author} {\bibfnamefont {E.}~\bibnamefont {{Carmona}}}, \bibinfo {author} {\bibfnamefont {A.}~\bibnamefont {{Carosi}}}, \bibinfo {author} {\bibfnamefont {P.}~\bibnamefont {{Colin}}}, \bibinfo {author} {\bibfnamefont {E.}~\bibnamefont {{Colombo}}}, \bibinfo {author} {\bibfnamefont {J.~L.}\ \bibnamefont {{Contreras}}}, \bibinfo {author} {\bibfnamefont {J.}~\bibnamefont {{Cortina}}}, \bibinfo {author} {\bibfnamefont {L.}~\bibnamefont {{Cossio}}}, \bibinfo {author} {\bibfnamefont {S.}~\bibnamefont {{Covino}}}, \bibinfo {author} {\bibfnamefont {F.}~\bibnamefont {{Dazzi}}}, \bibinfo {author} {\bibfnamefont {A.}~\bibnamefont {{De Angelis}}}, \bibinfo {author} {\bibfnamefont {E.}~\bibnamefont {{De Cea del Pozo}}}, \bibinfo {author} {\bibfnamefont {B.}~\bibnamefont {{De
  Lotto}}}, \bibinfo {author} {\bibfnamefont {C.}~\bibnamefont {{Delgado Mendez}}}, \bibinfo {author} {\bibfnamefont {A.}~\bibnamefont {{Diago Ortega}}}, \bibinfo {author} {\bibfnamefont {M.}~\bibnamefont {{Doert}}}, \bibinfo {author} {\bibfnamefont {A.}~\bibnamefont {{Dom{\'\i}nguez}}}, \bibinfo {author} {\bibfnamefont {D.}~\bibnamefont {{Dominis Prester}}}, \bibinfo {author} {\bibfnamefont {D.}~\bibnamefont {{Dorner}}}, \bibinfo {author} {\bibfnamefont {M.}~\bibnamefont {{Doro}}}, \bibinfo {author} {\bibfnamefont {D.}~\bibnamefont {{Elsaesser}}}, \bibinfo {author} {\bibfnamefont {D.}~\bibnamefont {{Ferenc}}}, \bibinfo {author} {\bibfnamefont {M.~V.}\ \bibnamefont {{Fonseca}}}, \bibinfo {author} {\bibfnamefont {L.}~\bibnamefont {{Font}}}, \bibinfo {author} {\bibfnamefont {C.}~\bibnamefont {{Fruck}}}, \bibinfo {author} {\bibfnamefont {R.~J.}\ \bibnamefont {{Garc{\'\i}a L{\'o}pez}}}, \bibinfo {author} {\bibfnamefont {M.}~\bibnamefont {{Garczarczyk}}}, \bibinfo {author} {\bibfnamefont {D.}~\bibnamefont
  {{Garrido}}}, \bibinfo {author} {\bibfnamefont {G.}~\bibnamefont {{Giavitto}}}, \bibinfo {author} {\bibfnamefont {N.}~\bibnamefont {{Godinovi{\'c}}}}, \bibinfo {author} {\bibfnamefont {D.}~\bibnamefont {{Hadasch}}}, \bibinfo {author} {\bibfnamefont {D.}~\bibnamefont {{H{\"a}fner}}}, \bibinfo {author} {\bibfnamefont {A.}~\bibnamefont {{Herrero}}}, \bibinfo {author} {\bibfnamefont {D.}~\bibnamefont {{Hildebrand}}}, \bibinfo {author} {\bibfnamefont {D.}~\bibnamefont {{H{\"o}hne-M{\"o}nch}}}, \bibinfo {author} {\bibfnamefont {J.}~\bibnamefont {{Hose}}}, \bibinfo {author} {\bibfnamefont {D.}~\bibnamefont {{Hrupec}}}, \bibinfo {author} {\bibfnamefont {B.}~\bibnamefont {{Huber}}}, \bibinfo {author} {\bibfnamefont {T.}~\bibnamefont {{Jogler}}}, \bibinfo {author} {\bibfnamefont {S.}~\bibnamefont {{Klepser}}}, \bibinfo {author} {\bibfnamefont {T.}~\bibnamefont {{Kr{\"a}henb{\"u}hl}}}, \bibinfo {author} {\bibfnamefont {J.}~\bibnamefont {{Krause}}}, \bibinfo {author} {\bibfnamefont {A.}~\bibnamefont {{La Barbera}}},
  \bibinfo {author} {\bibfnamefont {D.}~\bibnamefont {{Lelas}}}, \bibinfo {author} {\bibfnamefont {E.}~\bibnamefont {{Leonardo}}}, \bibinfo {author} {\bibfnamefont {E.}~\bibnamefont {{Lindfors}}}, \bibinfo {author} {\bibfnamefont {S.}~\bibnamefont {{Lombardi}}}, \bibinfo {author} {\bibfnamefont {M.}~\bibnamefont {{L{\'o}pez}}}, \bibinfo {author} {\bibfnamefont {E.}~\bibnamefont {{Lorenz}}}, \bibinfo {author} {\bibfnamefont {M.}~\bibnamefont {{Makariev}}}, \bibinfo {author} {\bibfnamefont {G.}~\bibnamefont {{Maneva}}}, \bibinfo {author} {\bibfnamefont {N.}~\bibnamefont {{Mankuzhiyil}}}, \bibinfo {author} {\bibfnamefont {K.}~\bibnamefont {{Mannheim}}}, \bibinfo {author} {\bibfnamefont {L.}~\bibnamefont {{Maraschi}}}, \bibinfo {author} {\bibfnamefont {M.}~\bibnamefont {{Mariotti}}}, \bibinfo {author} {\bibfnamefont {M.}~\bibnamefont {{Mart{\'\i}nez}}}, \bibinfo {author} {\bibfnamefont {D.}~\bibnamefont {{Mazin}}}, \bibinfo {author} {\bibfnamefont {M.}~\bibnamefont {{Meucci}}}, \bibinfo {author} {\bibfnamefont
  {J.~M.}\ \bibnamefont {{Miranda}}}, \bibinfo {author} {\bibfnamefont {R.}~\bibnamefont {{Mirzoyan}}}, \bibinfo {author} {\bibfnamefont {H.}~\bibnamefont {{Miyamoto}}}, \bibinfo {author} {\bibfnamefont {J.}~\bibnamefont {{Mold{\'o}n}}}, \bibinfo {author} {\bibfnamefont {A.}~\bibnamefont {{Moralejo}}}, \bibinfo {author} {\bibfnamefont {P.}~\bibnamefont {{Munar}}}, \bibinfo {author} {\bibfnamefont {D.}~\bibnamefont {{Nieto}}}, \bibinfo {author} {\bibfnamefont {K.}~\bibnamefont {{Nilsson}}}, \bibinfo {author} {\bibfnamefont {R.}~\bibnamefont {{Orito}}}, \bibinfo {author} {\bibfnamefont {I.}~\bibnamefont {{Oya}}}, \bibinfo {author} {\bibfnamefont {D.}~\bibnamefont {{Paneque}}}, \bibinfo {author} {\bibfnamefont {R.}~\bibnamefont {{Paoletti}}}, \bibinfo {author} {\bibfnamefont {S.}~\bibnamefont {{Pardo}}}, \bibinfo {author} {\bibfnamefont {J.~M.}\ \bibnamefont {{Paredes}}}, \bibinfo {author} {\bibfnamefont {S.}~\bibnamefont {{Partini}}}, \bibinfo {author} {\bibfnamefont {M.}~\bibnamefont {{Pasanen}}}, \bibinfo
  {author} {\bibfnamefont {F.}~\bibnamefont {{Pauss}}}, \bibinfo {author} {\bibfnamefont {M.~A.}\ \bibnamefont {{Perez-Torres}}}, \bibinfo {author} {\bibfnamefont {M.}~\bibnamefont {{Persic}}}, \bibinfo {author} {\bibfnamefont {L.}~\bibnamefont {{Peruzzo}}}, \bibinfo {author} {\bibfnamefont {M.}~\bibnamefont {{Pilia}}}, \bibinfo {author} {\bibfnamefont {J.}~\bibnamefont {{Pochon}}}, \bibinfo {author} {\bibfnamefont {F.}~\bibnamefont {{Prada}}}, \bibinfo {author} {\bibfnamefont {P.~G.}\ \bibnamefont {{Prada Moroni}}}, \bibinfo {author} {\bibfnamefont {E.}~\bibnamefont {{Prandini}}}, \bibinfo {author} {\bibfnamefont {I.}~\bibnamefont {{Puljak}}}, \bibinfo {author} {\bibfnamefont {I.}~\bibnamefont {{Reichardt}}}, \bibinfo {author} {\bibfnamefont {R.}~\bibnamefont {{Reinthal}}}, \bibinfo {author} {\bibfnamefont {W.}~\bibnamefont {{Rhode}}}, \bibinfo {author} {\bibfnamefont {M.}~\bibnamefont {{Rib{\'o}}}}, \bibinfo {author} {\bibfnamefont {J.}~\bibnamefont {{Rico}}}, \bibinfo {author} {\bibfnamefont
  {S.}~\bibnamefont {{R{\"u}gamer}}}, \bibinfo {author} {\bibfnamefont {A.}~\bibnamefont {{Saggion}}}, \bibinfo {author} {\bibfnamefont {K.}~\bibnamefont {{Saito}}}, \bibinfo {author} {\bibfnamefont {T.~Y.}\ \bibnamefont {{Saito}}}, \bibinfo {author} {\bibfnamefont {M.}~\bibnamefont {{Salvati}}}, \bibinfo {author} {\bibfnamefont {K.}~\bibnamefont {{Satalecka}}}, \bibinfo {author} {\bibfnamefont {V.}~\bibnamefont {{Scalzotto}}}, \bibinfo {author} {\bibfnamefont {V.}~\bibnamefont {{Scapin}}}, \bibinfo {author} {\bibfnamefont {C.}~\bibnamefont {{Schultz}}}, \bibinfo {author} {\bibfnamefont {T.}~\bibnamefont {{Schweizer}}}, \bibinfo {author} {\bibfnamefont {M.}~\bibnamefont {{Shayduk}}}, \bibinfo {author} {\bibfnamefont {S.~N.}\ \bibnamefont {{Shore}}}, \bibinfo {author} {\bibfnamefont {A.}~\bibnamefont {{Sillanp{\"a}{\"a}}}}, \bibinfo {author} {\bibfnamefont {J.}~\bibnamefont {{Sitarek}}}, \bibinfo {author} {\bibfnamefont {D.}~\bibnamefont {{Sobczynska}}}, \bibinfo {author} {\bibfnamefont {F.}~\bibnamefont
  {{Spanier}}}, \bibinfo {author} {\bibfnamefont {S.}~\bibnamefont {{Spiro}}}, \bibinfo {author} {\bibfnamefont {A.}~\bibnamefont {{Stamerra}}}, \bibinfo {author} {\bibfnamefont {B.}~\bibnamefont {{Steinke}}}, \bibinfo {author} {\bibfnamefont {J.}~\bibnamefont {{Storz}}}, \bibinfo {author} {\bibfnamefont {N.}~\bibnamefont {{Strah}}}, \bibinfo {author} {\bibfnamefont {T.}~\bibnamefont {{Suri{\'c}}}}, \bibinfo {author} {\bibfnamefont {L.}~\bibnamefont {{Takalo}}}, \bibinfo {author} {\bibfnamefont {H.}~\bibnamefont {{Takami}}}, \bibinfo {author} {\bibfnamefont {F.}~\bibnamefont {{Tavecchio}}}, \bibinfo {author} {\bibfnamefont {P.}~\bibnamefont {{Temnikov}}}, \bibinfo {author} {\bibfnamefont {T.}~\bibnamefont {{Terzi{\'c}}}}, \bibinfo {author} {\bibfnamefont {D.}~\bibnamefont {{Tescaro}}}, \bibinfo {author} {\bibfnamefont {M.}~\bibnamefont {{Teshima}}}, \bibinfo {author} {\bibfnamefont {M.}~\bibnamefont {{Thom}}}, \bibinfo {author} {\bibfnamefont {O.}~\bibnamefont {{Tibolla}}}, \bibinfo {author} {\bibfnamefont
  {D.~F.}\ \bibnamefont {{Torres}}}, \bibinfo {author} {\bibfnamefont {A.}~\bibnamefont {{Treves}}}, \bibinfo {author} {\bibfnamefont {H.}~\bibnamefont {{Vankov}}}, \bibinfo {author} {\bibfnamefont {P.}~\bibnamefont {{Vogler}}}, \bibinfo {author} {\bibfnamefont {R.~M.}\ \bibnamefont {{Wagner}}}, \bibinfo {author} {\bibfnamefont {Q.}~\bibnamefont {{Weitzel}}}, \bibinfo {author} {\bibfnamefont {V.}~\bibnamefont {{Zabalza}}}, \bibinfo {author} {\bibfnamefont {F.}~\bibnamefont {{Zandanel}}}, \bibinfo {author} {\bibfnamefont {R.}~\bibnamefont {{Zanin}}}, \bibinfo {author} {\bibnamefont {{MAGIC Collaboration}}}, \bibinfo {author} {\bibfnamefont {T.}~\bibnamefont {{Arlen}}}, \bibinfo {author} {\bibfnamefont {T.}~\bibnamefont {{Aune}}}, \bibinfo {author} {\bibfnamefont {M.}~\bibnamefont {{Beilicke}}}, \bibinfo {author} {\bibfnamefont {W.}~\bibnamefont {{Benbow}}}, \bibinfo {author} {\bibfnamefont {A.}~\bibnamefont {{Bouvier}}}, \bibinfo {author} {\bibfnamefont {S.~M.}\ \bibnamefont {{Bradbury}}}, \bibinfo {author}
  {\bibfnamefont {J.~H.}\ \bibnamefont {{Buckley}}}, \bibinfo {author} {\bibfnamefont {V.}~\bibnamefont {{Bugaev}}}, \bibinfo {author} {\bibfnamefont {K.}~\bibnamefont {{Byrum}}}, \bibinfo {author} {\bibfnamefont {A.}~\bibnamefont {{Cannon}}}, \bibinfo {author} {\bibfnamefont {A.}~\bibnamefont {{Cesarini}}}, \bibinfo {author} {\bibfnamefont {L.}~\bibnamefont {{Ciupik}}}, \bibinfo {author} {\bibfnamefont {M.~P.}\ \bibnamefont {{Connolly}}}, \bibinfo {author} {\bibfnamefont {W.}~\bibnamefont {{Cui}}}, \bibinfo {author} {\bibfnamefont {R.}~\bibnamefont {{Dickherber}}}, \bibinfo {author} {\bibfnamefont {C.}~\bibnamefont {{Duke}}}, \bibinfo {author} {\bibfnamefont {M.}~\bibnamefont {{Errando}}}, \bibinfo {author} {\bibfnamefont {A.}~\bibnamefont {{Falcone}}}, \bibinfo {author} {\bibfnamefont {J.~P.}\ \bibnamefont {{Finley}}}, \bibinfo {author} {\bibfnamefont {G.}~\bibnamefont {{Finnegan}}}, \bibinfo {author} {\bibfnamefont {L.}~\bibnamefont {{Fortson}}}, \bibinfo {author} {\bibfnamefont {A.}~\bibnamefont
  {{Furniss}}}, \bibinfo {author} {\bibfnamefont {N.}~\bibnamefont {{Galante}}}, \bibinfo {author} {\bibfnamefont {D.}~\bibnamefont {{Gall}}}, \bibinfo {author} {\bibfnamefont {S.}~\bibnamefont {{Godambe}}}, \bibinfo {author} {\bibfnamefont {S.}~\bibnamefont {{Griffin}}}, \bibinfo {author} {\bibfnamefont {J.}~\bibnamefont {{Grube}}}, \bibinfo {author} {\bibfnamefont {G.}~\bibnamefont {{Gyuk}}}, \bibinfo {author} {\bibfnamefont {D.}~\bibnamefont {{Hanna}}}, \bibinfo {author} {\bibfnamefont {J.}~\bibnamefont {{Holder}}}, \bibinfo {author} {\bibfnamefont {H.}~\bibnamefont {{Huan}}}, \bibinfo {author} {\bibfnamefont {C.~M.}\ \bibnamefont {{Hui}}}, \bibinfo {author} {\bibfnamefont {P.}~\bibnamefont {{Kaaret}}}, \bibinfo {author} {\bibfnamefont {N.}~\bibnamefont {{Karlsson}}}, \bibinfo {author} {\bibfnamefont {M.}~\bibnamefont {{Kertzman}}}, \bibinfo {author} {\bibfnamefont {Y.}~\bibnamefont {{Khassen}}}, \bibinfo {author} {\bibfnamefont {D.}~\bibnamefont {{Kieda}}}, \bibinfo {author} {\bibfnamefont
  {H.}~\bibnamefont {{Krawczynski}}}, \bibinfo {author} {\bibfnamefont {F.}~\bibnamefont {{Krennrich}}}, \bibinfo {author} {\bibfnamefont {M.~J.}\ \bibnamefont {{Lang}}}, \bibinfo {author} {\bibfnamefont {S.}~\bibnamefont {{LeBohec}}}, \bibinfo {author} {\bibfnamefont {G.}~\bibnamefont {{Maier}}}, \bibinfo {author} {\bibfnamefont {S.}~\bibnamefont {{McArthur}}}, \bibinfo {author} {\bibfnamefont {A.}~\bibnamefont {{McCann}}}, \bibinfo {author} {\bibfnamefont {P.}~\bibnamefont {{Moriarty}}}, \bibinfo {author} {\bibfnamefont {R.}~\bibnamefont {{Mukherjee}}}, \bibinfo {author} {\bibfnamefont {P.~D.}\ \bibnamefont {{Nu{\~n}ez}}}, \bibinfo {author} {\bibfnamefont {R.~A.}\ \bibnamefont {{Ong}}}, \bibinfo {author} {\bibfnamefont {M.}~\bibnamefont {{Orr}}}, \bibinfo {author} {\bibfnamefont {A.~N.}\ \bibnamefont {{Otte}}}, \bibinfo {author} {\bibfnamefont {N.}~\bibnamefont {{Park}}}, \bibinfo {author} {\bibfnamefont {J.~S.}\ \bibnamefont {{Perkins}}}, \bibinfo {author} {\bibfnamefont {A.}~\bibnamefont {{Pichel}}},
  \bibinfo {author} {\bibfnamefont {M.}~\bibnamefont {{Pohl}}}, \bibinfo {author} {\bibfnamefont {H.}~\bibnamefont {{Prokoph}}}, \bibinfo {author} {\bibfnamefont {K.}~\bibnamefont {{Ragan}}}, \bibinfo {author} {\bibfnamefont {L.~C.}\ \bibnamefont {{Reyes}}}, \bibinfo {author} {\bibfnamefont {P.~T.}\ \bibnamefont {{Reynolds}}}, \bibinfo {author} {\bibfnamefont {E.}~\bibnamefont {{Roache}}}, \bibinfo {author} {\bibfnamefont {H.~J.}\ \bibnamefont {{Rose}}}, \bibinfo {author} {\bibfnamefont {J.}~\bibnamefont {{Ruppel}}}, \bibinfo {author} {\bibfnamefont {M.}~\bibnamefont {{Schroedter}}}, \bibinfo {author} {\bibfnamefont {G.~H.}\ \bibnamefont {{Sembroski}}}, \bibinfo {author} {\bibfnamefont {G.~D.}\ \bibnamefont {{{\c{S}}ent{\"u}rk}}}, \bibinfo {author} {\bibfnamefont {I.}~\bibnamefont {{Telezhinsky}}}, \bibinfo {author} {\bibfnamefont {G.}~\bibnamefont {{Te{\v{s}}i{\'c}}}}, \bibinfo {author} {\bibfnamefont {M.}~\bibnamefont {{Theiling}}}, \bibinfo {author} {\bibfnamefont {S.}~\bibnamefont {{Thibadeau}}}, \bibinfo
  {author} {\bibfnamefont {A.}~\bibnamefont {{Varlotta}}}, \bibinfo {author} {\bibfnamefont {V.~V.}\ \bibnamefont {{Vassiliev}}}, \bibinfo {author} {\bibfnamefont {M.}~\bibnamefont {{Vivier}}}, \bibinfo {author} {\bibfnamefont {S.~P.}\ \bibnamefont {{Wakely}}}, \bibinfo {author} {\bibfnamefont {T.~C.}\ \bibnamefont {{Weekes}}}, \bibinfo {author} {\bibfnamefont {D.~A.}\ \bibnamefont {{Williams}}}, \bibinfo {author} {\bibfnamefont {B.}~\bibnamefont {{Zitzer}}}, \bibinfo {author} {\bibnamefont {{VERITAS Collaboration}}}, \bibinfo {author} {\bibfnamefont {U.}~\bibnamefont {{Barres de Almeida}}}, \bibinfo {author} {\bibfnamefont {M.}~\bibnamefont {{Cara}}}, \bibinfo {author} {\bibfnamefont {C.}~\bibnamefont {{Casadio}}}, \bibinfo {author} {\bibfnamefont {C.~C.}\ \bibnamefont {{Cheung}}}, \bibinfo {author} {\bibfnamefont {W.}~\bibnamefont {{McConville}}}, \bibinfo {author} {\bibfnamefont {F.}~\bibnamefont {{Davies}}}, \bibinfo {author} {\bibfnamefont {A.}~\bibnamefont {{Doi}}}, \bibinfo {author} {\bibfnamefont
  {G.}~\bibnamefont {{Giovannini}}}, \bibinfo {author} {\bibfnamefont {M.}~\bibnamefont {{Giroletti}}}, \bibinfo {author} {\bibfnamefont {K.}~\bibnamefont {{Hada}}}, \bibinfo {author} {\bibfnamefont {P.}~\bibnamefont {{Hardee}}}, \bibinfo {author} {\bibfnamefont {D.~E.}\ \bibnamefont {{Harris}}}, \bibinfo {author} {\bibfnamefont {W.}~\bibnamefont {{Junor}}}, \bibinfo {author} {\bibfnamefont {M.}~\bibnamefont {{Kino}}}, \bibinfo {author} {\bibfnamefont {N.~P.}\ \bibnamefont {{Lee}}}, \bibinfo {author} {\bibfnamefont {C.}~\bibnamefont {{Ly}}}, \bibinfo {author} {\bibfnamefont {J.}~\bibnamefont {{Madrid}}}, \bibinfo {author} {\bibfnamefont {F.}~\bibnamefont {{Massaro}}}, \bibinfo {author} {\bibfnamefont {C.~G.}\ \bibnamefont {{Mundell}}}, \bibinfo {author} {\bibfnamefont {H.}~\bibnamefont {{Nagai}}}, \bibinfo {author} {\bibfnamefont {E.~S.}\ \bibnamefont {{Perlman}}}, \bibinfo {author} {\bibfnamefont {I.~A.}\ \bibnamefont {{Steele}}}, \bibinfo {author} {\bibfnamefont {R.~C.}\ \bibnamefont {{Walker}}},\ and\
  \bibinfo {author} {\bibfnamefont {D.~L.}\ \bibnamefont {{Wood}}},\ }\href {https://doi.org/10.1088/0004-637X/746/2/151} {\bibfield  {journal} {\bibinfo  {journal} {ApJ}\ }\textbf {\bibinfo {volume} {746}},\ \bibinfo {eid} {151} (\bibinfo {year} {2012})},\ \Eprint {https://arxiv.org/abs/1111.5341} {arXiv:1111.5341 [astro-ph.CO]} \BibitemShut {NoStop}%
\bibitem [{\citenamefont {{Algaba}}\ \emph {et~al.}(2024)\citenamefont {{Algaba}}, \citenamefont {{Balokovi{\'c}}}, \citenamefont {{Chandra}}, \citenamefont {{Cheong}}, \citenamefont {{Cui}}, \citenamefont {{D'Ammando}}, \citenamefont {{Falcone}}, \citenamefont {{Ford}}, \citenamefont {{Giroletti}}, \citenamefont {{Goddi}}, \citenamefont {{Gurwell}}, \citenamefont {{Hada}}, \citenamefont {{Haggard}}, \citenamefont {{Jorstad}}, \citenamefont {{Kaur}}, \citenamefont {{Kawashima}}, \citenamefont {{Kerby}}, \citenamefont {{Kim}}, \citenamefont {{Kino}}, \citenamefont {{Kravchenko}}, \citenamefont {{Lee}}, \citenamefont {{Lu}}, \citenamefont {{Markoff}}, \citenamefont {{Michail}}, \citenamefont {{Neilsen}}, \citenamefont {{Nowak}}, \citenamefont {{Principe}}, \citenamefont {{Ramakrishnan}}, \citenamefont {{Ripperda}}, \citenamefont {{Sasada}}, \citenamefont {{Savchenko}}, \citenamefont {{Sheridan}}, \citenamefont {{Akiyama}}, \citenamefont {{Alberdi}}, \citenamefont {{Alef}}, \citenamefont {{Anantua}},
  \citenamefont {{Asada}}, \citenamefont {{Azulay}}, \citenamefont {{Bach}}, \citenamefont {{Baczko}}, \citenamefont {{Ball}}, \citenamefont {{Bandyopadhyay}}, \citenamefont {{Barrett}}, \citenamefont {{Baub{\"o}ck}}, \citenamefont {{Benson}}, \citenamefont {{Bintley}}, \citenamefont {{lackburn}}, \citenamefont {{Blundell}}, \citenamefont {{Bouman}}, \citenamefont {{Bower}}, \citenamefont {{Boyce}}, \citenamefont {{Bremer}}, \citenamefont {{Brissenden}}, \citenamefont {{Britzen}}, \citenamefont {{Broderick}}, \citenamefont {{Broguiere}}, \citenamefont {{Bronzwaer}}, \citenamefont {{Bustamante}}, \citenamefont {{Carlstrom}}, \citenamefont {{Chael}}, \citenamefont {{Chan}}, \citenamefont {{Chang}}, \citenamefont {{Chatterjee}}, \citenamefont {{Chatterjee}}, \citenamefont {{Chen}}, \citenamefont {{Chen}}, \citenamefont {{Cheng}}, \citenamefont {{Cho}}, \citenamefont {{Christian}}, \citenamefont {{Conroy}}, \citenamefont {{Conway}}, \citenamefont {{Crawford}}, \citenamefont {{Crew}}, \citenamefont
  {{Cruz-Osorio}}, \citenamefont {{Dahale}}, \citenamefont {{Davelaar}}, \citenamefont {{De Laurentis}}, \citenamefont {{Deane}}, \citenamefont {{Dempsey}}, \citenamefont {{Desvignes}}, \citenamefont {{Dexter}}, \citenamefont {{Dhruv}}, \citenamefont {{Dihingia}}, \citenamefont {{Doeleman}}, \citenamefont {{Dzib}}, \citenamefont {{Eatough}}, \citenamefont {{Emami}}, \citenamefont {{Falcke}}, \citenamefont {{Farah}}, \citenamefont {{Fish}}, \citenamefont {{Fomalont}}, \citenamefont {{Ford}}, \citenamefont {{Foschi}}, \citenamefont {{Fraga-Encinas}}, \citenamefont {{Freeman}}, \citenamefont {{Friberg}}, \citenamefont {{Fromm}}, \citenamefont {{Fuentes}}, \citenamefont {{Galison}}, \citenamefont {{Gammie}}, \citenamefont {{Garc{\'\i}a}}, \citenamefont {{Gentaz}}, \citenamefont {{Georgiev}}, \citenamefont {{Gold}}, \citenamefont {{G{\'o}mez-Ruiz}}, \citenamefont {{G{\'o}mez}}, \citenamefont {{Gu}}, \citenamefont {{Hesper}}, \citenamefont {{Heumann}}, \citenamefont {{Ho}}, \citenamefont {{Ho}}, \citenamefont
  {{Honma}}, \citenamefont {{Huang}}, \citenamefont {{Huang}}, \citenamefont {{Hughes}}, \citenamefont {{Ikeda}}, \citenamefont {{Impellizzeri}}, \citenamefont {{Inoue}}, \citenamefont {{Issaoun}}, \citenamefont {{James}}, \citenamefont {{Jannuzi}}, \citenamefont {{Janssen}}, \citenamefont {{Jeter}}, \citenamefont {{Jiang}}, \citenamefont {{Jim{\'e}nez-Rosales}}, \citenamefont {{Johnson}}, \citenamefont {{Jones}}, \citenamefont {{Joshi}}, \citenamefont {{Jung}}, \citenamefont {{Karuppusamy}}, \citenamefont {{Keating}}, \citenamefont {{Kettenis}}, \citenamefont {{Kim}}, \citenamefont {{Kim}}, \citenamefont {{Kim}}, \citenamefont {{Koay}}, \citenamefont {{Kocherlakota}}, \citenamefont {{Kofuji}}, \citenamefont {{Koch}}, \citenamefont {{Koyama}}, \citenamefont {{Kramer}}, \citenamefont {{Kramer}}, \citenamefont {{Kramer}}, \citenamefont {{Krichbaum}}, \citenamefont {{Kuo}}, \citenamefont {{La Bella}}, \citenamefont {{Levis}}, \citenamefont {{Li}}, \citenamefont {{Lico}}, \citenamefont {{Lindahl}}, \citenamefont
  {{Lindqvist}}, \citenamefont {{Lisakov}}, \citenamefont {{Liu}}, \citenamefont {{Liu}}, \citenamefont {{Liuzzo}}, \citenamefont {{Lo}}, \citenamefont {{Lobanov}}, \citenamefont {{Loinard}}, \citenamefont {{Lonsdale}}, \citenamefont {{Lowitz}}, \citenamefont {{MacDonald}}, \citenamefont {{Mao}}, \citenamefont {{Marchili}}, \citenamefont {{Marrone}}, \citenamefont {{Marscher}}, \citenamefont {{Mart{\'\i}-Vidal}}, \citenamefont {{Matsushita}}, \citenamefont {{Matthews}}, \citenamefont {{Medeiros}}, \citenamefont {{Menten}}, \citenamefont {{Mizuno}}, \citenamefont {{Mizuno}}, \citenamefont {{Montgomery}}, \citenamefont {{Moran}}, \citenamefont {{Moriyama}}, \citenamefont {{Moscibrodzka}}, \citenamefont {{Mulaudzi}}, \citenamefont {{M{\"u}ller}}, \citenamefont {{M{\"u}ller}}, \citenamefont {{Mus}}, \citenamefont {{Musoke}}, \citenamefont {{Myserlis}}, \citenamefont {{Nagai}}, \citenamefont {{Nagar}}, \citenamefont {{Nair}}, \citenamefont {{Nakamura}}, \citenamefont {{Narayanan}}, \citenamefont {{Natarajan}},
  \citenamefont {{Nathanail}}, \citenamefont {{Navarro Fuentes}}, \citenamefont {{Ni}}, \citenamefont {{Oh}}, \citenamefont {{Okino}}, \citenamefont {{Olivares}}, \citenamefont {{Oyama}}, \citenamefont {{{\"O}zel}}, \citenamefont {{Palumbo}}, \citenamefont {{Filippos Paraschos}}, \citenamefont {{Park}},\ and\ \citenamefont {{Parsons}}}]{Algaba24}%
  \BibitemOpen
  \bibfield  {author} {\bibinfo {author} {\bibfnamefont {J.~C.}\ \bibnamefont {{Algaba}}}, \bibinfo {author} {\bibfnamefont {M.}~\bibnamefont {{Balokovi{\'c}}}}, \bibinfo {author} {\bibfnamefont {S.}~\bibnamefont {{Chandra}}}, \bibinfo {author} {\bibfnamefont {W.~Y.}\ \bibnamefont {{Cheong}}}, \bibinfo {author} {\bibfnamefont {Y.~Z.}\ \bibnamefont {{Cui}}}, \bibinfo {author} {\bibfnamefont {F.}~\bibnamefont {{D'Ammando}}}, \bibinfo {author} {\bibfnamefont {A.~D.}\ \bibnamefont {{Falcone}}}, \bibinfo {author} {\bibfnamefont {N.~M.}\ \bibnamefont {{Ford}}}, \bibinfo {author} {\bibfnamefont {M.}~\bibnamefont {{Giroletti}}}, \bibinfo {author} {\bibfnamefont {C.}~\bibnamefont {{Goddi}}}, \bibinfo {author} {\bibfnamefont {M.~A.}\ \bibnamefont {{Gurwell}}}, \bibinfo {author} {\bibfnamefont {K.}~\bibnamefont {{Hada}}}, \bibinfo {author} {\bibfnamefont {D.}~\bibnamefont {{Haggard}}}, \bibinfo {author} {\bibfnamefont {S.}~\bibnamefont {{Jorstad}}}, \bibinfo {author} {\bibfnamefont {A.}~\bibnamefont {{Kaur}}}, \bibinfo
  {author} {\bibfnamefont {T.}~\bibnamefont {{Kawashima}}}, \bibinfo {author} {\bibfnamefont {S.}~\bibnamefont {{Kerby}}}, \bibinfo {author} {\bibfnamefont {J.~Y.}\ \bibnamefont {{Kim}}}, \bibinfo {author} {\bibfnamefont {M.}~\bibnamefont {{Kino}}}, \bibinfo {author} {\bibfnamefont {E.~V.}\ \bibnamefont {{Kravchenko}}}, \bibinfo {author} {\bibfnamefont {S.~S.}\ \bibnamefont {{Lee}}}, \bibinfo {author} {\bibfnamefont {R.~S.}\ \bibnamefont {{Lu}}}, \bibinfo {author} {\bibfnamefont {S.}~\bibnamefont {{Markoff}}}, \bibinfo {author} {\bibfnamefont {J.}~\bibnamefont {{Michail}}}, \bibinfo {author} {\bibfnamefont {J.}~\bibnamefont {{Neilsen}}}, \bibinfo {author} {\bibfnamefont {M.~A.}\ \bibnamefont {{Nowak}}}, \bibinfo {author} {\bibfnamefont {G.}~\bibnamefont {{Principe}}}, \bibinfo {author} {\bibfnamefont {V.}~\bibnamefont {{Ramakrishnan}}}, \bibinfo {author} {\bibfnamefont {B.}~\bibnamefont {{Ripperda}}}, \bibinfo {author} {\bibfnamefont {M.}~\bibnamefont {{Sasada}}}, \bibinfo {author} {\bibfnamefont {S.~S.}\
  \bibnamefont {{Savchenko}}}, \bibinfo {author} {\bibfnamefont {C.}~\bibnamefont {{Sheridan}}}, \bibinfo {author} {\bibfnamefont {K.}~\bibnamefont {{Akiyama}}}, \bibinfo {author} {\bibfnamefont {A.}~\bibnamefont {{Alberdi}}}, \bibinfo {author} {\bibfnamefont {W.}~\bibnamefont {{Alef}}}, \bibinfo {author} {\bibfnamefont {R.}~\bibnamefont {{Anantua}}}, \bibinfo {author} {\bibfnamefont {K.}~\bibnamefont {{Asada}}}, \bibinfo {author} {\bibfnamefont {R.}~\bibnamefont {{Azulay}}}, \bibinfo {author} {\bibfnamefont {U.}~\bibnamefont {{Bach}}}, \bibinfo {author} {\bibfnamefont {A.~K.}\ \bibnamefont {{Baczko}}}, \bibinfo {author} {\bibfnamefont {D.}~\bibnamefont {{Ball}}}, \bibinfo {author} {\bibfnamefont {B.}~\bibnamefont {{Bandyopadhyay}}}, \bibinfo {author} {\bibfnamefont {J.}~\bibnamefont {{Barrett}}}, \bibinfo {author} {\bibfnamefont {M.}~\bibnamefont {{Baub{\"o}ck}}}, \bibinfo {author} {\bibfnamefont {B.~A.}\ \bibnamefont {{Benson}}}, \bibinfo {author} {\bibfnamefont {D.}~\bibnamefont {{Bintley}}}, \bibinfo
  {author} {\bibfnamefont {L.}~\bibnamefont {{lackburn}}}, \bibinfo {author} {\bibfnamefont {R.}~\bibnamefont {{Blundell}}}, \bibinfo {author} {\bibfnamefont {K.~L.}\ \bibnamefont {{Bouman}}}, \bibinfo {author} {\bibfnamefont {G.~C.}\ \bibnamefont {{Bower}}}, \bibinfo {author} {\bibfnamefont {H.}~\bibnamefont {{Boyce}}}, \bibinfo {author} {\bibfnamefont {M.}~\bibnamefont {{Bremer}}}, \bibinfo {author} {\bibfnamefont {R.}~\bibnamefont {{Brissenden}}}, \bibinfo {author} {\bibfnamefont {S.}~\bibnamefont {{Britzen}}}, \bibinfo {author} {\bibfnamefont {A.~E.}\ \bibnamefont {{Broderick}}}, \bibinfo {author} {\bibfnamefont {D.}~\bibnamefont {{Broguiere}}}, \bibinfo {author} {\bibfnamefont {T.}~\bibnamefont {{Bronzwaer}}}, \bibinfo {author} {\bibfnamefont {S.}~\bibnamefont {{Bustamante}}}, \bibinfo {author} {\bibfnamefont {J.~E.}\ \bibnamefont {{Carlstrom}}}, \bibinfo {author} {\bibfnamefont {A.}~\bibnamefont {{Chael}}}, \bibinfo {author} {\bibfnamefont {C.~k.}\ \bibnamefont {{Chan}}}, \bibinfo {author}
  {\bibfnamefont {D.~O.}\ \bibnamefont {{Chang}}}, \bibinfo {author} {\bibfnamefont {K.}~\bibnamefont {{Chatterjee}}}, \bibinfo {author} {\bibfnamefont {S.}~\bibnamefont {{Chatterjee}}}, \bibinfo {author} {\bibfnamefont {M.~T.}\ \bibnamefont {{Chen}}}, \bibinfo {author} {\bibfnamefont {Y.}~\bibnamefont {{Chen}}}, \bibinfo {author} {\bibfnamefont {X.}~\bibnamefont {{Cheng}}}, \bibinfo {author} {\bibfnamefont {I.}~\bibnamefont {{Cho}}}, \bibinfo {author} {\bibfnamefont {P.}~\bibnamefont {{Christian}}}, \bibinfo {author} {\bibfnamefont {N.~S.}\ \bibnamefont {{Conroy}}}, \bibinfo {author} {\bibfnamefont {J.~E.}\ \bibnamefont {{Conway}}}, \bibinfo {author} {\bibfnamefont {T.~M.}\ \bibnamefont {{Crawford}}}, \bibinfo {author} {\bibfnamefont {G.~B.}\ \bibnamefont {{Crew}}}, \bibinfo {author} {\bibfnamefont {A.}~\bibnamefont {{Cruz-Osorio}}}, \bibinfo {author} {\bibfnamefont {R.}~\bibnamefont {{Dahale}}}, \bibinfo {author} {\bibfnamefont {J.}~\bibnamefont {{Davelaar}}}, \bibinfo {author} {\bibfnamefont
  {M.}~\bibnamefont {{De Laurentis}}}, \bibinfo {author} {\bibfnamefont {R.}~\bibnamefont {{Deane}}}, \bibinfo {author} {\bibfnamefont {J.}~\bibnamefont {{Dempsey}}}, \bibinfo {author} {\bibfnamefont {G.}~\bibnamefont {{Desvignes}}}, \bibinfo {author} {\bibfnamefont {J.}~\bibnamefont {{Dexter}}}, \bibinfo {author} {\bibfnamefont {V.}~\bibnamefont {{Dhruv}}}, \bibinfo {author} {\bibfnamefont {I.~K.}\ \bibnamefont {{Dihingia}}}, \bibinfo {author} {\bibfnamefont {S.~S.}\ \bibnamefont {{Doeleman}}}, \bibinfo {author} {\bibfnamefont {S.~A.}\ \bibnamefont {{Dzib}}}, \bibinfo {author} {\bibfnamefont {R.~P.}\ \bibnamefont {{Eatough}}}, \bibinfo {author} {\bibfnamefont {R.}~\bibnamefont {{Emami}}}, \bibinfo {author} {\bibfnamefont {H.}~\bibnamefont {{Falcke}}}, \bibinfo {author} {\bibfnamefont {J.}~\bibnamefont {{Farah}}}, \bibinfo {author} {\bibfnamefont {V.~L.}\ \bibnamefont {{Fish}}}, \bibinfo {author} {\bibfnamefont {E.}~\bibnamefont {{Fomalont}}}, \bibinfo {author} {\bibfnamefont {H.~A.}\ \bibnamefont {{Ford}}},
  \bibinfo {author} {\bibfnamefont {M.}~\bibnamefont {{Foschi}}}, \bibinfo {author} {\bibfnamefont {R.}~\bibnamefont {{Fraga-Encinas}}}, \bibinfo {author} {\bibfnamefont {W.~T.}\ \bibnamefont {{Freeman}}}, \bibinfo {author} {\bibfnamefont {P.}~\bibnamefont {{Friberg}}}, \bibinfo {author} {\bibfnamefont {C.~M.}\ \bibnamefont {{Fromm}}}, \bibinfo {author} {\bibfnamefont {A.}~\bibnamefont {{Fuentes}}}, \bibinfo {author} {\bibfnamefont {P.}~\bibnamefont {{Galison}}}, \bibinfo {author} {\bibfnamefont {C.~F.}\ \bibnamefont {{Gammie}}}, \bibinfo {author} {\bibfnamefont {R.}~\bibnamefont {{Garc{\'\i}a}}}, \bibinfo {author} {\bibfnamefont {O.}~\bibnamefont {{Gentaz}}}, \bibinfo {author} {\bibfnamefont {B.}~\bibnamefont {{Georgiev}}}, \bibinfo {author} {\bibfnamefont {R.}~\bibnamefont {{Gold}}}, \bibinfo {author} {\bibfnamefont {A.~I.}\ \bibnamefont {{G{\'o}mez-Ruiz}}}, \bibinfo {author} {\bibfnamefont {J.~L.}\ \bibnamefont {{G{\'o}mez}}}, \bibinfo {author} {\bibfnamefont {M.}~\bibnamefont {{Gu}}}, \bibinfo {author}
  {\bibfnamefont {R.}~\bibnamefont {{Hesper}}}, \bibinfo {author} {\bibfnamefont {D.}~\bibnamefont {{Heumann}}}, \bibinfo {author} {\bibfnamefont {L.~C.}\ \bibnamefont {{Ho}}}, \bibinfo {author} {\bibfnamefont {P.}~\bibnamefont {{Ho}}}, \bibinfo {author} {\bibfnamefont {M.}~\bibnamefont {{Honma}}}, \bibinfo {author} {\bibfnamefont {C.~W.~L.}\ \bibnamefont {{Huang}}}, \bibinfo {author} {\bibfnamefont {L.}~\bibnamefont {{Huang}}}, \bibinfo {author} {\bibfnamefont {D.~H.}\ \bibnamefont {{Hughes}}}, \bibinfo {author} {\bibfnamefont {S.}~\bibnamefont {{Ikeda}}}, \bibinfo {author} {\bibfnamefont {C.~M.~V.}\ \bibnamefont {{Impellizzeri}}}, \bibinfo {author} {\bibfnamefont {M.}~\bibnamefont {{Inoue}}}, \bibinfo {author} {\bibfnamefont {S.}~\bibnamefont {{Issaoun}}}, \bibinfo {author} {\bibfnamefont {D.~J.}\ \bibnamefont {{James}}}, \bibinfo {author} {\bibfnamefont {B.~T.}\ \bibnamefont {{Jannuzi}}}, \bibinfo {author} {\bibfnamefont {M.}~\bibnamefont {{Janssen}}}, \bibinfo {author} {\bibfnamefont {B.}~\bibnamefont
  {{Jeter}}}, \bibinfo {author} {\bibfnamefont {W.}~\bibnamefont {{Jiang}}}, \bibinfo {author} {\bibfnamefont {A.}~\bibnamefont {{Jim{\'e}nez-Rosales}}}, \bibinfo {author} {\bibfnamefont {M.~D.}\ \bibnamefont {{Johnson}}}, \bibinfo {author} {\bibfnamefont {A.~C.}\ \bibnamefont {{Jones}}}, \bibinfo {author} {\bibfnamefont {A.~V.}\ \bibnamefont {{Joshi}}}, \bibinfo {author} {\bibfnamefont {T.}~\bibnamefont {{Jung}}}, \bibinfo {author} {\bibfnamefont {R.}~\bibnamefont {{Karuppusamy}}}, \bibinfo {author} {\bibfnamefont {G.~K.}\ \bibnamefont {{Keating}}}, \bibinfo {author} {\bibfnamefont {M.}~\bibnamefont {{Kettenis}}}, \bibinfo {author} {\bibfnamefont {D.~J.}\ \bibnamefont {{Kim}}}, \bibinfo {author} {\bibfnamefont {J.}~\bibnamefont {{Kim}}}, \bibinfo {author} {\bibfnamefont {J.}~\bibnamefont {{Kim}}}, \bibinfo {author} {\bibfnamefont {J.~Y.}\ \bibnamefont {{Koay}}}, \bibinfo {author} {\bibfnamefont {P.}~\bibnamefont {{Kocherlakota}}}, \bibinfo {author} {\bibfnamefont {Y.}~\bibnamefont {{Kofuji}}}, \bibinfo
  {author} {\bibfnamefont {P.~M.}\ \bibnamefont {{Koch}}}, \bibinfo {author} {\bibfnamefont {S.}~\bibnamefont {{Koyama}}}, \bibinfo {author} {\bibfnamefont {C.}~\bibnamefont {{Kramer}}}, \bibinfo {author} {\bibfnamefont {J.~A.}\ \bibnamefont {{Kramer}}}, \bibinfo {author} {\bibfnamefont {M.}~\bibnamefont {{Kramer}}}, \bibinfo {author} {\bibfnamefont {T.~P.}\ \bibnamefont {{Krichbaum}}}, \bibinfo {author} {\bibfnamefont {C.~Y.}\ \bibnamefont {{Kuo}}}, \bibinfo {author} {\bibfnamefont {N.}~\bibnamefont {{La Bella}}}, \bibinfo {author} {\bibfnamefont {A.}~\bibnamefont {{Levis}}}, \bibinfo {author} {\bibfnamefont {Z.}~\bibnamefont {{Li}}}, \bibinfo {author} {\bibfnamefont {R.}~\bibnamefont {{Lico}}}, \bibinfo {author} {\bibfnamefont {G.}~\bibnamefont {{Lindahl}}}, \bibinfo {author} {\bibfnamefont {M.}~\bibnamefont {{Lindqvist}}}, \bibinfo {author} {\bibfnamefont {M.}~\bibnamefont {{Lisakov}}}, \bibinfo {author} {\bibfnamefont {J.}~\bibnamefont {{Liu}}}, \bibinfo {author} {\bibfnamefont {K.}~\bibnamefont {{Liu}}},
  \bibinfo {author} {\bibfnamefont {E.}~\bibnamefont {{Liuzzo}}}, \bibinfo {author} {\bibfnamefont {W.~P.}\ \bibnamefont {{Lo}}}, \bibinfo {author} {\bibfnamefont {A.~P.}\ \bibnamefont {{Lobanov}}}, \bibinfo {author} {\bibfnamefont {L.}~\bibnamefont {{Loinard}}}, \bibinfo {author} {\bibfnamefont {C.~J.}\ \bibnamefont {{Lonsdale}}}, \bibinfo {author} {\bibfnamefont {A.~E.}\ \bibnamefont {{Lowitz}}}, \bibinfo {author} {\bibfnamefont {N.~R.}\ \bibnamefont {{MacDonald}}}, \bibinfo {author} {\bibfnamefont {J.}~\bibnamefont {{Mao}}}, \bibinfo {author} {\bibfnamefont {N.}~\bibnamefont {{Marchili}}}, \bibinfo {author} {\bibfnamefont {D.~P.}\ \bibnamefont {{Marrone}}}, \bibinfo {author} {\bibfnamefont {A.~P.}\ \bibnamefont {{Marscher}}}, \bibinfo {author} {\bibfnamefont {I.}~\bibnamefont {{Mart{\'\i}-Vidal}}}, \bibinfo {author} {\bibfnamefont {S.}~\bibnamefont {{Matsushita}}}, \bibinfo {author} {\bibfnamefont {L.~D.}\ \bibnamefont {{Matthews}}}, \bibinfo {author} {\bibfnamefont {L.}~\bibnamefont {{Medeiros}}},
  \bibinfo {author} {\bibfnamefont {K.~M.}\ \bibnamefont {{Menten}}}, \bibinfo {author} {\bibfnamefont {I.}~\bibnamefont {{Mizuno}}}, \bibinfo {author} {\bibfnamefont {Y.}~\bibnamefont {{Mizuno}}}, \bibinfo {author} {\bibfnamefont {J.}~\bibnamefont {{Montgomery}}}, \bibinfo {author} {\bibfnamefont {J.~M.}\ \bibnamefont {{Moran}}}, \bibinfo {author} {\bibfnamefont {K.}~\bibnamefont {{Moriyama}}}, \bibinfo {author} {\bibfnamefont {M.}~\bibnamefont {{Moscibrodzka}}}, \bibinfo {author} {\bibfnamefont {W.}~\bibnamefont {{Mulaudzi}}}, \bibinfo {author} {\bibfnamefont {C.}~\bibnamefont {{M{\"u}ller}}}, \bibinfo {author} {\bibfnamefont {H.}~\bibnamefont {{M{\"u}ller}}}, \bibinfo {author} {\bibfnamefont {A.}~\bibnamefont {{Mus}}}, \bibinfo {author} {\bibfnamefont {G.}~\bibnamefont {{Musoke}}}, \bibinfo {author} {\bibfnamefont {I.}~\bibnamefont {{Myserlis}}}, \bibinfo {author} {\bibfnamefont {H.}~\bibnamefont {{Nagai}}}, \bibinfo {author} {\bibfnamefont {N.~M.}\ \bibnamefont {{Nagar}}}, \bibinfo {author} {\bibfnamefont
  {D.~G.}\ \bibnamefont {{Nair}}}, \bibinfo {author} {\bibfnamefont {M.}~\bibnamefont {{Nakamura}}}, \bibinfo {author} {\bibfnamefont {G.}~\bibnamefont {{Narayanan}}}, \bibinfo {author} {\bibfnamefont {I.}~\bibnamefont {{Natarajan}}}, \bibinfo {author} {\bibfnamefont {A.}~\bibnamefont {{Nathanail}}}, \bibinfo {author} {\bibfnamefont {S.}~\bibnamefont {{Navarro Fuentes}}}, \bibinfo {author} {\bibfnamefont {C.}~\bibnamefont {{Ni}}}, \bibinfo {author} {\bibfnamefont {J.}~\bibnamefont {{Oh}}}, \bibinfo {author} {\bibfnamefont {H.}~\bibnamefont {{Okino}}}, \bibinfo {author} {\bibfnamefont {H.}~\bibnamefont {{Olivares}}}, \bibinfo {author} {\bibfnamefont {T.}~\bibnamefont {{Oyama}}}, \bibinfo {author} {\bibfnamefont {F.}~\bibnamefont {{{\"O}zel}}}, \bibinfo {author} {\bibfnamefont {D.~C.~M.}\ \bibnamefont {{Palumbo}}}, \bibinfo {author} {\bibfnamefont {G.}~\bibnamefont {{Filippos Paraschos}}}, \bibinfo {author} {\bibfnamefont {J.}~\bibnamefont {{Park}}},\ and\ \bibinfo {author} {\bibfnamefont {H.}~\bibnamefont
  {{Parsons}}},\ }\href {https://doi.org/10.1051/0004-6361/202450497} {\bibfield  {journal} {\bibinfo  {journal} {A\&A}\ }\textbf {\bibinfo {volume} {692}},\ \bibinfo {eid} {A140} (\bibinfo {year} {2024})},\ \Eprint {https://arxiv.org/abs/2404.17623} {arXiv:2404.17623 [astro-ph.HE]} \BibitemShut {NoStop}%
\bibitem [{\citenamefont {{Cao}}\ \emph {et~al.}(2024)\citenamefont {{Cao}}, \citenamefont {{Aharonian}}, \citenamefont {{Axikegu}}, \citenamefont {{Bai}}, \citenamefont {{Bao}}, \citenamefont {{Bastieri}}, \citenamefont {{Bi}}, \citenamefont {{Bi}}, \citenamefont {{Bian}}, \citenamefont {{Bukevich}}, \citenamefont {{Cao}}, \citenamefont {{Cao}}, \citenamefont {{Cao}}, \citenamefont {{Chang}}, \citenamefont {{Chang}}, \citenamefont {{Chen}}, \citenamefont {{Chen}}, \citenamefont {{Chen}}, \citenamefont {{Chen}}, \citenamefont {{Chen}}, \citenamefont {{Chen}}, \citenamefont {{Chen}}, \citenamefont {{Chen}}, \citenamefont {{Chen}}, \citenamefont {{Chen}}, \citenamefont {{Chen}}, \citenamefont {{Chen}}, \citenamefont {{Chen}}, \citenamefont {{Chen}}, \citenamefont {{Cheng}}, \citenamefont {{Cheng}}, \citenamefont {{Chu}}, \citenamefont {{Cui}}, \citenamefont {{Cui}}, \citenamefont {{Cui}}, \citenamefont {{Cui}}, \citenamefont {{Dai}}, \citenamefont {{Dai}}, \citenamefont {{Dai}}, \citenamefont {{Danzengluobu}},
  \citenamefont {{Dong}}, \citenamefont {{Duan}}, \citenamefont {{Fan}}, \citenamefont {{Fan}}, \citenamefont {{Fang}}, \citenamefont {{Fang}}, \citenamefont {{Fang}}, \citenamefont {{Feng}}, \citenamefont {{Feng}}, \citenamefont {{Feng}}, \citenamefont {{Feng}}, \citenamefont {{Feng}}, \citenamefont {{Feng}}, \citenamefont {{Feng}}, \citenamefont {{Gabici}}, \citenamefont {{Gao}}, \citenamefont {{Gao}}, \citenamefont {{Gao}}, \citenamefont {{Gao}}, \citenamefont {{Gao}}, \citenamefont {{Ge}}, \citenamefont {{Ge}}, \citenamefont {{Geng}}, \citenamefont {{Giacinti}}, \citenamefont {{Gong}}, \citenamefont {{Gou}}, \citenamefont {{Gu}}, \citenamefont {{Guo}}, \citenamefont {{Guo}}, \citenamefont {{Guo}}, \citenamefont {{Guo}}, \citenamefont {{Guo}}, \citenamefont {{Han}}, \citenamefont {{Hannuksela}}, \citenamefont {{Hasan}}, \citenamefont {{He}}, \citenamefont {{He}}, \citenamefont {{He}}, \citenamefont {{He}}, \citenamefont {{Hor}}, \citenamefont {{Hou}}, \citenamefont {{Hou}}, \citenamefont {{Hou}},
  \citenamefont {{Hu}}, \citenamefont {{Hu}}, \citenamefont {{Hu}}, \citenamefont {{Huang}}, \citenamefont {{Huang}}, \citenamefont {{Huang}}, \citenamefont {{Huang}}, \citenamefont {{Huang}}, \citenamefont {{Huang}}, \citenamefont {{Huang}}, \citenamefont {{Huang}}, \citenamefont {{Ji}}, \citenamefont {{Jia}}, \citenamefont {{Jia}}, \citenamefont {{Jiang}}, \citenamefont {{Jiang}}, \citenamefont {{Jiang}}, \citenamefont {{Jiang}}, \citenamefont {{Jin}}, \citenamefont {{Kang}}, \citenamefont {{Karpikov}}, \citenamefont {{Khangulyan}}, \citenamefont {{Kuleshov}}, \citenamefont {{Kurinov}}, \citenamefont {{Li}}, \citenamefont {{Li}}, \citenamefont {{Li}}, \citenamefont {{Li}}, \citenamefont {{Li}}, \citenamefont {{Li}}, \citenamefont {{Li}}, \citenamefont {{Li}}, \citenamefont {{Li}}, \citenamefont {{Li}}, \citenamefont {{Li}}, \citenamefont {{Li}}, \citenamefont {{Li}}, \citenamefont {{Li}}, \citenamefont {{Li}}, \citenamefont {{Li}}, \citenamefont {{Li}}, \citenamefont {{Li}}, \citenamefont {{Li}},
  \citenamefont {{Liang}}, \citenamefont {{Liang}}, \citenamefont {{Lin}}, \citenamefont {{Liu}}, \citenamefont {{Liu}}, \citenamefont {{Liu}}, \citenamefont {{Liu}}, \citenamefont {{Liu}}, \citenamefont {{Liu}}, \citenamefont {{Liu}}, \citenamefont {{Liu}}, \citenamefont {{Liu}}, \citenamefont {{Liu}}, \citenamefont {{Liu}}, \citenamefont {{Liu}}, \citenamefont {{Liu}}, \citenamefont {{Liu}}, \citenamefont {{Luo}}, \citenamefont {{Luo}}, \citenamefont {{Lv}}, \citenamefont {{Ma}}, \citenamefont {{Ma}}, \citenamefont {{Ma}}, \citenamefont {{Mao}}, \citenamefont {{Min}}, \citenamefont {{Mitthumsiri}}, \citenamefont {{Mu}}, \citenamefont {{Nan}}, \citenamefont {{Neronov}}, \citenamefont {{Ng}}, \citenamefont {{Ou}}, \citenamefont {{Pattarakijwanich}}, \citenamefont {{Pei}}, \citenamefont {{Qi}}, \citenamefont {{Qi}}, \citenamefont {{Qiao}}, \citenamefont {{Qin}}, \citenamefont {{Raza}}, \citenamefont {{Ruffolo}}, \citenamefont {{S{\'a}iz}}, \citenamefont {{Saeed}}, \citenamefont {{Semikoz}}, \citenamefont
  {{Shao}}, \citenamefont {{Shchegolev}}, \citenamefont {{Sheng}}, \citenamefont {{Shu}}, \citenamefont {{Song}}, \citenamefont {{Stenkin}}, \citenamefont {{Stepanov}}, \citenamefont {{Su}}, \citenamefont {{Sun}}, \citenamefont {{Sun}}, \citenamefont {{Sun}}, \citenamefont {{Sun}}, \citenamefont {{Takata}}, \citenamefont {{Tam}}, \citenamefont {{Tang}}, \citenamefont {{Tang}}, \citenamefont {{Tang}}, \citenamefont {{Tian}}, \citenamefont {{Wan}}, \citenamefont {{Wang}}, \citenamefont {{Wang}}, \citenamefont {{Wang}}, \citenamefont {{Wang}}, \citenamefont {{Wang}}, \citenamefont {{Wang}}, \citenamefont {{Wang}}, \citenamefont {{Wang}}, \citenamefont {{Wang}}, \citenamefont {{Wang}}, \citenamefont {{Wang}}, \citenamefont {{Wang}},\ and\ \citenamefont {{Wang}}}]{Cao24}%
  \BibitemOpen
  \bibfield  {author} {\bibinfo {author} {\bibfnamefont {Z.}~\bibnamefont {{Cao}}}, \bibinfo {author} {\bibfnamefont {F.}~\bibnamefont {{Aharonian}}}, \bibinfo {author} {\bibnamefont {{Axikegu}}}, \bibinfo {author} {\bibfnamefont {Y.~X.}\ \bibnamefont {{Bai}}}, \bibinfo {author} {\bibfnamefont {Y.~W.}\ \bibnamefont {{Bao}}}, \bibinfo {author} {\bibfnamefont {D.}~\bibnamefont {{Bastieri}}}, \bibinfo {author} {\bibfnamefont {X.~J.}\ \bibnamefont {{Bi}}}, \bibinfo {author} {\bibfnamefont {Y.~J.}\ \bibnamefont {{Bi}}}, \bibinfo {author} {\bibfnamefont {W.}~\bibnamefont {{Bian}}}, \bibinfo {author} {\bibfnamefont {A.~V.}\ \bibnamefont {{Bukevich}}}, \bibinfo {author} {\bibfnamefont {Q.}~\bibnamefont {{Cao}}}, \bibinfo {author} {\bibfnamefont {W.~Y.}\ \bibnamefont {{Cao}}}, \bibinfo {author} {\bibfnamefont {Z.}~\bibnamefont {{Cao}}}, \bibinfo {author} {\bibfnamefont {J.}~\bibnamefont {{Chang}}}, \bibinfo {author} {\bibfnamefont {J.~F.}\ \bibnamefont {{Chang}}}, \bibinfo {author} {\bibfnamefont {A.~M.}\ \bibnamefont
  {{Chen}}}, \bibinfo {author} {\bibfnamefont {E.~S.}\ \bibnamefont {{Chen}}}, \bibinfo {author} {\bibfnamefont {H.~X.}\ \bibnamefont {{Chen}}}, \bibinfo {author} {\bibfnamefont {L.}~\bibnamefont {{Chen}}}, \bibinfo {author} {\bibfnamefont {L.}~\bibnamefont {{Chen}}}, \bibinfo {author} {\bibfnamefont {L.}~\bibnamefont {{Chen}}}, \bibinfo {author} {\bibfnamefont {M.~J.}\ \bibnamefont {{Chen}}}, \bibinfo {author} {\bibfnamefont {M.~L.}\ \bibnamefont {{Chen}}}, \bibinfo {author} {\bibfnamefont {Q.~H.}\ \bibnamefont {{Chen}}}, \bibinfo {author} {\bibfnamefont {S.}~\bibnamefont {{Chen}}}, \bibinfo {author} {\bibfnamefont {S.~H.}\ \bibnamefont {{Chen}}}, \bibinfo {author} {\bibfnamefont {S.~Z.}\ \bibnamefont {{Chen}}}, \bibinfo {author} {\bibfnamefont {T.~L.}\ \bibnamefont {{Chen}}}, \bibinfo {author} {\bibfnamefont {Y.}~\bibnamefont {{Chen}}}, \bibinfo {author} {\bibfnamefont {N.}~\bibnamefont {{Cheng}}}, \bibinfo {author} {\bibfnamefont {Y.~D.}\ \bibnamefont {{Cheng}}}, \bibinfo {author} {\bibfnamefont {M.~C.}\
  \bibnamefont {{Chu}}}, \bibinfo {author} {\bibfnamefont {M.~Y.}\ \bibnamefont {{Cui}}}, \bibinfo {author} {\bibfnamefont {S.~W.}\ \bibnamefont {{Cui}}}, \bibinfo {author} {\bibfnamefont {X.~H.}\ \bibnamefont {{Cui}}}, \bibinfo {author} {\bibfnamefont {Y.~D.}\ \bibnamefont {{Cui}}}, \bibinfo {author} {\bibfnamefont {B.~Z.}\ \bibnamefont {{Dai}}}, \bibinfo {author} {\bibfnamefont {H.~L.}\ \bibnamefont {{Dai}}}, \bibinfo {author} {\bibfnamefont {Z.~G.}\ \bibnamefont {{Dai}}}, \bibinfo {author} {\bibnamefont {{Danzengluobu}}}, \bibinfo {author} {\bibfnamefont {X.~Q.}\ \bibnamefont {{Dong}}}, \bibinfo {author} {\bibfnamefont {K.~K.}\ \bibnamefont {{Duan}}}, \bibinfo {author} {\bibfnamefont {J.~H.}\ \bibnamefont {{Fan}}}, \bibinfo {author} {\bibfnamefont {Y.~Z.}\ \bibnamefont {{Fan}}}, \bibinfo {author} {\bibfnamefont {J.}~\bibnamefont {{Fang}}}, \bibinfo {author} {\bibfnamefont {J.~H.}\ \bibnamefont {{Fang}}}, \bibinfo {author} {\bibfnamefont {K.}~\bibnamefont {{Fang}}}, \bibinfo {author} {\bibfnamefont {C.~F.}\
  \bibnamefont {{Feng}}}, \bibinfo {author} {\bibfnamefont {H.}~\bibnamefont {{Feng}}}, \bibinfo {author} {\bibfnamefont {L.}~\bibnamefont {{Feng}}}, \bibinfo {author} {\bibfnamefont {S.~H.}\ \bibnamefont {{Feng}}}, \bibinfo {author} {\bibfnamefont {X.~T.}\ \bibnamefont {{Feng}}}, \bibinfo {author} {\bibfnamefont {Y.}~\bibnamefont {{Feng}}}, \bibinfo {author} {\bibfnamefont {Y.~L.}\ \bibnamefont {{Feng}}}, \bibinfo {author} {\bibfnamefont {S.}~\bibnamefont {{Gabici}}}, \bibinfo {author} {\bibfnamefont {B.}~\bibnamefont {{Gao}}}, \bibinfo {author} {\bibfnamefont {C.~D.}\ \bibnamefont {{Gao}}}, \bibinfo {author} {\bibfnamefont {Q.}~\bibnamefont {{Gao}}}, \bibinfo {author} {\bibfnamefont {W.}~\bibnamefont {{Gao}}}, \bibinfo {author} {\bibfnamefont {W.~K.}\ \bibnamefont {{Gao}}}, \bibinfo {author} {\bibfnamefont {M.~M.}\ \bibnamefont {{Ge}}}, \bibinfo {author} {\bibfnamefont {T.~T.}\ \bibnamefont {{Ge}}}, \bibinfo {author} {\bibfnamefont {L.~S.}\ \bibnamefont {{Geng}}}, \bibinfo {author} {\bibfnamefont
  {G.}~\bibnamefont {{Giacinti}}}, \bibinfo {author} {\bibfnamefont {G.~H.}\ \bibnamefont {{Gong}}}, \bibinfo {author} {\bibfnamefont {Q.~B.}\ \bibnamefont {{Gou}}}, \bibinfo {author} {\bibfnamefont {M.~H.}\ \bibnamefont {{Gu}}}, \bibinfo {author} {\bibfnamefont {F.~L.}\ \bibnamefont {{Guo}}}, \bibinfo {author} {\bibfnamefont {J.}~\bibnamefont {{Guo}}}, \bibinfo {author} {\bibfnamefont {X.~L.}\ \bibnamefont {{Guo}}}, \bibinfo {author} {\bibfnamefont {Y.~Q.}\ \bibnamefont {{Guo}}}, \bibinfo {author} {\bibfnamefont {Y.~Y.}\ \bibnamefont {{Guo}}}, \bibinfo {author} {\bibfnamefont {Y.~A.}\ \bibnamefont {{Han}}}, \bibinfo {author} {\bibfnamefont {O.~A.}\ \bibnamefont {{Hannuksela}}}, \bibinfo {author} {\bibfnamefont {M.}~\bibnamefont {{Hasan}}}, \bibinfo {author} {\bibfnamefont {H.~H.}\ \bibnamefont {{He}}}, \bibinfo {author} {\bibfnamefont {H.~N.}\ \bibnamefont {{He}}}, \bibinfo {author} {\bibfnamefont {J.~Y.}\ \bibnamefont {{He}}}, \bibinfo {author} {\bibfnamefont {Y.}~\bibnamefont {{He}}}, \bibinfo {author}
  {\bibfnamefont {Y.~K.}\ \bibnamefont {{Hor}}}, \bibinfo {author} {\bibfnamefont {B.~W.}\ \bibnamefont {{Hou}}}, \bibinfo {author} {\bibfnamefont {C.}~\bibnamefont {{Hou}}}, \bibinfo {author} {\bibfnamefont {X.}~\bibnamefont {{Hou}}}, \bibinfo {author} {\bibfnamefont {H.~B.}\ \bibnamefont {{Hu}}}, \bibinfo {author} {\bibfnamefont {Q.}~\bibnamefont {{Hu}}}, \bibinfo {author} {\bibfnamefont {S.~C.}\ \bibnamefont {{Hu}}}, \bibinfo {author} {\bibfnamefont {C.}~\bibnamefont {{Huang}}}, \bibinfo {author} {\bibfnamefont {D.~H.}\ \bibnamefont {{Huang}}}, \bibinfo {author} {\bibfnamefont {T.~Q.}\ \bibnamefont {{Huang}}}, \bibinfo {author} {\bibfnamefont {W.~J.}\ \bibnamefont {{Huang}}}, \bibinfo {author} {\bibfnamefont {X.~T.}\ \bibnamefont {{Huang}}}, \bibinfo {author} {\bibfnamefont {X.~Y.}\ \bibnamefont {{Huang}}}, \bibinfo {author} {\bibfnamefont {Y.}~\bibnamefont {{Huang}}}, \bibinfo {author} {\bibfnamefont {Y.~Y.}\ \bibnamefont {{Huang}}}, \bibinfo {author} {\bibfnamefont {X.~L.}\ \bibnamefont {{Ji}}}, \bibinfo
  {author} {\bibfnamefont {H.~Y.}\ \bibnamefont {{Jia}}}, \bibinfo {author} {\bibfnamefont {K.}~\bibnamefont {{Jia}}}, \bibinfo {author} {\bibfnamefont {H.~B.}\ \bibnamefont {{Jiang}}}, \bibinfo {author} {\bibfnamefont {K.}~\bibnamefont {{Jiang}}}, \bibinfo {author} {\bibfnamefont {X.~W.}\ \bibnamefont {{Jiang}}}, \bibinfo {author} {\bibfnamefont {Z.~J.}\ \bibnamefont {{Jiang}}}, \bibinfo {author} {\bibfnamefont {M.}~\bibnamefont {{Jin}}}, \bibinfo {author} {\bibfnamefont {M.~M.}\ \bibnamefont {{Kang}}}, \bibinfo {author} {\bibfnamefont {I.}~\bibnamefont {{Karpikov}}}, \bibinfo {author} {\bibfnamefont {D.}~\bibnamefont {{Khangulyan}}}, \bibinfo {author} {\bibfnamefont {D.}~\bibnamefont {{Kuleshov}}}, \bibinfo {author} {\bibfnamefont {K.}~\bibnamefont {{Kurinov}}}, \bibinfo {author} {\bibfnamefont {B.~B.}\ \bibnamefont {{Li}}}, \bibinfo {author} {\bibfnamefont {C.~M.}\ \bibnamefont {{Li}}}, \bibinfo {author} {\bibfnamefont {C.}~\bibnamefont {{Li}}}, \bibinfo {author} {\bibfnamefont {C.}~\bibnamefont {{Li}}},
  \bibinfo {author} {\bibfnamefont {D.}~\bibnamefont {{Li}}}, \bibinfo {author} {\bibfnamefont {F.}~\bibnamefont {{Li}}}, \bibinfo {author} {\bibfnamefont {H.~B.}\ \bibnamefont {{Li}}}, \bibinfo {author} {\bibfnamefont {H.~C.}\ \bibnamefont {{Li}}}, \bibinfo {author} {\bibfnamefont {J.}~\bibnamefont {{Li}}}, \bibinfo {author} {\bibfnamefont {J.}~\bibnamefont {{Li}}}, \bibinfo {author} {\bibfnamefont {K.}~\bibnamefont {{Li}}}, \bibinfo {author} {\bibfnamefont {S.~D.}\ \bibnamefont {{Li}}}, \bibinfo {author} {\bibfnamefont {W.~L.}\ \bibnamefont {{Li}}}, \bibinfo {author} {\bibfnamefont {W.~L.}\ \bibnamefont {{Li}}}, \bibinfo {author} {\bibfnamefont {X.~R.}\ \bibnamefont {{Li}}}, \bibinfo {author} {\bibfnamefont {X.}~\bibnamefont {{Li}}}, \bibinfo {author} {\bibfnamefont {Y.~Z.}\ \bibnamefont {{Li}}}, \bibinfo {author} {\bibfnamefont {Z.}~\bibnamefont {{Li}}}, \bibinfo {author} {\bibfnamefont {Z.}~\bibnamefont {{Li}}}, \bibinfo {author} {\bibfnamefont {E.~W.}\ \bibnamefont {{Liang}}}, \bibinfo {author}
  {\bibfnamefont {Y.~F.}\ \bibnamefont {{Liang}}}, \bibinfo {author} {\bibfnamefont {S.~J.}\ \bibnamefont {{Lin}}}, \bibinfo {author} {\bibfnamefont {B.}~\bibnamefont {{Liu}}}, \bibinfo {author} {\bibfnamefont {C.}~\bibnamefont {{Liu}}}, \bibinfo {author} {\bibfnamefont {D.}~\bibnamefont {{Liu}}}, \bibinfo {author} {\bibfnamefont {D.~B.}\ \bibnamefont {{Liu}}}, \bibinfo {author} {\bibfnamefont {H.}~\bibnamefont {{Liu}}}, \bibinfo {author} {\bibfnamefont {H.~D.}\ \bibnamefont {{Liu}}}, \bibinfo {author} {\bibfnamefont {J.}~\bibnamefont {{Liu}}}, \bibinfo {author} {\bibfnamefont {J.~L.}\ \bibnamefont {{Liu}}}, \bibinfo {author} {\bibfnamefont {M.~Y.}\ \bibnamefont {{Liu}}}, \bibinfo {author} {\bibfnamefont {R.~Y.}\ \bibnamefont {{Liu}}}, \bibinfo {author} {\bibfnamefont {S.~M.}\ \bibnamefont {{Liu}}}, \bibinfo {author} {\bibfnamefont {W.}~\bibnamefont {{Liu}}}, \bibinfo {author} {\bibfnamefont {Y.}~\bibnamefont {{Liu}}}, \bibinfo {author} {\bibfnamefont {Y.~N.}\ \bibnamefont {{Liu}}}, \bibinfo {author}
  {\bibfnamefont {Q.}~\bibnamefont {{Luo}}}, \bibinfo {author} {\bibfnamefont {Y.}~\bibnamefont {{Luo}}}, \bibinfo {author} {\bibfnamefont {H.~K.}\ \bibnamefont {{Lv}}}, \bibinfo {author} {\bibfnamefont {B.~Q.}\ \bibnamefont {{Ma}}}, \bibinfo {author} {\bibfnamefont {L.~L.}\ \bibnamefont {{Ma}}}, \bibinfo {author} {\bibfnamefont {X.~H.}\ \bibnamefont {{Ma}}}, \bibinfo {author} {\bibfnamefont {J.~R.}\ \bibnamefont {{Mao}}}, \bibinfo {author} {\bibfnamefont {Z.}~\bibnamefont {{Min}}}, \bibinfo {author} {\bibfnamefont {W.}~\bibnamefont {{Mitthumsiri}}}, \bibinfo {author} {\bibfnamefont {H.~J.}\ \bibnamefont {{Mu}}}, \bibinfo {author} {\bibfnamefont {Y.~C.}\ \bibnamefont {{Nan}}}, \bibinfo {author} {\bibfnamefont {A.}~\bibnamefont {{Neronov}}}, \bibinfo {author} {\bibfnamefont {K.~C.~Y.}\ \bibnamefont {{Ng}}}, \bibinfo {author} {\bibfnamefont {L.~J.}\ \bibnamefont {{Ou}}}, \bibinfo {author} {\bibfnamefont {P.}~\bibnamefont {{Pattarakijwanich}}}, \bibinfo {author} {\bibfnamefont {Z.~Y.}\ \bibnamefont {{Pei}}},
  \bibinfo {author} {\bibfnamefont {J.~C.}\ \bibnamefont {{Qi}}}, \bibinfo {author} {\bibfnamefont {M.~Y.}\ \bibnamefont {{Qi}}}, \bibinfo {author} {\bibfnamefont {B.~Q.}\ \bibnamefont {{Qiao}}}, \bibinfo {author} {\bibfnamefont {J.~J.}\ \bibnamefont {{Qin}}}, \bibinfo {author} {\bibfnamefont {A.}~\bibnamefont {{Raza}}}, \bibinfo {author} {\bibfnamefont {D.}~\bibnamefont {{Ruffolo}}}, \bibinfo {author} {\bibfnamefont {A.}~\bibnamefont {{S{\'a}iz}}}, \bibinfo {author} {\bibfnamefont {M.}~\bibnamefont {{Saeed}}}, \bibinfo {author} {\bibfnamefont {D.}~\bibnamefont {{Semikoz}}}, \bibinfo {author} {\bibfnamefont {L.}~\bibnamefont {{Shao}}}, \bibinfo {author} {\bibfnamefont {O.}~\bibnamefont {{Shchegolev}}}, \bibinfo {author} {\bibfnamefont {X.~D.}\ \bibnamefont {{Sheng}}}, \bibinfo {author} {\bibfnamefont {F.~W.}\ \bibnamefont {{Shu}}}, \bibinfo {author} {\bibfnamefont {H.~C.}\ \bibnamefont {{Song}}}, \bibinfo {author} {\bibfnamefont {Y.~V.}\ \bibnamefont {{Stenkin}}}, \bibinfo {author} {\bibfnamefont
  {V.}~\bibnamefont {{Stepanov}}}, \bibinfo {author} {\bibfnamefont {Y.}~\bibnamefont {{Su}}}, \bibinfo {author} {\bibfnamefont {D.~X.}\ \bibnamefont {{Sun}}}, \bibinfo {author} {\bibfnamefont {Q.~N.}\ \bibnamefont {{Sun}}}, \bibinfo {author} {\bibfnamefont {X.~N.}\ \bibnamefont {{Sun}}}, \bibinfo {author} {\bibfnamefont {Z.~B.}\ \bibnamefont {{Sun}}}, \bibinfo {author} {\bibfnamefont {J.}~\bibnamefont {{Takata}}}, \bibinfo {author} {\bibfnamefont {P.~H.~T.}\ \bibnamefont {{Tam}}}, \bibinfo {author} {\bibfnamefont {Q.~W.}\ \bibnamefont {{Tang}}}, \bibinfo {author} {\bibfnamefont {R.}~\bibnamefont {{Tang}}}, \bibinfo {author} {\bibfnamefont {Z.~B.}\ \bibnamefont {{Tang}}}, \bibinfo {author} {\bibfnamefont {W.~W.}\ \bibnamefont {{Tian}}}, \bibinfo {author} {\bibfnamefont {L.~H.}\ \bibnamefont {{Wan}}}, \bibinfo {author} {\bibfnamefont {C.}~\bibnamefont {{Wang}}}, \bibinfo {author} {\bibfnamefont {C.~B.}\ \bibnamefont {{Wang}}}, \bibinfo {author} {\bibfnamefont {G.~W.}\ \bibnamefont {{Wang}}}, \bibinfo {author}
  {\bibfnamefont {H.~G.}\ \bibnamefont {{Wang}}}, \bibinfo {author} {\bibfnamefont {H.~H.}\ \bibnamefont {{Wang}}}, \bibinfo {author} {\bibfnamefont {J.~C.}\ \bibnamefont {{Wang}}}, \bibinfo {author} {\bibfnamefont {K.}~\bibnamefont {{Wang}}}, \bibinfo {author} {\bibfnamefont {K.}~\bibnamefont {{Wang}}}, \bibinfo {author} {\bibfnamefont {L.~P.}\ \bibnamefont {{Wang}}}, \bibinfo {author} {\bibfnamefont {L.~Y.}\ \bibnamefont {{Wang}}}, \bibinfo {author} {\bibfnamefont {P.~H.}\ \bibnamefont {{Wang}}}, \bibinfo {author} {\bibfnamefont {R.}~\bibnamefont {{Wang}}},\ and\ \bibinfo {author} {\bibfnamefont {W.}~\bibnamefont {{Wang}}},\ }\href {https://doi.org/10.3847/2041-8213/ad8921} {\bibfield  {journal} {\bibinfo  {journal} {ApJL}\ }\textbf {\bibinfo {volume} {975}},\ \bibinfo {eid} {L44} (\bibinfo {year} {2024})},\ \Eprint {https://arxiv.org/abs/2410.15353} {arXiv:2410.15353 [astro-ph.HE]} \BibitemShut {NoStop}%
\bibitem [{\citenamefont {{Koide}}(2003)}]{Koide03}%
  \BibitemOpen
  \bibfield  {author} {\bibinfo {author} {\bibfnamefont {S.}~\bibnamefont {{Koide}}},\ }\href {https://doi.org/10.1103/PhysRevD.67.104010} {\bibfield  {journal} {\bibinfo  {journal} {\prd}\ }\textbf {\bibinfo {volume} {67}},\ \bibinfo {eid} {104010} (\bibinfo {year} {2003})}\BibitemShut {NoStop}%
\bibitem [{\citenamefont {{Komissarov}}(2004)}]{Komissarov04}%
  \BibitemOpen
  \bibfield  {author} {\bibinfo {author} {\bibfnamefont {S.~S.}\ \bibnamefont {{Komissarov}}},\ }\href {https://doi.org/10.1111/j.1365-2966.2004.07598.x} {\bibfield  {journal} {\bibinfo  {journal} {MNRAS}\ }\textbf {\bibinfo {volume} {350}},\ \bibinfo {pages} {427} (\bibinfo {year} {2004})}\BibitemShut {NoStop}%
\bibitem [{\citenamefont {{Komissarov}}\ and\ \citenamefont {{McKinney}}(2007)}]{Komissarov&McKinney07}%
  \BibitemOpen
  \bibfield  {author} {\bibinfo {author} {\bibfnamefont {S.~S.}\ \bibnamefont {{Komissarov}}}\ and\ \bibinfo {author} {\bibfnamefont {J.~C.}\ \bibnamefont {{McKinney}}},\ }\href {https://doi.org/10.1111/j.1745-3933.2007.00301.x} {\bibfield  {journal} {\bibinfo  {journal} {MNRAS}\ }\textbf {\bibinfo {volume} {377}},\ \bibinfo {pages} {L49} (\bibinfo {year} {2007})},\ \Eprint {https://arxiv.org/abs/astro-ph/0702269} {arXiv:astro-ph/0702269 [astro-ph]} \BibitemShut {NoStop}%
\bibitem [{\citenamefont {{Ripperda}}\ \emph {et~al.}(2020)\citenamefont {{Ripperda}}, \citenamefont {{Bacchini}},\ and\ \citenamefont {{Philippov}}}]{Ripperda20}%
  \BibitemOpen
  \bibfield  {author} {\bibinfo {author} {\bibfnamefont {B.}~\bibnamefont {{Ripperda}}}, \bibinfo {author} {\bibfnamefont {F.}~\bibnamefont {{Bacchini}}},\ and\ \bibinfo {author} {\bibfnamefont {A.~A.}\ \bibnamefont {{Philippov}}},\ }\href {https://doi.org/10.3847/1538-4357/ababab} {\bibfield  {journal} {\bibinfo  {journal} {ApJ}\ }\textbf {\bibinfo {volume} {900}},\ \bibinfo {eid} {100} (\bibinfo {year} {2020})},\ \Eprint {https://arxiv.org/abs/2003.04330} {arXiv:2003.04330 [astro-ph.HE]} \BibitemShut {NoStop}%
\bibitem [{\citenamefont {{Zhou}}\ \emph {et~al.}(2026)\citenamefont {{Zhou}}, \citenamefont {{Huang}}, \citenamefont {{Li}}, \citenamefont {{Zhang}}, \citenamefont {{Hou}}, \citenamefont {{Guo}},\ and\ \citenamefont {{Chen}}}]{Zhou25}%
  \BibitemOpen
  \bibfield  {author} {\bibinfo {author} {\bibfnamefont {F.}~\bibnamefont {{Zhou}}}, \bibinfo {author} {\bibfnamefont {J.}~\bibnamefont {{Huang}}}, \bibinfo {author} {\bibfnamefont {Y.}~\bibnamefont {{Li}}}, \bibinfo {author} {\bibfnamefont {Z.}~\bibnamefont {{Zhang}}}, \bibinfo {author} {\bibfnamefont {Y.}~\bibnamefont {{Hou}}}, \bibinfo {author} {\bibfnamefont {M.}~\bibnamefont {{Guo}}},\ and\ \bibinfo {author} {\bibfnamefont {B.}~\bibnamefont {{Chen}}},\ }\href {https://doi.org/10.3847/1538-4357/ae5e67} {\bibfield  {journal} {\bibinfo  {journal} {ApJ}\ }\textbf {\bibinfo {volume} {1002}},\ \bibinfo {eid} {152} (\bibinfo {year} {2026})},\ \Eprint {https://arxiv.org/abs/2512.06803} {arXiv:2512.06803 [astro-ph.HE]} \BibitemShut {NoStop}%
\bibitem [{\citenamefont {{Hakobyan}}\ \emph {et~al.}(2023)\citenamefont {{Hakobyan}}, \citenamefont {{Ripperda}},\ and\ \citenamefont {{Philippov}}}]{Hakobyan23}%
  \BibitemOpen
  \bibfield  {author} {\bibinfo {author} {\bibfnamefont {H.}~\bibnamefont {{Hakobyan}}}, \bibinfo {author} {\bibfnamefont {B.}~\bibnamefont {{Ripperda}}},\ and\ \bibinfo {author} {\bibfnamefont {A.~A.}\ \bibnamefont {{Philippov}}},\ }\href {https://doi.org/10.3847/2041-8213/acb264} {\bibfield  {journal} {\bibinfo  {journal} {ApJL}\ }\textbf {\bibinfo {volume} {943}},\ \bibinfo {eid} {L29} (\bibinfo {year} {2023})},\ \Eprint {https://arxiv.org/abs/2209.02105} {arXiv:2209.02105 [astro-ph.HE]} \BibitemShut {NoStop}%
\bibitem [{\citenamefont {{Hakobyan}}\ \emph {et~al.}(2025)\citenamefont {{Hakobyan}}, \citenamefont {{Levinson}}, \citenamefont {{Sironi}}, \citenamefont {{Philippov}},\ and\ \citenamefont {{Ripperda}}}]{Hakobyan25}%
  \BibitemOpen
  \bibfield  {author} {\bibinfo {author} {\bibfnamefont {H.}~\bibnamefont {{Hakobyan}}}, \bibinfo {author} {\bibfnamefont {A.}~\bibnamefont {{Levinson}}}, \bibinfo {author} {\bibfnamefont {L.}~\bibnamefont {{Sironi}}}, \bibinfo {author} {\bibfnamefont {A.}~\bibnamefont {{Philippov}}},\ and\ \bibinfo {author} {\bibfnamefont {B.}~\bibnamefont {{Ripperda}}},\ }\href {https://doi.org/10.3847/2041-8213/ae286a} {\bibfield  {journal} {\bibinfo  {journal} {ApJL}\ }\textbf {\bibinfo {volume} {995}},\ \bibinfo {eid} {L73} (\bibinfo {year} {2025})},\ \Eprint {https://arxiv.org/abs/2507.14002} {arXiv:2507.14002 [astro-ph.HE]} \BibitemShut {NoStop}%
\bibitem [{\citenamefont {{Solanki}}\ \emph {et~al.}(2025)\citenamefont {{Solanki}}, \citenamefont {{Davelaar}}, \citenamefont {{Ripperda}},\ and\ \citenamefont {{Philippov}}}]{Solanki25}%
  \BibitemOpen
  \bibfield  {author} {\bibinfo {author} {\bibfnamefont {S.}~\bibnamefont {{Solanki}}}, \bibinfo {author} {\bibfnamefont {J.}~\bibnamefont {{Davelaar}}}, \bibinfo {author} {\bibfnamefont {B.}~\bibnamefont {{Ripperda}}},\ and\ \bibinfo {author} {\bibfnamefont {A.}~\bibnamefont {{Philippov}}},\ }\href {https://doi.org/10.3847/1538-4357/adcba9} {\bibfield  {journal} {\bibinfo  {journal} {ApJ}\ }\textbf {\bibinfo {volume} {985}},\ \bibinfo {eid} {147} (\bibinfo {year} {2025})},\ \Eprint {https://arxiv.org/abs/2504.16298} {arXiv:2504.16298 [astro-ph.HE]} \BibitemShut {NoStop}%
\bibitem [{\citenamefont {{Werner}}\ \emph {et~al.}(2019)\citenamefont {{Werner}}, \citenamefont {{Philippov}},\ and\ \citenamefont {{Uzdensky}}}]{Werner19}%
  \BibitemOpen
  \bibfield  {author} {\bibinfo {author} {\bibfnamefont {G.~R.}\ \bibnamefont {{Werner}}}, \bibinfo {author} {\bibfnamefont {A.~A.}\ \bibnamefont {{Philippov}}},\ and\ \bibinfo {author} {\bibfnamefont {D.~A.}\ \bibnamefont {{Uzdensky}}},\ }\href {https://doi.org/10.1093/mnrasl/sly157} {\bibfield  {journal} {\bibinfo  {journal} {MNRAS}\ }\textbf {\bibinfo {volume} {482}},\ \bibinfo {pages} {L60} (\bibinfo {year} {2019})},\ \Eprint {https://arxiv.org/abs/1805.01910} {arXiv:1805.01910 [astro-ph.HE]} \BibitemShut {NoStop}%
\bibitem [{\citenamefont {{Sironi}}\ and\ \citenamefont {{Beloborodov}}(2020)}]{Sironi20}%
  \BibitemOpen
  \bibfield  {author} {\bibinfo {author} {\bibfnamefont {L.}~\bibnamefont {{Sironi}}}\ and\ \bibinfo {author} {\bibfnamefont {A.~M.}\ \bibnamefont {{Beloborodov}}},\ }\href {https://doi.org/10.3847/1538-4357/aba622} {\bibfield  {journal} {\bibinfo  {journal} {ApJ}\ }\textbf {\bibinfo {volume} {899}},\ \bibinfo {eid} {52} (\bibinfo {year} {2020})},\ \Eprint {https://arxiv.org/abs/1908.08138} {arXiv:1908.08138 [astro-ph.HE]} \BibitemShut {NoStop}%
\bibitem [{\citenamefont {{Guo}}\ \emph {et~al.}(2021)\citenamefont {{Guo}}, \citenamefont {{Li}}, \citenamefont {{Daughton}}, \citenamefont {{Li}}, \citenamefont {{Kilian}}, \citenamefont {{Liu}}, \citenamefont {{Zhang}},\ and\ \citenamefont {{Zhang}}}]{Guo21}%
  \BibitemOpen
  \bibfield  {author} {\bibinfo {author} {\bibfnamefont {F.}~\bibnamefont {{Guo}}}, \bibinfo {author} {\bibfnamefont {X.}~\bibnamefont {{Li}}}, \bibinfo {author} {\bibfnamefont {W.}~\bibnamefont {{Daughton}}}, \bibinfo {author} {\bibfnamefont {H.}~\bibnamefont {{Li}}}, \bibinfo {author} {\bibfnamefont {P.}~\bibnamefont {{Kilian}}}, \bibinfo {author} {\bibfnamefont {Y.-H.}\ \bibnamefont {{Liu}}}, \bibinfo {author} {\bibfnamefont {Q.}~\bibnamefont {{Zhang}}},\ and\ \bibinfo {author} {\bibfnamefont {H.}~\bibnamefont {{Zhang}}},\ }\href {https://doi.org/10.3847/1538-4357/ac0918} {\bibfield  {journal} {\bibinfo  {journal} {ApJ}\ }\textbf {\bibinfo {volume} {919}},\ \bibinfo {eid} {111} (\bibinfo {year} {2021})},\ \Eprint {https://arxiv.org/abs/2008.02743} {arXiv:2008.02743 [astro-ph.HE]} \BibitemShut {NoStop}%
\bibitem [{\citenamefont {{Schoeffler}}\ \emph {et~al.}(2023)\citenamefont {{Schoeffler}}, \citenamefont {{Grismayer}}, \citenamefont {{Uzdensky}},\ and\ \citenamefont {{Silva}}}]{Schoeffler23}%
  \BibitemOpen
  \bibfield  {author} {\bibinfo {author} {\bibfnamefont {K.~M.}\ \bibnamefont {{Schoeffler}}}, \bibinfo {author} {\bibfnamefont {T.}~\bibnamefont {{Grismayer}}}, \bibinfo {author} {\bibfnamefont {D.}~\bibnamefont {{Uzdensky}}},\ and\ \bibinfo {author} {\bibfnamefont {L.~O.}\ \bibnamefont {{Silva}}},\ }\href {https://doi.org/10.1093/mnras/stad1588} {\bibfield  {journal} {\bibinfo  {journal} {MNRAS}\ }\textbf {\bibinfo {volume} {523}},\ \bibinfo {pages} {3812} (\bibinfo {year} {2023})},\ \Eprint {https://arxiv.org/abs/2303.16643} {arXiv:2303.16643 [astro-ph.HE]} \BibitemShut {NoStop}%
\bibitem [{\citenamefont {{Landau}}\ and\ \citenamefont {{Lifshitz}}(1975)}]{Landau75}%
  \BibitemOpen
  \bibfield  {author} {\bibinfo {author} {\bibfnamefont {L.~D.}\ \bibnamefont {{Landau}}}\ and\ \bibinfo {author} {\bibfnamefont {E.~M.}\ \bibnamefont {{Lifshitz}}},\ }\href@noop {} {\emph {\bibinfo {title} {{The classical theory of fields}}}}\ (\bibinfo {year} {1975})\BibitemShut {NoStop}%
\bibitem [{\citenamefont {{MacDonald}}\ and\ \citenamefont {{Thorne}}(1982)}]{Macdonald82}%
  \BibitemOpen
  \bibfield  {author} {\bibinfo {author} {\bibfnamefont {D.}~\bibnamefont {{MacDonald}}}\ and\ \bibinfo {author} {\bibfnamefont {K.~S.}\ \bibnamefont {{Thorne}}},\ }\href {https://doi.org/10.1093/mnras/198.2.345} {\bibfield  {journal} {\bibinfo  {journal} {MNRAS}\ }\textbf {\bibinfo {volume} {198}},\ \bibinfo {pages} {345} (\bibinfo {year} {1982})}\BibitemShut {NoStop}%
\bibitem [{\citenamefont {{Landau}}\ and\ \citenamefont {{Lifshitz}}(1971)}]{Landau71}%
  \BibitemOpen
  \bibfield  {author} {\bibinfo {author} {\bibfnamefont {L.~D.}\ \bibnamefont {{Landau}}}\ and\ \bibinfo {author} {\bibfnamefont {E.~M.}\ \bibnamefont {{Lifshitz}}},\ }\href@noop {} {\emph {\bibinfo {title} {{The classical theory of fields}}}}\ (\bibinfo {year} {1971})\BibitemShut {NoStop}%
\bibitem [{\citenamefont {Hughes}\ \emph {et~al.}(1994)\citenamefont {Hughes}, \citenamefont {Keeton}, \citenamefont {Walker}, \citenamefont {Walsh}, \citenamefont {Shapiro},\ and\ \citenamefont {Teukolsky}}]{Hughes94}%
  \BibitemOpen
  \bibfield  {author} {\bibinfo {author} {\bibfnamefont {S.~A.}\ \bibnamefont {Hughes}}, \bibinfo {author} {\bibfnamefont {C.~R.}\ \bibnamefont {Keeton}}, \bibinfo {author} {\bibfnamefont {P.}~\bibnamefont {Walker}}, \bibinfo {author} {\bibfnamefont {K.~T.}\ \bibnamefont {Walsh}}, \bibinfo {author} {\bibfnamefont {S.~L.}\ \bibnamefont {Shapiro}},\ and\ \bibinfo {author} {\bibfnamefont {S.~A.}\ \bibnamefont {Teukolsky}},\ }\href {https://doi.org/10.1103/PhysRevD.49.4004} {\bibfield  {journal} {\bibinfo  {journal} {Phys. Rev. D}\ }\textbf {\bibinfo {volume} {49}},\ \bibinfo {pages} {4004} (\bibinfo {year} {1994})}\BibitemShut {NoStop}%
\bibitem [{\citenamefont {{Bardeen}}\ \emph {et~al.}(1972)\citenamefont {{Bardeen}}, \citenamefont {{Press}},\ and\ \citenamefont {{Teukolsky}}}]{bardeen72}%
  \BibitemOpen
  \bibfield  {author} {\bibinfo {author} {\bibfnamefont {J.~M.}\ \bibnamefont {{Bardeen}}}, \bibinfo {author} {\bibfnamefont {W.~H.}\ \bibnamefont {{Press}}},\ and\ \bibinfo {author} {\bibfnamefont {S.~A.}\ \bibnamefont {{Teukolsky}}},\ }\href {https://doi.org/10.1086/151796} {\bibfield  {journal} {\bibinfo  {journal} {ApJ}\ }\textbf {\bibinfo {volume} {178}},\ \bibinfo {pages} {347} (\bibinfo {year} {1972})}\BibitemShut {NoStop}%
\bibitem [{\citenamefont {{Thorne}}\ \emph {et~al.}(1986)\citenamefont {{Thorne}}, \citenamefont {{Price}},\ and\ \citenamefont {{MacDonald}}}]{Thorne86}%
  \BibitemOpen
  \bibfield  {author} {\bibinfo {author} {\bibfnamefont {K.~S.}\ \bibnamefont {{Thorne}}}, \bibinfo {author} {\bibfnamefont {R.~H.}\ \bibnamefont {{Price}}},\ and\ \bibinfo {author} {\bibfnamefont {D.~A.}\ \bibnamefont {{MacDonald}}},\ }\href@noop {} {\emph {\bibinfo {title} {{Black holes: The membrane paradigm}}}}\ (\bibinfo {year} {1986})\BibitemShut {NoStop}%
\bibitem [{\citenamefont {{Goldreich}}\ and\ \citenamefont {{Julian}}(1969)}]{GJ69}%
  \BibitemOpen
  \bibfield  {author} {\bibinfo {author} {\bibfnamefont {P.}~\bibnamefont {{Goldreich}}}\ and\ \bibinfo {author} {\bibfnamefont {W.~H.}\ \bibnamefont {{Julian}}},\ }\href {https://doi.org/10.1086/150119} {\bibfield  {journal} {\bibinfo  {journal} {ApJ}\ }\textbf {\bibinfo {volume} {157}},\ \bibinfo {pages} {869} (\bibinfo {year} {1969})}\BibitemShut {NoStop}%
\bibitem [{\citenamefont {{Wald}}(1974)}]{Wald74}%
  \BibitemOpen
  \bibfield  {author} {\bibinfo {author} {\bibfnamefont {R.~M.}\ \bibnamefont {{Wald}}},\ }\href {https://doi.org/10.1103/PhysRevD.10.1680} {\bibfield  {journal} {\bibinfo  {journal} {\prd}\ }\textbf {\bibinfo {volume} {10}},\ \bibinfo {pages} {1680} (\bibinfo {year} {1974})}\BibitemShut {NoStop}%
\bibitem [{\citenamefont {Berenger}(1994)}]{Berenger94}%
  \BibitemOpen
  \bibfield  {author} {\bibinfo {author} {\bibfnamefont {J.-P.}\ \bibnamefont {Berenger}},\ }\href {https://doi.org/https://doi.org/10.1006/jcph.1994.1159} {\bibfield  {journal} {\bibinfo  {journal} {Journal of Computational Physics}\ }\textbf {\bibinfo {volume} {114}},\ \bibinfo {pages} {185} (\bibinfo {year} {1994})}\BibitemShut {NoStop}%
\bibitem [{\citenamefont {{Cerutti}}\ \emph {et~al.}(2025)\citenamefont {{Cerutti}}, \citenamefont {{Figueiredo}},\ and\ \citenamefont {{Dubus}}}]{Cerutti25}%
  \BibitemOpen
  \bibfield  {author} {\bibinfo {author} {\bibfnamefont {B.}~\bibnamefont {{Cerutti}}}, \bibinfo {author} {\bibfnamefont {E.}~\bibnamefont {{Figueiredo}}},\ and\ \bibinfo {author} {\bibfnamefont {G.}~\bibnamefont {{Dubus}}},\ }\href {https://doi.org/10.1051/0004-6361/202451948} {\bibfield  {journal} {\bibinfo  {journal} {A\&A}\ }\textbf {\bibinfo {volume} {695}},\ \bibinfo {eid} {A93} (\bibinfo {year} {2025})},\ \Eprint {https://arxiv.org/abs/2412.02307} {arXiv:2412.02307 [astro-ph.HE]} \BibitemShut {NoStop}%
\bibitem [{\citenamefont {{Kuo}}\ \emph {et~al.}(2014)\citenamefont {{Kuo}}, \citenamefont {{Asada}}, \citenamefont {{Rao}}, \citenamefont {{Nakamura}}, \citenamefont {{Algaba}}, \citenamefont {{Liu}}, \citenamefont {{Inoue}}, \citenamefont {{Koch}}, \citenamefont {{Ho}}, \citenamefont {{Matsushita}}, \citenamefont {{Pu}}, \citenamefont {{Akiyama}}, \citenamefont {{Nishioka}},\ and\ \citenamefont {{Pradel}}}]{Kuo14}%
  \BibitemOpen
  \bibfield  {author} {\bibinfo {author} {\bibfnamefont {C.~Y.}\ \bibnamefont {{Kuo}}}, \bibinfo {author} {\bibfnamefont {K.}~\bibnamefont {{Asada}}}, \bibinfo {author} {\bibfnamefont {R.}~\bibnamefont {{Rao}}}, \bibinfo {author} {\bibfnamefont {M.}~\bibnamefont {{Nakamura}}}, \bibinfo {author} {\bibfnamefont {J.~C.}\ \bibnamefont {{Algaba}}}, \bibinfo {author} {\bibfnamefont {H.~B.}\ \bibnamefont {{Liu}}}, \bibinfo {author} {\bibfnamefont {M.}~\bibnamefont {{Inoue}}}, \bibinfo {author} {\bibfnamefont {P.~M.}\ \bibnamefont {{Koch}}}, \bibinfo {author} {\bibfnamefont {P.~T.~P.}\ \bibnamefont {{Ho}}}, \bibinfo {author} {\bibfnamefont {S.}~\bibnamefont {{Matsushita}}}, \bibinfo {author} {\bibfnamefont {H.-Y.}\ \bibnamefont {{Pu}}}, \bibinfo {author} {\bibfnamefont {K.}~\bibnamefont {{Akiyama}}}, \bibinfo {author} {\bibfnamefont {H.}~\bibnamefont {{Nishioka}}},\ and\ \bibinfo {author} {\bibfnamefont {N.}~\bibnamefont {{Pradel}}},\ }\href {https://doi.org/10.1088/2041-8205/783/2/L33} {\bibfield  {journal}
  {\bibinfo  {journal} {ApJL}\ }\textbf {\bibinfo {volume} {783}},\ \bibinfo {eid} {L33} (\bibinfo {year} {2014})},\ \Eprint {https://arxiv.org/abs/1402.5238} {arXiv:1402.5238 [astro-ph.GA]} \BibitemShut {NoStop}%
\bibitem [{\citenamefont {{Event Horizon Telescope Collaboration}}\ \emph {et~al.}(2021)\citenamefont {{Event Horizon Telescope Collaboration}}, \citenamefont {{Akiyama}}, \citenamefont {{Algaba}}, \citenamefont {{Alberdi}}, \citenamefont {{Alef}}, \citenamefont {{Anantua}}, \citenamefont {{Asada}}, \citenamefont {{Azulay}}, \citenamefont {{Baczko}}, \citenamefont {{Ball}}, \citenamefont {{Balokovi{\'c}}}, \citenamefont {{Barrett}}, \citenamefont {{Benson}}, \citenamefont {{Bintley}}, \citenamefont {{Blackburn}}, \citenamefont {{Blundell}}, \citenamefont {{Boland}}, \citenamefont {{Bouman}}, \citenamefont {{Bower}}, \citenamefont {{Boyce}}, \citenamefont {{Bremer}}, \citenamefont {{Brinkerink}}, \citenamefont {{Brissenden}}, \citenamefont {{Britzen}}, \citenamefont {{Broderick}}, \citenamefont {{Broguiere}}, \citenamefont {{Bronzwaer}}, \citenamefont {{Byun}}, \citenamefont {{Carlstrom}}, \citenamefont {{Chael}}, \citenamefont {{Chan}}, \citenamefont {{Chatterjee}}, \citenamefont {{Chatterjee}}, \citenamefont
  {{Chen}}, \citenamefont {{Chen}}, \citenamefont {{Chesler}}, \citenamefont {{Cho}}, \citenamefont {{Christian}}, \citenamefont {{Conway}}, \citenamefont {{Cordes}}, \citenamefont {{Crawford}}, \citenamefont {{Crew}}, \citenamefont {{Cruz-Osorio}}, \citenamefont {{Cui}}, \citenamefont {{Davelaar}}, \citenamefont {{De Laurentis}}, \citenamefont {{Deane}}, \citenamefont {{Dempsey}}, \citenamefont {{Desvignes}}, \citenamefont {{Dexter}}, \citenamefont {{Doeleman}}, \citenamefont {{Eatough}}, \citenamefont {{Falcke}}, \citenamefont {{Farah}}, \citenamefont {{Fish}}, \citenamefont {{Fomalont}}, \citenamefont {{Ford}}, \citenamefont {{Fraga-Encinas}}, \citenamefont {{Friberg}}, \citenamefont {{Fromm}}, \citenamefont {{Fuentes}}, \citenamefont {{Galison}}, \citenamefont {{Gammie}}, \citenamefont {{Garc{\'\i}a}}, \citenamefont {{Gelles}}, \citenamefont {{Gentaz}}, \citenamefont {{Georgiev}}, \citenamefont {{Goddi}}, \citenamefont {{Gold}}, \citenamefont {{G{\'o}mez}}, \citenamefont {{G{\'o}mez-Ruiz}}, \citenamefont
  {{Gu}}, \citenamefont {{Gurwell}}, \citenamefont {{Hada}}, \citenamefont {{Haggard}}, \citenamefont {{Hecht}}, \citenamefont {{Hesper}}, \citenamefont {{Himwich}}, \citenamefont {{Ho}}, \citenamefont {{Ho}}, \citenamefont {{Honma}}, \citenamefont {{Huang}}, \citenamefont {{Huang}}, \citenamefont {{Hughes}}, \citenamefont {{Ikeda}}, \citenamefont {{Inoue}}, \citenamefont {{Issaoun}}, \citenamefont {{James}}, \citenamefont {{Jannuzi}}, \citenamefont {{Janssen}}, \citenamefont {{Jeter}}, \citenamefont {{Jiang}}, \citenamefont {{Jimenez-Rosales}}, \citenamefont {{Johnson}}, \citenamefont {{Jorstad}}, \citenamefont {{Jung}}, \citenamefont {{Karami}}, \citenamefont {{Karuppusamy}}, \citenamefont {{Kawashima}}, \citenamefont {{Keating}}, \citenamefont {{Kettenis}}, \citenamefont {{Kim}}, \citenamefont {{Kim}}, \citenamefont {{Kim}}, \citenamefont {{Kim}}, \citenamefont {{Kino}}, \citenamefont {{Koay}}, \citenamefont {{Kofuji}}, \citenamefont {{Koch}}, \citenamefont {{Koyama}}, \citenamefont {{Kramer}},
  \citenamefont {{Kramer}}, \citenamefont {{Krichbaum}}, \citenamefont {{Kuo}}, \citenamefont {{Lauer}}, \citenamefont {{Lee}}, \citenamefont {{Levis}}, \citenamefont {{Li}}, \citenamefont {{Li}}, \citenamefont {{Lindqvist}}, \citenamefont {{Lico}}, \citenamefont {{Lindahl}}, \citenamefont {{Liu}}, \citenamefont {{Liu}}, \citenamefont {{Liuzzo}}, \citenamefont {{Lo}}, \citenamefont {{Lobanov}}, \citenamefont {{Loinard}}, \citenamefont {{Lonsdale}}, \citenamefont {{Lu}}, \citenamefont {{MacDonald}}, \citenamefont {{Mao}}, \citenamefont {{Marchili}}, \citenamefont {{Markoff}}, \citenamefont {{Marrone}}, \citenamefont {{Marscher}}, \citenamefont {{Mart{\'\i}-Vidal}}, \citenamefont {{Matsushita}}, \citenamefont {{Matthews}}, \citenamefont {{Medeiros}}, \citenamefont {{Menten}}, \citenamefont {{Mizuno}}, \citenamefont {{Mizuno}}, \citenamefont {{Moran}}, \citenamefont {{Moriyama}}, \citenamefont {{Moscibrodzka}}, \citenamefont {{M{\"u}ller}}, \citenamefont {{Musoke}}, \citenamefont {{Mus Mej{\'\i}as}},
  \citenamefont {{Michalik}}, \citenamefont {{Nadolski}}, \citenamefont {{Nagai}}, \citenamefont {{Nagar}}, \citenamefont {{Nakamura}}, \citenamefont {{Narayan}}, \citenamefont {{Narayanan}}, \citenamefont {{Natarajan}}, \citenamefont {{Nathanail}}, \citenamefont {{Neilsen}}, \citenamefont {{Neri}}, \citenamefont {{Ni}}, \citenamefont {{Noutsos}}, \citenamefont {{Nowak}}, \citenamefont {{Okino}}, \citenamefont {{Olivares}}, \citenamefont {{Ortiz-Le{\'o}n}}, \citenamefont {{Oyama}}, \citenamefont {{{\"O}zel}}, \citenamefont {{Palumbo}}, \citenamefont {{Park}}, \citenamefont {{Patel}}, \citenamefont {{Pen}}, \citenamefont {{Pesce}}, \citenamefont {{Pi{\'e}tu}}, \citenamefont {{Plambeck}}, \citenamefont {{PopStefanija}}, \citenamefont {{Porth}}, \citenamefont {{P{\"o}tzl}}, \citenamefont {{Prather}}, \citenamefont {{Preciado-L{\'o}pez}}, \citenamefont {{Psaltis}}, \citenamefont {{Pu}}, \citenamefont {{Ramakrishnan}}, \citenamefont {{Rao}}, \citenamefont {{Rawlings}}, \citenamefont {{Raymond}}, \citenamefont
  {{Rezzolla}}, \citenamefont {{Ricarte}}, \citenamefont {{Ripperda}}, \citenamefont {{Roelofs}}, \citenamefont {{Rogers}}, \citenamefont {{Ros}}, \citenamefont {{Rose}}, \citenamefont {{Roshanineshat}}, \citenamefont {{Rottmann}}, \citenamefont {{Roy}}, \citenamefont {{Ruszczyk}}, \citenamefont {{Rygl}}, \citenamefont {{S{\'a}nchez}},\ and\ \citenamefont {{S{\'a}nchez-Arguelles}}}]{EHT21_mag}%
  \BibitemOpen
  \bibfield  {author} {\bibinfo {author} {\bibnamefont {{Event Horizon Telescope Collaboration}}}, \bibinfo {author} {\bibfnamefont {K.}~\bibnamefont {{Akiyama}}}, \bibinfo {author} {\bibfnamefont {J.~C.}\ \bibnamefont {{Algaba}}}, \bibinfo {author} {\bibfnamefont {A.}~\bibnamefont {{Alberdi}}}, \bibinfo {author} {\bibfnamefont {W.}~\bibnamefont {{Alef}}}, \bibinfo {author} {\bibfnamefont {R.}~\bibnamefont {{Anantua}}}, \bibinfo {author} {\bibfnamefont {K.}~\bibnamefont {{Asada}}}, \bibinfo {author} {\bibfnamefont {R.}~\bibnamefont {{Azulay}}}, \bibinfo {author} {\bibfnamefont {A.-K.}\ \bibnamefont {{Baczko}}}, \bibinfo {author} {\bibfnamefont {D.}~\bibnamefont {{Ball}}}, \bibinfo {author} {\bibfnamefont {M.}~\bibnamefont {{Balokovi{\'c}}}}, \bibinfo {author} {\bibfnamefont {J.}~\bibnamefont {{Barrett}}}, \bibinfo {author} {\bibfnamefont {B.~A.}\ \bibnamefont {{Benson}}}, \bibinfo {author} {\bibfnamefont {D.}~\bibnamefont {{Bintley}}}, \bibinfo {author} {\bibfnamefont {L.}~\bibnamefont {{Blackburn}}},
  \bibinfo {author} {\bibfnamefont {R.}~\bibnamefont {{Blundell}}}, \bibinfo {author} {\bibfnamefont {W.}~\bibnamefont {{Boland}}}, \bibinfo {author} {\bibfnamefont {K.~L.}\ \bibnamefont {{Bouman}}}, \bibinfo {author} {\bibfnamefont {G.~C.}\ \bibnamefont {{Bower}}}, \bibinfo {author} {\bibfnamefont {H.}~\bibnamefont {{Boyce}}}, \bibinfo {author} {\bibfnamefont {M.}~\bibnamefont {{Bremer}}}, \bibinfo {author} {\bibfnamefont {C.~D.}\ \bibnamefont {{Brinkerink}}}, \bibinfo {author} {\bibfnamefont {R.}~\bibnamefont {{Brissenden}}}, \bibinfo {author} {\bibfnamefont {S.}~\bibnamefont {{Britzen}}}, \bibinfo {author} {\bibfnamefont {A.~E.}\ \bibnamefont {{Broderick}}}, \bibinfo {author} {\bibfnamefont {D.}~\bibnamefont {{Broguiere}}}, \bibinfo {author} {\bibfnamefont {T.}~\bibnamefont {{Bronzwaer}}}, \bibinfo {author} {\bibfnamefont {D.-Y.}\ \bibnamefont {{Byun}}}, \bibinfo {author} {\bibfnamefont {J.~E.}\ \bibnamefont {{Carlstrom}}}, \bibinfo {author} {\bibfnamefont {A.}~\bibnamefont {{Chael}}}, \bibinfo {author}
  {\bibfnamefont {C.-k.}\ \bibnamefont {{Chan}}}, \bibinfo {author} {\bibfnamefont {S.}~\bibnamefont {{Chatterjee}}}, \bibinfo {author} {\bibfnamefont {K.}~\bibnamefont {{Chatterjee}}}, \bibinfo {author} {\bibfnamefont {M.-T.}\ \bibnamefont {{Chen}}}, \bibinfo {author} {\bibfnamefont {Y.}~\bibnamefont {{Chen}}}, \bibinfo {author} {\bibfnamefont {P.~M.}\ \bibnamefont {{Chesler}}}, \bibinfo {author} {\bibfnamefont {I.}~\bibnamefont {{Cho}}}, \bibinfo {author} {\bibfnamefont {P.}~\bibnamefont {{Christian}}}, \bibinfo {author} {\bibfnamefont {J.~E.}\ \bibnamefont {{Conway}}}, \bibinfo {author} {\bibfnamefont {J.~M.}\ \bibnamefont {{Cordes}}}, \bibinfo {author} {\bibfnamefont {T.~M.}\ \bibnamefont {{Crawford}}}, \bibinfo {author} {\bibfnamefont {G.~B.}\ \bibnamefont {{Crew}}}, \bibinfo {author} {\bibfnamefont {A.}~\bibnamefont {{Cruz-Osorio}}}, \bibinfo {author} {\bibfnamefont {Y.}~\bibnamefont {{Cui}}}, \bibinfo {author} {\bibfnamefont {J.}~\bibnamefont {{Davelaar}}}, \bibinfo {author} {\bibfnamefont
  {M.}~\bibnamefont {{De Laurentis}}}, \bibinfo {author} {\bibfnamefont {R.}~\bibnamefont {{Deane}}}, \bibinfo {author} {\bibfnamefont {J.}~\bibnamefont {{Dempsey}}}, \bibinfo {author} {\bibfnamefont {G.}~\bibnamefont {{Desvignes}}}, \bibinfo {author} {\bibfnamefont {J.}~\bibnamefont {{Dexter}}}, \bibinfo {author} {\bibfnamefont {S.~S.}\ \bibnamefont {{Doeleman}}}, \bibinfo {author} {\bibfnamefont {R.~P.}\ \bibnamefont {{Eatough}}}, \bibinfo {author} {\bibfnamefont {H.}~\bibnamefont {{Falcke}}}, \bibinfo {author} {\bibfnamefont {J.}~\bibnamefont {{Farah}}}, \bibinfo {author} {\bibfnamefont {V.~L.}\ \bibnamefont {{Fish}}}, \bibinfo {author} {\bibfnamefont {E.}~\bibnamefont {{Fomalont}}}, \bibinfo {author} {\bibfnamefont {H.~A.}\ \bibnamefont {{Ford}}}, \bibinfo {author} {\bibfnamefont {R.}~\bibnamefont {{Fraga-Encinas}}}, \bibinfo {author} {\bibfnamefont {P.}~\bibnamefont {{Friberg}}}, \bibinfo {author} {\bibfnamefont {C.~M.}\ \bibnamefont {{Fromm}}}, \bibinfo {author} {\bibfnamefont {A.}~\bibnamefont
  {{Fuentes}}}, \bibinfo {author} {\bibfnamefont {P.}~\bibnamefont {{Galison}}}, \bibinfo {author} {\bibfnamefont {C.~F.}\ \bibnamefont {{Gammie}}}, \bibinfo {author} {\bibfnamefont {R.}~\bibnamefont {{Garc{\'\i}a}}}, \bibinfo {author} {\bibfnamefont {Z.}~\bibnamefont {{Gelles}}}, \bibinfo {author} {\bibfnamefont {O.}~\bibnamefont {{Gentaz}}}, \bibinfo {author} {\bibfnamefont {B.}~\bibnamefont {{Georgiev}}}, \bibinfo {author} {\bibfnamefont {C.}~\bibnamefont {{Goddi}}}, \bibinfo {author} {\bibfnamefont {R.}~\bibnamefont {{Gold}}}, \bibinfo {author} {\bibfnamefont {J.~L.}\ \bibnamefont {{G{\'o}mez}}}, \bibinfo {author} {\bibfnamefont {A.~I.}\ \bibnamefont {{G{\'o}mez-Ruiz}}}, \bibinfo {author} {\bibfnamefont {M.}~\bibnamefont {{Gu}}}, \bibinfo {author} {\bibfnamefont {M.}~\bibnamefont {{Gurwell}}}, \bibinfo {author} {\bibfnamefont {K.}~\bibnamefont {{Hada}}}, \bibinfo {author} {\bibfnamefont {D.}~\bibnamefont {{Haggard}}}, \bibinfo {author} {\bibfnamefont {M.~H.}\ \bibnamefont {{Hecht}}}, \bibinfo {author}
  {\bibfnamefont {R.}~\bibnamefont {{Hesper}}}, \bibinfo {author} {\bibfnamefont {E.}~\bibnamefont {{Himwich}}}, \bibinfo {author} {\bibfnamefont {L.~C.}\ \bibnamefont {{Ho}}}, \bibinfo {author} {\bibfnamefont {P.}~\bibnamefont {{Ho}}}, \bibinfo {author} {\bibfnamefont {M.}~\bibnamefont {{Honma}}}, \bibinfo {author} {\bibfnamefont {C.-W.~L.}\ \bibnamefont {{Huang}}}, \bibinfo {author} {\bibfnamefont {L.}~\bibnamefont {{Huang}}}, \bibinfo {author} {\bibfnamefont {D.~H.}\ \bibnamefont {{Hughes}}}, \bibinfo {author} {\bibfnamefont {S.}~\bibnamefont {{Ikeda}}}, \bibinfo {author} {\bibfnamefont {M.}~\bibnamefont {{Inoue}}}, \bibinfo {author} {\bibfnamefont {S.}~\bibnamefont {{Issaoun}}}, \bibinfo {author} {\bibfnamefont {D.~J.}\ \bibnamefont {{James}}}, \bibinfo {author} {\bibfnamefont {B.~T.}\ \bibnamefont {{Jannuzi}}}, \bibinfo {author} {\bibfnamefont {M.}~\bibnamefont {{Janssen}}}, \bibinfo {author} {\bibfnamefont {B.}~\bibnamefont {{Jeter}}}, \bibinfo {author} {\bibfnamefont {W.}~\bibnamefont {{Jiang}}},
  \bibinfo {author} {\bibfnamefont {A.}~\bibnamefont {{Jimenez-Rosales}}}, \bibinfo {author} {\bibfnamefont {M.~D.}\ \bibnamefont {{Johnson}}}, \bibinfo {author} {\bibfnamefont {S.}~\bibnamefont {{Jorstad}}}, \bibinfo {author} {\bibfnamefont {T.}~\bibnamefont {{Jung}}}, \bibinfo {author} {\bibfnamefont {M.}~\bibnamefont {{Karami}}}, \bibinfo {author} {\bibfnamefont {R.}~\bibnamefont {{Karuppusamy}}}, \bibinfo {author} {\bibfnamefont {T.}~\bibnamefont {{Kawashima}}}, \bibinfo {author} {\bibfnamefont {G.~K.}\ \bibnamefont {{Keating}}}, \bibinfo {author} {\bibfnamefont {M.}~\bibnamefont {{Kettenis}}}, \bibinfo {author} {\bibfnamefont {D.-J.}\ \bibnamefont {{Kim}}}, \bibinfo {author} {\bibfnamefont {J.-Y.}\ \bibnamefont {{Kim}}}, \bibinfo {author} {\bibfnamefont {J.}~\bibnamefont {{Kim}}}, \bibinfo {author} {\bibfnamefont {J.}~\bibnamefont {{Kim}}}, \bibinfo {author} {\bibfnamefont {M.}~\bibnamefont {{Kino}}}, \bibinfo {author} {\bibfnamefont {J.~Y.}\ \bibnamefont {{Koay}}}, \bibinfo {author} {\bibfnamefont
  {Y.}~\bibnamefont {{Kofuji}}}, \bibinfo {author} {\bibfnamefont {P.~M.}\ \bibnamefont {{Koch}}}, \bibinfo {author} {\bibfnamefont {S.}~\bibnamefont {{Koyama}}}, \bibinfo {author} {\bibfnamefont {M.}~\bibnamefont {{Kramer}}}, \bibinfo {author} {\bibfnamefont {C.}~\bibnamefont {{Kramer}}}, \bibinfo {author} {\bibfnamefont {T.~P.}\ \bibnamefont {{Krichbaum}}}, \bibinfo {author} {\bibfnamefont {C.-Y.}\ \bibnamefont {{Kuo}}}, \bibinfo {author} {\bibfnamefont {T.~R.}\ \bibnamefont {{Lauer}}}, \bibinfo {author} {\bibfnamefont {S.-S.}\ \bibnamefont {{Lee}}}, \bibinfo {author} {\bibfnamefont {A.}~\bibnamefont {{Levis}}}, \bibinfo {author} {\bibfnamefont {Y.-R.}\ \bibnamefont {{Li}}}, \bibinfo {author} {\bibfnamefont {Z.}~\bibnamefont {{Li}}}, \bibinfo {author} {\bibfnamefont {M.}~\bibnamefont {{Lindqvist}}}, \bibinfo {author} {\bibfnamefont {R.}~\bibnamefont {{Lico}}}, \bibinfo {author} {\bibfnamefont {G.}~\bibnamefont {{Lindahl}}}, \bibinfo {author} {\bibfnamefont {J.}~\bibnamefont {{Liu}}}, \bibinfo {author}
  {\bibfnamefont {K.}~\bibnamefont {{Liu}}}, \bibinfo {author} {\bibfnamefont {E.}~\bibnamefont {{Liuzzo}}}, \bibinfo {author} {\bibfnamefont {W.-P.}\ \bibnamefont {{Lo}}}, \bibinfo {author} {\bibfnamefont {A.~P.}\ \bibnamefont {{Lobanov}}}, \bibinfo {author} {\bibfnamefont {L.}~\bibnamefont {{Loinard}}}, \bibinfo {author} {\bibfnamefont {C.}~\bibnamefont {{Lonsdale}}}, \bibinfo {author} {\bibfnamefont {R.-S.}\ \bibnamefont {{Lu}}}, \bibinfo {author} {\bibfnamefont {N.~R.}\ \bibnamefont {{MacDonald}}}, \bibinfo {author} {\bibfnamefont {J.}~\bibnamefont {{Mao}}}, \bibinfo {author} {\bibfnamefont {N.}~\bibnamefont {{Marchili}}}, \bibinfo {author} {\bibfnamefont {S.}~\bibnamefont {{Markoff}}}, \bibinfo {author} {\bibfnamefont {D.~P.}\ \bibnamefont {{Marrone}}}, \bibinfo {author} {\bibfnamefont {A.~P.}\ \bibnamefont {{Marscher}}}, \bibinfo {author} {\bibfnamefont {I.}~\bibnamefont {{Mart{\'\i}-Vidal}}}, \bibinfo {author} {\bibfnamefont {S.}~\bibnamefont {{Matsushita}}}, \bibinfo {author} {\bibfnamefont {L.~D.}\
  \bibnamefont {{Matthews}}}, \bibinfo {author} {\bibfnamefont {L.}~\bibnamefont {{Medeiros}}}, \bibinfo {author} {\bibfnamefont {K.~M.}\ \bibnamefont {{Menten}}}, \bibinfo {author} {\bibfnamefont {I.}~\bibnamefont {{Mizuno}}}, \bibinfo {author} {\bibfnamefont {Y.}~\bibnamefont {{Mizuno}}}, \bibinfo {author} {\bibfnamefont {J.~M.}\ \bibnamefont {{Moran}}}, \bibinfo {author} {\bibfnamefont {K.}~\bibnamefont {{Moriyama}}}, \bibinfo {author} {\bibfnamefont {M.}~\bibnamefont {{Moscibrodzka}}}, \bibinfo {author} {\bibfnamefont {C.}~\bibnamefont {{M{\"u}ller}}}, \bibinfo {author} {\bibfnamefont {G.}~\bibnamefont {{Musoke}}}, \bibinfo {author} {\bibfnamefont {A.}~\bibnamefont {{Mus Mej{\'\i}as}}}, \bibinfo {author} {\bibfnamefont {D.}~\bibnamefont {{Michalik}}}, \bibinfo {author} {\bibfnamefont {A.}~\bibnamefont {{Nadolski}}}, \bibinfo {author} {\bibfnamefont {H.}~\bibnamefont {{Nagai}}}, \bibinfo {author} {\bibfnamefont {N.~M.}\ \bibnamefont {{Nagar}}}, \bibinfo {author} {\bibfnamefont {M.}~\bibnamefont
  {{Nakamura}}}, \bibinfo {author} {\bibfnamefont {R.}~\bibnamefont {{Narayan}}}, \bibinfo {author} {\bibfnamefont {G.}~\bibnamefont {{Narayanan}}}, \bibinfo {author} {\bibfnamefont {I.}~\bibnamefont {{Natarajan}}}, \bibinfo {author} {\bibfnamefont {A.}~\bibnamefont {{Nathanail}}}, \bibinfo {author} {\bibfnamefont {J.}~\bibnamefont {{Neilsen}}}, \bibinfo {author} {\bibfnamefont {R.}~\bibnamefont {{Neri}}}, \bibinfo {author} {\bibfnamefont {C.}~\bibnamefont {{Ni}}}, \bibinfo {author} {\bibfnamefont {A.}~\bibnamefont {{Noutsos}}}, \bibinfo {author} {\bibfnamefont {M.~A.}\ \bibnamefont {{Nowak}}}, \bibinfo {author} {\bibfnamefont {H.}~\bibnamefont {{Okino}}}, \bibinfo {author} {\bibfnamefont {H.}~\bibnamefont {{Olivares}}}, \bibinfo {author} {\bibfnamefont {G.~N.}\ \bibnamefont {{Ortiz-Le{\'o}n}}}, \bibinfo {author} {\bibfnamefont {T.}~\bibnamefont {{Oyama}}}, \bibinfo {author} {\bibfnamefont {F.}~\bibnamefont {{{\"O}zel}}}, \bibinfo {author} {\bibfnamefont {D.~C.~M.}\ \bibnamefont {{Palumbo}}}, \bibinfo
  {author} {\bibfnamefont {J.}~\bibnamefont {{Park}}}, \bibinfo {author} {\bibfnamefont {N.}~\bibnamefont {{Patel}}}, \bibinfo {author} {\bibfnamefont {U.-L.}\ \bibnamefont {{Pen}}}, \bibinfo {author} {\bibfnamefont {D.~W.}\ \bibnamefont {{Pesce}}}, \bibinfo {author} {\bibfnamefont {V.}~\bibnamefont {{Pi{\'e}tu}}}, \bibinfo {author} {\bibfnamefont {R.}~\bibnamefont {{Plambeck}}}, \bibinfo {author} {\bibfnamefont {A.}~\bibnamefont {{PopStefanija}}}, \bibinfo {author} {\bibfnamefont {O.}~\bibnamefont {{Porth}}}, \bibinfo {author} {\bibfnamefont {F.~M.}\ \bibnamefont {{P{\"o}tzl}}}, \bibinfo {author} {\bibfnamefont {B.}~\bibnamefont {{Prather}}}, \bibinfo {author} {\bibfnamefont {J.~A.}\ \bibnamefont {{Preciado-L{\'o}pez}}}, \bibinfo {author} {\bibfnamefont {D.}~\bibnamefont {{Psaltis}}}, \bibinfo {author} {\bibfnamefont {H.-Y.}\ \bibnamefont {{Pu}}}, \bibinfo {author} {\bibfnamefont {V.}~\bibnamefont {{Ramakrishnan}}}, \bibinfo {author} {\bibfnamefont {R.}~\bibnamefont {{Rao}}}, \bibinfo {author} {\bibfnamefont
  {M.~G.}\ \bibnamefont {{Rawlings}}}, \bibinfo {author} {\bibfnamefont {A.~W.}\ \bibnamefont {{Raymond}}}, \bibinfo {author} {\bibfnamefont {L.}~\bibnamefont {{Rezzolla}}}, \bibinfo {author} {\bibfnamefont {A.}~\bibnamefont {{Ricarte}}}, \bibinfo {author} {\bibfnamefont {B.}~\bibnamefont {{Ripperda}}}, \bibinfo {author} {\bibfnamefont {F.}~\bibnamefont {{Roelofs}}}, \bibinfo {author} {\bibfnamefont {A.}~\bibnamefont {{Rogers}}}, \bibinfo {author} {\bibfnamefont {E.}~\bibnamefont {{Ros}}}, \bibinfo {author} {\bibfnamefont {M.}~\bibnamefont {{Rose}}}, \bibinfo {author} {\bibfnamefont {A.}~\bibnamefont {{Roshanineshat}}}, \bibinfo {author} {\bibfnamefont {H.}~\bibnamefont {{Rottmann}}}, \bibinfo {author} {\bibfnamefont {A.~L.}\ \bibnamefont {{Roy}}}, \bibinfo {author} {\bibfnamefont {C.}~\bibnamefont {{Ruszczyk}}}, \bibinfo {author} {\bibfnamefont {K.~L.~J.}\ \bibnamefont {{Rygl}}}, \bibinfo {author} {\bibfnamefont {S.}~\bibnamefont {{S{\'a}nchez}}},\ and\ \bibinfo {author} {\bibfnamefont {D.}~\bibnamefont
  {{S{\'a}nchez-Arguelles}}},\ }\href {https://doi.org/10.3847/2041-8213/abe4de} {\bibfield  {journal} {\bibinfo  {journal} {ApJL}\ }\textbf {\bibinfo {volume} {910}},\ \bibinfo {eid} {L13} (\bibinfo {year} {2021})},\ \Eprint {https://arxiv.org/abs/2105.01173} {arXiv:2105.01173 [astro-ph.HE]} \BibitemShut {NoStop}%
\bibitem [{\citenamefont {{Event Horizon Telescope Collaboration}}\ \emph {et~al.}(2019)\citenamefont {{Event Horizon Telescope Collaboration}}, \citenamefont {{Akiyama}}, \citenamefont {{Alberdi}}, \citenamefont {{Alef}}, \citenamefont {{Asada}}, \citenamefont {{Azulay}}, \citenamefont {{Baczko}}, \citenamefont {{Ball}}, \citenamefont {{Balokovi{\'c}}}, \citenamefont {{Barrett}}, \citenamefont {{Bintley}}, \citenamefont {{Blackburn}}, \citenamefont {{Boland}}, \citenamefont {{Bouman}}, \citenamefont {{Bower}}, \citenamefont {{Bremer}}, \citenamefont {{Brinkerink}}, \citenamefont {{Brissenden}}, \citenamefont {{Britzen}}, \citenamefont {{Broderick}}, \citenamefont {{Broguiere}}, \citenamefont {{Bronzwaer}}, \citenamefont {{Byun}}, \citenamefont {{Carlstrom}}, \citenamefont {{Chael}}, \citenamefont {{Chan}}, \citenamefont {{Chatterjee}}, \citenamefont {{Chatterjee}}, \citenamefont {{Chen}}, \citenamefont {{Chen}}, \citenamefont {{Cho}}, \citenamefont {{Christian}}, \citenamefont {{Conway}}, \citenamefont
  {{Cordes}}, \citenamefont {{Crew}}, \citenamefont {{Cui}}, \citenamefont {{Davelaar}}, \citenamefont {{De Laurentis}}, \citenamefont {{Deane}}, \citenamefont {{Dempsey}}, \citenamefont {{Desvignes}}, \citenamefont {{Dexter}}, \citenamefont {{Doeleman}}, \citenamefont {{Eatough}}, \citenamefont {{Falcke}}, \citenamefont {{Fish}}, \citenamefont {{Fomalont}}, \citenamefont {{Fraga-Encinas}}, \citenamefont {{Freeman}}, \citenamefont {{Friberg}}, \citenamefont {{Fromm}}, \citenamefont {{G{\'o}mez}}, \citenamefont {{Galison}}, \citenamefont {{Gammie}}, \citenamefont {{Garc{\'\i}a}}, \citenamefont {{Gentaz}}, \citenamefont {{Georgiev}}, \citenamefont {{Goddi}}, \citenamefont {{Gold}}, \citenamefont {{Gu}}, \citenamefont {{Gurwell}}, \citenamefont {{Hada}}, \citenamefont {{Hecht}}, \citenamefont {{Hesper}}, \citenamefont {{Ho}}, \citenamefont {{Ho}}, \citenamefont {{Honma}}, \citenamefont {{Huang}}, \citenamefont {{Huang}}, \citenamefont {{Hughes}}, \citenamefont {{Ikeda}}, \citenamefont {{Inoue}}, \citenamefont
  {{Issaoun}}, \citenamefont {{James}}, \citenamefont {{Jannuzi}}, \citenamefont {{Janssen}}, \citenamefont {{Jeter}}, \citenamefont {{Jiang}}, \citenamefont {{Johnson}}, \citenamefont {{Jorstad}}, \citenamefont {{Jung}}, \citenamefont {{Karami}}, \citenamefont {{Karuppusamy}}, \citenamefont {{Kawashima}}, \citenamefont {{Keating}}, \citenamefont {{Kettenis}}, \citenamefont {{Kim}}, \citenamefont {{Kim}}, \citenamefont {{Kim}}, \citenamefont {{Kino}}, \citenamefont {{Koay}}, \citenamefont {{Koch}}, \citenamefont {{Koyama}}, \citenamefont {{Kramer}}, \citenamefont {{Kramer}}, \citenamefont {{Krichbaum}}, \citenamefont {{Kuo}}, \citenamefont {{Lauer}}, \citenamefont {{Lee}}, \citenamefont {{Li}}, \citenamefont {{Li}}, \citenamefont {{Lindqvist}}, \citenamefont {{Liu}}, \citenamefont {{Liuzzo}}, \citenamefont {{Lo}}, \citenamefont {{Lobanov}}, \citenamefont {{Loinard}}, \citenamefont {{Lonsdale}}, \citenamefont {{Lu}}, \citenamefont {{MacDonald}}, \citenamefont {{Mao}}, \citenamefont {{Markoff}}, \citenamefont
  {{Marrone}}, \citenamefont {{Marscher}}, \citenamefont {{Mart{\'\i}-Vidal}}, \citenamefont {{Matsushita}}, \citenamefont {{Matthews}}, \citenamefont {{Medeiros}}, \citenamefont {{Menten}}, \citenamefont {{Mizuno}}, \citenamefont {{Mizuno}}, \citenamefont {{Moran}}, \citenamefont {{Moriyama}}, \citenamefont {{Moscibrodzka}}, \citenamefont {{M{\"u}ller}}, \citenamefont {{Nagai}}, \citenamefont {{Nagar}}, \citenamefont {{Nakamura}}, \citenamefont {{Narayan}}, \citenamefont {{Narayanan}}, \citenamefont {{Natarajan}}, \citenamefont {{Neri}}, \citenamefont {{Ni}}, \citenamefont {{Noutsos}}, \citenamefont {{Okino}}, \citenamefont {{Olivares}}, \citenamefont {{Ortiz-Le{\'o}n}}, \citenamefont {{Oyama}}, \citenamefont {{{\"O}zel}}, \citenamefont {{Palumbo}}, \citenamefont {{Patel}}, \citenamefont {{Pen}}, \citenamefont {{Pesce}}, \citenamefont {{Pi{\'e}tu}}, \citenamefont {{Plambeck}}, \citenamefont {{PopStefanija}}, \citenamefont {{Porth}}, \citenamefont {{Prather}}, \citenamefont {{Preciado-L{\'o}pez}},
  \citenamefont {{Psaltis}}, \citenamefont {{Pu}}, \citenamefont {{Ramakrishnan}}, \citenamefont {{Rao}}, \citenamefont {{Rawlings}}, \citenamefont {{Raymond}}, \citenamefont {{Rezzolla}}, \citenamefont {{Ripperda}}, \citenamefont {{Roelofs}}, \citenamefont {{Rogers}}, \citenamefont {{Ros}}, \citenamefont {{Rose}}, \citenamefont {{Roshanineshat}}, \citenamefont {{Rottmann}}, \citenamefont {{Roy}}, \citenamefont {{Ruszczyk}}, \citenamefont {{Ryan}}, \citenamefont {{Rygl}}, \citenamefont {{S{\'a}nchez}}, \citenamefont {{S{\'a}nchez-Arguelles}}, \citenamefont {{Sasada}}, \citenamefont {{Savolainen}}, \citenamefont {{Schloerb}}, \citenamefont {{Schuster}}, \citenamefont {{Shao}}, \citenamefont {{Shen}}, \citenamefont {{Small}}, \citenamefont {{Sohn}}, \citenamefont {{SooHoo}}, \citenamefont {{Tazaki}}, \citenamefont {{Tiede}}, \citenamefont {{Tilanus}}, \citenamefont {{Titus}}, \citenamefont {{Toma}}, \citenamefont {{Torne}}, \citenamefont {{Trent}}, \citenamefont {{Trippe}}, \citenamefont {{Tsuda}},
  \citenamefont {{van Bemmel}}, \citenamefont {{van Langevelde}}, \citenamefont {{van Rossum}}, \citenamefont {{Wagner}}, \citenamefont {{Wardle}}, \citenamefont {{Weintroub}}, \citenamefont {{Wex}}, \citenamefont {{Wharton}}, \citenamefont {{Wielgus}}, \citenamefont {{Wong}}, \citenamefont {{Wu}}, \citenamefont {{Young}},\ and\ \citenamefont {{Young}}}]{EHT19}%
  \BibitemOpen
  \bibfield  {author} {\bibinfo {author} {\bibnamefont {{Event Horizon Telescope Collaboration}}}, \bibinfo {author} {\bibfnamefont {K.}~\bibnamefont {{Akiyama}}}, \bibinfo {author} {\bibfnamefont {A.}~\bibnamefont {{Alberdi}}}, \bibinfo {author} {\bibfnamefont {W.}~\bibnamefont {{Alef}}}, \bibinfo {author} {\bibfnamefont {K.}~\bibnamefont {{Asada}}}, \bibinfo {author} {\bibfnamefont {R.}~\bibnamefont {{Azulay}}}, \bibinfo {author} {\bibfnamefont {A.-K.}\ \bibnamefont {{Baczko}}}, \bibinfo {author} {\bibfnamefont {D.}~\bibnamefont {{Ball}}}, \bibinfo {author} {\bibfnamefont {M.}~\bibnamefont {{Balokovi{\'c}}}}, \bibinfo {author} {\bibfnamefont {J.}~\bibnamefont {{Barrett}}}, \bibinfo {author} {\bibfnamefont {D.}~\bibnamefont {{Bintley}}}, \bibinfo {author} {\bibfnamefont {L.}~\bibnamefont {{Blackburn}}}, \bibinfo {author} {\bibfnamefont {W.}~\bibnamefont {{Boland}}}, \bibinfo {author} {\bibfnamefont {K.~L.}\ \bibnamefont {{Bouman}}}, \bibinfo {author} {\bibfnamefont {G.~C.}\ \bibnamefont {{Bower}}}, \bibinfo
  {author} {\bibfnamefont {M.}~\bibnamefont {{Bremer}}}, \bibinfo {author} {\bibfnamefont {C.~D.}\ \bibnamefont {{Brinkerink}}}, \bibinfo {author} {\bibfnamefont {R.}~\bibnamefont {{Brissenden}}}, \bibinfo {author} {\bibfnamefont {S.}~\bibnamefont {{Britzen}}}, \bibinfo {author} {\bibfnamefont {A.~E.}\ \bibnamefont {{Broderick}}}, \bibinfo {author} {\bibfnamefont {D.}~\bibnamefont {{Broguiere}}}, \bibinfo {author} {\bibfnamefont {T.}~\bibnamefont {{Bronzwaer}}}, \bibinfo {author} {\bibfnamefont {D.-Y.}\ \bibnamefont {{Byun}}}, \bibinfo {author} {\bibfnamefont {J.~E.}\ \bibnamefont {{Carlstrom}}}, \bibinfo {author} {\bibfnamefont {A.}~\bibnamefont {{Chael}}}, \bibinfo {author} {\bibfnamefont {C.-k.}\ \bibnamefont {{Chan}}}, \bibinfo {author} {\bibfnamefont {S.}~\bibnamefont {{Chatterjee}}}, \bibinfo {author} {\bibfnamefont {K.}~\bibnamefont {{Chatterjee}}}, \bibinfo {author} {\bibfnamefont {M.-T.}\ \bibnamefont {{Chen}}}, \bibinfo {author} {\bibfnamefont {Y.}~\bibnamefont {{Chen}}}, \bibinfo {author}
  {\bibfnamefont {I.}~\bibnamefont {{Cho}}}, \bibinfo {author} {\bibfnamefont {P.}~\bibnamefont {{Christian}}}, \bibinfo {author} {\bibfnamefont {J.~E.}\ \bibnamefont {{Conway}}}, \bibinfo {author} {\bibfnamefont {J.~M.}\ \bibnamefont {{Cordes}}}, \bibinfo {author} {\bibfnamefont {G.~B.}\ \bibnamefont {{Crew}}}, \bibinfo {author} {\bibfnamefont {Y.}~\bibnamefont {{Cui}}}, \bibinfo {author} {\bibfnamefont {J.}~\bibnamefont {{Davelaar}}}, \bibinfo {author} {\bibfnamefont {M.}~\bibnamefont {{De Laurentis}}}, \bibinfo {author} {\bibfnamefont {R.}~\bibnamefont {{Deane}}}, \bibinfo {author} {\bibfnamefont {J.}~\bibnamefont {{Dempsey}}}, \bibinfo {author} {\bibfnamefont {G.}~\bibnamefont {{Desvignes}}}, \bibinfo {author} {\bibfnamefont {J.}~\bibnamefont {{Dexter}}}, \bibinfo {author} {\bibfnamefont {S.~S.}\ \bibnamefont {{Doeleman}}}, \bibinfo {author} {\bibfnamefont {R.~P.}\ \bibnamefont {{Eatough}}}, \bibinfo {author} {\bibfnamefont {H.}~\bibnamefont {{Falcke}}}, \bibinfo {author} {\bibfnamefont {V.~L.}\
  \bibnamefont {{Fish}}}, \bibinfo {author} {\bibfnamefont {E.}~\bibnamefont {{Fomalont}}}, \bibinfo {author} {\bibfnamefont {R.}~\bibnamefont {{Fraga-Encinas}}}, \bibinfo {author} {\bibfnamefont {W.~T.}\ \bibnamefont {{Freeman}}}, \bibinfo {author} {\bibfnamefont {P.}~\bibnamefont {{Friberg}}}, \bibinfo {author} {\bibfnamefont {C.~M.}\ \bibnamefont {{Fromm}}}, \bibinfo {author} {\bibfnamefont {J.~L.}\ \bibnamefont {{G{\'o}mez}}}, \bibinfo {author} {\bibfnamefont {P.}~\bibnamefont {{Galison}}}, \bibinfo {author} {\bibfnamefont {C.~F.}\ \bibnamefont {{Gammie}}}, \bibinfo {author} {\bibfnamefont {R.}~\bibnamefont {{Garc{\'\i}a}}}, \bibinfo {author} {\bibfnamefont {O.}~\bibnamefont {{Gentaz}}}, \bibinfo {author} {\bibfnamefont {B.}~\bibnamefont {{Georgiev}}}, \bibinfo {author} {\bibfnamefont {C.}~\bibnamefont {{Goddi}}}, \bibinfo {author} {\bibfnamefont {R.}~\bibnamefont {{Gold}}}, \bibinfo {author} {\bibfnamefont {M.}~\bibnamefont {{Gu}}}, \bibinfo {author} {\bibfnamefont {M.}~\bibnamefont {{Gurwell}}},
  \bibinfo {author} {\bibfnamefont {K.}~\bibnamefont {{Hada}}}, \bibinfo {author} {\bibfnamefont {M.~H.}\ \bibnamefont {{Hecht}}}, \bibinfo {author} {\bibfnamefont {R.}~\bibnamefont {{Hesper}}}, \bibinfo {author} {\bibfnamefont {L.~C.}\ \bibnamefont {{Ho}}}, \bibinfo {author} {\bibfnamefont {P.}~\bibnamefont {{Ho}}}, \bibinfo {author} {\bibfnamefont {M.}~\bibnamefont {{Honma}}}, \bibinfo {author} {\bibfnamefont {C.-W.~L.}\ \bibnamefont {{Huang}}}, \bibinfo {author} {\bibfnamefont {L.}~\bibnamefont {{Huang}}}, \bibinfo {author} {\bibfnamefont {D.~H.}\ \bibnamefont {{Hughes}}}, \bibinfo {author} {\bibfnamefont {S.}~\bibnamefont {{Ikeda}}}, \bibinfo {author} {\bibfnamefont {M.}~\bibnamefont {{Inoue}}}, \bibinfo {author} {\bibfnamefont {S.}~\bibnamefont {{Issaoun}}}, \bibinfo {author} {\bibfnamefont {D.~J.}\ \bibnamefont {{James}}}, \bibinfo {author} {\bibfnamefont {B.~T.}\ \bibnamefont {{Jannuzi}}}, \bibinfo {author} {\bibfnamefont {M.}~\bibnamefont {{Janssen}}}, \bibinfo {author} {\bibfnamefont
  {B.}~\bibnamefont {{Jeter}}}, \bibinfo {author} {\bibfnamefont {W.}~\bibnamefont {{Jiang}}}, \bibinfo {author} {\bibfnamefont {M.~D.}\ \bibnamefont {{Johnson}}}, \bibinfo {author} {\bibfnamefont {S.}~\bibnamefont {{Jorstad}}}, \bibinfo {author} {\bibfnamefont {T.}~\bibnamefont {{Jung}}}, \bibinfo {author} {\bibfnamefont {M.}~\bibnamefont {{Karami}}}, \bibinfo {author} {\bibfnamefont {R.}~\bibnamefont {{Karuppusamy}}}, \bibinfo {author} {\bibfnamefont {T.}~\bibnamefont {{Kawashima}}}, \bibinfo {author} {\bibfnamefont {G.~K.}\ \bibnamefont {{Keating}}}, \bibinfo {author} {\bibfnamefont {M.}~\bibnamefont {{Kettenis}}}, \bibinfo {author} {\bibfnamefont {J.-Y.}\ \bibnamefont {{Kim}}}, \bibinfo {author} {\bibfnamefont {J.}~\bibnamefont {{Kim}}}, \bibinfo {author} {\bibfnamefont {J.}~\bibnamefont {{Kim}}}, \bibinfo {author} {\bibfnamefont {M.}~\bibnamefont {{Kino}}}, \bibinfo {author} {\bibfnamefont {J.~Y.}\ \bibnamefont {{Koay}}}, \bibinfo {author} {\bibfnamefont {P.~M.}\ \bibnamefont {{Koch}}}, \bibinfo {author}
  {\bibfnamefont {S.}~\bibnamefont {{Koyama}}}, \bibinfo {author} {\bibfnamefont {M.}~\bibnamefont {{Kramer}}}, \bibinfo {author} {\bibfnamefont {C.}~\bibnamefont {{Kramer}}}, \bibinfo {author} {\bibfnamefont {T.~P.}\ \bibnamefont {{Krichbaum}}}, \bibinfo {author} {\bibfnamefont {C.-Y.}\ \bibnamefont {{Kuo}}}, \bibinfo {author} {\bibfnamefont {T.~R.}\ \bibnamefont {{Lauer}}}, \bibinfo {author} {\bibfnamefont {S.-S.}\ \bibnamefont {{Lee}}}, \bibinfo {author} {\bibfnamefont {Y.-R.}\ \bibnamefont {{Li}}}, \bibinfo {author} {\bibfnamefont {Z.}~\bibnamefont {{Li}}}, \bibinfo {author} {\bibfnamefont {M.}~\bibnamefont {{Lindqvist}}}, \bibinfo {author} {\bibfnamefont {K.}~\bibnamefont {{Liu}}}, \bibinfo {author} {\bibfnamefont {E.}~\bibnamefont {{Liuzzo}}}, \bibinfo {author} {\bibfnamefont {W.-P.}\ \bibnamefont {{Lo}}}, \bibinfo {author} {\bibfnamefont {A.~P.}\ \bibnamefont {{Lobanov}}}, \bibinfo {author} {\bibfnamefont {L.}~\bibnamefont {{Loinard}}}, \bibinfo {author} {\bibfnamefont {C.}~\bibnamefont {{Lonsdale}}},
  \bibinfo {author} {\bibfnamefont {R.-S.}\ \bibnamefont {{Lu}}}, \bibinfo {author} {\bibfnamefont {N.~R.}\ \bibnamefont {{MacDonald}}}, \bibinfo {author} {\bibfnamefont {J.}~\bibnamefont {{Mao}}}, \bibinfo {author} {\bibfnamefont {S.}~\bibnamefont {{Markoff}}}, \bibinfo {author} {\bibfnamefont {D.~P.}\ \bibnamefont {{Marrone}}}, \bibinfo {author} {\bibfnamefont {A.~P.}\ \bibnamefont {{Marscher}}}, \bibinfo {author} {\bibfnamefont {I.}~\bibnamefont {{Mart{\'\i}-Vidal}}}, \bibinfo {author} {\bibfnamefont {S.}~\bibnamefont {{Matsushita}}}, \bibinfo {author} {\bibfnamefont {L.~D.}\ \bibnamefont {{Matthews}}}, \bibinfo {author} {\bibfnamefont {L.}~\bibnamefont {{Medeiros}}}, \bibinfo {author} {\bibfnamefont {K.~M.}\ \bibnamefont {{Menten}}}, \bibinfo {author} {\bibfnamefont {Y.}~\bibnamefont {{Mizuno}}}, \bibinfo {author} {\bibfnamefont {I.}~\bibnamefont {{Mizuno}}}, \bibinfo {author} {\bibfnamefont {J.~M.}\ \bibnamefont {{Moran}}}, \bibinfo {author} {\bibfnamefont {K.}~\bibnamefont {{Moriyama}}}, \bibinfo
  {author} {\bibfnamefont {M.}~\bibnamefont {{Moscibrodzka}}}, \bibinfo {author} {\bibfnamefont {C.}~\bibnamefont {{M{\"u}ller}}}, \bibinfo {author} {\bibfnamefont {H.}~\bibnamefont {{Nagai}}}, \bibinfo {author} {\bibfnamefont {N.~M.}\ \bibnamefont {{Nagar}}}, \bibinfo {author} {\bibfnamefont {M.}~\bibnamefont {{Nakamura}}}, \bibinfo {author} {\bibfnamefont {R.}~\bibnamefont {{Narayan}}}, \bibinfo {author} {\bibfnamefont {G.}~\bibnamefont {{Narayanan}}}, \bibinfo {author} {\bibfnamefont {I.}~\bibnamefont {{Natarajan}}}, \bibinfo {author} {\bibfnamefont {R.}~\bibnamefont {{Neri}}}, \bibinfo {author} {\bibfnamefont {C.}~\bibnamefont {{Ni}}}, \bibinfo {author} {\bibfnamefont {A.}~\bibnamefont {{Noutsos}}}, \bibinfo {author} {\bibfnamefont {H.}~\bibnamefont {{Okino}}}, \bibinfo {author} {\bibfnamefont {H.}~\bibnamefont {{Olivares}}}, \bibinfo {author} {\bibfnamefont {G.~N.}\ \bibnamefont {{Ortiz-Le{\'o}n}}}, \bibinfo {author} {\bibfnamefont {T.}~\bibnamefont {{Oyama}}}, \bibinfo {author} {\bibfnamefont
  {F.}~\bibnamefont {{{\"O}zel}}}, \bibinfo {author} {\bibfnamefont {D.~C.~M.}\ \bibnamefont {{Palumbo}}}, \bibinfo {author} {\bibfnamefont {N.}~\bibnamefont {{Patel}}}, \bibinfo {author} {\bibfnamefont {U.-L.}\ \bibnamefont {{Pen}}}, \bibinfo {author} {\bibfnamefont {D.~W.}\ \bibnamefont {{Pesce}}}, \bibinfo {author} {\bibfnamefont {V.}~\bibnamefont {{Pi{\'e}tu}}}, \bibinfo {author} {\bibfnamefont {R.}~\bibnamefont {{Plambeck}}}, \bibinfo {author} {\bibfnamefont {A.}~\bibnamefont {{PopStefanija}}}, \bibinfo {author} {\bibfnamefont {O.}~\bibnamefont {{Porth}}}, \bibinfo {author} {\bibfnamefont {B.}~\bibnamefont {{Prather}}}, \bibinfo {author} {\bibfnamefont {J.~A.}\ \bibnamefont {{Preciado-L{\'o}pez}}}, \bibinfo {author} {\bibfnamefont {D.}~\bibnamefont {{Psaltis}}}, \bibinfo {author} {\bibfnamefont {H.-Y.}\ \bibnamefont {{Pu}}}, \bibinfo {author} {\bibfnamefont {V.}~\bibnamefont {{Ramakrishnan}}}, \bibinfo {author} {\bibfnamefont {R.}~\bibnamefont {{Rao}}}, \bibinfo {author} {\bibfnamefont {M.~G.}\
  \bibnamefont {{Rawlings}}}, \bibinfo {author} {\bibfnamefont {A.~W.}\ \bibnamefont {{Raymond}}}, \bibinfo {author} {\bibfnamefont {L.}~\bibnamefont {{Rezzolla}}}, \bibinfo {author} {\bibfnamefont {B.}~\bibnamefont {{Ripperda}}}, \bibinfo {author} {\bibfnamefont {F.}~\bibnamefont {{Roelofs}}}, \bibinfo {author} {\bibfnamefont {A.}~\bibnamefont {{Rogers}}}, \bibinfo {author} {\bibfnamefont {E.}~\bibnamefont {{Ros}}}, \bibinfo {author} {\bibfnamefont {M.}~\bibnamefont {{Rose}}}, \bibinfo {author} {\bibfnamefont {A.}~\bibnamefont {{Roshanineshat}}}, \bibinfo {author} {\bibfnamefont {H.}~\bibnamefont {{Rottmann}}}, \bibinfo {author} {\bibfnamefont {A.~L.}\ \bibnamefont {{Roy}}}, \bibinfo {author} {\bibfnamefont {C.}~\bibnamefont {{Ruszczyk}}}, \bibinfo {author} {\bibfnamefont {B.~R.}\ \bibnamefont {{Ryan}}}, \bibinfo {author} {\bibfnamefont {K.~L.~J.}\ \bibnamefont {{Rygl}}}, \bibinfo {author} {\bibfnamefont {S.}~\bibnamefont {{S{\'a}nchez}}}, \bibinfo {author} {\bibfnamefont {D.}~\bibnamefont
  {{S{\'a}nchez-Arguelles}}}, \bibinfo {author} {\bibfnamefont {M.}~\bibnamefont {{Sasada}}}, \bibinfo {author} {\bibfnamefont {T.}~\bibnamefont {{Savolainen}}}, \bibinfo {author} {\bibfnamefont {F.~P.}\ \bibnamefont {{Schloerb}}}, \bibinfo {author} {\bibfnamefont {K.-F.}\ \bibnamefont {{Schuster}}}, \bibinfo {author} {\bibfnamefont {L.}~\bibnamefont {{Shao}}}, \bibinfo {author} {\bibfnamefont {Z.}~\bibnamefont {{Shen}}}, \bibinfo {author} {\bibfnamefont {D.}~\bibnamefont {{Small}}}, \bibinfo {author} {\bibfnamefont {B.~W.}\ \bibnamefont {{Sohn}}}, \bibinfo {author} {\bibfnamefont {J.}~\bibnamefont {{SooHoo}}}, \bibinfo {author} {\bibfnamefont {F.}~\bibnamefont {{Tazaki}}}, \bibinfo {author} {\bibfnamefont {P.}~\bibnamefont {{Tiede}}}, \bibinfo {author} {\bibfnamefont {R.~P.~J.}\ \bibnamefont {{Tilanus}}}, \bibinfo {author} {\bibfnamefont {M.}~\bibnamefont {{Titus}}}, \bibinfo {author} {\bibfnamefont {K.}~\bibnamefont {{Toma}}}, \bibinfo {author} {\bibfnamefont {P.}~\bibnamefont {{Torne}}}, \bibinfo {author}
  {\bibfnamefont {T.}~\bibnamefont {{Trent}}}, \bibinfo {author} {\bibfnamefont {S.}~\bibnamefont {{Trippe}}}, \bibinfo {author} {\bibfnamefont {S.}~\bibnamefont {{Tsuda}}}, \bibinfo {author} {\bibfnamefont {I.}~\bibnamefont {{van Bemmel}}}, \bibinfo {author} {\bibfnamefont {H.~J.}\ \bibnamefont {{van Langevelde}}}, \bibinfo {author} {\bibfnamefont {D.~R.}\ \bibnamefont {{van Rossum}}}, \bibinfo {author} {\bibfnamefont {J.}~\bibnamefont {{Wagner}}}, \bibinfo {author} {\bibfnamefont {J.}~\bibnamefont {{Wardle}}}, \bibinfo {author} {\bibfnamefont {J.}~\bibnamefont {{Weintroub}}}, \bibinfo {author} {\bibfnamefont {N.}~\bibnamefont {{Wex}}}, \bibinfo {author} {\bibfnamefont {R.}~\bibnamefont {{Wharton}}}, \bibinfo {author} {\bibfnamefont {M.}~\bibnamefont {{Wielgus}}}, \bibinfo {author} {\bibfnamefont {G.~N.}\ \bibnamefont {{Wong}}}, \bibinfo {author} {\bibfnamefont {Q.}~\bibnamefont {{Wu}}}, \bibinfo {author} {\bibfnamefont {K.}~\bibnamefont {{Young}}},\ and\ \bibinfo {author} {\bibfnamefont {A.}~\bibnamefont
  {{Young}}},\ }\href {https://doi.org/10.3847/2041-8213/ab0ec7} {\bibfield  {journal} {\bibinfo  {journal} {ApJL}\ }\textbf {\bibinfo {volume} {875}},\ \bibinfo {eid} {L1} (\bibinfo {year} {2019})},\ \Eprint {https://arxiv.org/abs/1906.11238} {arXiv:1906.11238 [astro-ph.GA]} \BibitemShut {NoStop}%
\bibitem [{\citenamefont {{The Event Horizon Telescope Collaboration}}\ \emph {et~al.}(2024)\citenamefont {{The Event Horizon Telescope Collaboration}}, \citenamefont {Akiyama}, \citenamefont {Alberdi}, \citenamefont {Alef}, \citenamefont {Algaba}, \citenamefont {Anantua}, \citenamefont {Asada}, \citenamefont {Azulay}, \citenamefont {Bach}, \citenamefont {Baczko}, \citenamefont {Ball}, \citenamefont {Baloković}, \citenamefont {Bandyopadhyay}, \citenamefont {Barrett}, \citenamefont {Bauböck}, \citenamefont {Benson}, \citenamefont {Bintley}, \citenamefont {Blackburn}, \citenamefont {Blundell}, \citenamefont {Bouman}, \citenamefont {Bower}, \citenamefont {Boyce}, \citenamefont {Bremer}, \citenamefont {Brinkerink}, \citenamefont {Brissenden}, \citenamefont {Britzen}, \citenamefont {Broderick}, \citenamefont {Broguiere}, \citenamefont {Bronzwaer}, \citenamefont {Bustamante}, \citenamefont {Byun}, \citenamefont {Carlstrom}, \citenamefont {Ceccobello}, \citenamefont {Chael}, \citenamefont {Chan}, \citenamefont
  {Chang}, \citenamefont {Chatterjee}, \citenamefont {Chatterjee}, \citenamefont {Chen}, \citenamefont {Chen}, \citenamefont {Cheng}, \citenamefont {Cho}, \citenamefont {Christian}, \citenamefont {Conroy}, \citenamefont {Conway}, \citenamefont {Cordes}, \citenamefont {Crawford}, \citenamefont {Crew}, \citenamefont {Cruz-Osorio}, \citenamefont {Cui}, \citenamefont {Dahale}, \citenamefont {Davelaar}, \citenamefont {De~Laurentis}, \citenamefont {Deane}, \citenamefont {Dempsey}, \citenamefont {Desvignes}, \citenamefont {Dexter}, \citenamefont {Dhruv}, \citenamefont {Dihingia}, \citenamefont {Doeleman}, \citenamefont {Dougall}, \citenamefont {Dzib}, \citenamefont {Eatough}, \citenamefont {Emami}, \citenamefont {Falcke}, \citenamefont {Farah}, \citenamefont {Fish}, \citenamefont {Fomalont}, \citenamefont {Ford}, \citenamefont {Foschi}, \citenamefont {Fraga-Encinas}, \citenamefont {Freeman}, \citenamefont {Friberg}, \citenamefont {Fromm}, \citenamefont {Fuentes}, \citenamefont {Galison}, \citenamefont {Gammie},
  \citenamefont {García}, \citenamefont {Gentaz}, \citenamefont {Georgiev}, \citenamefont {Goddi}, \citenamefont {Gold}, \citenamefont {Gómez-Ruiz}, \citenamefont {Gómez}, \citenamefont {Gu}, \citenamefont {Gurwell}, \citenamefont {Hada}, \citenamefont {Haggard}, \citenamefont {Haworth}, \citenamefont {Hecht}, \citenamefont {Hesper}, \citenamefont {Heumann}, \citenamefont {Ho}, \citenamefont {Ho}, \citenamefont {Honma}, \citenamefont {Huang}, \citenamefont {Huang}, \citenamefont {Hughes}, \citenamefont {Ikeda}, \citenamefont {Impellizzeri}, \citenamefont {Inoue}, \citenamefont {Issaoun}, \citenamefont {James}, \citenamefont {Jannuzi}, \citenamefont {Janssen}, \citenamefont {Jeter}, \citenamefont {Jiang}, \citenamefont {Jiménez-Rosales}, \citenamefont {Johnson}, \citenamefont {Jorstad}, \citenamefont {Joshi}, \citenamefont {Jung}, \citenamefont {Karami}, \citenamefont {Karuppusamy}, \citenamefont {Kawashima}, \citenamefont {Keating}, \citenamefont {Kettenis}, \citenamefont {Kim}, \citenamefont {Kim},
  \citenamefont {Kim}, \citenamefont {Kim}, \citenamefont {Kino}, \citenamefont {Koay}, \citenamefont {Kocherlakota}, \citenamefont {Kofuji}, \citenamefont {Koch}, \citenamefont {Koyama}, \citenamefont {Kramer}, \citenamefont {Kramer}, \citenamefont {Kramer}, \citenamefont {Krichbaum}, \citenamefont {Kuo}, \citenamefont {La~Bella}, \citenamefont {Lauer}, \citenamefont {Lee}, \citenamefont {Lee}, \citenamefont {Leung}, \citenamefont {Levis}, \citenamefont {Li}, \citenamefont {Lico}, \citenamefont {Lindahl}, \citenamefont {Lindqvist}, \citenamefont {Lisakov}, \citenamefont {Liu}, \citenamefont {Liu}, \citenamefont {Liuzzo}, \citenamefont {Lo}, \citenamefont {Lobanov}, \citenamefont {Loinard}, \citenamefont {Lonsdale}, \citenamefont {Lowitz}, \citenamefont {Lu}, \citenamefont {MacDonald}, \citenamefont {Mao}, \citenamefont {Marchili}, \citenamefont {Markoff}, \citenamefont {Marrone}, \citenamefont {Marscher}, \citenamefont {Martí-Vidal}, \citenamefont {Matsushita}, \citenamefont {Matthews}, \citenamefont
  {Medeiros}, \citenamefont {Menten}, \citenamefont {Michalik}, \citenamefont {Mizuno}, \citenamefont {Mizuno}, \citenamefont {Moran}, \citenamefont {Moriyama}, \citenamefont {Moscibrodzka}, \citenamefont {Mulaudzi}, \citenamefont {Müller}, \citenamefont {Müller}, \citenamefont {Mus}, \citenamefont {Musoke}, \citenamefont {Myserlis}, \citenamefont {Nadolski}, \citenamefont {Nagai}, \citenamefont {Nagar}, \citenamefont {Nakamura}, \citenamefont {Narayanan}, \citenamefont {Natarajan}, \citenamefont {Nathanail}, \citenamefont {Navarro~Fuentes}, \citenamefont {Neilsen}, \citenamefont {Neri}, \citenamefont {Ni}, \citenamefont {Noutsos}, \citenamefont {Nowak}, \citenamefont {Oh}, \citenamefont {Okino}, \citenamefont {Olivares}, \citenamefont {Ortiz-León}, \citenamefont {Oyama}, \citenamefont {Özel}, \citenamefont {Palumbo}, \citenamefont {Paraschos}, \citenamefont {Park}, \citenamefont {Parsons}, \citenamefont {Patel}, \citenamefont {Pen}, \citenamefont {Pesce}, \citenamefont {Piétu}, \citenamefont {Plambeck},
  \citenamefont {PopStefanija}, \citenamefont {Porth}, \citenamefont {Pötzl}, \citenamefont {Prather}, \citenamefont {Preciado-López}, \citenamefont {Psaltis}, \citenamefont {Pu}, \citenamefont {Ramakrishnan}, \citenamefont {Rao}, \citenamefont {Rawlings}, \citenamefont {Raymond}, \citenamefont {Rezzolla}, \citenamefont {Ricarte}, \citenamefont {Ripperda}, \citenamefont {Roelofs}, \citenamefont {Rogers}, \citenamefont {Romero-Cañizales}, \citenamefont {Ros}, \citenamefont {Roshanineshat}, \citenamefont {Rottmann}, \citenamefont {Roy}, \citenamefont {Ruiz}, \citenamefont {Ruszczyk}, \citenamefont {Rygl}, \citenamefont {Sánchez}, \citenamefont {Sánchez-Argüelles}, \citenamefont {Sánchez-Portal}, \citenamefont {Sasada}, \citenamefont {Satapathy}, \citenamefont {Savolainen}, \citenamefont {Schloerb}, \citenamefont {Schonfeld}, \citenamefont {Schuster}, \citenamefont {Shao}, \citenamefont {Shen}, \citenamefont {Small}, \citenamefont {Sohn}, \citenamefont {SooHoo}, \citenamefont {Sosapanta~Salas},
  \citenamefont {Souccar}, \citenamefont {Stanway}, \citenamefont {Sun}, \citenamefont {Tazaki}, \citenamefont {Tetarenko}, \citenamefont {Tiede}, \citenamefont {Tilanus}, \citenamefont {Titus}, \citenamefont {Torne}, \citenamefont {Toscano}, \citenamefont {Traianou}, \citenamefont {Trent}, \citenamefont {Trippe}, \citenamefont {Turk}, \citenamefont {van Bemmel}, \citenamefont {van Langevelde}, \citenamefont {van Rossum}, \citenamefont {Vos}, \citenamefont {Wagner}, \citenamefont {Ward-Thompson}, \citenamefont {Wardle}, \citenamefont {Washington}, \citenamefont {Weintroub}, \citenamefont {Wharton}, \citenamefont {Wielgus}, \citenamefont {Wiik}, \citenamefont {Witzel}, \citenamefont {Wondrak}, \citenamefont {Wong}, \citenamefont {Wu}, \citenamefont {Yadlapalli}, \citenamefont {Yamaguchi}, \citenamefont {Yfantis}, \citenamefont {Yoon}, \citenamefont {Young}, \citenamefont {Young}, \citenamefont {Younsi}, \citenamefont {Yu}, \citenamefont {Yuan}, \citenamefont {Yuan}, \citenamefont {Zensus}, \citenamefont
  {Zhang}, \citenamefont {Zhao}, \citenamefont {Zhao},\ and\ \citenamefont {Najafi-Ziyazi}}]{EHT_SgrA_spin}%
  \BibitemOpen
  \bibfield  {author} {\bibinfo {author} {\bibnamefont {{The Event Horizon Telescope Collaboration}}}, \bibinfo {author} {\bibfnamefont {K.}~\bibnamefont {Akiyama}}, \bibinfo {author} {\bibfnamefont {A.}~\bibnamefont {Alberdi}}, \bibinfo {author} {\bibfnamefont {W.}~\bibnamefont {Alef}}, \bibinfo {author} {\bibfnamefont {J.~C.}\ \bibnamefont {Algaba}}, \bibinfo {author} {\bibfnamefont {R.}~\bibnamefont {Anantua}}, \bibinfo {author} {\bibfnamefont {K.}~\bibnamefont {Asada}}, \bibinfo {author} {\bibfnamefont {R.}~\bibnamefont {Azulay}}, \bibinfo {author} {\bibfnamefont {U.}~\bibnamefont {Bach}}, \bibinfo {author} {\bibfnamefont {A.-K.}\ \bibnamefont {Baczko}}, \bibinfo {author} {\bibfnamefont {D.}~\bibnamefont {Ball}}, \bibinfo {author} {\bibfnamefont {M.}~\bibnamefont {Baloković}}, \bibinfo {author} {\bibfnamefont {B.}~\bibnamefont {Bandyopadhyay}}, \bibinfo {author} {\bibfnamefont {J.}~\bibnamefont {Barrett}}, \bibinfo {author} {\bibfnamefont {M.}~\bibnamefont {Bauböck}}, \bibinfo {author} {\bibfnamefont
  {B.~A.}\ \bibnamefont {Benson}}, \bibinfo {author} {\bibfnamefont {D.}~\bibnamefont {Bintley}}, \bibinfo {author} {\bibfnamefont {L.}~\bibnamefont {Blackburn}}, \bibinfo {author} {\bibfnamefont {R.}~\bibnamefont {Blundell}}, \bibinfo {author} {\bibfnamefont {K.~L.}\ \bibnamefont {Bouman}}, \bibinfo {author} {\bibfnamefont {G.~C.}\ \bibnamefont {Bower}}, \bibinfo {author} {\bibfnamefont {H.}~\bibnamefont {Boyce}}, \bibinfo {author} {\bibfnamefont {M.}~\bibnamefont {Bremer}}, \bibinfo {author} {\bibfnamefont {C.~D.}\ \bibnamefont {Brinkerink}}, \bibinfo {author} {\bibfnamefont {R.}~\bibnamefont {Brissenden}}, \bibinfo {author} {\bibfnamefont {S.}~\bibnamefont {Britzen}}, \bibinfo {author} {\bibfnamefont {A.~E.}\ \bibnamefont {Broderick}}, \bibinfo {author} {\bibfnamefont {D.}~\bibnamefont {Broguiere}}, \bibinfo {author} {\bibfnamefont {T.}~\bibnamefont {Bronzwaer}}, \bibinfo {author} {\bibfnamefont {S.}~\bibnamefont {Bustamante}}, \bibinfo {author} {\bibfnamefont {D.-Y.}\ \bibnamefont {Byun}}, \bibinfo
  {author} {\bibfnamefont {J.~E.}\ \bibnamefont {Carlstrom}}, \bibinfo {author} {\bibfnamefont {C.}~\bibnamefont {Ceccobello}}, \bibinfo {author} {\bibfnamefont {A.}~\bibnamefont {Chael}}, \bibinfo {author} {\bibfnamefont {C.-k.}\ \bibnamefont {Chan}}, \bibinfo {author} {\bibfnamefont {D.~O.}\ \bibnamefont {Chang}}, \bibinfo {author} {\bibfnamefont {K.}~\bibnamefont {Chatterjee}}, \bibinfo {author} {\bibfnamefont {S.}~\bibnamefont {Chatterjee}}, \bibinfo {author} {\bibfnamefont {M.-T.}\ \bibnamefont {Chen}}, \bibinfo {author} {\bibfnamefont {Y.}~\bibnamefont {Chen}}, \bibinfo {author} {\bibfnamefont {X.}~\bibnamefont {Cheng}}, \bibinfo {author} {\bibfnamefont {I.}~\bibnamefont {Cho}}, \bibinfo {author} {\bibfnamefont {P.}~\bibnamefont {Christian}}, \bibinfo {author} {\bibfnamefont {N.~S.}\ \bibnamefont {Conroy}}, \bibinfo {author} {\bibfnamefont {J.~E.}\ \bibnamefont {Conway}}, \bibinfo {author} {\bibfnamefont {J.~M.}\ \bibnamefont {Cordes}}, \bibinfo {author} {\bibfnamefont {T.~M.}\ \bibnamefont {Crawford}},
  \bibinfo {author} {\bibfnamefont {G.~B.}\ \bibnamefont {Crew}}, \bibinfo {author} {\bibfnamefont {A.}~\bibnamefont {Cruz-Osorio}}, \bibinfo {author} {\bibfnamefont {Y.}~\bibnamefont {Cui}}, \bibinfo {author} {\bibfnamefont {R.}~\bibnamefont {Dahale}}, \bibinfo {author} {\bibfnamefont {J.}~\bibnamefont {Davelaar}}, \bibinfo {author} {\bibfnamefont {M.}~\bibnamefont {De~Laurentis}}, \bibinfo {author} {\bibfnamefont {R.}~\bibnamefont {Deane}}, \bibinfo {author} {\bibfnamefont {J.}~\bibnamefont {Dempsey}}, \bibinfo {author} {\bibfnamefont {G.}~\bibnamefont {Desvignes}}, \bibinfo {author} {\bibfnamefont {J.}~\bibnamefont {Dexter}}, \bibinfo {author} {\bibfnamefont {V.}~\bibnamefont {Dhruv}}, \bibinfo {author} {\bibfnamefont {I.~K.}\ \bibnamefont {Dihingia}}, \bibinfo {author} {\bibfnamefont {S.~S.}\ \bibnamefont {Doeleman}}, \bibinfo {author} {\bibfnamefont {S.}~\bibnamefont {Dougall}}, \bibinfo {author} {\bibfnamefont {S.~A.}\ \bibnamefont {Dzib}}, \bibinfo {author} {\bibfnamefont {R.~P.}\ \bibnamefont
  {Eatough}}, \bibinfo {author} {\bibfnamefont {R.}~\bibnamefont {Emami}}, \bibinfo {author} {\bibfnamefont {H.}~\bibnamefont {Falcke}}, \bibinfo {author} {\bibfnamefont {J.}~\bibnamefont {Farah}}, \bibinfo {author} {\bibfnamefont {V.~L.}\ \bibnamefont {Fish}}, \bibinfo {author} {\bibfnamefont {E.}~\bibnamefont {Fomalont}}, \bibinfo {author} {\bibfnamefont {H.~A.}\ \bibnamefont {Ford}}, \bibinfo {author} {\bibfnamefont {M.}~\bibnamefont {Foschi}}, \bibinfo {author} {\bibfnamefont {R.}~\bibnamefont {Fraga-Encinas}}, \bibinfo {author} {\bibfnamefont {W.~T.}\ \bibnamefont {Freeman}}, \bibinfo {author} {\bibfnamefont {P.}~\bibnamefont {Friberg}}, \bibinfo {author} {\bibfnamefont {C.~M.}\ \bibnamefont {Fromm}}, \bibinfo {author} {\bibfnamefont {A.}~\bibnamefont {Fuentes}}, \bibinfo {author} {\bibfnamefont {P.}~\bibnamefont {Galison}}, \bibinfo {author} {\bibfnamefont {C.~F.}\ \bibnamefont {Gammie}}, \bibinfo {author} {\bibfnamefont {R.}~\bibnamefont {García}}, \bibinfo {author} {\bibfnamefont {O.}~\bibnamefont
  {Gentaz}}, \bibinfo {author} {\bibfnamefont {B.}~\bibnamefont {Georgiev}}, \bibinfo {author} {\bibfnamefont {C.}~\bibnamefont {Goddi}}, \bibinfo {author} {\bibfnamefont {R.}~\bibnamefont {Gold}}, \bibinfo {author} {\bibfnamefont {A.~I.}\ \bibnamefont {Gómez-Ruiz}}, \bibinfo {author} {\bibfnamefont {J.~L.}\ \bibnamefont {Gómez}}, \bibinfo {author} {\bibfnamefont {M.}~\bibnamefont {Gu}}, \bibinfo {author} {\bibfnamefont {M.}~\bibnamefont {Gurwell}}, \bibinfo {author} {\bibfnamefont {K.}~\bibnamefont {Hada}}, \bibinfo {author} {\bibfnamefont {D.}~\bibnamefont {Haggard}}, \bibinfo {author} {\bibfnamefont {K.}~\bibnamefont {Haworth}}, \bibinfo {author} {\bibfnamefont {M.~H.}\ \bibnamefont {Hecht}}, \bibinfo {author} {\bibfnamefont {R.}~\bibnamefont {Hesper}}, \bibinfo {author} {\bibfnamefont {D.}~\bibnamefont {Heumann}}, \bibinfo {author} {\bibfnamefont {L.~C.}\ \bibnamefont {Ho}}, \bibinfo {author} {\bibfnamefont {P.}~\bibnamefont {Ho}}, \bibinfo {author} {\bibfnamefont {M.}~\bibnamefont {Honma}}, \bibinfo
  {author} {\bibfnamefont {C.-W.~L.}\ \bibnamefont {Huang}}, \bibinfo {author} {\bibfnamefont {L.}~\bibnamefont {Huang}}, \bibinfo {author} {\bibfnamefont {D.~H.}\ \bibnamefont {Hughes}}, \bibinfo {author} {\bibfnamefont {S.}~\bibnamefont {Ikeda}}, \bibinfo {author} {\bibfnamefont {C.~M.~V.}\ \bibnamefont {Impellizzeri}}, \bibinfo {author} {\bibfnamefont {M.}~\bibnamefont {Inoue}}, \bibinfo {author} {\bibfnamefont {S.}~\bibnamefont {Issaoun}}, \bibinfo {author} {\bibfnamefont {D.~J.}\ \bibnamefont {James}}, \bibinfo {author} {\bibfnamefont {B.~T.}\ \bibnamefont {Jannuzi}}, \bibinfo {author} {\bibfnamefont {M.}~\bibnamefont {Janssen}}, \bibinfo {author} {\bibfnamefont {B.}~\bibnamefont {Jeter}}, \bibinfo {author} {\bibfnamefont {W.}~\bibnamefont {Jiang}}, \bibinfo {author} {\bibfnamefont {A.}~\bibnamefont {Jiménez-Rosales}}, \bibinfo {author} {\bibfnamefont {M.~D.}\ \bibnamefont {Johnson}}, \bibinfo {author} {\bibfnamefont {S.}~\bibnamefont {Jorstad}}, \bibinfo {author} {\bibfnamefont {A.~V.}\ \bibnamefont
  {Joshi}}, \bibinfo {author} {\bibfnamefont {T.}~\bibnamefont {Jung}}, \bibinfo {author} {\bibfnamefont {M.}~\bibnamefont {Karami}}, \bibinfo {author} {\bibfnamefont {R.}~\bibnamefont {Karuppusamy}}, \bibinfo {author} {\bibfnamefont {T.}~\bibnamefont {Kawashima}}, \bibinfo {author} {\bibfnamefont {G.~K.}\ \bibnamefont {Keating}}, \bibinfo {author} {\bibfnamefont {M.}~\bibnamefont {Kettenis}}, \bibinfo {author} {\bibfnamefont {D.-J.}\ \bibnamefont {Kim}}, \bibinfo {author} {\bibfnamefont {J.-Y.}\ \bibnamefont {Kim}}, \bibinfo {author} {\bibfnamefont {J.}~\bibnamefont {Kim}}, \bibinfo {author} {\bibfnamefont {J.}~\bibnamefont {Kim}}, \bibinfo {author} {\bibfnamefont {M.}~\bibnamefont {Kino}}, \bibinfo {author} {\bibfnamefont {J.~Y.}\ \bibnamefont {Koay}}, \bibinfo {author} {\bibfnamefont {P.}~\bibnamefont {Kocherlakota}}, \bibinfo {author} {\bibfnamefont {Y.}~\bibnamefont {Kofuji}}, \bibinfo {author} {\bibfnamefont {P.~M.}\ \bibnamefont {Koch}}, \bibinfo {author} {\bibfnamefont {S.}~\bibnamefont {Koyama}},
  \bibinfo {author} {\bibfnamefont {C.}~\bibnamefont {Kramer}}, \bibinfo {author} {\bibfnamefont {J.~A.}\ \bibnamefont {Kramer}}, \bibinfo {author} {\bibfnamefont {M.}~\bibnamefont {Kramer}}, \bibinfo {author} {\bibfnamefont {T.~P.}\ \bibnamefont {Krichbaum}}, \bibinfo {author} {\bibfnamefont {C.-Y.}\ \bibnamefont {Kuo}}, \bibinfo {author} {\bibfnamefont {N.}~\bibnamefont {La~Bella}}, \bibinfo {author} {\bibfnamefont {T.~R.}\ \bibnamefont {Lauer}}, \bibinfo {author} {\bibfnamefont {D.}~\bibnamefont {Lee}}, \bibinfo {author} {\bibfnamefont {S.-S.}\ \bibnamefont {Lee}}, \bibinfo {author} {\bibfnamefont {P.~K.}\ \bibnamefont {Leung}}, \bibinfo {author} {\bibfnamefont {A.}~\bibnamefont {Levis}}, \bibinfo {author} {\bibfnamefont {Z.}~\bibnamefont {Li}}, \bibinfo {author} {\bibfnamefont {R.}~\bibnamefont {Lico}}, \bibinfo {author} {\bibfnamefont {G.}~\bibnamefont {Lindahl}}, \bibinfo {author} {\bibfnamefont {M.}~\bibnamefont {Lindqvist}}, \bibinfo {author} {\bibfnamefont {M.}~\bibnamefont {Lisakov}}, \bibinfo
  {author} {\bibfnamefont {J.}~\bibnamefont {Liu}}, \bibinfo {author} {\bibfnamefont {K.}~\bibnamefont {Liu}}, \bibinfo {author} {\bibfnamefont {E.}~\bibnamefont {Liuzzo}}, \bibinfo {author} {\bibfnamefont {W.-P.}\ \bibnamefont {Lo}}, \bibinfo {author} {\bibfnamefont {A.~P.}\ \bibnamefont {Lobanov}}, \bibinfo {author} {\bibfnamefont {L.}~\bibnamefont {Loinard}}, \bibinfo {author} {\bibfnamefont {C.~J.}\ \bibnamefont {Lonsdale}}, \bibinfo {author} {\bibfnamefont {A.~E.}\ \bibnamefont {Lowitz}}, \bibinfo {author} {\bibfnamefont {R.-S.}\ \bibnamefont {Lu}}, \bibinfo {author} {\bibfnamefont {N.~R.}\ \bibnamefont {MacDonald}}, \bibinfo {author} {\bibfnamefont {J.}~\bibnamefont {Mao}}, \bibinfo {author} {\bibfnamefont {N.}~\bibnamefont {Marchili}}, \bibinfo {author} {\bibfnamefont {S.}~\bibnamefont {Markoff}}, \bibinfo {author} {\bibfnamefont {D.~P.}\ \bibnamefont {Marrone}}, \bibinfo {author} {\bibfnamefont {A.~P.}\ \bibnamefont {Marscher}}, \bibinfo {author} {\bibfnamefont {I.}~\bibnamefont {Martí-Vidal}},
  \bibinfo {author} {\bibfnamefont {S.}~\bibnamefont {Matsushita}}, \bibinfo {author} {\bibfnamefont {L.~D.}\ \bibnamefont {Matthews}}, \bibinfo {author} {\bibfnamefont {L.}~\bibnamefont {Medeiros}}, \bibinfo {author} {\bibfnamefont {K.~M.}\ \bibnamefont {Menten}}, \bibinfo {author} {\bibfnamefont {D.}~\bibnamefont {Michalik}}, \bibinfo {author} {\bibfnamefont {I.}~\bibnamefont {Mizuno}}, \bibinfo {author} {\bibfnamefont {Y.}~\bibnamefont {Mizuno}}, \bibinfo {author} {\bibfnamefont {J.~M.}\ \bibnamefont {Moran}}, \bibinfo {author} {\bibfnamefont {K.}~\bibnamefont {Moriyama}}, \bibinfo {author} {\bibfnamefont {M.}~\bibnamefont {Moscibrodzka}}, \bibinfo {author} {\bibfnamefont {W.}~\bibnamefont {Mulaudzi}}, \bibinfo {author} {\bibfnamefont {C.}~\bibnamefont {Müller}}, \bibinfo {author} {\bibfnamefont {H.}~\bibnamefont {Müller}}, \bibinfo {author} {\bibfnamefont {A.}~\bibnamefont {Mus}}, \bibinfo {author} {\bibfnamefont {G.}~\bibnamefont {Musoke}}, \bibinfo {author} {\bibfnamefont {I.}~\bibnamefont
  {Myserlis}}, \bibinfo {author} {\bibfnamefont {A.}~\bibnamefont {Nadolski}}, \bibinfo {author} {\bibfnamefont {H.}~\bibnamefont {Nagai}}, \bibinfo {author} {\bibfnamefont {N.~M.}\ \bibnamefont {Nagar}}, \bibinfo {author} {\bibfnamefont {M.}~\bibnamefont {Nakamura}}, \bibinfo {author} {\bibfnamefont {G.}~\bibnamefont {Narayanan}}, \bibinfo {author} {\bibfnamefont {I.}~\bibnamefont {Natarajan}}, \bibinfo {author} {\bibfnamefont {A.}~\bibnamefont {Nathanail}}, \bibinfo {author} {\bibfnamefont {S.}~\bibnamefont {Navarro~Fuentes}}, \bibinfo {author} {\bibfnamefont {J.}~\bibnamefont {Neilsen}}, \bibinfo {author} {\bibfnamefont {R.}~\bibnamefont {Neri}}, \bibinfo {author} {\bibfnamefont {C.}~\bibnamefont {Ni}}, \bibinfo {author} {\bibfnamefont {A.}~\bibnamefont {Noutsos}}, \bibinfo {author} {\bibfnamefont {M.~A.}\ \bibnamefont {Nowak}}, \bibinfo {author} {\bibfnamefont {J.}~\bibnamefont {Oh}}, \bibinfo {author} {\bibfnamefont {H.}~\bibnamefont {Okino}}, \bibinfo {author} {\bibfnamefont {H.}~\bibnamefont
  {Olivares}}, \bibinfo {author} {\bibfnamefont {G.~N.}\ \bibnamefont {Ortiz-León}}, \bibinfo {author} {\bibfnamefont {T.}~\bibnamefont {Oyama}}, \bibinfo {author} {\bibfnamefont {F.}~\bibnamefont {Özel}}, \bibinfo {author} {\bibfnamefont {D.~C.~M.}\ \bibnamefont {Palumbo}}, \bibinfo {author} {\bibfnamefont {G.~F.}\ \bibnamefont {Paraschos}}, \bibinfo {author} {\bibfnamefont {J.}~\bibnamefont {Park}}, \bibinfo {author} {\bibfnamefont {H.}~\bibnamefont {Parsons}}, \bibinfo {author} {\bibfnamefont {N.}~\bibnamefont {Patel}}, \bibinfo {author} {\bibfnamefont {U.-L.}\ \bibnamefont {Pen}}, \bibinfo {author} {\bibfnamefont {D.~W.}\ \bibnamefont {Pesce}}, \bibinfo {author} {\bibfnamefont {V.}~\bibnamefont {Piétu}}, \bibinfo {author} {\bibfnamefont {R.}~\bibnamefont {Plambeck}}, \bibinfo {author} {\bibfnamefont {A.}~\bibnamefont {PopStefanija}}, \bibinfo {author} {\bibfnamefont {O.}~\bibnamefont {Porth}}, \bibinfo {author} {\bibfnamefont {F.~M.}\ \bibnamefont {Pötzl}}, \bibinfo {author} {\bibfnamefont
  {B.}~\bibnamefont {Prather}}, \bibinfo {author} {\bibfnamefont {J.~A.}\ \bibnamefont {Preciado-López}}, \bibinfo {author} {\bibfnamefont {D.}~\bibnamefont {Psaltis}}, \bibinfo {author} {\bibfnamefont {H.-Y.}\ \bibnamefont {Pu}}, \bibinfo {author} {\bibfnamefont {V.}~\bibnamefont {Ramakrishnan}}, \bibinfo {author} {\bibfnamefont {R.}~\bibnamefont {Rao}}, \bibinfo {author} {\bibfnamefont {M.~G.}\ \bibnamefont {Rawlings}}, \bibinfo {author} {\bibfnamefont {A.~W.}\ \bibnamefont {Raymond}}, \bibinfo {author} {\bibfnamefont {L.}~\bibnamefont {Rezzolla}}, \bibinfo {author} {\bibfnamefont {A.}~\bibnamefont {Ricarte}}, \bibinfo {author} {\bibfnamefont {B.}~\bibnamefont {Ripperda}}, \bibinfo {author} {\bibfnamefont {F.}~\bibnamefont {Roelofs}}, \bibinfo {author} {\bibfnamefont {A.}~\bibnamefont {Rogers}}, \bibinfo {author} {\bibfnamefont {C.}~\bibnamefont {Romero-Cañizales}}, \bibinfo {author} {\bibfnamefont {E.}~\bibnamefont {Ros}}, \bibinfo {author} {\bibfnamefont {A.}~\bibnamefont {Roshanineshat}}, \bibinfo
  {author} {\bibfnamefont {H.}~\bibnamefont {Rottmann}}, \bibinfo {author} {\bibfnamefont {A.~L.}\ \bibnamefont {Roy}}, \bibinfo {author} {\bibfnamefont {I.}~\bibnamefont {Ruiz}}, \bibinfo {author} {\bibfnamefont {C.}~\bibnamefont {Ruszczyk}}, \bibinfo {author} {\bibfnamefont {K.~L.~J.}\ \bibnamefont {Rygl}}, \bibinfo {author} {\bibfnamefont {S.}~\bibnamefont {Sánchez}}, \bibinfo {author} {\bibfnamefont {D.}~\bibnamefont {Sánchez-Argüelles}}, \bibinfo {author} {\bibfnamefont {M.}~\bibnamefont {Sánchez-Portal}}, \bibinfo {author} {\bibfnamefont {M.}~\bibnamefont {Sasada}}, \bibinfo {author} {\bibfnamefont {K.}~\bibnamefont {Satapathy}}, \bibinfo {author} {\bibfnamefont {T.}~\bibnamefont {Savolainen}}, \bibinfo {author} {\bibfnamefont {F.~P.}\ \bibnamefont {Schloerb}}, \bibinfo {author} {\bibfnamefont {J.}~\bibnamefont {Schonfeld}}, \bibinfo {author} {\bibfnamefont {K.-F.}\ \bibnamefont {Schuster}}, \bibinfo {author} {\bibfnamefont {L.}~\bibnamefont {Shao}}, \bibinfo {author} {\bibfnamefont
  {Z.}~\bibnamefont {Shen}}, \bibinfo {author} {\bibfnamefont {D.}~\bibnamefont {Small}}, \bibinfo {author} {\bibfnamefont {B.~W.}\ \bibnamefont {Sohn}}, \bibinfo {author} {\bibfnamefont {J.}~\bibnamefont {SooHoo}}, \bibinfo {author} {\bibfnamefont {L.~D.}\ \bibnamefont {Sosapanta~Salas}}, \bibinfo {author} {\bibfnamefont {K.}~\bibnamefont {Souccar}}, \bibinfo {author} {\bibfnamefont {J.~S.}\ \bibnamefont {Stanway}}, \bibinfo {author} {\bibfnamefont {H.}~\bibnamefont {Sun}}, \bibinfo {author} {\bibfnamefont {F.}~\bibnamefont {Tazaki}}, \bibinfo {author} {\bibfnamefont {A.~J.}\ \bibnamefont {Tetarenko}}, \bibinfo {author} {\bibfnamefont {P.}~\bibnamefont {Tiede}}, \bibinfo {author} {\bibfnamefont {R.~P.~J.}\ \bibnamefont {Tilanus}}, \bibinfo {author} {\bibfnamefont {M.}~\bibnamefont {Titus}}, \bibinfo {author} {\bibfnamefont {P.}~\bibnamefont {Torne}}, \bibinfo {author} {\bibfnamefont {T.}~\bibnamefont {Toscano}}, \bibinfo {author} {\bibfnamefont {E.}~\bibnamefont {Traianou}}, \bibinfo {author} {\bibfnamefont
  {T.}~\bibnamefont {Trent}}, \bibinfo {author} {\bibfnamefont {S.}~\bibnamefont {Trippe}}, \bibinfo {author} {\bibfnamefont {M.}~\bibnamefont {Turk}}, \bibinfo {author} {\bibfnamefont {I.}~\bibnamefont {van Bemmel}}, \bibinfo {author} {\bibfnamefont {H.~J.}\ \bibnamefont {van Langevelde}}, \bibinfo {author} {\bibfnamefont {D.~R.}\ \bibnamefont {van Rossum}}, \bibinfo {author} {\bibfnamefont {J.}~\bibnamefont {Vos}}, \bibinfo {author} {\bibfnamefont {J.}~\bibnamefont {Wagner}}, \bibinfo {author} {\bibfnamefont {D.}~\bibnamefont {Ward-Thompson}}, \bibinfo {author} {\bibfnamefont {J.}~\bibnamefont {Wardle}}, \bibinfo {author} {\bibfnamefont {J.~E.}\ \bibnamefont {Washington}}, \bibinfo {author} {\bibfnamefont {J.}~\bibnamefont {Weintroub}}, \bibinfo {author} {\bibfnamefont {R.}~\bibnamefont {Wharton}}, \bibinfo {author} {\bibfnamefont {M.}~\bibnamefont {Wielgus}}, \bibinfo {author} {\bibfnamefont {K.}~\bibnamefont {Wiik}}, \bibinfo {author} {\bibfnamefont {G.}~\bibnamefont {Witzel}}, \bibinfo {author}
  {\bibfnamefont {M.~F.}\ \bibnamefont {Wondrak}}, \bibinfo {author} {\bibfnamefont {G.~N.}\ \bibnamefont {Wong}}, \bibinfo {author} {\bibfnamefont {Q.}~\bibnamefont {Wu}}, \bibinfo {author} {\bibfnamefont {N.}~\bibnamefont {Yadlapalli}}, \bibinfo {author} {\bibfnamefont {P.}~\bibnamefont {Yamaguchi}}, \bibinfo {author} {\bibfnamefont {A.}~\bibnamefont {Yfantis}}, \bibinfo {author} {\bibfnamefont {D.}~\bibnamefont {Yoon}}, \bibinfo {author} {\bibfnamefont {A.}~\bibnamefont {Young}}, \bibinfo {author} {\bibfnamefont {K.}~\bibnamefont {Young}}, \bibinfo {author} {\bibfnamefont {Z.}~\bibnamefont {Younsi}}, \bibinfo {author} {\bibfnamefont {W.}~\bibnamefont {Yu}}, \bibinfo {author} {\bibfnamefont {F.}~\bibnamefont {Yuan}}, \bibinfo {author} {\bibfnamefont {Y.-F.}\ \bibnamefont {Yuan}}, \bibinfo {author} {\bibfnamefont {J.~A.}\ \bibnamefont {Zensus}}, \bibinfo {author} {\bibfnamefont {S.}~\bibnamefont {Zhang}}, \bibinfo {author} {\bibfnamefont {G.-Y.}\ \bibnamefont {Zhao}}, \bibinfo {author} {\bibfnamefont
  {S.-S.}\ \bibnamefont {Zhao}},\ and\ \bibinfo {author} {\bibfnamefont {M.}~\bibnamefont {Najafi-Ziyazi}},\ }\href {https://doi.org/10.3847/2041-8213/ad2df1} {\bibfield  {journal} {\bibinfo  {journal} {The Astrophysical Journal Letters}\ }\textbf {\bibinfo {volume} {964}},\ \bibinfo {pages} {L26} (\bibinfo {year} {2024})}\BibitemShut {NoStop}%
\bibitem [{\citenamefont {{Event Horizon Telescope Collaboration}}\ \emph {et~al.}(2022)\citenamefont {{Event Horizon Telescope Collaboration}}, \citenamefont {{Akiyama}}, \citenamefont {{Alberdi}}, \citenamefont {{Alef}}, \citenamefont {{Algaba}}, \citenamefont {{Anantua}}, \citenamefont {{Asada}}, \citenamefont {{Azulay}}, \citenamefont {{Bach}}, \citenamefont {{Baczko}}, \citenamefont {{Ball}}, \citenamefont {{Balokovi{\'c}}}, \citenamefont {{Barrett}}, \citenamefont {{Baub{\"o}ck}}, \citenamefont {{Benson}}, \citenamefont {{Bintley}}, \citenamefont {{Blackburn}}, \citenamefont {{Blundell}}, \citenamefont {{Bouman}}, \citenamefont {{Bower}}, \citenamefont {{Boyce}}, \citenamefont {{Bremer}}, \citenamefont {{Brinkerink}}, \citenamefont {{Brissenden}}, \citenamefont {{Britzen}}, \citenamefont {{Broderick}}, \citenamefont {{Broguiere}}, \citenamefont {{Bronzwaer}}, \citenamefont {{Bustamante}}, \citenamefont {{Byun}}, \citenamefont {{Carlstrom}}, \citenamefont {{Ceccobello}}, \citenamefont {{Chael}},
  \citenamefont {{Chan}}, \citenamefont {{Chatterjee}}, \citenamefont {{Chatterjee}}, \citenamefont {{Chen}}, \citenamefont {{Chen}}, \citenamefont {{Cheng}}, \citenamefont {{Cho}}, \citenamefont {{Christian}}, \citenamefont {{Conroy}}, \citenamefont {{Conway}}, \citenamefont {{Cordes}}, \citenamefont {{Crawford}}, \citenamefont {{Crew}}, \citenamefont {{Cruz-Osorio}}, \citenamefont {{Cui}}, \citenamefont {{Davelaar}}, \citenamefont {{De Laurentis}}, \citenamefont {{Deane}}, \citenamefont {{Dempsey}}, \citenamefont {{Desvignes}}, \citenamefont {{Dexter}}, \citenamefont {{Dhruv}}, \citenamefont {{Doeleman}}, \citenamefont {{Dougal}}, \citenamefont {{Dzib}}, \citenamefont {{Eatough}}, \citenamefont {{Emami}}, \citenamefont {{Falcke}}, \citenamefont {{Farah}}, \citenamefont {{Fish}}, \citenamefont {{Fomalont}}, \citenamefont {{Ford}}, \citenamefont {{Fraga-Encinas}}, \citenamefont {{Freeman}}, \citenamefont {{Friberg}}, \citenamefont {{Fromm}}, \citenamefont {{Fuentes}}, \citenamefont {{Galison}}, \citenamefont
  {{Gammie}}, \citenamefont {{Garc{\'\i}a}}, \citenamefont {{Gentaz}}, \citenamefont {{Georgiev}}, \citenamefont {{Goddi}}, \citenamefont {{Gold}}, \citenamefont {{G{\'o}mez-Ruiz}}, \citenamefont {{G{\'o}mez}}, \citenamefont {{Gu}}, \citenamefont {{Gurwell}}, \citenamefont {{Hada}}, \citenamefont {{Haggard}}, \citenamefont {{Haworth}}, \citenamefont {{Hecht}}, \citenamefont {{Hesper}}, \citenamefont {{Heumann}}, \citenamefont {{Ho}}, \citenamefont {{Ho}}, \citenamefont {{Honma}}, \citenamefont {{Huang}}, \citenamefont {{Huang}}, \citenamefont {{Hughes}}, \citenamefont {{Ikeda}}, \citenamefont {{Impellizzeri}}, \citenamefont {{Inoue}}, \citenamefont {{Issaoun}}, \citenamefont {{James}}, \citenamefont {{Jannuzi}}, \citenamefont {{Janssen}}, \citenamefont {{Jeter}}, \citenamefont {{Jiang}}, \citenamefont {{Jim{\'e}nez-Rosales}}, \citenamefont {{Johnson}}, \citenamefont {{Jorstad}}, \citenamefont {{Joshi}}, \citenamefont {{Jung}}, \citenamefont {{Karami}}, \citenamefont {{Karuppusamy}}, \citenamefont
  {{Kawashima}}, \citenamefont {{Keating}}, \citenamefont {{Kettenis}}, \citenamefont {{Kim}}, \citenamefont {{Kim}}, \citenamefont {{Kim}}, \citenamefont {{Kim}}, \citenamefont {{Kino}}, \citenamefont {{Koay}}, \citenamefont {{Kocherlakota}}, \citenamefont {{Kofuji}}, \citenamefont {{Koch}}, \citenamefont {{Koyama}}, \citenamefont {{Kramer}}, \citenamefont {{Kramer}}, \citenamefont {{Krichbaum}}, \citenamefont {{Kuo}}, \citenamefont {{La Bella}}, \citenamefont {{Lauer}}, \citenamefont {{Lee}}, \citenamefont {{Lee}}, \citenamefont {{Leung}}, \citenamefont {{Levis}}, \citenamefont {{Li}}, \citenamefont {{Lico}}, \citenamefont {{Lindahl}}, \citenamefont {{Lindqvist}}, \citenamefont {{Lisakov}}, \citenamefont {{Liu}}, \citenamefont {{Liu}}, \citenamefont {{Liuzzo}}, \citenamefont {{Lo}}, \citenamefont {{Lobanov}}, \citenamefont {{Loinard}}, \citenamefont {{Lonsdale}}, \citenamefont {{Lu}}, \citenamefont {{Mao}}, \citenamefont {{Marchili}}, \citenamefont {{Markoff}}, \citenamefont {{Marrone}}, \citenamefont
  {{Marscher}}, \citenamefont {{Mart{\'\i}-Vidal}}, \citenamefont {{Matsushita}}, \citenamefont {{Matthews}}, \citenamefont {{Medeiros}}, \citenamefont {{Menten}}, \citenamefont {{Michalik}}, \citenamefont {{Mizuno}}, \citenamefont {{Mizuno}}, \citenamefont {{Moran}}, \citenamefont {{Moriyama}}, \citenamefont {{Moscibrodzka}}, \citenamefont {{M{\"u}ller}}, \citenamefont {{Mus}}, \citenamefont {{Musoke}}, \citenamefont {{Myserlis}}, \citenamefont {{Nadolski}}, \citenamefont {{Nagai}}, \citenamefont {{Nagar}}, \citenamefont {{Nakamura}}, \citenamefont {{Narayan}}, \citenamefont {{Narayanan}}, \citenamefont {{Natarajan}}, \citenamefont {{Nathanail}}, \citenamefont {{Fuentes}}, \citenamefont {{Neilsen}}, \citenamefont {{Neri}}, \citenamefont {{Ni}}, \citenamefont {{Noutsos}}, \citenamefont {{Nowak}}, \citenamefont {{Oh}}, \citenamefont {{Okino}}, \citenamefont {{Olivares}}, \citenamefont {{Ortiz-Le{\'o}n}}, \citenamefont {{Oyama}}, \citenamefont {{{\"O}zel}}, \citenamefont {{Palumbo}}, \citenamefont
  {{Paraschos}}, \citenamefont {{Park}}, \citenamefont {{Parsons}}, \citenamefont {{Patel}}, \citenamefont {{Pen}}, \citenamefont {{Pesce}}, \citenamefont {{Pi{\'e}tu}}, \citenamefont {{Plambeck}}, \citenamefont {{PopStefanija}}, \citenamefont {{Porth}}, \citenamefont {{P{\"o}tzl}}, \citenamefont {{Prather}}, \citenamefont {{Preciado-L{\'o}pez}},\ and\ \citenamefont {{Psaltis}}}]{EHT_SgrA1}%
  \BibitemOpen
  \bibfield  {author} {\bibinfo {author} {\bibnamefont {{Event Horizon Telescope Collaboration}}}, \bibinfo {author} {\bibfnamefont {K.}~\bibnamefont {{Akiyama}}}, \bibinfo {author} {\bibfnamefont {A.}~\bibnamefont {{Alberdi}}}, \bibinfo {author} {\bibfnamefont {W.}~\bibnamefont {{Alef}}}, \bibinfo {author} {\bibfnamefont {J.~C.}\ \bibnamefont {{Algaba}}}, \bibinfo {author} {\bibfnamefont {R.}~\bibnamefont {{Anantua}}}, \bibinfo {author} {\bibfnamefont {K.}~\bibnamefont {{Asada}}}, \bibinfo {author} {\bibfnamefont {R.}~\bibnamefont {{Azulay}}}, \bibinfo {author} {\bibfnamefont {U.}~\bibnamefont {{Bach}}}, \bibinfo {author} {\bibfnamefont {A.-K.}\ \bibnamefont {{Baczko}}}, \bibinfo {author} {\bibfnamefont {D.}~\bibnamefont {{Ball}}}, \bibinfo {author} {\bibfnamefont {M.}~\bibnamefont {{Balokovi{\'c}}}}, \bibinfo {author} {\bibfnamefont {J.}~\bibnamefont {{Barrett}}}, \bibinfo {author} {\bibfnamefont {M.}~\bibnamefont {{Baub{\"o}ck}}}, \bibinfo {author} {\bibfnamefont {B.~A.}\ \bibnamefont {{Benson}}}, \bibinfo
  {author} {\bibfnamefont {D.}~\bibnamefont {{Bintley}}}, \bibinfo {author} {\bibfnamefont {L.}~\bibnamefont {{Blackburn}}}, \bibinfo {author} {\bibfnamefont {R.}~\bibnamefont {{Blundell}}}, \bibinfo {author} {\bibfnamefont {K.~L.}\ \bibnamefont {{Bouman}}}, \bibinfo {author} {\bibfnamefont {G.~C.}\ \bibnamefont {{Bower}}}, \bibinfo {author} {\bibfnamefont {H.}~\bibnamefont {{Boyce}}}, \bibinfo {author} {\bibfnamefont {M.}~\bibnamefont {{Bremer}}}, \bibinfo {author} {\bibfnamefont {C.~D.}\ \bibnamefont {{Brinkerink}}}, \bibinfo {author} {\bibfnamefont {R.}~\bibnamefont {{Brissenden}}}, \bibinfo {author} {\bibfnamefont {S.}~\bibnamefont {{Britzen}}}, \bibinfo {author} {\bibfnamefont {A.~E.}\ \bibnamefont {{Broderick}}}, \bibinfo {author} {\bibfnamefont {D.}~\bibnamefont {{Broguiere}}}, \bibinfo {author} {\bibfnamefont {T.}~\bibnamefont {{Bronzwaer}}}, \bibinfo {author} {\bibfnamefont {S.}~\bibnamefont {{Bustamante}}}, \bibinfo {author} {\bibfnamefont {D.-Y.}\ \bibnamefont {{Byun}}}, \bibinfo {author}
  {\bibfnamefont {J.~E.}\ \bibnamefont {{Carlstrom}}}, \bibinfo {author} {\bibfnamefont {C.}~\bibnamefont {{Ceccobello}}}, \bibinfo {author} {\bibfnamefont {A.}~\bibnamefont {{Chael}}}, \bibinfo {author} {\bibfnamefont {C.-k.}\ \bibnamefont {{Chan}}}, \bibinfo {author} {\bibfnamefont {K.}~\bibnamefont {{Chatterjee}}}, \bibinfo {author} {\bibfnamefont {S.}~\bibnamefont {{Chatterjee}}}, \bibinfo {author} {\bibfnamefont {M.-T.}\ \bibnamefont {{Chen}}}, \bibinfo {author} {\bibfnamefont {Y.}~\bibnamefont {{Chen}}}, \bibinfo {author} {\bibfnamefont {X.}~\bibnamefont {{Cheng}}}, \bibinfo {author} {\bibfnamefont {I.}~\bibnamefont {{Cho}}}, \bibinfo {author} {\bibfnamefont {P.}~\bibnamefont {{Christian}}}, \bibinfo {author} {\bibfnamefont {N.~S.}\ \bibnamefont {{Conroy}}}, \bibinfo {author} {\bibfnamefont {J.~E.}\ \bibnamefont {{Conway}}}, \bibinfo {author} {\bibfnamefont {J.~M.}\ \bibnamefont {{Cordes}}}, \bibinfo {author} {\bibfnamefont {T.~M.}\ \bibnamefont {{Crawford}}}, \bibinfo {author} {\bibfnamefont {G.~B.}\
  \bibnamefont {{Crew}}}, \bibinfo {author} {\bibfnamefont {A.}~\bibnamefont {{Cruz-Osorio}}}, \bibinfo {author} {\bibfnamefont {Y.}~\bibnamefont {{Cui}}}, \bibinfo {author} {\bibfnamefont {J.}~\bibnamefont {{Davelaar}}}, \bibinfo {author} {\bibfnamefont {M.}~\bibnamefont {{De Laurentis}}}, \bibinfo {author} {\bibfnamefont {R.}~\bibnamefont {{Deane}}}, \bibinfo {author} {\bibfnamefont {J.}~\bibnamefont {{Dempsey}}}, \bibinfo {author} {\bibfnamefont {G.}~\bibnamefont {{Desvignes}}}, \bibinfo {author} {\bibfnamefont {J.}~\bibnamefont {{Dexter}}}, \bibinfo {author} {\bibfnamefont {V.}~\bibnamefont {{Dhruv}}}, \bibinfo {author} {\bibfnamefont {S.~S.}\ \bibnamefont {{Doeleman}}}, \bibinfo {author} {\bibfnamefont {S.}~\bibnamefont {{Dougal}}}, \bibinfo {author} {\bibfnamefont {S.~A.}\ \bibnamefont {{Dzib}}}, \bibinfo {author} {\bibfnamefont {R.~P.}\ \bibnamefont {{Eatough}}}, \bibinfo {author} {\bibfnamefont {R.}~\bibnamefont {{Emami}}}, \bibinfo {author} {\bibfnamefont {H.}~\bibnamefont {{Falcke}}}, \bibinfo
  {author} {\bibfnamefont {J.}~\bibnamefont {{Farah}}}, \bibinfo {author} {\bibfnamefont {V.~L.}\ \bibnamefont {{Fish}}}, \bibinfo {author} {\bibfnamefont {E.}~\bibnamefont {{Fomalont}}}, \bibinfo {author} {\bibfnamefont {H.~A.}\ \bibnamefont {{Ford}}}, \bibinfo {author} {\bibfnamefont {R.}~\bibnamefont {{Fraga-Encinas}}}, \bibinfo {author} {\bibfnamefont {W.~T.}\ \bibnamefont {{Freeman}}}, \bibinfo {author} {\bibfnamefont {P.}~\bibnamefont {{Friberg}}}, \bibinfo {author} {\bibfnamefont {C.~M.}\ \bibnamefont {{Fromm}}}, \bibinfo {author} {\bibfnamefont {A.}~\bibnamefont {{Fuentes}}}, \bibinfo {author} {\bibfnamefont {P.}~\bibnamefont {{Galison}}}, \bibinfo {author} {\bibfnamefont {C.~F.}\ \bibnamefont {{Gammie}}}, \bibinfo {author} {\bibfnamefont {R.}~\bibnamefont {{Garc{\'\i}a}}}, \bibinfo {author} {\bibfnamefont {O.}~\bibnamefont {{Gentaz}}}, \bibinfo {author} {\bibfnamefont {B.}~\bibnamefont {{Georgiev}}}, \bibinfo {author} {\bibfnamefont {C.}~\bibnamefont {{Goddi}}}, \bibinfo {author} {\bibfnamefont
  {R.}~\bibnamefont {{Gold}}}, \bibinfo {author} {\bibfnamefont {A.~I.}\ \bibnamefont {{G{\'o}mez-Ruiz}}}, \bibinfo {author} {\bibfnamefont {J.~L.}\ \bibnamefont {{G{\'o}mez}}}, \bibinfo {author} {\bibfnamefont {M.}~\bibnamefont {{Gu}}}, \bibinfo {author} {\bibfnamefont {M.}~\bibnamefont {{Gurwell}}}, \bibinfo {author} {\bibfnamefont {K.}~\bibnamefont {{Hada}}}, \bibinfo {author} {\bibfnamefont {D.}~\bibnamefont {{Haggard}}}, \bibinfo {author} {\bibfnamefont {K.}~\bibnamefont {{Haworth}}}, \bibinfo {author} {\bibfnamefont {M.~H.}\ \bibnamefont {{Hecht}}}, \bibinfo {author} {\bibfnamefont {R.}~\bibnamefont {{Hesper}}}, \bibinfo {author} {\bibfnamefont {D.}~\bibnamefont {{Heumann}}}, \bibinfo {author} {\bibfnamefont {L.~C.}\ \bibnamefont {{Ho}}}, \bibinfo {author} {\bibfnamefont {P.}~\bibnamefont {{Ho}}}, \bibinfo {author} {\bibfnamefont {M.}~\bibnamefont {{Honma}}}, \bibinfo {author} {\bibfnamefont {C.-W.~L.}\ \bibnamefont {{Huang}}}, \bibinfo {author} {\bibfnamefont {L.}~\bibnamefont {{Huang}}}, \bibinfo
  {author} {\bibfnamefont {D.~H.}\ \bibnamefont {{Hughes}}}, \bibinfo {author} {\bibfnamefont {S.}~\bibnamefont {{Ikeda}}}, \bibinfo {author} {\bibfnamefont {C.~M.~V.}\ \bibnamefont {{Impellizzeri}}}, \bibinfo {author} {\bibfnamefont {M.}~\bibnamefont {{Inoue}}}, \bibinfo {author} {\bibfnamefont {S.}~\bibnamefont {{Issaoun}}}, \bibinfo {author} {\bibfnamefont {D.~J.}\ \bibnamefont {{James}}}, \bibinfo {author} {\bibfnamefont {B.~T.}\ \bibnamefont {{Jannuzi}}}, \bibinfo {author} {\bibfnamefont {M.}~\bibnamefont {{Janssen}}}, \bibinfo {author} {\bibfnamefont {B.}~\bibnamefont {{Jeter}}}, \bibinfo {author} {\bibfnamefont {W.}~\bibnamefont {{Jiang}}}, \bibinfo {author} {\bibfnamefont {A.}~\bibnamefont {{Jim{\'e}nez-Rosales}}}, \bibinfo {author} {\bibfnamefont {M.~D.}\ \bibnamefont {{Johnson}}}, \bibinfo {author} {\bibfnamefont {S.}~\bibnamefont {{Jorstad}}}, \bibinfo {author} {\bibfnamefont {A.~V.}\ \bibnamefont {{Joshi}}}, \bibinfo {author} {\bibfnamefont {T.}~\bibnamefont {{Jung}}}, \bibinfo {author}
  {\bibfnamefont {M.}~\bibnamefont {{Karami}}}, \bibinfo {author} {\bibfnamefont {R.}~\bibnamefont {{Karuppusamy}}}, \bibinfo {author} {\bibfnamefont {T.}~\bibnamefont {{Kawashima}}}, \bibinfo {author} {\bibfnamefont {G.~K.}\ \bibnamefont {{Keating}}}, \bibinfo {author} {\bibfnamefont {M.}~\bibnamefont {{Kettenis}}}, \bibinfo {author} {\bibfnamefont {D.-J.}\ \bibnamefont {{Kim}}}, \bibinfo {author} {\bibfnamefont {J.-Y.}\ \bibnamefont {{Kim}}}, \bibinfo {author} {\bibfnamefont {J.}~\bibnamefont {{Kim}}}, \bibinfo {author} {\bibfnamefont {J.}~\bibnamefont {{Kim}}}, \bibinfo {author} {\bibfnamefont {M.}~\bibnamefont {{Kino}}}, \bibinfo {author} {\bibfnamefont {J.~Y.}\ \bibnamefont {{Koay}}}, \bibinfo {author} {\bibfnamefont {P.}~\bibnamefont {{Kocherlakota}}}, \bibinfo {author} {\bibfnamefont {Y.}~\bibnamefont {{Kofuji}}}, \bibinfo {author} {\bibfnamefont {P.~M.}\ \bibnamefont {{Koch}}}, \bibinfo {author} {\bibfnamefont {S.}~\bibnamefont {{Koyama}}}, \bibinfo {author} {\bibfnamefont {C.}~\bibnamefont
  {{Kramer}}}, \bibinfo {author} {\bibfnamefont {M.}~\bibnamefont {{Kramer}}}, \bibinfo {author} {\bibfnamefont {T.~P.}\ \bibnamefont {{Krichbaum}}}, \bibinfo {author} {\bibfnamefont {C.-Y.}\ \bibnamefont {{Kuo}}}, \bibinfo {author} {\bibfnamefont {N.}~\bibnamefont {{La Bella}}}, \bibinfo {author} {\bibfnamefont {T.~R.}\ \bibnamefont {{Lauer}}}, \bibinfo {author} {\bibfnamefont {D.}~\bibnamefont {{Lee}}}, \bibinfo {author} {\bibfnamefont {S.-S.}\ \bibnamefont {{Lee}}}, \bibinfo {author} {\bibfnamefont {P.~K.}\ \bibnamefont {{Leung}}}, \bibinfo {author} {\bibfnamefont {A.}~\bibnamefont {{Levis}}}, \bibinfo {author} {\bibfnamefont {Z.}~\bibnamefont {{Li}}}, \bibinfo {author} {\bibfnamefont {R.}~\bibnamefont {{Lico}}}, \bibinfo {author} {\bibfnamefont {G.}~\bibnamefont {{Lindahl}}}, \bibinfo {author} {\bibfnamefont {M.}~\bibnamefont {{Lindqvist}}}, \bibinfo {author} {\bibfnamefont {M.}~\bibnamefont {{Lisakov}}}, \bibinfo {author} {\bibfnamefont {J.}~\bibnamefont {{Liu}}}, \bibinfo {author} {\bibfnamefont
  {K.}~\bibnamefont {{Liu}}}, \bibinfo {author} {\bibfnamefont {E.}~\bibnamefont {{Liuzzo}}}, \bibinfo {author} {\bibfnamefont {W.-P.}\ \bibnamefont {{Lo}}}, \bibinfo {author} {\bibfnamefont {A.~P.}\ \bibnamefont {{Lobanov}}}, \bibinfo {author} {\bibfnamefont {L.}~\bibnamefont {{Loinard}}}, \bibinfo {author} {\bibfnamefont {C.~J.}\ \bibnamefont {{Lonsdale}}}, \bibinfo {author} {\bibfnamefont {R.-S.}\ \bibnamefont {{Lu}}}, \bibinfo {author} {\bibfnamefont {J.}~\bibnamefont {{Mao}}}, \bibinfo {author} {\bibfnamefont {N.}~\bibnamefont {{Marchili}}}, \bibinfo {author} {\bibfnamefont {S.}~\bibnamefont {{Markoff}}}, \bibinfo {author} {\bibfnamefont {D.~P.}\ \bibnamefont {{Marrone}}}, \bibinfo {author} {\bibfnamefont {A.~P.}\ \bibnamefont {{Marscher}}}, \bibinfo {author} {\bibfnamefont {I.}~\bibnamefont {{Mart{\'\i}-Vidal}}}, \bibinfo {author} {\bibfnamefont {S.}~\bibnamefont {{Matsushita}}}, \bibinfo {author} {\bibfnamefont {L.~D.}\ \bibnamefont {{Matthews}}}, \bibinfo {author} {\bibfnamefont {L.}~\bibnamefont
  {{Medeiros}}}, \bibinfo {author} {\bibfnamefont {K.~M.}\ \bibnamefont {{Menten}}}, \bibinfo {author} {\bibfnamefont {D.}~\bibnamefont {{Michalik}}}, \bibinfo {author} {\bibfnamefont {I.}~\bibnamefont {{Mizuno}}}, \bibinfo {author} {\bibfnamefont {Y.}~\bibnamefont {{Mizuno}}}, \bibinfo {author} {\bibfnamefont {J.~M.}\ \bibnamefont {{Moran}}}, \bibinfo {author} {\bibfnamefont {K.}~\bibnamefont {{Moriyama}}}, \bibinfo {author} {\bibfnamefont {M.}~\bibnamefont {{Moscibrodzka}}}, \bibinfo {author} {\bibfnamefont {C.}~\bibnamefont {{M{\"u}ller}}}, \bibinfo {author} {\bibfnamefont {A.}~\bibnamefont {{Mus}}}, \bibinfo {author} {\bibfnamefont {G.}~\bibnamefont {{Musoke}}}, \bibinfo {author} {\bibfnamefont {I.}~\bibnamefont {{Myserlis}}}, \bibinfo {author} {\bibfnamefont {A.}~\bibnamefont {{Nadolski}}}, \bibinfo {author} {\bibfnamefont {H.}~\bibnamefont {{Nagai}}}, \bibinfo {author} {\bibfnamefont {N.~M.}\ \bibnamefont {{Nagar}}}, \bibinfo {author} {\bibfnamefont {M.}~\bibnamefont {{Nakamura}}}, \bibinfo {author}
  {\bibfnamefont {R.}~\bibnamefont {{Narayan}}}, \bibinfo {author} {\bibfnamefont {G.}~\bibnamefont {{Narayanan}}}, \bibinfo {author} {\bibfnamefont {I.}~\bibnamefont {{Natarajan}}}, \bibinfo {author} {\bibfnamefont {A.}~\bibnamefont {{Nathanail}}}, \bibinfo {author} {\bibfnamefont {S.~N.}\ \bibnamefont {{Fuentes}}}, \bibinfo {author} {\bibfnamefont {J.}~\bibnamefont {{Neilsen}}}, \bibinfo {author} {\bibfnamefont {R.}~\bibnamefont {{Neri}}}, \bibinfo {author} {\bibfnamefont {C.}~\bibnamefont {{Ni}}}, \bibinfo {author} {\bibfnamefont {A.}~\bibnamefont {{Noutsos}}}, \bibinfo {author} {\bibfnamefont {M.~A.}\ \bibnamefont {{Nowak}}}, \bibinfo {author} {\bibfnamefont {J.}~\bibnamefont {{Oh}}}, \bibinfo {author} {\bibfnamefont {H.}~\bibnamefont {{Okino}}}, \bibinfo {author} {\bibfnamefont {H.}~\bibnamefont {{Olivares}}}, \bibinfo {author} {\bibfnamefont {G.~N.}\ \bibnamefont {{Ortiz-Le{\'o}n}}}, \bibinfo {author} {\bibfnamefont {T.}~\bibnamefont {{Oyama}}}, \bibinfo {author} {\bibfnamefont {F.}~\bibnamefont
  {{{\"O}zel}}}, \bibinfo {author} {\bibfnamefont {D.~C.~M.}\ \bibnamefont {{Palumbo}}}, \bibinfo {author} {\bibfnamefont {G.~F.}\ \bibnamefont {{Paraschos}}}, \bibinfo {author} {\bibfnamefont {J.}~\bibnamefont {{Park}}}, \bibinfo {author} {\bibfnamefont {H.}~\bibnamefont {{Parsons}}}, \bibinfo {author} {\bibfnamefont {N.}~\bibnamefont {{Patel}}}, \bibinfo {author} {\bibfnamefont {U.-L.}\ \bibnamefont {{Pen}}}, \bibinfo {author} {\bibfnamefont {D.~W.}\ \bibnamefont {{Pesce}}}, \bibinfo {author} {\bibfnamefont {V.}~\bibnamefont {{Pi{\'e}tu}}}, \bibinfo {author} {\bibfnamefont {R.}~\bibnamefont {{Plambeck}}}, \bibinfo {author} {\bibfnamefont {A.}~\bibnamefont {{PopStefanija}}}, \bibinfo {author} {\bibfnamefont {O.}~\bibnamefont {{Porth}}}, \bibinfo {author} {\bibfnamefont {F.~M.}\ \bibnamefont {{P{\"o}tzl}}}, \bibinfo {author} {\bibfnamefont {B.}~\bibnamefont {{Prather}}}, \bibinfo {author} {\bibfnamefont {J.~A.}\ \bibnamefont {{Preciado-L{\'o}pez}}},\ and\ \bibinfo {author} {\bibfnamefont {D.}~\bibnamefont
  {{Psaltis}}},\ }\href {https://doi.org/10.3847/2041-8213/ac6674} {\bibfield  {journal} {\bibinfo  {journal} {ApJL}\ }\textbf {\bibinfo {volume} {930}},\ \bibinfo {eid} {L12} (\bibinfo {year} {2022})}\BibitemShut {NoStop}%
\bibitem [{\citenamefont {{Toma}}\ and\ \citenamefont {{Takahara}}(2014)}]{Toma14}%
  \BibitemOpen
  \bibfield  {author} {\bibinfo {author} {\bibfnamefont {K.}~\bibnamefont {{Toma}}}\ and\ \bibinfo {author} {\bibfnamefont {F.}~\bibnamefont {{Takahara}}},\ }\href {https://doi.org/10.1093/mnras/stu1053} {\bibfield  {journal} {\bibinfo  {journal} {MNRAS}\ }\textbf {\bibinfo {volume} {442}},\ \bibinfo {pages} {2855} (\bibinfo {year} {2014})},\ \Eprint {https://arxiv.org/abs/1405.7437} {arXiv:1405.7437 [astro-ph.HE]} \BibitemShut {NoStop}%
\bibitem [{\citenamefont {{Coppi}}\ and\ \citenamefont {{Blandford}}(1990)}]{Coppi90}%
  \BibitemOpen
  \bibfield  {author} {\bibinfo {author} {\bibfnamefont {P.~S.}\ \bibnamefont {{Coppi}}}\ and\ \bibinfo {author} {\bibfnamefont {R.~D.}\ \bibnamefont {{Blandford}}},\ }\href {https://doi.org/10.1093/mnras/245.3.453} {\bibfield  {journal} {\bibinfo  {journal} {MNRAS}\ }\textbf {\bibinfo {volume} {245}},\ \bibinfo {pages} {453} (\bibinfo {year} {1990})}\BibitemShut {NoStop}%
\bibitem [{\citenamefont {{Kolo{\v{s}}}}\ \emph {et~al.}(2026)\citenamefont {{Kolo{\v{s}}}}, \citenamefont {{Abdulkhamidov}}, \citenamefont {{Tursunov}}, \citenamefont {{Rueda}},\ and\ \citenamefont {{Cerutti}}}]{Kolos26}%
  \BibitemOpen
  \bibfield  {author} {\bibinfo {author} {\bibfnamefont {M.}~\bibnamefont {{Kolo{\v{s}}}}}, \bibinfo {author} {\bibfnamefont {F.}~\bibnamefont {{Abdulkhamidov}}}, \bibinfo {author} {\bibfnamefont {A.}~\bibnamefont {{Tursunov}}}, \bibinfo {author} {\bibfnamefont {J.~A.}\ \bibnamefont {{Rueda}}},\ and\ \bibinfo {author} {\bibfnamefont {B.}~\bibnamefont {{Cerutti}}},\ }\href {https://doi.org/10.48550/arXiv.2609.16447} {\bibfield  {journal} {\bibinfo  {journal} {arXiv e-prints}\ ,\ \bibinfo {eid} {arXiv:2609.16447}} (\bibinfo {year} {2026})},\ \Eprint {https://arxiv.org/abs/2609.16447} {arXiv:2609.16447 [gr-qc]} \BibitemShut {NoStop}%
\bibitem [{\citenamefont {{Prieto}}\ \emph {et~al.}(2016)\citenamefont {{Prieto}}, \citenamefont {{Fern{\'a}ndez-Ontiveros}}, \citenamefont {{Markoff}}, \citenamefont {{Espada}},\ and\ \citenamefont {{Gonz{\'a}lez-Mart{\'\i}n}}}]{Prieto16}%
  \BibitemOpen
  \bibfield  {author} {\bibinfo {author} {\bibfnamefont {M.~A.}\ \bibnamefont {{Prieto}}}, \bibinfo {author} {\bibfnamefont {J.~A.}\ \bibnamefont {{Fern{\'a}ndez-Ontiveros}}}, \bibinfo {author} {\bibfnamefont {S.}~\bibnamefont {{Markoff}}}, \bibinfo {author} {\bibfnamefont {D.}~\bibnamefont {{Espada}}},\ and\ \bibinfo {author} {\bibfnamefont {O.}~\bibnamefont {{Gonz{\'a}lez-Mart{\'\i}n}}},\ }\href {https://doi.org/10.1093/mnras/stw166} {\bibfield  {journal} {\bibinfo  {journal} {MNRAS}\ }\textbf {\bibinfo {volume} {457}},\ \bibinfo {pages} {3801} (\bibinfo {year} {2016})},\ \Eprint {https://arxiv.org/abs/1508.02302} {arXiv:1508.02302 [astro-ph.GA]} \BibitemShut {NoStop}%
\bibitem [{\citenamefont {{Biretta}}\ \emph {et~al.}(1999)\citenamefont {{Biretta}}, \citenamefont {{Sparks}},\ and\ \citenamefont {{Macchetto}}}]{Biretta99}%
  \BibitemOpen
  \bibfield  {author} {\bibinfo {author} {\bibfnamefont {J.~A.}\ \bibnamefont {{Biretta}}}, \bibinfo {author} {\bibfnamefont {W.~B.}\ \bibnamefont {{Sparks}}},\ and\ \bibinfo {author} {\bibfnamefont {F.}~\bibnamefont {{Macchetto}}},\ }\href {https://doi.org/10.1086/307499} {\bibfield  {journal} {\bibinfo  {journal} {ApJ}\ }\textbf {\bibinfo {volume} {520}},\ \bibinfo {pages} {621} (\bibinfo {year} {1999})}\BibitemShut {NoStop}%
\bibitem [{\citenamefont {{Perlman}}\ \emph {et~al.}(2011)\citenamefont {{Perlman}}, \citenamefont {{Adams}}, \citenamefont {{Cara}}, \citenamefont {{Bourque}}, \citenamefont {{Harris}}, \citenamefont {{Madrid}}, \citenamefont {{Simons}}, \citenamefont {{Clausen-Brown}}, \citenamefont {{Cheung}}, \citenamefont {{Stawarz}}, \citenamefont {{Georganopoulos}}, \citenamefont {{Sparks}},\ and\ \citenamefont {{Biretta}}}]{Perlman11}%
  \BibitemOpen
  \bibfield  {author} {\bibinfo {author} {\bibfnamefont {E.~S.}\ \bibnamefont {{Perlman}}}, \bibinfo {author} {\bibfnamefont {S.~C.}\ \bibnamefont {{Adams}}}, \bibinfo {author} {\bibfnamefont {M.}~\bibnamefont {{Cara}}}, \bibinfo {author} {\bibfnamefont {M.}~\bibnamefont {{Bourque}}}, \bibinfo {author} {\bibfnamefont {D.~E.}\ \bibnamefont {{Harris}}}, \bibinfo {author} {\bibfnamefont {J.~P.}\ \bibnamefont {{Madrid}}}, \bibinfo {author} {\bibfnamefont {R.~C.}\ \bibnamefont {{Simons}}}, \bibinfo {author} {\bibfnamefont {E.}~\bibnamefont {{Clausen-Brown}}}, \bibinfo {author} {\bibfnamefont {C.~C.}\ \bibnamefont {{Cheung}}}, \bibinfo {author} {\bibfnamefont {L.}~\bibnamefont {{Stawarz}}}, \bibinfo {author} {\bibfnamefont {M.}~\bibnamefont {{Georganopoulos}}}, \bibinfo {author} {\bibfnamefont {W.~B.}\ \bibnamefont {{Sparks}}},\ and\ \bibinfo {author} {\bibfnamefont {J.~A.}\ \bibnamefont {{Biretta}}},\ }\href {https://doi.org/10.1088/0004-637X/743/2/119} {\bibfield  {journal} {\bibinfo  {journal} {ApJ}\ }\textbf
  {\bibinfo {volume} {743}},\ \bibinfo {eid} {119} (\bibinfo {year} {2011})},\ \Eprint {https://arxiv.org/abs/1109.6252} {arXiv:1109.6252 [astro-ph.HE]} \BibitemShut {NoStop}%
\bibitem [{\citenamefont {{Hada}}\ \emph {et~al.}(2013)\citenamefont {{Hada}}, \citenamefont {{Kino}}, \citenamefont {{Doi}}, \citenamefont {{Nagai}}, \citenamefont {{Honma}}, \citenamefont {{Hagiwara}}, \citenamefont {{Giroletti}}, \citenamefont {{Giovannini}},\ and\ \citenamefont {{Kawaguchi}}}]{Hada13}%
  \BibitemOpen
  \bibfield  {author} {\bibinfo {author} {\bibfnamefont {K.}~\bibnamefont {{Hada}}}, \bibinfo {author} {\bibfnamefont {M.}~\bibnamefont {{Kino}}}, \bibinfo {author} {\bibfnamefont {A.}~\bibnamefont {{Doi}}}, \bibinfo {author} {\bibfnamefont {H.}~\bibnamefont {{Nagai}}}, \bibinfo {author} {\bibfnamefont {M.}~\bibnamefont {{Honma}}}, \bibinfo {author} {\bibfnamefont {Y.}~\bibnamefont {{Hagiwara}}}, \bibinfo {author} {\bibfnamefont {M.}~\bibnamefont {{Giroletti}}}, \bibinfo {author} {\bibfnamefont {G.}~\bibnamefont {{Giovannini}}},\ and\ \bibinfo {author} {\bibfnamefont {N.}~\bibnamefont {{Kawaguchi}}},\ }\href {https://doi.org/10.1088/0004-637X/775/1/70} {\bibfield  {journal} {\bibinfo  {journal} {ApJ}\ }\textbf {\bibinfo {volume} {775}},\ \bibinfo {eid} {70} (\bibinfo {year} {2013})},\ \Eprint {https://arxiv.org/abs/1308.1411} {arXiv:1308.1411 [astro-ph.CO]} \BibitemShut {NoStop}%
\bibitem [{\citenamefont {{Jia}}\ \emph {et~al.}(2023)\citenamefont {{Jia}}, \citenamefont {{Ripperda}}, \citenamefont {{Quataert}}, \citenamefont {{White}}, \citenamefont {{Chatterjee}}, \citenamefont {{Philippov}},\ and\ \citenamefont {{Liska}}}]{Jia23}%
  \BibitemOpen
  \bibfield  {author} {\bibinfo {author} {\bibfnamefont {H.}~\bibnamefont {{Jia}}}, \bibinfo {author} {\bibfnamefont {B.}~\bibnamefont {{Ripperda}}}, \bibinfo {author} {\bibfnamefont {E.}~\bibnamefont {{Quataert}}}, \bibinfo {author} {\bibfnamefont {C.~J.}\ \bibnamefont {{White}}}, \bibinfo {author} {\bibfnamefont {K.}~\bibnamefont {{Chatterjee}}}, \bibinfo {author} {\bibfnamefont {A.}~\bibnamefont {{Philippov}}},\ and\ \bibinfo {author} {\bibfnamefont {M.}~\bibnamefont {{Liska}}},\ }\href {https://doi.org/10.1093/mnras/stad2935} {\bibfield  {journal} {\bibinfo  {journal} {MNRAS}\ }\textbf {\bibinfo {volume} {526}},\ \bibinfo {pages} {2924} (\bibinfo {year} {2023})},\ \Eprint {https://arxiv.org/abs/2301.09014} {arXiv:2301.09014 [astro-ph.HE]} \BibitemShut {NoStop}%
\bibitem [{\citenamefont {{Levinson}}\ and\ \citenamefont {{Rieger}}(2011)}]{Levinson11}%
  \BibitemOpen
  \bibfield  {author} {\bibinfo {author} {\bibfnamefont {A.}~\bibnamefont {{Levinson}}}\ and\ \bibinfo {author} {\bibfnamefont {F.}~\bibnamefont {{Rieger}}},\ }\href {https://doi.org/10.1088/0004-637X/730/2/123} {\bibfield  {journal} {\bibinfo  {journal} {ApJ}\ }\textbf {\bibinfo {volume} {730}},\ \bibinfo {eid} {123} (\bibinfo {year} {2011})},\ \Eprint {https://arxiv.org/abs/1011.5319} {arXiv:1011.5319 [astro-ph.HE]} \BibitemShut {NoStop}%
\bibitem [{\citenamefont {{Hirotani}}\ and\ \citenamefont {{Pu}}(2016)}]{Hirotani16}%
  \BibitemOpen
  \bibfield  {author} {\bibinfo {author} {\bibfnamefont {K.}~\bibnamefont {{Hirotani}}}\ and\ \bibinfo {author} {\bibfnamefont {H.-Y.}\ \bibnamefont {{Pu}}},\ }\href {https://doi.org/10.3847/0004-637X/818/1/50} {\bibfield  {journal} {\bibinfo  {journal} {ApJ}\ }\textbf {\bibinfo {volume} {818}},\ \bibinfo {eid} {50} (\bibinfo {year} {2016})},\ \Eprint {https://arxiv.org/abs/1512.05026} {arXiv:1512.05026 [astro-ph.HE]} \BibitemShut {NoStop}%
\bibitem [{\citenamefont {Katsoulakos}\ and\ \citenamefont {Rieger}(2018)}]{Katsoulakos18}%
  \BibitemOpen
  \bibfield  {author} {\bibinfo {author} {\bibfnamefont {G.}~\bibnamefont {Katsoulakos}}\ and\ \bibinfo {author} {\bibfnamefont {F.~M.}\ \bibnamefont {Rieger}},\ }\href {https://doi.org/10.3847/1538-4357/aaa003} {\bibfield  {journal} {\bibinfo  {journal} {The Astrophysical Journal}\ }\textbf {\bibinfo {volume} {852}},\ \bibinfo {pages} {112} (\bibinfo {year} {2018})}\BibitemShut {NoStop}%
\bibitem [{\citenamefont {{Chernoglazov}}\ \emph {et~al.}(2023)\citenamefont {{Chernoglazov}}, \citenamefont {{Hakobyan}},\ and\ \citenamefont {{Philippov}}}]{Chernoglazov23}%
  \BibitemOpen
  \bibfield  {author} {\bibinfo {author} {\bibfnamefont {A.}~\bibnamefont {{Chernoglazov}}}, \bibinfo {author} {\bibfnamefont {H.}~\bibnamefont {{Hakobyan}}},\ and\ \bibinfo {author} {\bibfnamefont {A.}~\bibnamefont {{Philippov}}},\ }\href {https://doi.org/10.3847/1538-4357/acffc6} {\bibfield  {journal} {\bibinfo  {journal} {ApJ}\ }\textbf {\bibinfo {volume} {959}},\ \bibinfo {eid} {122} (\bibinfo {year} {2023})},\ \Eprint {https://arxiv.org/abs/2305.02348} {arXiv:2305.02348 [astro-ph.HE]} \BibitemShut {NoStop}%
\bibitem [{\citenamefont {French}\ \emph {et~al.}(2023)\citenamefont {French}, \citenamefont {Guo}, \citenamefont {Zhang},\ and\ \citenamefont {Uzdensky}}]{French23}%
  \BibitemOpen
  \bibfield  {author} {\bibinfo {author} {\bibfnamefont {O.}~\bibnamefont {French}}, \bibinfo {author} {\bibfnamefont {F.}~\bibnamefont {Guo}}, \bibinfo {author} {\bibfnamefont {Q.}~\bibnamefont {Zhang}},\ and\ \bibinfo {author} {\bibfnamefont {D.~A.}\ \bibnamefont {Uzdensky}},\ }\href {https://doi.org/10.3847/1538-4357/acb7dd} {\bibfield  {journal} {\bibinfo  {journal} {The Astrophysical Journal}\ }\textbf {\bibinfo {volume} {948}},\ \bibinfo {pages} {19} (\bibinfo {year} {2023})}\BibitemShut {NoStop}%
\bibitem [{\citenamefont {{Figueiredo}}\ \emph {et~al.}(2024)\citenamefont {{Figueiredo}}, \citenamefont {{Cerutti}}, \citenamefont {{Mehlhaff}},\ and\ \citenamefont {{Scepi}}}]{Figueiredo24}%
  \BibitemOpen
  \bibfield  {author} {\bibinfo {author} {\bibfnamefont {E.}~\bibnamefont {{Figueiredo}}}, \bibinfo {author} {\bibfnamefont {B.}~\bibnamefont {{Cerutti}}}, \bibinfo {author} {\bibfnamefont {J.}~\bibnamefont {{Mehlhaff}}},\ and\ \bibinfo {author} {\bibfnamefont {N.}~\bibnamefont {{Scepi}}},\ }\href {https://doi.org/10.1051/0004-6361/202451229} {\bibfield  {journal} {\bibinfo  {journal} {A\&A}\ }\textbf {\bibinfo {volume} {690}},\ \bibinfo {eid} {A389} (\bibinfo {year} {2024})},\ \Eprint {https://arxiv.org/abs/2409.13495} {arXiv:2409.13495 [astro-ph.HE]} \BibitemShut {NoStop}%
\bibitem [{\citenamefont {Peery}\ \emph {et~al.}(2024)\citenamefont {Peery}, \citenamefont {Liu},\ and\ \citenamefont {Li}}]{Peery24}%
  \BibitemOpen
  \bibfield  {author} {\bibinfo {author} {\bibfnamefont {S.}~\bibnamefont {Peery}}, \bibinfo {author} {\bibfnamefont {Y.-H.}\ \bibnamefont {Liu}},\ and\ \bibinfo {author} {\bibfnamefont {X.}~\bibnamefont {Li}},\ }\href {https://doi.org/10.3847/1538-4357/ad24e8} {\bibfield  {journal} {\bibinfo  {journal} {The Astrophysical Journal}\ }\textbf {\bibinfo {volume} {964}},\ \bibinfo {pages} {144} (\bibinfo {year} {2024})}\BibitemShut {NoStop}%
\bibitem [{\citenamefont {{Chen}}\ \emph {et~al.}(2023)\citenamefont {{Chen}}, \citenamefont {{Uzdensky}},\ and\ \citenamefont {{Dexter}}}]{Chen23}%
  \BibitemOpen
  \bibfield  {author} {\bibinfo {author} {\bibfnamefont {A.~Y.}\ \bibnamefont {{Chen}}}, \bibinfo {author} {\bibfnamefont {D.}~\bibnamefont {{Uzdensky}}},\ and\ \bibinfo {author} {\bibfnamefont {J.}~\bibnamefont {{Dexter}}},\ }\href {https://doi.org/10.3847/1538-4357/acb68a} {\bibfield  {journal} {\bibinfo  {journal} {ApJ}\ }\textbf {\bibinfo {volume} {944}},\ \bibinfo {eid} {173} (\bibinfo {year} {2023})},\ \Eprint {https://arxiv.org/abs/2209.03249} {arXiv:2209.03249 [astro-ph.HE]} \BibitemShut {NoStop}%
\bibitem [{\citenamefont {{Kuze}}\ \emph {et~al.}(2025)\citenamefont {{Kuze}}, \citenamefont {{Kimura}},\ and\ \citenamefont {{Fang}}}]{Kuze25}%
  \BibitemOpen
  \bibfield  {author} {\bibinfo {author} {\bibfnamefont {R.}~\bibnamefont {{Kuze}}}, \bibinfo {author} {\bibfnamefont {S.~S.}\ \bibnamefont {{Kimura}}},\ and\ \bibinfo {author} {\bibfnamefont {K.}~\bibnamefont {{Fang}}},\ }\href {https://doi.org/10.3847/1538-4357/adcc1b} {\bibfield  {journal} {\bibinfo  {journal} {ApJ}\ }\textbf {\bibinfo {volume} {985}},\ \bibinfo {eid} {139} (\bibinfo {year} {2025})},\ \Eprint {https://arxiv.org/abs/2501.17467} {arXiv:2501.17467 [astro-ph.HE]} \BibitemShut {NoStop}%
\bibitem [{\citenamefont {{Oikawa}}\ \emph {et~al.}(2026)\citenamefont {{Oikawa}}, \citenamefont {{Toma}},\ and\ \citenamefont {{Kimura}}}]{Oikawa26}%
  \BibitemOpen
  \bibfield  {author} {\bibinfo {author} {\bibfnamefont {R.}~\bibnamefont {{Oikawa}}}, \bibinfo {author} {\bibfnamefont {K.}~\bibnamefont {{Toma}}},\ and\ \bibinfo {author} {\bibfnamefont {S.~S.}\ \bibnamefont {{Kimura}}},\ }\href {https://doi.org/10.3847/1538-4357/ae969b} {\bibfield  {journal} {\bibinfo  {journal} {ApJ}\ }\textbf {\bibinfo {volume} {1008}},\ \bibinfo {eid} {185} (\bibinfo {year} {2026})},\ \Eprint {https://arxiv.org/abs/2605.01002} {arXiv:2605.01002 [astro-ph.HE]} \BibitemShut {NoStop}%
\bibitem [{\citenamefont {{Wong}}\ \emph {et~al.}(2021)\citenamefont {{Wong}}, \citenamefont {{Ryan}},\ and\ \citenamefont {{Gammie}}}]{Wong21}%
  \BibitemOpen
  \bibfield  {author} {\bibinfo {author} {\bibfnamefont {G.~N.}\ \bibnamefont {{Wong}}}, \bibinfo {author} {\bibfnamefont {B.~R.}\ \bibnamefont {{Ryan}}},\ and\ \bibinfo {author} {\bibfnamefont {C.~F.}\ \bibnamefont {{Gammie}}},\ }\href {https://doi.org/10.3847/1538-4357/abd0f9} {\bibfield  {journal} {\bibinfo  {journal} {ApJ}\ }\textbf {\bibinfo {volume} {907}},\ \bibinfo {eid} {73} (\bibinfo {year} {2021})},\ \Eprint {https://arxiv.org/abs/2012.04658} {arXiv:2012.04658 [astro-ph.HE]} \BibitemShut {NoStop}%
\bibitem [{\citenamefont {{Spada}}\ \emph {et~al.}(2001)\citenamefont {{Spada}}, \citenamefont {{Ghisellini}}, \citenamefont {{Lazzati}},\ and\ \citenamefont {{Celotti}}}]{Spada01}%
  \BibitemOpen
  \bibfield  {author} {\bibinfo {author} {\bibfnamefont {M.}~\bibnamefont {{Spada}}}, \bibinfo {author} {\bibfnamefont {G.}~\bibnamefont {{Ghisellini}}}, \bibinfo {author} {\bibfnamefont {D.}~\bibnamefont {{Lazzati}}},\ and\ \bibinfo {author} {\bibfnamefont {A.}~\bibnamefont {{Celotti}}},\ }\href {https://doi.org/10.1046/j.1365-8711.2001.04557.x} {\bibfield  {journal} {\bibinfo  {journal} {MNRAS}\ }\textbf {\bibinfo {volume} {325}},\ \bibinfo {pages} {1559} (\bibinfo {year} {2001})},\ \Eprint {https://arxiv.org/abs/astro-ph/0103424} {arXiv:astro-ph/0103424 [astro-ph]} \BibitemShut {NoStop}%
\bibitem [{\citenamefont {{Baring}}\ \emph {et~al.}(2017)\citenamefont {{Baring}}, \citenamefont {{B{\"o}ttcher}},\ and\ \citenamefont {{Summerlin}}}]{Baring17}%
  \BibitemOpen
  \bibfield  {author} {\bibinfo {author} {\bibfnamefont {M.~G.}\ \bibnamefont {{Baring}}}, \bibinfo {author} {\bibfnamefont {M.}~\bibnamefont {{B{\"o}ttcher}}},\ and\ \bibinfo {author} {\bibfnamefont {E.~J.}\ \bibnamefont {{Summerlin}}},\ }\href {https://doi.org/10.1093/mnras/stw2344} {\bibfield  {journal} {\bibinfo  {journal} {MNRAS}\ }\textbf {\bibinfo {volume} {464}},\ \bibinfo {pages} {4875} (\bibinfo {year} {2017})},\ \Eprint {https://arxiv.org/abs/1609.03899} {arXiv:1609.03899 [astro-ph.HE]} \BibitemShut {NoStop}%
\bibitem [{\citenamefont {{B{\"o}ttcher}}\ and\ \citenamefont {{Baring}}(2019)}]{Bottcher19}%
  \BibitemOpen
  \bibfield  {author} {\bibinfo {author} {\bibfnamefont {M.}~\bibnamefont {{B{\"o}ttcher}}}\ and\ \bibinfo {author} {\bibfnamefont {M.~G.}\ \bibnamefont {{Baring}}},\ }\href {https://doi.org/10.3847/1538-4357/ab552a} {\bibfield  {journal} {\bibinfo  {journal} {ApJ}\ }\textbf {\bibinfo {volume} {887}},\ \bibinfo {eid} {133} (\bibinfo {year} {2019})},\ \Eprint {https://arxiv.org/abs/1911.02834} {arXiv:1911.02834 [astro-ph.HE]} \BibitemShut {NoStop}%
\bibitem [{\citenamefont {{Kimura}}\ and\ \citenamefont {{Toma}}(2020)}]{Kimura20}%
  \BibitemOpen
  \bibfield  {author} {\bibinfo {author} {\bibfnamefont {S.~S.}\ \bibnamefont {{Kimura}}}\ and\ \bibinfo {author} {\bibfnamefont {K.}~\bibnamefont {{Toma}}},\ }\href {https://doi.org/10.3847/1538-4357/abc343} {\bibfield  {journal} {\bibinfo  {journal} {ApJ}\ }\textbf {\bibinfo {volume} {905}},\ \bibinfo {eid} {178} (\bibinfo {year} {2020})},\ \Eprint {https://arxiv.org/abs/2003.13173} {arXiv:2003.13173 [astro-ph.HE]} \BibitemShut {NoStop}%
\bibitem [{\citenamefont {{Kuze}}\ \emph {et~al.}(2024)\citenamefont {{Kuze}}, \citenamefont {{Kimura}},\ and\ \citenamefont {{Toma}}}]{Kuze24}%
  \BibitemOpen
  \bibfield  {author} {\bibinfo {author} {\bibfnamefont {R.}~\bibnamefont {{Kuze}}}, \bibinfo {author} {\bibfnamefont {S.~S.}\ \bibnamefont {{Kimura}}},\ and\ \bibinfo {author} {\bibfnamefont {K.}~\bibnamefont {{Toma}}},\ }\href {https://doi.org/10.3847/1538-4357/ad88f4} {\bibfield  {journal} {\bibinfo  {journal} {ApJ}\ }\textbf {\bibinfo {volume} {977}},\ \bibinfo {eid} {22} (\bibinfo {year} {2024})}\BibitemShut {NoStop}%
\bibitem [{\citenamefont {{Chow}}\ \emph {et~al.}(2026)\citenamefont {{Chow}}, \citenamefont {{Sironi}}, \citenamefont {{Ripperda}},\ and\ \citenamefont {{Levinson}}}]{Chow26}%
  \BibitemOpen
  \bibfield  {author} {\bibinfo {author} {\bibfnamefont {A.}~\bibnamefont {{Chow}}}, \bibinfo {author} {\bibfnamefont {L.}~\bibnamefont {{Sironi}}}, \bibinfo {author} {\bibfnamefont {B.}~\bibnamefont {{Ripperda}}},\ and\ \bibinfo {author} {\bibfnamefont {A.}~\bibnamefont {{Levinson}}},\ }\href {https://doi.org/10.1093/mnras/stag1631} {\bibfield  {journal} {\bibinfo  {journal} {MNRAS}\ }\textbf {\bibinfo {volume} {552}},\ \bibinfo {eid} {stag1631} (\bibinfo {year} {2026})},\ \Eprint {https://arxiv.org/abs/2603.10122} {arXiv:2603.10122 [astro-ph.HE]} \BibitemShut {NoStop}%
\bibitem [{\citenamefont {{Kimura}}\ \emph {et~al.}(2022)\citenamefont {{Kimura}}, \citenamefont {{Toma}}, \citenamefont {{Noda}},\ and\ \citenamefont {{Hada}}}]{Kimura22}%
  \BibitemOpen
  \bibfield  {author} {\bibinfo {author} {\bibfnamefont {S.~S.}\ \bibnamefont {{Kimura}}}, \bibinfo {author} {\bibfnamefont {K.}~\bibnamefont {{Toma}}}, \bibinfo {author} {\bibfnamefont {H.}~\bibnamefont {{Noda}}},\ and\ \bibinfo {author} {\bibfnamefont {K.}~\bibnamefont {{Hada}}},\ }\href {https://doi.org/10.3847/2041-8213/ac8d5a} {\bibfield  {journal} {\bibinfo  {journal} {ApJL}\ }\textbf {\bibinfo {volume} {937}},\ \bibinfo {eid} {L34} (\bibinfo {year} {2022})},\ \Eprint {https://arxiv.org/abs/2208.01882} {arXiv:2208.01882 [astro-ph.HE]} \BibitemShut {NoStop}%
\bibitem [{\citenamefont {{Nalewajko}}\ \emph {et~al.}(2026)\citenamefont {{Nalewajko}}, \citenamefont {{Kapusta}}, \citenamefont {{Ripperda}},\ and\ \citenamefont {{Philippov}}}]{Nalewajko26}%
  \BibitemOpen
  \bibfield  {author} {\bibinfo {author} {\bibfnamefont {K.}~\bibnamefont {{Nalewajko}}}, \bibinfo {author} {\bibfnamefont {M.}~\bibnamefont {{Kapusta}}}, \bibinfo {author} {\bibfnamefont {B.}~\bibnamefont {{Ripperda}}},\ and\ \bibinfo {author} {\bibfnamefont {A.}~\bibnamefont {{Philippov}}},\ }\href {https://doi.org/10.3847/1538-4357/ae6653} {\bibfield  {journal} {\bibinfo  {journal} {ApJ}\ }\textbf {\bibinfo {volume} {1003}},\ \bibinfo {eid} {174} (\bibinfo {year} {2026})},\ \Eprint {https://arxiv.org/abs/2604.25996} {arXiv:2604.25996 [astro-ph.HE]} \BibitemShut {NoStop}%
\bibitem [{\citenamefont {{Thorne}}\ and\ \citenamefont {{MacDonald}}(1982)}]{Thorne82}%
  \BibitemOpen
  \bibfield  {author} {\bibinfo {author} {\bibfnamefont {K.~S.}\ \bibnamefont {{Thorne}}}\ and\ \bibinfo {author} {\bibfnamefont {D.}~\bibnamefont {{MacDonald}}},\ }\href {https://doi.org/10.1093/mnras/198.2.339} {\bibfield  {journal} {\bibinfo  {journal} {MNRAS}\ }\textbf {\bibinfo {volume} {198}},\ \bibinfo {pages} {339} (\bibinfo {year} {1982})}\BibitemShut {NoStop}%
\bibitem [{\citenamefont {{Garofalo}}\ and\ \citenamefont {{Meier}}(2010)}]{Garofalo10}%
  \BibitemOpen
  \bibfield  {author} {\bibinfo {author} {\bibfnamefont {D.}~\bibnamefont {{Garofalo}}}\ and\ \bibinfo {author} {\bibfnamefont {D.~L.}\ \bibnamefont {{Meier}}},\ }\href {https://doi.org/10.1111/j.1365-2966.2010.16815.x} {\bibfield  {journal} {\bibinfo  {journal} {MNRAS}\ }\textbf {\bibinfo {volume} {406}},\ \bibinfo {pages} {2047} (\bibinfo {year} {2010})},\ \Eprint {https://arxiv.org/abs/1004.1600} {arXiv:1004.1600 [astro-ph.HE]} \BibitemShut {NoStop}%
\bibitem [{\citenamefont {{Komissarov}}(2005)}]{Komissarov05}%
  \BibitemOpen
  \bibfield  {author} {\bibinfo {author} {\bibfnamefont {S.~S.}\ \bibnamefont {{Komissarov}}},\ }\href {https://doi.org/10.1111/j.1365-2966.2005.08974.x} {\bibfield  {journal} {\bibinfo  {journal} {MNRAS}\ }\textbf {\bibinfo {volume} {359}},\ \bibinfo {pages} {801} (\bibinfo {year} {2005})},\ \Eprint {https://arxiv.org/abs/astro-ph/0501599} {arXiv:astro-ph/0501599 [astro-ph]} \BibitemShut {NoStop}%
\bibitem [{\citenamefont {{Cheng}}\ \emph {et~al.}(1986)\citenamefont {{Cheng}}, \citenamefont {{Ho}},\ and\ \citenamefont {{Ruderman}}}]{Cheng86}%
  \BibitemOpen
  \bibfield  {author} {\bibinfo {author} {\bibfnamefont {K.~S.}\ \bibnamefont {{Cheng}}}, \bibinfo {author} {\bibfnamefont {C.}~\bibnamefont {{Ho}}},\ and\ \bibinfo {author} {\bibfnamefont {M.}~\bibnamefont {{Ruderman}}},\ }\href {https://doi.org/10.1086/163829} {\bibfield  {journal} {\bibinfo  {journal} {ApJ}\ }\textbf {\bibinfo {volume} {300}},\ \bibinfo {pages} {500} (\bibinfo {year} {1986})}\BibitemShut {NoStop}%
\bibitem [{\citenamefont {{Romani}}(1996)}]{Romani96}%
  \BibitemOpen
  \bibfield  {author} {\bibinfo {author} {\bibfnamefont {R.~W.}\ \bibnamefont {{Romani}}},\ }\href {https://doi.org/10.1086/177878} {\bibfield  {journal} {\bibinfo  {journal} {ApJ}\ }\textbf {\bibinfo {volume} {470}},\ \bibinfo {pages} {469} (\bibinfo {year} {1996})}\BibitemShut {NoStop}%
\end{thebibliography}
